\documentclass[12pt,a4paper]{article}

\usepackage{ifthen} 
\newboolean{pdflatex}
\setboolean{pdflatex}{true} 

\newboolean{articletitles}
\setboolean{articletitles}{true} 

\newboolean{uprightparticles}
\setboolean{uprightparticles}{false} 

\newboolean{firstbigtable}
\setboolean{firstbigtable}{true}

\newboolean{summaryonly}
\setboolean{summaryonly}{false}

\usepackage{microtype}
\usepackage{lineno}  
\usepackage{xspace} 

\usepackage{graphicx}  
\usepackage{color}
\usepackage{colortbl}
\usepackage{rotating}
\usepackage{subfigure}

\graphicspath{{./figs/}} 

\usepackage{amsmath} 
\usepackage{amssymb}
\usepackage{amsfonts}
\usepackage{upgreek} 

\newcommand*\patchAmsMathEnvironmentForLineno[1]{%
\expandafter\let\csname old#1\expandafter\endcsname\csname #1\endcsname
\expandafter\let\csname oldend#1\expandafter\endcsname\csname
end#1\endcsname
 \renewenvironment{#1}%
   {\linenomath\csname old#1\endcsname}%
   {\csname oldend#1\endcsname\endlinenomath}%
}
\newcommand*\patchBothAmsMathEnvironmentsForLineno[1]{%
  \patchAmsMathEnvironmentForLineno{#1}%
  \patchAmsMathEnvironmentForLineno{#1*}%
}
\AtBeginDocument{%
\patchBothAmsMathEnvironmentsForLineno{equation}%
\patchBothAmsMathEnvironmentsForLineno{align}%
\patchBothAmsMathEnvironmentsForLineno{flalign}%
\patchBothAmsMathEnvironmentsForLineno{alignat}%
\patchBothAmsMathEnvironmentsForLineno{gather}%
\patchBothAmsMathEnvironmentsForLineno{multline}%
}

\usepackage{hyperref}    
\usepackage[all]{hypcap} 

\usepackage{cancel}

\def\lhcb {\mbox{LHCb}\xspace}
\def\ux85 {\mbox{UX85}\xspace}

\def\babar  {\mbox{BaBar}\xspace}
\def\belle  {\mbox{Belle}\xspace}

\def\cdf    {\mbox{CDF}\xspace}
\def\dzero  {\mbox{D0}\xspace}

\def\tevatron {Tevatron\xspace}

\ifthenelse{\boolean{uprightparticles}}%
{
 
 \def\Pgamma      {\ensuremath{\upgamma}\xspace}

 \def\Peta        {\ensuremath{\upeta}\xspace}

 \def\Pmu         {\ensuremath{\upmu}\xspace}                 
 \def\Pnu         {\ensuremath{\upnu}\xspace}                 
                  
 \def\Ppi         {\ensuremath{\uppi}\xspace}                 
                  
 \def\Prho        {\ensuremath{\uprho}\xspace}                 
                  
 \def\Ptau        {\ensuremath{\uptau}\xspace}                 
                  
 \def\Pphi        {\ensuremath{\upphi}\xspace}                 
                  
 \def\Pchi        {\ensuremath{\upchi}\xspace}                 
 \def\Ppsi        {\ensuremath{\uppsi}\xspace}                 
 \def\Pomega      {\ensuremath{\upomega}\xspace}                 

 \def\PDelta      {\ensuremath{\Delta}\xspace}                 
 \def\PXi      {\ensuremath{\Xi}\xspace}                 
 \def\PLambda      {\ensuremath{\Lambda}\xspace}                 
 \def\PSigma      {\ensuremath{\Sigma}\xspace}                 
 \def\POmega      {\ensuremath{\Omega}\xspace}                 
 \def\PUpsilon      {\ensuremath{\Upsilon}\xspace}                 
 
 \def\PB      {\ensuremath{\mathrm{B}}\xspace}                 
                  
 \def\PD      {\ensuremath{\mathrm{D}}\xspace}

 \def\PJ      {\ensuremath{\mathrm{J}}\xspace}                 
 \def\PK      {\ensuremath{\mathrm{K}}\xspace}

 \def\PP      {\ensuremath{\mathrm{P}}\xspace}

 \def\PV      {\ensuremath{\mathrm{V}}\xspace}                 
                  
 \def\PX      {\ensuremath{\mathrm{X}}\xspace}

 \def\Pb      {\ensuremath{\mathrm{b}}\xspace}                 
 \def\Pc      {\ensuremath{\mathrm{c}}\xspace}                 
 \def\Pd      {\ensuremath{\mathrm{d}}\xspace}                 
 \def\Pe      {\ensuremath{\mathrm{e}}\xspace}                 
                  
 \def\Pg      {\ensuremath{\mathrm{g}}\xspace}                 
                  
 \def\Pi      {\ensuremath{\mathrm{i}}\xspace}

 \def\Pq      {\ensuremath{\mathrm{q}}\xspace}                 
                  
 \def\Ps      {\ensuremath{\mathrm{s}}\xspace}                 
 \def\Pt      {\ensuremath{\mathrm{t}}\xspace}                 
 \def\Pu      {\ensuremath{\mathrm{u}}\xspace}

}
{
 
 \def\Pgamma      {\ensuremath{\gamma}\xspace}

 \def\Peta        {\ensuremath{\eta}\xspace}

 \def\Pmu         {\ensuremath{\mu}\xspace}                 
 \def\Pnu         {\ensuremath{\nu}\xspace}                 
                  
 \def\Ppi         {\ensuremath{\pi}\xspace}                 
                  
 \def\Prho        {\ensuremath{\rho}\xspace}                 
                  
 \def\Ptau        {\ensuremath{\tau}\xspace}                 
                  
 \def\Pphi        {\ensuremath{\phi}\xspace}                 
                  
 \def\Pchi        {\ensuremath{\chi}\xspace}                 
 \def\Ppsi        {\ensuremath{\psi}\xspace}                 
 \def\Pomega      {\ensuremath{\omega}\xspace}                 
 \mathchardef\PDelta="7101
 \mathchardef\PXi="7104
 \mathchardef\PLambda="7103
 \mathchardef\PSigma="7106
 \mathchardef\POmega="710A
 \mathchardef\PUpsilon="7107
                  
 \def\PB      {\ensuremath{B}\xspace}                 
                  
 \def\PD      {\ensuremath{D}\xspace}

 \def\PJ      {\ensuremath{J}\xspace}                 
 \def\PK      {\ensuremath{K}\xspace}

 \def\PP      {\ensuremath{P}\xspace}

 \def\PV      {\ensuremath{V}\xspace}                 
                  
 \def\PX      {\ensuremath{X}\xspace}

 \def\Pb      {\ensuremath{b}\xspace}                 
 \def\Pc      {\ensuremath{c}\xspace}                 
 \def\Pd      {\ensuremath{d}\xspace}                 
 \def\Pe      {\ensuremath{e}\xspace}                 
                  
 \def\Pg      {\ensuremath{g}\xspace}                 
                  
 \def\Pi      {\ensuremath{i}\xspace}

 \def\Pq      {\ensuremath{q}\xspace}                 
                  
 \def\Ps      {\ensuremath{s}\xspace}                 
 \def\Pt      {\ensuremath{t}\xspace}                 
 \def\Pu      {\ensuremath{u}\xspace}

}

\def\electron   {\ensuremath{\Pe}\xspace}
\def\en         {\ensuremath{\Pe^-}\xspace}   
\def\ep         {\ensuremath{\Pe^+}\xspace}
 
\def\epem       {\ensuremath{\Pe^+\Pe^-}\xspace}

\def\mmu        {\ensuremath{\Pmu}\xspace}
\def\mup        {\ensuremath{\Pmu^+}\xspace}
\def\mun        {\ensuremath{\Pmu^-}\xspace} 
\def\mumu       {\ensuremath{\Pmu^+\Pmu^-}\xspace}

\def\taup       {\ensuremath{\Ptau^+}\xspace}

\def\tautau     {\ensuremath{\Ptau^+\Ptau^-}\xspace}

\def\ellm       {\ensuremath{\ell^-}\xspace}
\def\ellp       {\ensuremath{\ell^+}\xspace}
\def\ellell     {\ensuremath{\ell^+ \ell^-}\xspace}

\def\neu        {\ensuremath{\Pnu}\xspace}
\def\neub       {\ensuremath{\overline{\Pnu}}\xspace}

\def\g      {\ensuremath{\Pgamma}\xspace}

\def\quark     {\ensuremath{\Pq}\xspace}
\def\quarkbar  {\ensuremath{\overline \quark}\xspace}
\def\qqbar     {\ensuremath{\quark\quarkbar}\xspace}
\def\uquark    {\ensuremath{\Pu}\xspace}
\def\uquarkbar {\ensuremath{\overline \uquark}\xspace}
\def\uubar     {\ensuremath{\uquark\uquarkbar}\xspace}
\def\dquark    {\ensuremath{\Pd}\xspace}

\def\squark    {\ensuremath{\Ps}\xspace}
\def\squarkbar {\ensuremath{\overline \squark}\xspace}
\def\ssbar     {\ensuremath{\squark\squarkbar}\xspace}
\def\cquark    {\ensuremath{\Pc}\xspace}
\def\cquarkbar {\ensuremath{\overline \cquark}\xspace}
\def\ccbar     {\ensuremath{\cquark\cquarkbar}\xspace}
\def\bquark    {\ensuremath{\Pb}\xspace}

\def\tquark    {\ensuremath{\Pt}\xspace}

\def\pion  {\ensuremath{\Ppi}\xspace}
\def\piz   {\ensuremath{\pion^0}\xspace}

\def\pip   {\ensuremath{\pion^+}\xspace}
\def\pim   {\ensuremath{\pion^-}\xspace}

\def\pipm  {\ensuremath{\pion^\pm}\xspace}

\def\kaon  {\ensuremath{\PK}\xspace}
  \def\Kbar  {\kern 0.2em\overline{\kern -0.2em \PK}{}\xspace}
\def\Kb    {\ensuremath{\Kbar}\xspace}
\def\Kz    {\ensuremath{\kaon^0}\xspace}
\def\Kzb   {\ensuremath{\Kbar^0}\xspace}
\def\KzKzb {\ensuremath{\Kz \kern -0.16em \Kzb}\xspace}
\def\Kp    {\ensuremath{\kaon^+}\xspace}
\def\Km    {\ensuremath{\kaon^-}\xspace}
\def\Kpm   {\ensuremath{\kaon^\pm}\xspace}
\def\Kmp   {\ensuremath{\kaon^\mp}\xspace}
\def\KpKm  {\ensuremath{\Kp \kern -0.16em \Km}\xspace}
\def\KS    {\ensuremath{\kaon^0_{\rm\scriptscriptstyle S}}\xspace} 
\def\KL    {\ensuremath{\kaon^0_{\rm\scriptscriptstyle L}}\xspace} 
\def\Kstarz  {\ensuremath{\kaon^{*0}}\xspace}
\def\Kstarzb {\ensuremath{\Kbar^{*0}}\xspace}
\def\Kstar   {\ensuremath{\kaon^*}\xspace}

\def\Kstarp  {\ensuremath{\kaon^{*+}}\xspace}
\def\Kstarm  {\ensuremath{\kaon^{*-}}\xspace}

  \def\Dbar    {\kern 0.2em\overline{\kern -0.2em \PD}{}\xspace}
\def\D       {\ensuremath{\PD}\xspace}
\def\Db      {\ensuremath{\Dbar}\xspace}
\def\Dz      {\ensuremath{\D^0}\xspace}
\def\Dzb     {\ensuremath{\Dbar^0}\xspace}
\def\DzDzb   {\ensuremath{\Dz {\kern -0.16em \Dzb}}\xspace}
\def\Dp      {\ensuremath{\D^+}\xspace}
\def\Dm      {\ensuremath{\D^-}\xspace}

\def\DpDm    {\ensuremath{\Dp {\kern -0.16em \Dm}}\xspace}

\def\Dstarp  {\ensuremath{\D^{*+}}\xspace}

\def\Ds      {\ensuremath{\D^+_\squark}\xspace}
\def\Dsp     {\ensuremath{\D^+_\squark}\xspace}
\def\Dsm     {\ensuremath{\D^-_\squark}\xspace}
\def\Dspm    {\ensuremath{\D^{\pm}_\squark}\xspace}
\def\Dsmp    {\ensuremath{\D^{\mp}_\squark}\xspace}

\def\B       {\ensuremath{\PB}\xspace}
\def\Bbar    {\kern 0.18em\overline{\kern -0.18em \PB}{}\xspace}
\def\Bb      {\ensuremath{\Bbar}\xspace}
 
\def\Bz      {\ensuremath{\B^0}\xspace}
\def\Bzb     {\ensuremath{\Bbar^0}\xspace}
\def\Bu      {\ensuremath{\B^+}\xspace}
\def\Bub     {\ensuremath{\B^-}\xspace}
\def\Bp      {\ensuremath{\Bu}\xspace}
\def\Bm      {\ensuremath{\Bub}\xspace}
\def\Bpm     {\ensuremath{\B^\pm}\xspace}

\def\Bd      {\ensuremath{\B^0}\xspace}
\def\Bs      {\ensuremath{\B^0_\squark}\xspace}
\def\Bsb     {\ensuremath{\Bbar^0_\squark}\xspace}
\def\Bdb     {\ensuremath{\Bbar^0}\xspace}
\def\Bc      {\ensuremath{\B_\cquark^+}\xspace}

\def\jpsi     {\ensuremath{{\PJ\mskip -3mu/\mskip -2mu\Ppsi\mskip 2mu}}\xspace}
\def\psitwos  {\ensuremath{\Ppsi{(2S)}}\xspace}

\def\etac     {\ensuremath{\Peta_\cquark}\xspace}
\def\chiczero {\ensuremath{\Pchi_{\cquark 0}}\xspace}
\def\chicone  {\ensuremath{\Pchi_{\cquark 1}}\xspace}
\def\chictwo  {\ensuremath{\Pchi_{\cquark 2}}\xspace}
  \def\Y#1S{\ensuremath{\PUpsilon{(#1S)}}\xspace}
\def\OneS  {\Y1S}
\def\TwoS  {\Y2S}
\def\ThreeS{\Y3S}

\def\nS {\ensuremath{\PUpsilon{(nS)}}\xspace}

\def\chicj  {\ensuremath{\Pchi_{cJ}}\xspace}
\def\chib  {\ensuremath{\Pchi_{b}}\xspace}
\def\hc  {\ensuremath{h_{c}}\xspace}
\def\etab     {\ensuremath{\Peta_\bquark}\xspace}
\def\chibzero {\ensuremath{\Pchi_{\bquark 0} (1P)}\xspace}

\def\chibtwo  {\ensuremath{\Pchi_{\bquark 2} (1P)}\xspace}

\def\L {\ensuremath{\PLambda}\xspace}
\def\Lbar {\ensuremath{\kern 0.1em\overline{\kern -0.1em\Lambda\kern -0.05em}\kern 0.05em{}}\xspace}

\def\Lb      {\ensuremath{\L^0_\bquark}\xspace}

\def\Lc      {\ensuremath{\L^+_\cquark}\xspace}

\def\BF         {{\ensuremath{\cal B}\xspace}}

\def\BR         {\BF}
\newcommand{\decay}[2]{\ensuremath{#1\!\to #2}\xspace}         
\def\ra                 {\ensuremath{\rightarrow}\xspace}
\def\to                 {\ensuremath{\rightarrow}\xspace}

\def\order   {\ensuremath{\mathcal{O}}\xspace}

\newcommand{\as}{\ensuremath{\alpha_{\scriptscriptstyle S}}\xspace}

\newcommand{\lqcd}{\ensuremath{\Lambda_{\mathrm{QCD}}}\xspace}
\def\qsq       {\ensuremath{q^2}\xspace}

\def\eps   {\ensuremath{\varepsilon}\xspace}

\def\CP                {\ensuremath{C\!P}\xspace}
\def\CPT               {\ensuremath{C\!PT}\xspace}

\newcommand{\dm}{\ensuremath{\Delta m}\xspace}

\newcommand{\DG}{\ensuremath{\Delta\Gamma}\xspace}
\newcommand{\DGs}{\ensuremath{\Delta\Gamma_{\squark}}\xspace}

\newcommand{\DGq}{\ensuremath{\Delta\Gamma_{\quark}}\xspace}
\newcommand{\Gq}{\ensuremath{\Gamma_{\quark}}\xspace}
\newcommand{\dmq}{\ensuremath{\Delta m_{\quark}}\xspace}

\newcommand{\ACP}{\ensuremath{{\cal A}_{\CP}}\xspace}

\def\BsToJPsiPhi  {\decay{\Bs}{\jpsi\phi}}

\def\BsToKK       {\decay{\Bs}{\Kp\Km}}

\def\AT#1     {\ensuremath{A_{\mathrm{T}}^{#1}}\xspace}           

\def\ctl       {\ensuremath{\cos{\theta_l}}\xspace}
\def\ctk       {\ensuremath{\cos{\theta_K}}\xspace}

\def\C#1      {\ensuremath{\mathcal{C}_{#1}}\xspace}                       
\def\Cp#1     {\ensuremath{\mathcal{C}_{#1}^{'}}\xspace}                    
\def\Ceff#1   {\ensuremath{\mathcal{C}_{#1}^{\mathrm{(eff)}}}\xspace}        
\def\Cpeff#1  {\ensuremath{\mathcal{C}_{#1}^{'\mathrm{(eff)}}}\xspace}       
\def\Ope#1    {\ensuremath{\mathcal{O}_{#1}}\xspace}                       
\def\Opep#1   {\ensuremath{\mathcal{O}_{#1}^{'}}\xspace}                    

\def\ycp        {\ensuremath{y_{\CP}}\xspace}
\def\agamma     {\ensuremath{A_{\Gamma}}\xspace}

\newcommand{\bra}[1]{\ensuremath{\langle #1|}}             
\newcommand{\ket}[1]{\ensuremath{|#1\rangle}}              

\newcommand{\tev}{\ensuremath{\mathrm{\,Te\kern -0.1em V}}\xspace}
\newcommand{\gev}{\ensuremath{\mathrm{\,Ge\kern -0.1em V}}\xspace}
\newcommand{\mev}{\ensuremath{\mathrm{\,Me\kern -0.1em V}}\xspace}
\newcommand{\kev}{\ensuremath{\mathrm{\,ke\kern -0.1em V}}\xspace}
\newcommand{\ev}{\ensuremath{\mathrm{\,e\kern -0.1em V}}\xspace}
\newcommand{\gevc}{\ensuremath{{\mathrm{\,Ge\kern -0.1em V\!/}c}}\xspace}
\newcommand{\mevc}{\ensuremath{{\mathrm{\,Me\kern -0.1em V\!/}c}}\xspace}
\newcommand{\gevcc}{\ensuremath{{\mathrm{\,Ge\kern -0.1em V\!/}c^2}}\xspace}
\newcommand{\gevgevcccc}{\ensuremath{{\mathrm{\,Ge\kern -0.1em V^2\!/}c^4}}\xspace}
\newcommand{\mevcc}{\ensuremath{{\mathrm{\,Me\kern -0.1em V\!/}c^2}}\xspace}

\def\cm   {\ensuremath{\rm \,cm}\xspace}

\def\mm   {\ensuremath{\rm \,mm}\xspace}

\def\mum  {\ensuremath{\,\upmu\rm m}\xspace}

\def\mub{\ensuremath{\rm \,\upmu b}\xspace}

\def\nb {\ensuremath{\rm \,nb}\xspace}

\def\pb {\ensuremath{\rm \,pb}\xspace}
\def\invpb {\ensuremath{\mbox{\,pb}^{-1}}\xspace}

\def\invfb   {\ensuremath{\mbox{\,fb}^{-1}}\xspace}

\def\sec  {\ensuremath{\rm {\,s}}\xspace}

\def\ns   {\ensuremath{{\rm \,ns}}\xspace}
\def\ps   {\ensuremath{{\rm \,ps}}\xspace}

\def\mhz  {\ensuremath{{\rm \,MHz}}\xspace}
\def\khz  {\ensuremath{{\rm \,kHz}}\xspace}

\def\invps{\ensuremath{{\rm \,ps^{-1}}}\xspace}

\def\order{{\ensuremath{\cal O}}\xspace}

\def\gsim{{~\raise.15em\hbox{$>$}\kern-.85em
          \lower.35em\hbox{$\sim$}~}\xspace}
\def\lsim{{~\raise.15em\hbox{$<$}\kern-.85em
          \lower.35em\hbox{$\sim$}~}\xspace}

\renewcommand{\Re}{\ensuremath{{\rm Re}}\xspace}
\renewcommand{\Im}{\ensuremath{{\rm Im}}\xspace}

\def\sqs   {\ensuremath{\protect\sqrt{s}}\xspace}

\def\pt         {\mbox{$p_{\rm T}$}\xspace}

\def\mrad{\ensuremath{\rm \,mrad}\xspace}
\def\rad{\ensuremath{\rm \,rad}\xspace}

\def\tell1  {TELL1\xspace}
\def\ukl1   {UKL1\xspace}

\newcommand{\eg}{\mbox{\itshape e.g.}\xspace}
\newcommand{\ie}{\mbox{\itshape i.e.}}

\newcommand{\etc}{\mbox{\itshape etc.}\xspace}

\newcommand{\SM}{{\rm SM}}

\newcommand{\slashed}{\slash \hspace{-0.23cm}}

\newcommand{\ACPcharm}{\ensuremath{{\cal A}_{\CP}}\xspace}
\newcommand{\ACPother}[1]{\ensuremath{a_{\CP}^{{#1}}}\xspace}
\newcommand{\ACPdir}{\ACPother{\ensuremath{\rm dir}}\xspace}

\def\Vudbare  {\ensuremath{V_{\uquark\dquark}}\xspace}
\def\Vcdbare  {\ensuremath{V_{\cquark\dquark}}\xspace}

\def\Vusbare  {\ensuremath{V_{\uquark\squark}}\xspace}
\def\Vcsbare  {\ensuremath{V_{\cquark\squark}}\xspace}

\def\Vubbare  {\ensuremath{V_{\uquark\bquark}}\xspace}
\def\Vcbbare  {\ensuremath{V_{\cquark\bquark}}\xspace}

\newcommand{\amplitude}[1]{\ensuremath{A({#1})}\xspace}
\newcommand{\ampbar}[1]{\ensuremath{{\kern 0.2em\overline{\kern -0.2em A}}({#1})}\xspace}
\newcommand{\ampsub}[1]{\ensuremath{A_{#1}}\xspace}
\newcommand{\ampbarsub}[1]{\ensuremath{{\kern 0.2em\overline{\kern -0.2em A}}_{#1}}\xspace}

\def\AfAfbar   {\ensuremath{\kern -0.2em \stackrel{\kern 0.2em \textsf{\fontsize{5pt}{1em}\selectfont(---)}}{A}_{\kern -0.3em f}\kern -0.3em}\xspace}
\def\dacp     {\ensuremath{\Delta \ACPcharm}\xspace}
\newcommand{\dacpdir}{\ensuremath{\Delta \ACPdir}\xspace}
\newcommand{\dacpother}[1]{\ensuremath{\Delta \ACPother{{#1}}}\xspace}

\newcommand{\Ks}{\ensuremath{K^0_S}\xspace}
\newcommand{\Ksmumu}{\ensuremath{\KS\to\mu^+\mu^-}\xspace}

\newcommand{\Bq}{\ensuremath{B^0_{(s)}}\xspace}

\newcommand{\Bsmumu}{\ensuremath{\Bs\to\mu^+\mu^-}\xspace}
\newcommand{\Bdmumu}{\ensuremath{\Bd\to\mu^+\mu^-}\xspace}

\newcommand{\Jpsi}{\ensuremath{\jpsi}\xspace}

\newcommand{\Bqmumu}{\ensuremath{\Bq\to\mu^+\mu^-}\xspace}

\newcommand{\BRof}[1]{\ensuremath{{\cal B}(#1)}\xspace}

\newcommand{\DzToKpi}{\decay{\Dz}{\Km\pip}}     
\newcommand{\DzTopiK}{\decay{\Dz}{\pim\Kp}}     
\newcommand{\DzToKK}{\decay{\Dz}{\Km\Kp}}
\newcommand{\DzTopipi}{\decay{\Dz}{\pim\pip}}
\newcommand{\DzbToKpi}{\decay{\Dzb}{\Kp\pim}}     
\newcommand{\DzbTopiK}{\decay{\Dzb}{\pip\Km}}     

\newcommand{\GeVc}{\ensuremath{\,{\rm GeV}/c}\xspace}

\newcommand{\figref}[1]{Fig.~\ref{#1}}

\newcommand{\secref}[1]{Sec.~\ref{#1}}

\usepackage{tabularx}
\usepackage{array}
\usepackage{multirow}
\usepackage{bbm}

\def \Gaml{{\Gamma_\ell}}

\def \apaL{{\ensuremath{A_\parallel^L}}\xspace}
\def \apaR{{\ensuremath{A_\parallel^R}}\xspace}

\def \apeL{{\ensuremath{A_\perp^L}}\xspace}
\def \apeR{{\ensuremath{A_\perp^R}}\xspace}

\usepackage{cite} 
\usepackage{mciteplus}

\begin{document}

\renewcommand{\thefootnote}{\fnsymbol{footnote}}
\setcounter{footnote}{1}

\begin{titlepage}
\pagenumbering{roman}

\vspace*{-1.5cm}
\centerline{\large EUROPEAN ORGANIZATION FOR NUCLEAR RESEARCH (CERN)}
\vspace*{1.5cm}
\hspace*{-0.5cm}
\begin{tabular*}{\linewidth}{lc@{\extracolsep{\fill}}r}
\ifthenelse{\boolean{pdflatex}}
{\vspace*{-2.7cm}\mbox{\!\!\!\includegraphics[width=.14\textwidth]{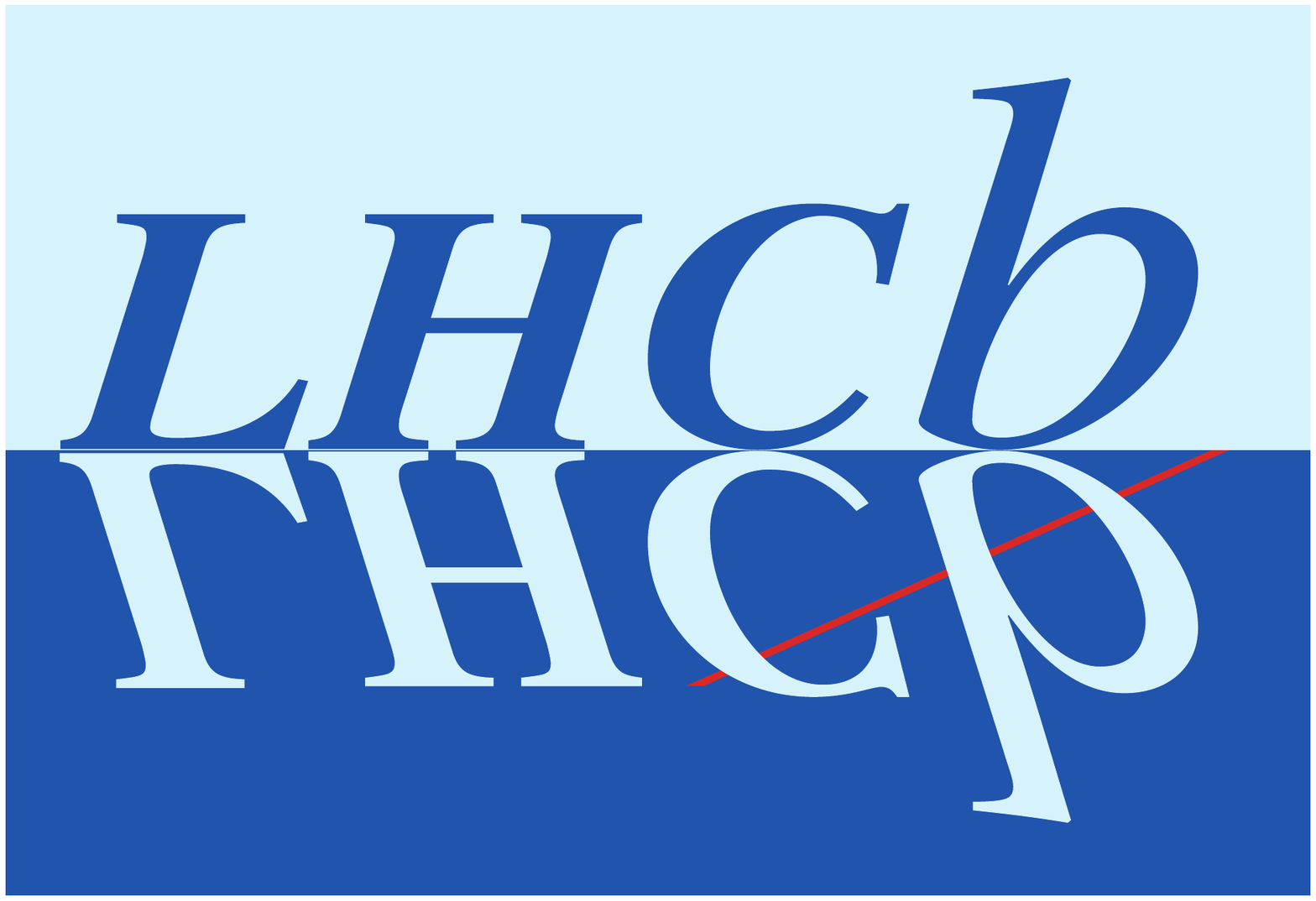}} & &}%
{\vspace*{-1.2cm}\mbox{\!\!\!\includegraphics[width=.12\textwidth]{figs/lhcb-logo.eps}} & &}%
\\
 & & CERN-PH-EP-2012-334 \\
 & & LHCb-PUB-2012-006, LHCb-PAPER-2012-031 \\
 & & \today  \\ 
 & & \\
\end{tabular*}

\vspace*{0.6cm}

\begin{center}
  {\bf
    {\huge
      Implications of LHCb measurements and future prospects
    }
  }
\end{center}

\vspace*{0.3cm}

\begin{center}
The LHCb collaboration\footnote{Authors are listed on the following pages.} \\
\vspace*{1ex}
and \\
\vspace*{1ex}
{\small 
A.~Bharucha\ifthenelse{\boolean{summaryonly}}{}{$^a$},
I.I.~Bigi\ifthenelse{\boolean{summaryonly}}{}{$^{b}$},
C.~Bobeth\ifthenelse{\boolean{summaryonly}}{}{$^c$},
M.~Bobrowski\ifthenelse{\boolean{summaryonly}}{}{$^{d}$},
J.~Brod\ifthenelse{\boolean{summaryonly}}{}{$^e$},
A.J.~Buras\ifthenelse{\boolean{summaryonly}}{}{$^{f}$},
C.T.H.~Davies\ifthenelse{\boolean{summaryonly}}{}{$^{g}$},
A.~Datta\ifthenelse{\boolean{summaryonly}}{}{$^h$},
C.~Delaunay\ifthenelse{\boolean{summaryonly}}{}{$^i$},
S.~Descotes-Genon\ifthenelse{\boolean{summaryonly}}{}{$^{j}$},
J.~Ellis\ifthenelse{\boolean{summaryonly}}{}{$^{i,k}$},
T.~Feldmann\ifthenelse{\boolean{summaryonly}}{}{$^{l}$},
R.~Fleischer\ifthenelse{\boolean{summaryonly}}{}{$^{m,n}$},
O.~Gedalia\ifthenelse{\boolean{summaryonly}}{}{$^{o}$},
J.~Girrbach\ifthenelse{\boolean{summaryonly}}{}{$^{f}$},
D.~Guadagnoli\ifthenelse{\boolean{summaryonly}}{}{$^p$},
G.~Hiller\ifthenelse{\boolean{summaryonly}}{}{$^{q}$},
Y.~Hochberg\ifthenelse{\boolean{summaryonly}}{}{$^{o}$},
T.~Hurth\ifthenelse{\boolean{summaryonly}}{}{$^{r}$},
G.~Isidori\ifthenelse{\boolean{summaryonly}}{}{$^{i,s}$},
S.~J\"{a}ger\ifthenelse{\boolean{summaryonly}}{}{$^{t}$},
M.~Jung\ifthenelse{\boolean{summaryonly}}{}{$^{q}$},
A.~Kagan\ifthenelse{\boolean{summaryonly}}{}{$^{e}$},
J.F.~Kamenik\ifthenelse{\boolean{summaryonly}}{}{$^{u,v}$},
A.~Lenz\ifthenelse{\boolean{summaryonly}}{}{$^{i,w}$},
Z.~Ligeti\ifthenelse{\boolean{summaryonly}}{}{$^{x}$},
D.~London\ifthenelse{\boolean{summaryonly}}{}{$^{y}$},
F.~Mahmoudi\ifthenelse{\boolean{summaryonly}}{}{$^{i,z}$},
J.~Matias\ifthenelse{\boolean{summaryonly}}{}{$^{aa}$},
S.~Nandi\ifthenelse{\boolean{summaryonly}}{}{$^{l}$},
Y.~Nir\ifthenelse{\boolean{summaryonly}}{}{$^{o}$},
P.~Paradisi\ifthenelse{\boolean{summaryonly}}{}{$^i$},
G.~Perez\ifthenelse{\boolean{summaryonly}}{}{$^{i,o}$},
A.A.~Petrov\ifthenelse{\boolean{summaryonly}}{}{$^{ab,ac}$},
R.~Rattazzi\ifthenelse{\boolean{summaryonly}}{}{$^{ad}$},
S.R.~Sharpe\ifthenelse{\boolean{summaryonly}}{}{$^{ae}$},
L.~Silvestrini\ifthenelse{\boolean{summaryonly}}{}{$^{af}$},
A.~Soni\ifthenelse{\boolean{summaryonly}}{}{$^{ag}$},
D.M.~Straub\ifthenelse{\boolean{summaryonly}}{}{$^{ah}$},
D.~van Dyk\ifthenelse{\boolean{summaryonly}}{}{$^{q}$},
J.~Virto\ifthenelse{\boolean{summaryonly}}{}{$^{aa}$},
Y.-M.~Wang\ifthenelse{\boolean{summaryonly}}{}{$^{l}$},
A.~Weiler\ifthenelse{\boolean{summaryonly}}{}{$^{ai}$},
J.~Zupan\ifthenelse{\boolean{summaryonly}}{}{$^{e}$} \\
\vspace*{1ex}
\ifthenelse{\boolean{summaryonly}}{}{
{\it 
$^a$Institut f\"{u}r Theoretische Physik, University of Hamburg, Hamburg, Germany \\
$^{b}$Department of Physics, University of Notre Dame du Lac, Notre Dame, USA \\
$^c$Technical University Munich, Excellence Cluster Universe, Garching, Germany \\
$^{d}$Karlsruhe Institute of Technology, Institut f{\"u}r Theoretische Teilchenphysik, Karlsruhe, Germany \\
$^e$Department of Physics, University of Cincinnati, Cincinnati, USA \\
$^{f}$TUM-Institute for Advanced Study, Garching, Germany \\
$^{g}$School of Physics and Astronomy, University of Glasgow, Glasgow, United Kingdom \\
$^h$Department of Physics and Astronomy, University of Mississippi, Oxford, USA \\
$^i$European Organization for Nuclear Research (CERN), Geneva, Switzerland \\
$^{j}$Laboratoire de Physique Th\'{e}orique, CNRS/Univ. Paris-Sud 11, Orsay, France \\
$^{k}$Physics Department, King’s College London, London, UK \\
$^{l}$Theoretische Elementarteilchenphysik, Naturwissenschaftlich Techn. Fakult\"{a}t, Universit\"{a}t Siegen, Siegen, Germany \\
$^{m}$Nikhef, Amsterdam, The Netherlands \\
$^{n}$Department of Physics and Astronomy, Vrije Universiteit Amsterdam, Amsterdam, The Netherlands \\
$^{o}$Department of Particle Physics and Astrophysics, Weizmann Institute of Science, Rehovot, Israel \\
$^p$LAPTh, Universit\'e de Savoie, CNRS/IN2P3, Annecy-le-Vieux, France \\
$^{q}$Institut f\"ur Physik, Technische Universit\"at Dortmund, Dortmund, Germany \\
$^{r}$Institute for Physics, Johannes Gutenberg University, Mainz, Germany \\
$^{s}$Laboratori Nazionali dell'INFN di Frascati, Frascati, Italy \\
$^{t}$Department of Physics \& Astronomy, University of Sussex, Brighton, UK \\
$^{u}$J. Stefan Institute, Ljubljana, Slovenia\\
$^{v}$Department of Physics, University of Ljubljana, Ljubljana, Slovenia \\
$^{w}$Institute for Particle Physics Phenomenology, Durham University, Durham, UK \\
$^{x}$Ernest Orlando Lawrence Berkeley National Laboratory, University of California, Berkeley, USA \\
$^{y}$Physique des Particules, Universit\'e de Montr\'eal, Montr\'eal, Canada \\
$^{z}$Clermont Universit{\'e}, Universit\'e Blaise Pascal, CNRS/IN2P3, Clermont-Ferrand, France \\ 
$^{aa}$Universitat Autonoma de Barcelona, Barcelona, Spain \\
$^{ab}$Department of Physics and Astronomy, Wayne State University, Detroit, USA \\
$^{ac}$Michigan Center for Theoretical Physics, University of Michigan, Ann Arbor, USA \\
$^{ad}$Institut de Th\'eorie des Ph\'enom\`enes Physiques, EPFL, Lausanne, Switzerland \\
$^{ae}$Physics Department, University of Washington, Seattle, USA \\
$^{af}$INFN, Sezione di Roma, Roma, Italy \\
$^{ag}$Department of Physics, Brookhaven National Laboratory, Upton, USA \\
$^{ah}$Scuola Normale Superiore and INFN, Pisa, Italy \\
$^{ai}$DESY, Hamburg, Germany\\
}
}
}

\end{center}

\vspace{\fill}

\begin{abstract}
  \noindent
  During 2011 the LHCb experiment at CERN collected $1.0 \invfb$ of $\sqrt{s} = 7 \tev$ $pp$ collisions.
  Due to the large heavy quark production cross-sections, these data provide unprecedented samples of heavy flavoured hadrons.
  The first results from LHCb have made a significant impact on the flavour physics landscape and have definitively proved the concept of a dedicated experiment in the forward region at a hadron collider.
  This document discusses the implications of these first measurements on classes of extensions to the Standard Model, bearing in mind the interplay with the results of searches for on-shell production of new particles at ATLAS and CMS.
  The physics potential of an upgrade to the LHCb detector, which would allow an order of magnitude more data to be collected, is emphasised.
\end{abstract}

\vspace*{0.5cm}

\begin{center}
  Submitted to Eur. Phys. J. C
\end{center}

\vspace{\fill}

{\footnotesize 
\centerline{\copyright~CERN on behalf of the \lhcb collaboration, license \href{http://creativecommons.org/licenses/by/3.0/}{CC-BY-3.0}.}}
\vspace*{2mm}

\end{titlepage}


\centerline{\large\bf LHCb collaboration}
\begin{flushleft}
\small
R.~Aaij$^{38}$, 
C.~Abellan~Beteta$^{33,n}$, 
A.~Adametz$^{11}$, 
B.~Adeva$^{34}$, 
M.~Adinolfi$^{43}$, 
C.~Adrover$^{6}$, 
A.~Affolder$^{49}$, 
Z.~Ajaltouni$^{5}$, 
J.~Albrecht$^{35}$, 
F.~Alessio$^{35}$
% ...

{\footnotesize \it
% ...
$ ^{5}$Clermont Universit\'{e}, Universit\'{e} Blaise Pascal, CNRS/IN2P3, LPC, Clermont-Ferrand, France\\
$ ^{6}$CPPM, Aix-Marseille Universit\'{e}, CNRS/IN2P3, Marseille, France\\
% ...
$ ^{11}$Physikalisches Institut, Ruprecht-Karls-Universit\"{a}t Heidelberg, Heidelberg, Germany\\
% ...
$ ^{33}$Universitat de Barcelona, Barcelona, Spain\\
$ ^{34}$Universidad de Santiago de Compostela, Santiago de Compostela, Spain\\
$ ^{35}$European Organization for Nuclear Research (CERN), Geneva, Switzerland\\
% ...
$ ^{38}$Nikhef National Institute for Subatomic Physics, Amsterdam, The Netherlands\\
% ...
$ ^{43}$H.H. Wills Physics Laboratory, University of Bristol, Bristol, United Kingdom\\
% ...
$ ^{49}$Oliver Lodge Laboratory, University of Liverpool, Liverpool, United Kingdom\\
% ...
$ ^{n}$LIFAELS, La Salle, Universitat Ramon Llull, Barcelona, Spain\\
% ...
}
\end{flushleft}

\cleardoublepage


\renewcommand{\thefootnote}{\arabic{footnote}}
\setcounter{footnote}{0}


\tableofcontents
\cleardoublepage


\pagestyle{plain} 
\setcounter{page}{1}
\pagenumbering{arabic}



\section{Introduction}
\label{sec:Introduction}

During 2011 the LHCb experiment~\cite{Alves:2008zz} at CERN collected $1.0 \invfb$ of $\sqrt{s} = 7 \tev$ $pp$ collisions.
Due to the large production cross-section, 
$\sigma(pp \to b\bar{b}X) = (89.6 \pm 6.4 \pm 15.5) \mub$ in the LHCb acceptance~\cite{LHCb-PAPER-2010-002}, with the comparable number for charm production about 20 times larger~\cite{LHCb-CONF-2010-013,LHCb-PAPER-2012-041},
these data provide unprecedented samples of heavy flavoured hadrons.
The first results from LHCb have made a significant impact on the flavour physics landscape and have definitively proved the concept of a flavour physics experiment in the forward region at a hadron collider.

The physics objectives of the first phase of LHCb were set out prior to the commencement of data taking in the ``roadmap document''~\cite{Adeva:2009ny}.
They centred on six main areas, in all of which LHCb has by now published its first results: 
(i) the tree-level determination of $\gamma$~\cite{LHCb-PAPER-2012-001,LHCb-PAPER-2012-027}, 
(ii) charmless two-body \B decays~\cite{LHCb-PAPER-2011-029,LHCb-PAPER-2012-002},
(iii) the measurement of mixing-induced \CP violation in $\Bs \to \jpsi \phi$~\cite{LHCb-PAPER-2011-021},
(iv) analysis of the decay $\Bs \to \mumu$~\cite{LHCb-PAPER-2011-004,LHCb-PAPER-2011-025,LHCb-PAPER-2012-007,LHCb-PAPER-2012-043}, 
(v) analysis of the decay $\Bz \to \Kstarz \mumu$~\cite{LHCb-PAPER-2011-020},
(vi) analysis of $\Bs \to \phi\gamma$ and other radiative \B decays~\cite{LHCb-PAPER-2011-042,LHCb-PAPER-2012-019}.\footnote{
  Throughout the document, the inclusion of charge conjugated modes is implied unless explicitly stated.
}
In addition, the search for \CP violation in the charm sector was established as a priority, and interesting results in this area have also been published~\cite{LHCb-PAPER-2011-023,LHCb-PAPER-2011-032}.

The results demonstrate the capability of LHCb to test the Standard Model (SM) and, potentially, to reveal new physics (NP) effects in the flavour sector.
This approach to search for NP is complementary to that used by the ATLAS and CMS experiments.  
While the high-\pt experiments search for on-shell production of new particles, LHCb can look for their effects in processes that are precisely predicted in the SM.
In particular, the SM has a highly distinctive flavour structure, with no tree-level flavour-changing neutral currents, and quark mixing described by the Cabibbo-Kobayashi-Maskawa (CKM) matrix~\cite{Cabibbo:1963yz,*Kobayashi:1973fv} which has a single source of \CP violation.  This structure is not necessarily replicated in extended models.
Historically, new particles have first been seen through their virtual effects since this approach allows one to probe mass scales beyond the energy frontier.  For example, the observation of \CP violation in the kaon system~\cite{Christenson:1964fg} was, in hindsight, the discovery of the third family of quarks, well before the observations of the bottom and top quarks.
Crucially, measurements of both high-\pt and flavour observables are necessary in order to decipher the nature of NP.

The early data also illustrated the potential for LHCb to expand its physics programme beyond these ``core'' measurements.  
In particular, the development of trigger algorithms that select events inclusively based on properties of $b$-hadron decays~\cite{LHCb-PUB-2011-016,LHCb-DP-2012-004} facilitates a much broader output than previously foreseen.
On the other hand, limitations imposed by the hardware trigger lead to a maximum instantaneous luminosity at which data can most effectively be collected (higher luminosity requires tighter trigger thresholds, so that there is no gain in yields, at least for channels that do not involve muons).  To overcome this limitation, an upgrade of the LHCb experiment has been proposed to be installed during the long shutdown of the LHC planned for 2018.  The upgraded detector will be read out at the maximum LHC bunch-crossing frequency of $40 \mhz$ so that the trigger can be fully implemented in software.
With such a flexible trigger strategy, the upgraded LHCb experiment can be considered as a general purpose detector in the forward region.

The Letter of Intent for the LHCb upgrade~\cite{CERN-LHCC-2011-001}, containing a detailed physics case, was submitted to the LHCC in March 2011 and was subsequently endorsed.  
Indeed, the LHCC viewed the physics case as ``compelling.''
Nevertheless, the LHCb collaboration continues to consider further possibilities to enhance the physics reach.
Moreover, given the strong motivation to exploit fully the flavour physics potential of the LHC, it is timely to update the estimated sensitivities for various key observables based on the latest available data.
These studies are described in this paper, and summarised in the framework technical design report for the LHCb upgrade~\cite{CERN-LHCC-2012-007}, submitted to the LHCC in June 2012 and endorsed in September 2012.


In the remainder of this introduction, a brief summary of the current LHCb detector is given, together with the common assumptions made to estimate the sensitivity achievable by the upgraded experiment.
Thereafter, the sections of the paper 
discuss rare charm and beauty decays in Sec.~\ref{sec:rare},
\CP violation in the \B system in Sec.~\ref{sec:B-CPV}
and mixing and \CP violation in the charm sector in Sec.~\ref{sec:charm}.
There are several other important topics, not covered in any of these sections, that can be studied at LHCb and its upgrade, and these are discussed in Sec.~\ref{sec:other}.
A summary is given in Sec.~\ref{sec:summary}.

\subsection{Current LHCb detector and performance}
\label{intro:det}

The \lhcb detector~\cite{Alves:2008zz} is a single-arm forward
spectrometer covering the \mbox{pseudorapidity} range $2<\eta <5$,
designed for the study of particles containing \bquark or \cquark
quarks. The detector includes a high precision tracking system
consisting of a silicon-strip vertex detector surrounding the $pp$
interaction region, a large-area silicon-strip detector located
upstream of a dipole magnet with a bending power of about
$4{\rm\,Tm}$, and three stations of silicon-strip detectors and straw
drift tubes placed downstream. The combined tracking system has a
momentum resolution $\Delta p/p$ that varies from 0.4\,\% at 5\gevc to
0.6\,\% at 100\gevc, and an impact parameter resolution of 20\mum for
tracks with high transverse momentum. Charged hadrons are identified
using two ring-imaging Cherenkov detectors. Photon, electron and
hadron candidates are identified by a calorimeter system consisting of
scintillating-pad and preshower detectors, an electromagnetic
calorimeter and a hadronic calorimeter. Muons are identified by a
system composed of alternating layers of iron and multiwire
proportional chambers. 
The trigger consists of a hardware stage, based
on information from the calorimeter and muon systems, followed by a
software stage which applies a full event reconstruction.


During 2011, the LHCb experiment collected $1.0 \invfb$ of integrated luminosity during the LHC $pp$ run at a centre-of-mass energy $\sqrt{s} = 7 \tev$. 
The majority of the data was recorded at an instantaneous luminosity of ${\cal L}_{\rm inst} = 3.5 \times 10^{32} \cm^{-2} \sec^{-1}$, nearly a factor of two above the LHCb design value, and with a pile-up rate (average number of visible interactions per crossing) of $\mu \sim 1.5$ (four times the nominal value, but below the rates of up to $\mu \sim 2.5$ seen in 2010). 
A luminosity levelling procedure, where the beams are displaced at the LHCb interaction region, allows LHCb to maintain an approximately constant luminosity throughout each LHC fill.
This procedure permitted reliable operation of the experiment and a stable trigger configuration throughout 2011.
The hardware stage of the trigger produced output at around 800 \khz, close to the nominal 1 \mhz, while the output of the software stage was around 3 \khz, above the nominal 2 \khz, divided roughly equally between channels with muons, $b$ decays to hadrons and charm decays.
During data taking, the magnet polarity was flipped at a frequency of about one cycle per month in order to collect equal sized data samples of both polarities for periods of stable running conditions.
Thanks to the excellent performance of the LHCb detector, the overall data taking efficiency exceeded 90\,\%.

\subsection{Assumptions for LHCb upgrade performance}
\label{intro:upgrade}

In the upgrade era, several important improvements compared to the current detector performance can be expected, as detailed in the framework TDR.
However, to be conservative, the sensitivity studies reported in this paper all assume detector performance as achieved during 2011 data taking.
The exception is in the trigger efficiency, where channels selected at hardware level by hadron, photon or electron triggers are expected to have their efficiencies double (channels selected by muon triggers are expected to have marginal gains, that have not been included in the extrapolations).
Several other assumptions are made:
\begin{itemize}
\item LHC collisions will be at $\sqrt{s} = 14 \tev$, with heavy flavour production cross-sections scaling linearly with $\sqrt{s}$;
\item the instantaneous luminosity\footnote{
  It is anticipated that any detectors that need replacement for the LHCb upgrade will be designed such that they can sustain a luminosity of ${\cal L}_{\rm inst} = 2 \times 10^{33} \cm^{-2} \sec^{-1}$~\cite{CERN-LHCC-2012-007}.
  Operation at instantaneous luminosities higher than the nominal value assumed for the estimations will allow the total data set to be accumulated in a shorter time.
} in LHCb will be ${\cal L}_{\rm inst} = 10^{33} \cm^{-2} \sec^{-1}$: this will be achieved with $25 \ns$ bunch crossings (compared to $50 \ns$ in 2011) and $\mu = 2$;
\item LHCb will change the polarity of its dipole magnet with similar frequency as in 2011/12 data taking, to approximately equalise the amount of data taken with each polarity for better control of certain potential systematic biases;
\item the integrated luminosity will be ${\cal L}_{\rm int} = 5 \invfb$ per year, and the experiment will run for 10 years to give a total sample of $50 \invfb$.
\end{itemize}

\clearpage

\section{Rare decays}
\label{sec:rare}

\subsection{Introduction}
\label{sec:rare:Introduction}

The term rare decay is used within this document to refer loosely to two classes of decays: 

\begin{itemize}
\item flavour-changing neutral current (FCNC) processes that are mediated by electroweak box and penguin type diagrams in the SM;
\item more exotic decays, including searches for lepton flavour or number violating decays of $B$ or $D$ mesons and for light scalar particles. 
\end{itemize}

\noindent
The first broad class of decays includes the rare radiative process
\decay{\Bs}{\phi\g} and rare leptonic and semileptonic decays
\decay{B^0_{(s)}}{\mumu} and \decay{\Bd}{\Kstarz\mumu}. These were
listed as priorities for the first phase of the \lhcb experiment in
the roadmap document~\cite{Adeva:2009ny}. In many well motivated new
physics models, new particles at the \tev scale can enter in diagrams
that compete with the SM processes, leading to modifications 
of branching fractions or angular distributions of the daughter
particles in these decays.   

For the second class of decay, there is either no SM contribution or
the SM contribution is vanishingly small and any signal would indicate
evidence for physics beyond the SM. Grouped in this class of decay are
searches for \gev scale new particles that might be directly produced
in $B$ or $D$ meson decays. 
This includes searches for light scalar particles 
and for $B$ meson decays to pairs of
same-charge leptons that can arise, for example, in models containing
Majorana neutrinos~\cite{Majorana:1937vz,Atre:2009rg,Cvetic:2010rw}.  

The focus of this section is on rare decays involving leptons or photons in the final states.
There are also several interesting rare decays involving hadronic final states that can be pursued at LHCb, such as $\Bp \to \Km\pip\pip$, $\Bp \to \Kp\Kp\pim$~\cite{Fajfer:2006av,Pirjol:2009vz}, $\Bs \to \phi \piz$ and $\Bs \to \phi\rho^0$~\cite{Hofer:2010ee}; however, these are not discussed in this document.

Section~\ref{sec:rare:operator} introduces the theoretical framework
(the operator product expansion) that is used when discussing rare
electroweak penguin processes. The observables and experimental
constraints coming from rare semileptonic, radiative and leptonic \B
decays are then discussed in Secs.~\ref{sec:rare:semileptonic},~\ref{sec:rare:radiative} and~\ref{sec:rare:leptonic} respectively. 
The implications of these experimental constraints for NP contributions are
discussed in Secs.~\ref{sec:rare:independent} and~\ref{sec:rare:dependent}. 
Possibilities with rare charm decays are then discussed in Sec.~\ref{sec:rare:charm},
and the potential of LHCb to search for rare kaon decays, lepton number and flavour violating decays, and for new light scalar particles is summarised in Secs.~\ref{sec:rare:kaon},~\ref{sec:rare:lepton} and~\ref{sec:rare:other} respectively.



\subsection{Model-independent analysis of new physics contributions to leptonic, semileptonic and radiative decays} 
\label{sec:rare:operator}

Contributions from physics beyond the SM to the observables in rare radiative, 
semileptonic and leptonic $B$ decays can be described by the modification of 
Wilson coefficients $C^{(\prime)}_i$ of local operators in an effective Hamiltonian of the form
\begin{equation}
\label{eq:Heff}
{\mathcal H}_{\text{eff}} = - \frac{4\,G_F}{\sqrt{2}} V_{tb}V_{tq}^* \frac{e^2}{16\pi^2}
\sum_i
(C_i O_i + C^\prime_i O^\prime_i) + \text{h.c.}~,
\end{equation}
where $q=d,s$, and where the primed operators indicate right-handed couplings. 
This framework is known as the operator product expansion, and is described in more detail in, \eg, Refs.~\cite{Uraltsev:1998bk,Buras:2005xt}.
In many concrete models, the operators that are most sensitive to NP are a subset of
\begin{align}
\label{eq:O7}
O_7^{(\prime)} &= \frac{m_b}{e}
(\bar{q} \sigma_{\mu \nu} P_{R(L)} b) F^{\mu \nu},
&
O_8^{(\prime)} &= \frac{g m_b}{e^2}
(\bar{q} \sigma_{\mu \nu} T^a P_{R(L)} b) G^{\mu \nu \, a},
\nonumber\\
O_9^{(\prime)} &= 
(\bar{q} \gamma_{\mu} P_{L(R)} b)(\bar{\ell} \gamma^\mu \ell)\,,
&
O_{10}^{(\prime)} &=
(\bar{q} \gamma_{\mu} P_{L(R)} b)( \bar{\ell} \gamma^\mu \gamma_5 \ell)\,,
\nonumber\\
O_S^{(\prime)} &= 
\frac{m_b}{m_{B_q}} (\bar{q} P_{R(L)} b)(  \bar{\ell} \ell)\,,
&
O_P^{(\prime)} &=
\frac{m_b}{m_{B_q}} (\bar{q} P_{R(L)} b)(  \bar{\ell} \gamma_5 \ell)\,,
\end{align}
which are customarily denoted as magnetic ($O_7^{(\prime)}$), chromomagnetic ($O_8^{(\prime)}$), semileptonic ($O_9^{(\prime)}$ and $O_{10}^{(\prime)}$), pseudoscalar ($O_P^{(\prime)}$) and scalar ($O_S^{(\prime)}$) operators.\footnote{
  In principle there are also tensor operators, $O_{T(5)} = (\bar{q}\sigma_{\mu\nu}b)(\bar{\ell}\sigma^{\mu\nu}(\gamma_5)\ell)$, which are relevant for some observables.
}
While the radiative $b\to q\gamma$ decays are sensitive only to 
the magnetic and chromomagnetic operators, semileptonic $b\to q\ell^+\ell^-$ 
decays are, in principle, sensitive to all these operators.\footnote{
  In radiative and semileptonic decays, the chromomagnetic operator $O_8$ enters at higher order in the strong coupling \as.
}

In the SM, models with minimal flavour violation (MFV)~\cite{Buras:2000dm,D'Ambrosio:2002ex} and models with a flavour symmetry relating the first two generations~\cite{Barbieri:2011ci}, the Wilson coefficients appearing in Eq.~(\ref{eq:Heff}) are equal for $q=d$ or $s$ and the ratio of amplitudes for $b\to d$ relative to $b \to s$ transitions is suppressed by $\left| V_{td}/V_{ts} \right|$. 
Due to this suppression, at the current level of experimental precision, constraints on decays with a $b\to d$ transition are much weaker than those on decays with a $b\to s$ transition for constraining $C_i^{(\prime)}$. 
In the future, precise measurements of $b \to d$ transitions will allow powerful tests to be made of this universality which could be violated by NP. 

The dependence on the Wilson coefficients, and the set of operators that can contribute, is different for different rare \B decays. 
In order to put the strongest constraints on the Wilson coefficients and to determine the room left for NP, it is therefore desirable to perform a combined analysis of all the available data on rare leptonic, semileptonic and radiative \B decays. 
A number of such analyses have recently been carried out for subsets of the Wilson coefficients~\cite{DescotesGenon:2011yn,Altmannshofer:2011gn,Bobeth:2011nj,Beaujean:2012uj,Altmannshofer:2012ir,Hurth:2012jn}. 

The theoretically cleanest branching ratios probing the $b\to s$ transition are the inclusive decays $B\to X_s\gamma$ and $B\to X_s\ell^+\ell^-$. 
In the former case, both the experimental measurement of the branching ratio and the SM expectation have uncertainties of about 7\,\%~\cite{HFAG,Misiak:2006zs}. 
In the latter case, semi-inclusive measurements at the $B$ factories still have errors at the 30\,\% level~\cite{HFAG}.  
At hadron colliders, the most promising modes to constrain NP are exclusive decays. 
In spite of the larger theory uncertainties on the branching fractions as compared to inclusive decays, the attainable experimental precision can lead to stringent constraints on the Wilson coefficients.
Moreover, beyond simple branching fraction measurements, exclusive decays offer
powerful probes of $C_7^{(\prime)}$, $C_9^{(\prime)}$ and $C_{10}^{(\prime)}$ through
angular and \CP-violating observables. 
The exclusive decays most sensitive to NP in $b\to s$ transitions are 
$B\to K^*\gamma$, $\Bs\to\mu^+\mu^-$, $B\to K\mu^+\mu^-$ and $B\to
K^*\mu^+\mu^-$. These decays are discussed in more detail below.  

\subsection{Rare semileptonic \B decays} 
\label{sec:rare:semileptonic}

The richest set of observables sensitive to NP are accessible through
rare semileptonic decays of \B mesons to a vector or pseudoscalar
meson and a pair of leptons. In particular the angular distribution of
$B\to K^*\mu^+\mu^-$ decays, discussed in Sec.~\ref{sec:BKll:pheno},
provides strong constraints on $C_{7}^{(\prime)}$, $C_{9}^{(\prime)}$
and $C_{10}^{(\prime)}$.  

\subsubsection{Theoretical treatment of rare semileptonic \decay{B}{M\ellell} decays}
\label{sec:rare:semileptonic:theory}

The theoretical treatment of exclusive rare semileptonic decays of the type \decay{B}{M\ellell} is possible in two kinematic regimes for the meson $M$: 
large recoil (corresponding to low dilepton invariant mass squared, \qsq) and small recoil (high \qsq). 
Calculations are difficult outside these regimes, in particular in the \qsq region close to the narrow \ccbar resonances (the \jpsi and \psitwos states).  

In the low $q^{2}$ region, these decays can be described by QCD-improved factorisation (QCDF)~\cite{Beneke:1999br,Beneke:2000ry} and the field theory formulation of soft-collinear effective theory (SCET)~\cite{Bauer:2000yr,Bauer:2001yt}. 
The combined limit of a heavy \bquark-quark and an energetic meson $M$, leads to the schematic form of the decay amplitude~\cite{Beneke:2001at,Beneke:2004dp}: 
\begin{align}
  \label{eq:factorisation}
  \mathcal T & 
   = C\, \xi + \phi_B \otimes T \otimes \phi_{M} 
     + \order(\lqcd/m_b)\,. 
\end{align}
which is accurate to leading order in $\lqcd/m_b$ and to all orders in \as. 
It factorises the calculation into process-independent non-perturbative quantities, $\B\to M$ form factors, $\xi$, and light cone distribution amplitudes (LCDAs), $\phi_{\B(M)}$, of the heavy (light) mesons, and perturbatively calculable quantities, $C$ and $T$ which are known to $\order(\as^1)$~\cite{Beneke:2001at,Beneke:2004dp}. 
Further, in the case that $M$ is a vector $V$ (pseudoscalar $P$), the seven (three) {\it a priori} independent $\B\to V$ ($\B\to P$) form factors reduce to two (one) universal {\it soft} form factors $\xi_{\bot,\|}$ ($\xi_P$) in QCDF/SCET~\cite{Charles:1998dr}. 
The factorisation formula Eq.~(\ref{eq:factorisation}) applies well in the dilepton mass range, $1 < q^2 < 6\gev^2$.\footnote{
  Light resonances at \qsq below $1\gev^2$ cannot be treated within QCDF, and their effects have to be estimated using other approaches. 
  In addition, the longitudinal amplitude in the QCDF/SCET approach generates a logarithmic divergence in the limit $q^2 \rightarrow 0$, indicating problems in the description below $1\gev^2$~\cite{Beneke:2001at}.
}

For \decay{\B}{\Kstar\ellell}, the three \Kstar spin amplitudes,
corresponding to longitudinal and transverse polarisations of the
\Kstar, are linear in the soft form factors
$\xi_{\bot,\|}$,  
\begin{align}
   A_{\perp,\parallel}^{L,R} & \propto C_{\perp}^{L,R}\, \xi_\perp, &
   A_{0}^{L,R} & \propto C_{\parallel}^{L,R}\, \xi_\parallel, &  
\end{align}
at leading order in $\lqcd/m_b$ and \as.  
The $C_{\perp,\parallel}^{L,R}$ are combinations of the Wilson coefficients ${\cal C}_{7,9,10}$ and the $L$ and $R$ indices refer to the chirality of the leptonic current.
Symmetry breaking corrections to these relationships of order \as are known~\cite{Beneke:2001at, Beneke:2004dp}. 
This simplification of the amplitudes as linear combinations of $C_{\perp,\parallel}^{L,R}$ and form factors, makes it possible to design a set of optimised observables in which any soft form factor dependence cancels out for all low dilepton masses~\qsq at leading order in \as and $\lqcd/m_b$~\cite{Egede:2008uy, Egede:2010zc, Matias:2012xw}, as discussed below in Sec.~\ref{sec:BKll:pheno}.  

Within the QCDF/SCET approach, a general, quantitative method to estimate the
important $\lqcd/m_b$ corrections to the heavy quark limit is missing. 
In semileptonic decays, a simple dimensional estimate of 10\,\% is often used, largely from matching of the soft form factors to the full-QCD form factors (see also Ref.~\cite{Khodjamirian:2010vf}).   


The high \qsq (low hadronic recoil) region, corresponds to dilepton
invariant masses above the two narrow resonances of $\jpsi$ and $\psitwos$, with $\qsq \gsim (14 - 15) \gev^2$. 
In this region, broad $\ccbar$-resonances are treated using a local operator product expansion~\cite{Grinstein:2004vb,Beylich:2011aq}. 
The operator product expansion (OPE) predicts small sub-leading corrections which are suppressed by either $(\lqcd/m_b)^2$~\cite{Beylich:2011aq} or $\as \lqcd/m_b$~\cite{Grinstein:2004vb} (depending on whether full QCD or subsequent matching on heavy quark effective theory in combination with form factor symmetries~\cite{Isgur:1989ed} is adopted). 
The sub-leading corrections to the amplitude have been estimated to be below $2$\,\%~\cite{Beylich:2011aq} and those due to form factor relations are suppressed numerically by $C_7 / C_9 \sim \order(0.1)$. 
Moreover, duality violating effects have been estimated within a model of resonances and found to be at the level of $2\,\%$ of the rate, if sufficiently large bins in \qsq are chosen~\cite{Beylich:2011aq}.  
Consequently, like the low \qsq region, this region is theoretically well under control. 

At high \qsq the heavy-to-light form factors are known only as extrapolations from light cone sum rules (LCSR) calculations at low \qsq. 
Results based on lattice calculations are being derived~\cite{Liu:2011raa}, and may play an important role in the near future in reducing the form factor uncertainties.  

\subsubsection{Angular distribution of \decay{\Bd}{\Kstarz\mumu} and \decay{\Bs}{\phi\mumu} decays}
\label{sec:BKll:pheno}

The physics opportunities of $\B\to V\ellell$ ($\ell=\electron, \mmu$, $V = \Kstar, \phi, \rho$) can be maximised through measurements of the angular distribution of the decay. 
Using the decay $\B \to \Kstar(\to \kaon\pion)\ellell$, with \Kstar on the mass shell, as an example, the angular distribution has the differential form~\cite{Kruger:1999xa,Altmannshofer:2008dz}
\begin{equation}
  \label{eq:differential decay rate}
  \frac{d^4\Gamma[\B \to \Kstar(\to \kaon\pion)\ellell]}
       {d\qsq\, d\ctl\, d\ctk\, d\phi} =
  \frac{9}{32\pi} \sum_i J_i(\qsq)\, g_i(\theta_l, \theta_K, \phi)\,,
\end{equation} 
with respect to \qsq and three decay angles $\theta_l$, $\theta_K$, and $\phi$. 
For the \Bd (\Bdb), $\theta_l$ is the angle between the \mup (\mun) and the opposite of the \Bd (\Bdb) direction in the dimuon rest frame, $\theta_K$ is the angle between the kaon and the direction opposite to the \B meson in the \Kstarz rest frame, and $\phi$ is the angle between the $\mumu$ and $\Kp\pim$ decay planes in the \B rest frame.
There are twelve angular terms appearing in the distribution and it is a long-term experimental goal to measure the coefficient functions $J_{i}(q^2)$ associated with these twelve terms, from which all other $\B\to \kaon^{(*)}\ellell$ observables can be derived.

In the SM, with massless leptons, the $J_i$ depend on bilinear products of six complex $K^*$ spin amplitudes $A_{\bot,\|,0}^{L,R}$,\footnote{
  Further amplitudes contribute in principle, but they are either suppressed by small lepton masses or originate from non-standard scalar/tensor operators.
}
such as
\begin{equation}
  \label{eq:J1s}
  J_{1s} = \frac{3}{4} \left[|\apeL|^2 + |\apaL|^2 + |\apeR|^2 + |\apaR|^2  \right]\,.
\end{equation}
The expressions for the eleven other $J_i$ terms are given for example in Refs.~\cite{Kruger:2005ep,Egede:2010zc}. 
Depending on the number of operators that are taken into account in the analysis, it is possible to relate some of the $J_i$ terms. 
The full derivation of these symmetries can be found in Ref.~\cite{Egede:2010zc}. 

When combining \B and \Bb decays, it is possible to form both \CP-averaged and \CP-asymmetric quantities: $S_i = (J_i + \bar{J_i})/[d(\Gamma + \bar{\Gamma})/d\qsq]$ and $A_i = (J_i - \bar{J_i})/[d(\Gamma + \bar{\Gamma})/d\qsq]$, from the
$J_i$~\cite{Bobeth:2008ij,Egede:2008uy,Altmannshofer:2008dz,Egede:2010zc,Bobeth:2010wg,Bobeth:2011gi,Kruger:2005ep}. 
The terms $J_{5,6,8,9}$ in the angular distribution are \CP-odd and, consequently, the associated \CP-asymmetry, $A_{5,6,8,9}$ can be extracted from an untagged analysis (making it possible for example to measure $A_{5,6,8,9}$ in \decay{\Bs}{\phi\mumu} decays). 
Moreover, the terms $J_{7,8,9}$ are $T$-odd and avoid the usual suppression of the corresponding \CP-asymmetries by small strong phases~\cite{Bobeth:2008ij}. 
The decay \decay{\Bd}{\Kstarz\mumu}, where the \Kstarz decays to $\Kp\pim$, is self-tagging (the flavour of the initial \B meson is determined from the decay products) and it is therefore possible to measure both the $A_i$ and $S_i$ for the twelve angular terms. 

In addition, a measurement of the $T$-odd \CP asymmetries, $A_7$, $A_8$ and $A_9$, which are zero in the SM and are not suppressed by small strong phases in the presence of NP, would be useful to constrain non-standard \CP violation. 
This is particularly true since the direct \CP asymmetry in the inclusive $B\to X_s\gamma$ decay is plagued by sizeable long-distance contributions and is therefore  not very useful as a constraint on NP~\cite{Benzke:2010tq}.  

\subsubsection{Strategies for analysis of $\Bd \to \Kstarz \ell^+\ell^-$ decays}

In $1.0\invfb$ of integrated luminosity, LHCb has collected the world's largest samples of \decay{\Bd}{\Kstarz\mumu} (with \decay{\Kstarz}{\Kp\pim}) and \decay{\Bs}{\phi\mumu}  decays, with around 900 and 80 signal candidates respectively reported in preliminary analyses~\cite{LHCb-CONF-2012-008,LHCb-CONF-2012-003}. 
These candidates are however sub-divided into six \qsq bins, following the binning scheme used in previous experiments~\cite{:2009zv}.
With the present statistics, the most populated \qsq bin contains $\sim 300$ \decay{\Bd}{\Kstarz\mumu} candidates which is not sufficient to perform a full angular analysis. 
The analyses are instead simplified by integrating  over two of the three angles or by applying a folding technique to the $\phi$ angle, $\phi \to \phi + \pi$ for $\phi < 0$, to cancel terms in the angular distribution.  

In the case of massless leptons, one finds:
\begin{equation}
 \label{eq:dGammadPhi}
    \frac{d\Gamma^\prime}{d\phi} =  \frac{\Gamma^\prime}{2\pi}\left(
      1 + S_3  \cos 2\phi + 
      A_{9} \sin 2\phi
    \right), 
\end{equation}
  \begin{equation}
    \label{kstarll-thetak}
    \frac{d\Gamma^\prime}{d\theta_K} = 
    \frac{3\Gamma^\prime}{4} \sin\theta_K \left(
      2  F_{\rm L} \cos^2 \theta_K +  (1- F_{\rm L})  \sin^2\theta_K
    \right),
\end{equation}
\begin{equation}
    \label{kstarll-thetal}
        \frac{d\Gamma^\prime}{d\theta_\ell} =  \Gamma^\prime \left(
      \frac{3}{4} F_{\rm L} \sin^2\theta_\ell + 
      \frac{3}{8} (1- F_{\rm L}) (1+\cos^2\theta_\ell) +
      A_{\rm FB}  \cos\theta_\ell 
    \right)\sin\theta_\ell.
\end{equation}
\noindent where $\Gamma^\prime = \Gamma + \bar{\Gamma}$. The
observables appear linearly in the expressions. Experimentally, the
fits are performed in bins of $q^2$ and the measured observables are
rate averaged over the $q^{2}$ bin. The observables appearing in the
angular projections are  the fraction of longitudinal polarisation of
the \Kstar, $F_{\rm L}$, the lepton system forward-backward asymmetry,
$A_{\rm FB}$, $S_{3}$ and $A_{9}$.   

\begin{figure}
\centering
\subfigure[]{\includegraphics[scale=0.34]{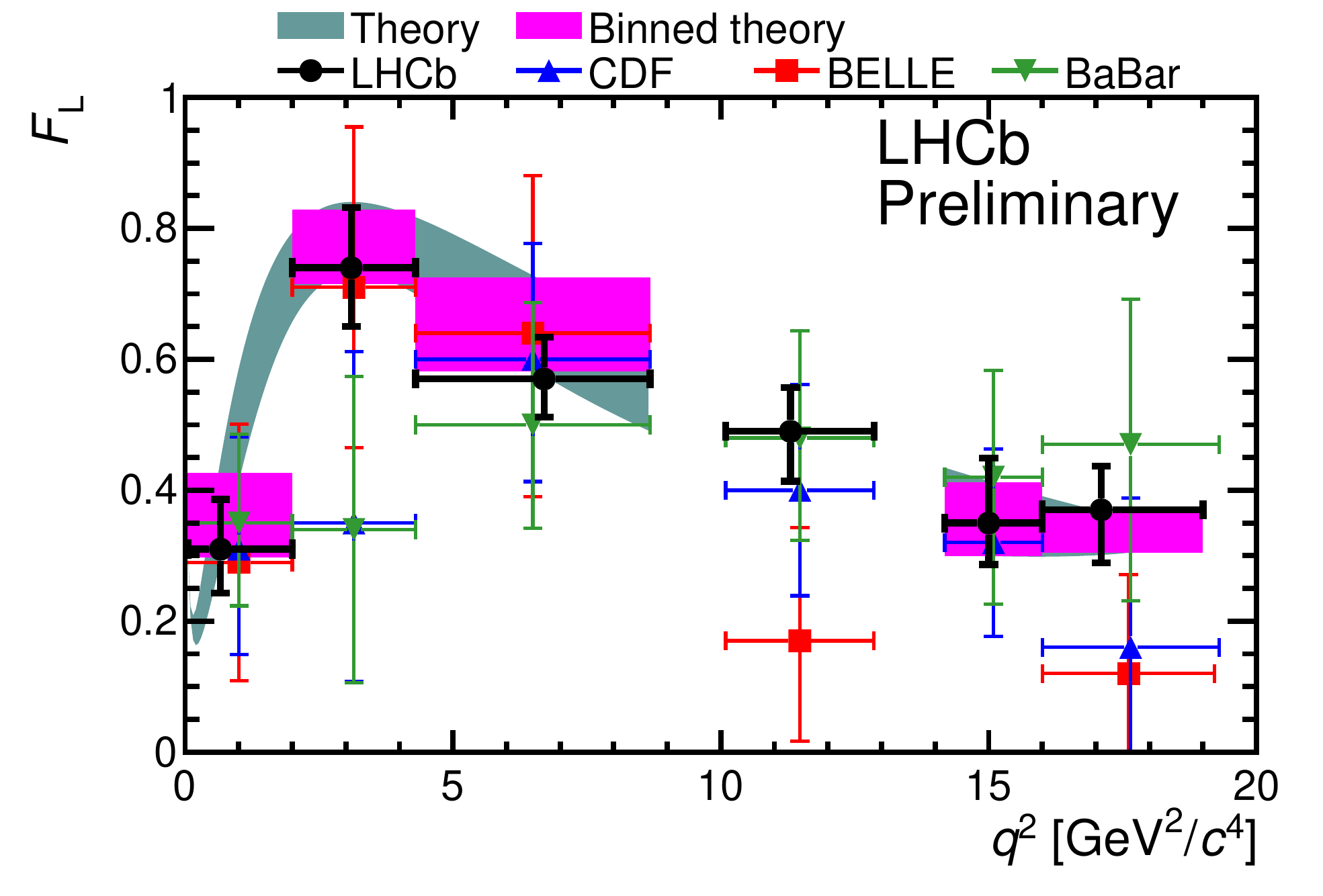}}
\subfigure[]{\includegraphics[scale=0.34]{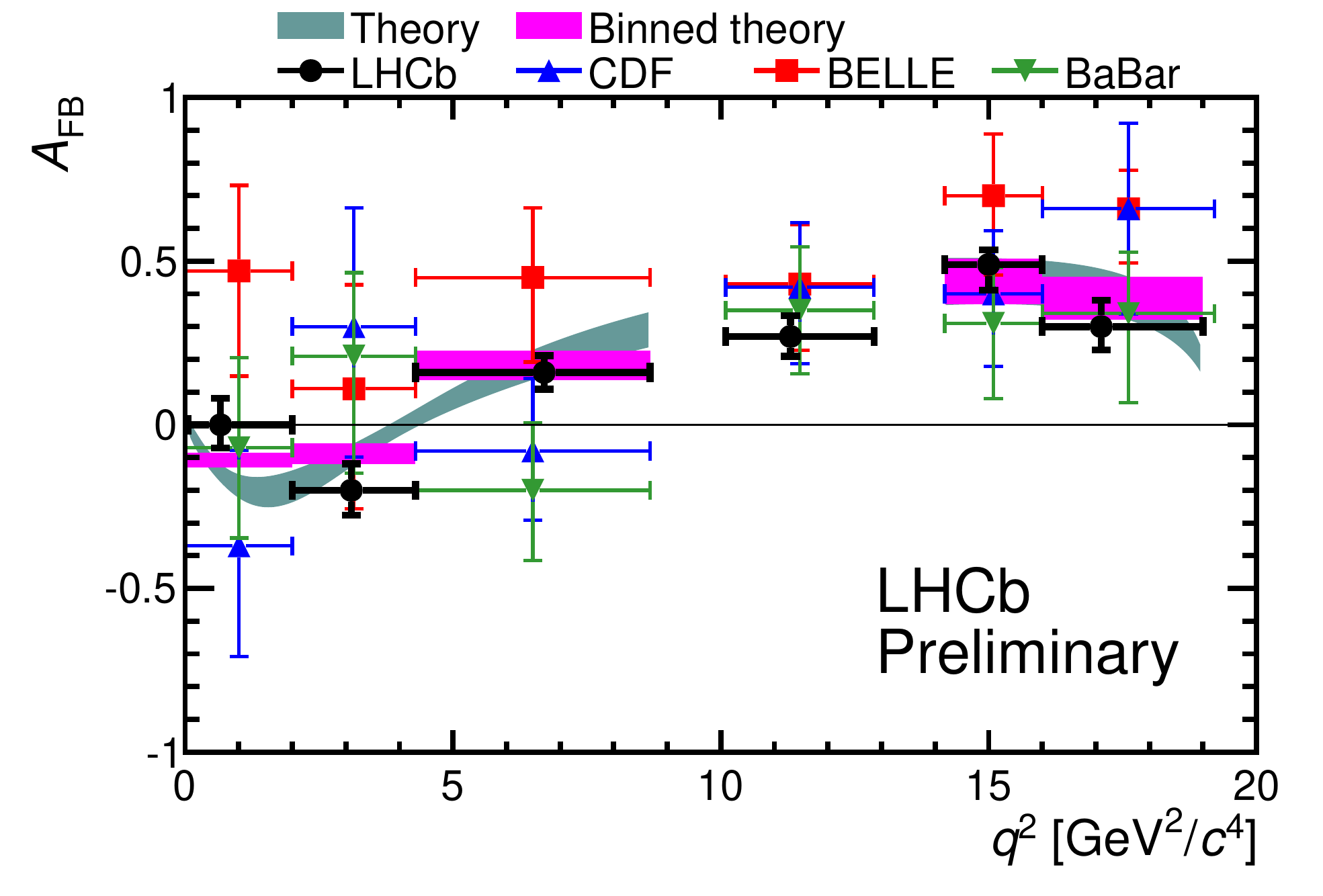}}
\subfigure[]{\includegraphics[scale=0.34]{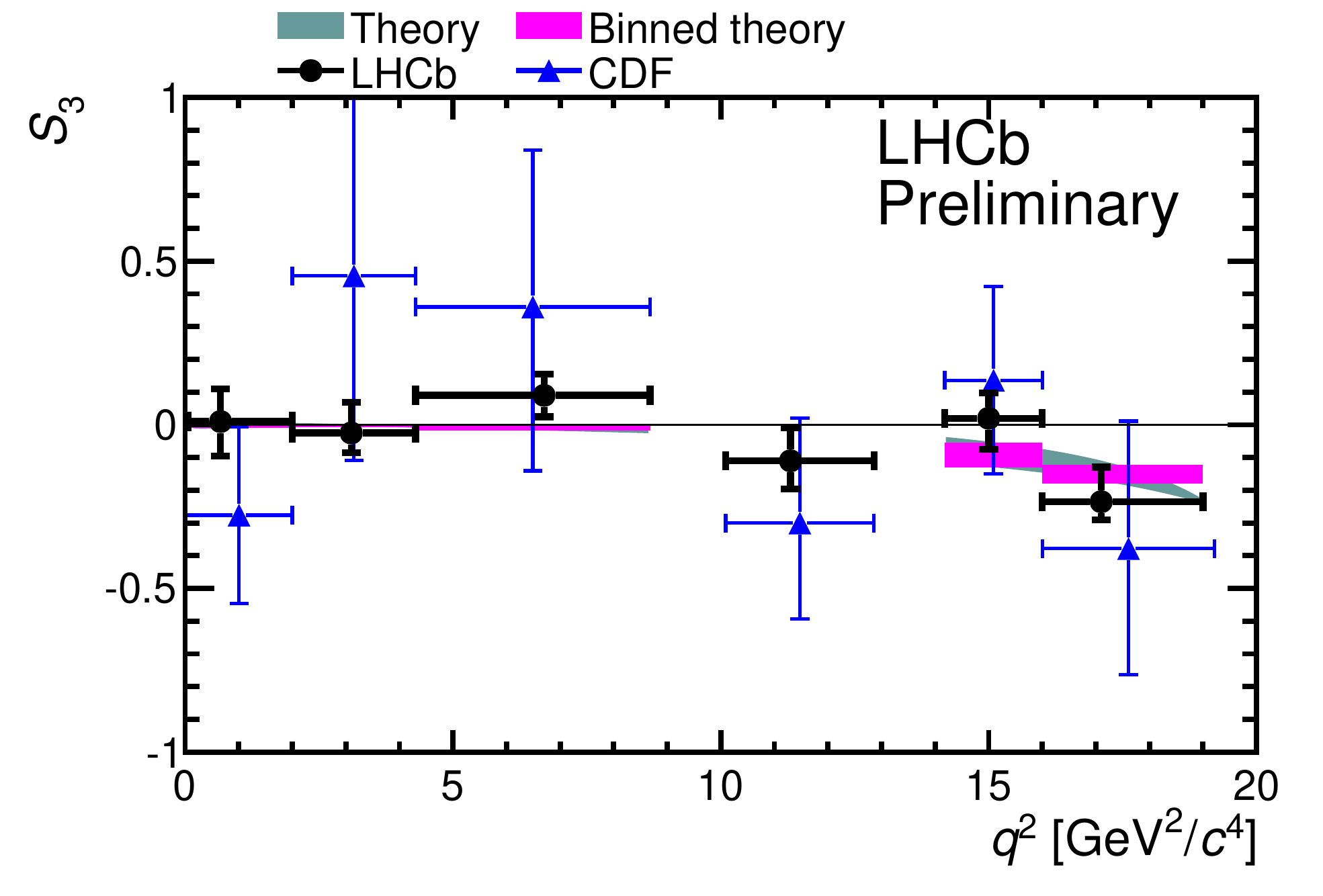}}
\subfigure[]{\includegraphics[scale=0.34]{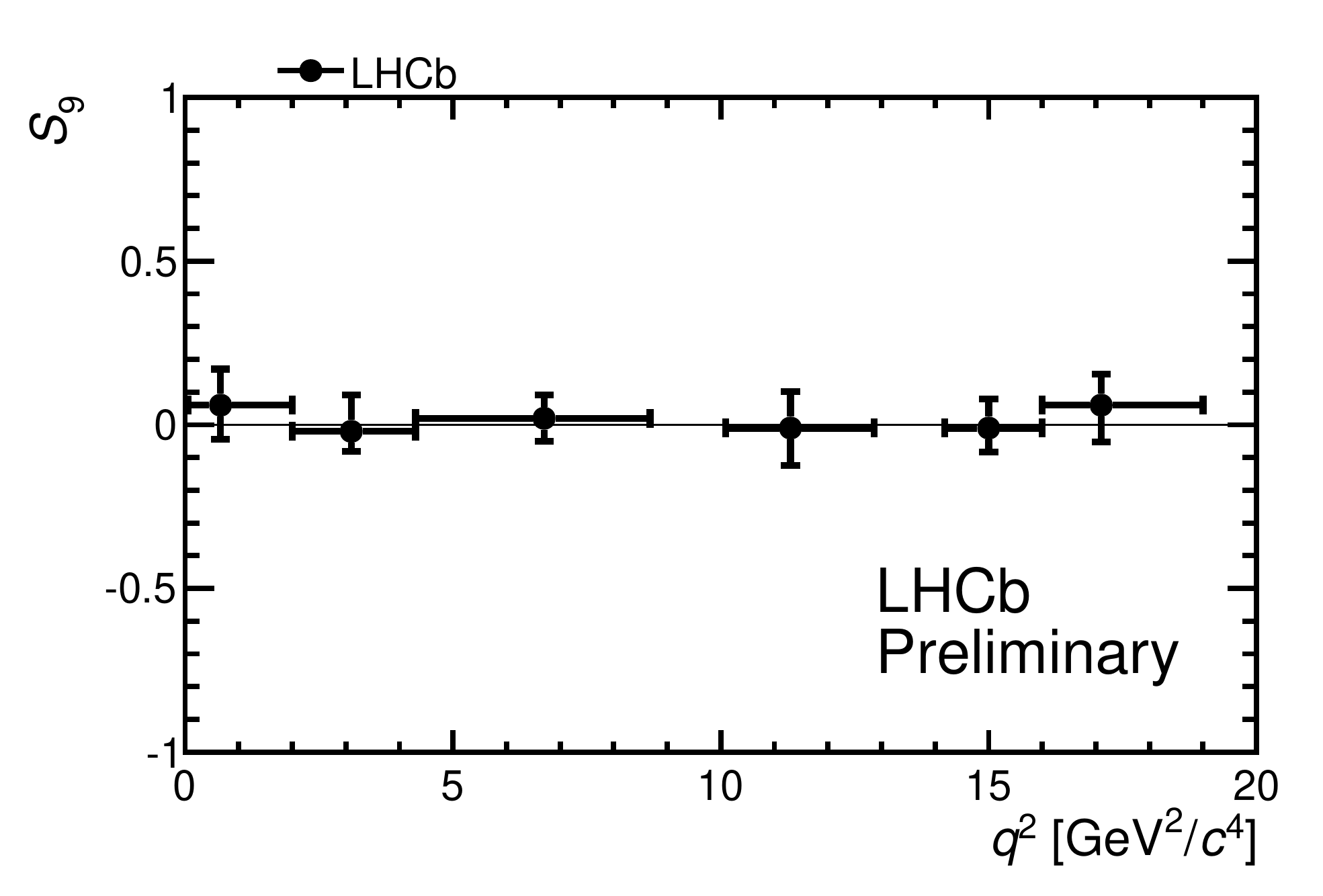}}
\caption{\small
  Summary of recent measurements of the angular observables (a) $F_{L}$, (b) $A_{\rm FB}$, (c) $S_{3}$ and (d) $S_{9}$ in \decay{\Bd}{\Kstarz\mumu} decays at LHCb, CDF and the \B factories~\cite{LHCb-CONF-2012-008}. 
  Descriptions of these observables are provided in the text (see Eqs.~(\ref{eq:dGammadPhi}),~(\ref{kstarll-thetak}) and~(\ref{kstarll-thetal}) and footnote~\ref{footnote:s9a9}). 
  The theory predictions at low- and high-dimuon invariant masses are indicated by the coloured bands and are also described in detail in the text.
}
 \label{fig:rare:bskstmm:anglular}
\end{figure}

The differential branching ratio, $A_{\rm FB}$ and $F_{\rm L}$ have been measured by the \B factories, CDF and LHCb~\cite{Aaltonen:2011ja,:2009zv,LHCb-CONF-2012-008}. 
The observable $S_3$ is related to the asymmetry between the parallel and perpendicular \Kstar spin amplitudes\footnote{
  The quantity $S_3 = (1-F_{\rm L})/2 \times A_T^{(2)}$ (in the massless case) allows access to one of the theoretically clean quantities, namely $A_T^{(2)}$. The observable $A_{T}^{(2)}$ is a theoretically cleaner observable than $S_{3}$ due to the cancellation of some of the form-factor dependence~\cite{DescotesGenon:2012zf}.
} 
is sensitive to right-handed operators ($C_7^\prime$) at low $q^{2}$, and is negligibly small in the SM. 
In the future, the decay \decay{\Bd}{\Kstarz\epem} could play an important role in constraining $C_7^\prime$ through $S_3$ since it allows one to probe to smaller values of $q^{2}$ than the \decay{\Bd}{\Kstarz\mumu} decay. 
First measurements have been performed by CDF and LHCb~\cite{Aaltonen:2011ja,LHCb-CONF-2012-008}.\footnote{
  \label{footnote:s9a9}
  Depending on the convention for the angle $\phi$, $d\Gamma^\prime/d\phi$ of Eq.~(\ref{eq:dGammadPhi}) can also depend on $S_9$, which is tiny in the SM and beyond.
  Note that, due to different angular conventions, the quantity $A_{\rm Im}$ reported in Ref.~\cite{LHCb-CONF-2012-008} corresponds to $S_9$, while $A_{\rm Im}$ in Ref.~\cite{Aaltonen:2011ja} corresponds to $A_9$.
}
The current experimental status of these \decay{\Bd}{\Kstarz\mumu} angular observables at \lhcb, the $B$ factories and CDF is shown in Fig.~\ref{fig:rare:bskstmm:anglular}.  
Improved measurements of these quantities would be useful to constrain the chirality-flipped Wilson coefficients ($C_7^{\prime}$, $C_9^{\prime}$ and $C_{10}^{\prime}$).

Whilst $A_{\rm FB}$ is not free from form-factor uncertainties at low $q^{2}$,  the value of the dilepton invariant mass $q^2_0$, for which the differential forward-backward asymmetry $A_{\rm FB}$ vanishes, can be predicted in a clean way.\footnote{
  In the QCDF approach at leading order in $\Lambda_{\rm QCD}/m_b$, the value of $q_0^2$ is free from hadronic uncertainties at order $\alpha_s^0$. 
  A dependence on the soft form factor and on the light-cone wave functions of the $B$ and $K^*$ mesons appears  only at order $\alpha_s^1$.
}  
The zero crossing-point is highly sensitive to the ratio of the two Wilson coefficients $C_7$ and $C_9$.  
In particular the model-independent upper bound on $\left| C_9 \right|$ implies $q_0^2 > 1.7 \gev^2/c^4$, which improves to $q_0^2 > 2.6 \gev^2/c^4$, assuming the sign of $C_7$ to be SM-like~\cite{Bobeth:2011nj}.
At next-to-leading order one finds~\cite{Beneke:2004dp}:\footnote{
  A recent determination of $q_0^2$ in \Bd decays gives $4.0 \pm 0.3 \gev^2/c^4$~\cite{Bobeth:2011nj}.  The shift with respect to Ref.~\cite{Beneke:2004dp} is of parametric origin and is driven in part by the choice of the renormalisation scale ($\mu = 4.2 \gev$ instead of $4.8 \gev$), but also due to differences in the implementation of higher ${\cal O}(\as)$ short-distance contributions.
}
\begin{equation}
  q_0^2[K^{*0}\ell^+\ell^-] = 4.36 \,^{+0.33}_{-0.31} \;\mbox{GeV}^2/c^{4},\;\;\;\;
  q_0^2[K^{*+}\ell^+\ell^-] = 4.15 \,^{+0.27}_{-0.27} \; \mbox{GeV}^2/c^{4}.
\end{equation}
\noindent where the first value is in good agreement with the recent preliminary result from \lhcb of $q_0^2 = 4.9\,^{+1.3}_{-1.1}\gev^{2}/c^{4}$~\cite{LHCb-CONF-2012-008} for the \decay{\Bd}{\Kstarz\mumu} decay.

It is possible to access information from other terms in the angular distribution by integrating over one of the angles and making an appropriate folding of the remaining two angles. 
From $\phi$ and $\theta_K$ only~\cite{Bharucha:2010bb} it is possible to extract: 
\begin{equation}
S_5 = -\frac{4}{3}\left[\int_{\pi/2}^{3\pi/2} - \int_{0}^{\pi/2} - \int_{3\pi/2}^{2\pi}\right]d\phi\left[ \int_0^1 - \int_{-1}^0  \right]d\cos\theta_K \frac{d^3(\Gamma-\bar\Gamma)}{dq^2\, d\cos\theta_Kd\phi} \bigg/ \frac{d(\Gamma+\bar\Gamma)}{dq^2}.
\end{equation}
Analogously to $A_{\rm FB}$, the zero-crossing point of $S_5$ has been shown to be theoretically clean. 
This observable is sensitive to the ratio of Wilson coefficients, 
$(C_7 + C_7^\prime)/(C_9  + \hat m_b (C_7 + C_7^\prime))$, and if measured would add complementary information to $A_{\rm FB}$ and $S_3$ about new right-handed currents.

\subsubsection{Theoretically clean observables in $\Bd \to \Kstarz \ell^+\ell^-$ decays}
\label{sec:rare:semileptonic:cleanobs}

By the time that $5\invfb$ of integrated luminosity is available at LHCb, it will be possible to exploit the complete NP sensitivity of the $B \rightarrow K^* \ell^+\ell^-$ both in the low- and high-$q^2$ regions, by performing a full angular analysis.
 The increasing size of the experimental samples makes it important to design 
optimised observables (by using specifically chosen combinations of the $J_i$) to reduce theoretical uncertainties.  In the low \qsq region, the linear dependence of the amplitudes on the soft form factors allows 
for a complete cancellation of the hadronic uncertainties due to the
form factors at leading order. This consequently increases the
sensitivity to the structure of NP models~\cite{Egede:2008uy, Egede:2010zc}. 


In the low $q^2$ region, the so-called transversity observables $A_T^{(i)}, i=2,3,4,5$ are an example set of observables that are constructed such that the soft form factor dependence cancels out at leading order. 
They represent the complete set of angular observables and are chosen to be highly sensitive to new  right-handed currents via $C_7^\prime$~\cite{Egede:2008uy,Egede:2010zc}. 
A second, complete, set of optimised angular observables was constructed (also in the cases of non-vanishing lepton masses and in the presence of scalar operators) in Ref.~\cite{Matias:2012xw}.  
Recently the effect of binning in $q^2$ on these observables has been considered~\cite{DescotesGenon:2012zf}. In these sets of observables, the unknown $\Lambda_{\rm QCD}/m_b$ corrections are estimated to be of order $10\,\%$ on the level of the spin amplitudes and represent the dominant source of theory uncertainty.

In general,  the angular observables  are shown to  offer high sensitivity to NP in the Wilson coefficients of the operators $O_7$, $O_9$, and $O_{10}$ and of  the chirally flipped operators~\cite{Bobeth:2008ij,Egede:2008uy,Altmannshofer:2008dz,Egede:2010zc}. 
In particular, the observables $S_3$, $A_9$ and the \CP-asymmetries $A_7$ and $A_8$ vanish at leading order in $\lqcd/m_b$ and \as in the SM operator basis~\cite{Bobeth:2008ij}. 
Importantly, this suppression is absent in extensions with non-vanishing chirality-flipped ${C}^\prime_{7,9,10}$, giving rise to contributions proportional to 
$\Re (C_i {C_j^{*}}^\prime)$ or $\Im (C_i {C_j^{*}}^\prime)$ and making these terms ideal probes of right-handed currents~\cite{Bobeth:2008ij,Egede:2008uy,Altmannshofer:2008dz,Egede:2010zc}.  \CP asymmetries are small in 
the SM,  because  the only \CP-violating phase affecting the decay is
doubly Cabibbo-suppressed, but can be significantly enhanced by NP
phases in $C_{9,10}$ and $C^{\prime}_{9,10}$, which at present are poorly
constrained. In a full angular analysis it can also be shown that
\CP-conserving observables provide 
indirect constraints on \CP-violating NP contributions~\cite{Egede:2010zc}.

At large \qsq, the dependence on the magnetic Wilson coefficients $C_7^{(\prime)}$ is suppressed, allowing, in turn, a cleaner extraction of semileptonic coefficients ($C_{9}^{(\prime)}$ and $C_{10}^{(\prime)}$). 
A set of transversity observables $H_T^{(i)}$, $i = 1,2,3$ have been designed to exploit the features of this kinematic region in order to have small hadronic uncertainties~\cite{Bobeth:2010wg}.
As a consequence of symmetry relations of the OPE~\cite{Bobeth:2010wg,Bobeth:2011gi,Bobeth:2011nj,Hambrock:2012dg}, at high \qsq, combinations of the angular observables $J_i$ can be formed within the SM operator basis (\ie\ with $C_i^\prime=0$), which depend:  
\begin{itemize}
\item only on short-distance quantities (\eg $H_T^{(2,3)}$) ; 
\item only on long-distance quantities ($F_{\rm L}$ and low $q^2$ optimised observables 
  $A_T^{(2,3)}$).
\end{itemize}
Deviations from these relations are due to small sub-leading corrections at order $(\lqcd/m_b)^2$ from the OPE. 

In the SM operator basis it is interesting to note that $A_T^{(2,3)}$,
which are highly sensitive to short distance contributions (from
$C_7^\prime$) at low $q^{2}$, instead become sensitive to long-distance
quantities (the ratio of form factors) at high $q^{2}$. The extraction
of form factor ratios is already possible with current data on $S_3$
($A_T^{(2)}$) and $F_{\rm L}$ and leads to a consistent picture between LCSR
calculations, lattice calculations and experimental data~\cite{Hambrock:2012dg,Beaujean:2012uj}. 
In the presence of chirality-flipped Wilson coefficients, these observables are no longer short-distance free, but are probes of right-handed currents~\cite{Altmannshofer:2012ir}.
At high $q^{2}$, the OPE framework predicts $H_T^{(2)} = H_T^{(3)}$ and $J_7 = J_8 = J_9 = 0$. 
Any deviation from these relationships, would indicate a problem with the OPE and the theoretical predictions in the high $q^2$ region. 

\subsubsection{\decay{\Bp}{\Kp\mumu} and \decay{\Bp}{\Kp\epem}} 
\label{sec:rare:semi:kmumu}

The branching fractions of $B^{0(+)} \to K^{0(+)}\mu^+\mu^-$ have been measured by \babar, Belle and CDF~\cite{Aaltonen:2011qs, :2009zv, :2012vw}. 
In $1.0 \invfb$ \lhcb observes 1250 \decay{\Bp}{\Kp\mumu} decays~\cite{LHCb-PAPER-2012-011}, and in the future will dominate measurements of these processes.

Since the $B\to K$ transition does not receive contributions from an axial vector current, the primed Wilson coefficients enter the $B^{0(+)}\to K^{0(+)}\mu^+\mu^-$ observables always in conjunction with their unprimed counterparts as $(C_i+C_i^\prime)$. 
This is in
contrast to the $B\to K^*\mu^+\mu^-$ decay and therefore provides complementary
constraints on the Wilson coefficients and their chirality-flipped counterparts.

An angular analysis of the $\mu^+\mu^-$ pair in the $B^{0(+)}\to K^{0(+)}\mu^+\mu^-$ decay would allow the measurement of two further observables, the forward-backward asymmetry $A_{\rm FB}$ and the so-called flat term $F_H$~\cite{Bobeth:2007dw}. 
The angular distribution of a \B meson decaying to a pseudoscalar
meson, $P$, and a pair of leptons involves just \qsq and a single angle
in the dilepton system, $\theta_l$~\cite{Bobeth:2007dw} 
\begin{equation}
  \label{eq:babar}
  \frac{1}{\Gaml} \frac{d\Gaml [\B \to P\ellell]}{d\!\ctl}
    = \frac{3}{4} (1 - F_{\rm H}) (1 - \cos^2\theta_l) 
    + \frac{1}{2} F_{\rm H} + A_{\rm FB} \ctl \, .
\end{equation} 

In the SM, the forward-backward asymmetry of the dilepton system is expected to be zero. 
Any non-zero forward-backward asymmetry would point to a contribution from new particles that extend the SM operator basis. 
Allowing for generic (pseudo-)scalar and tensor couplings, there is sizeable room for NP contributions in the range $|A_{\rm FB}| \lsim 15 \,\%$. 
The flat term, $F_{\rm H}/2$, that appears with $A_{\rm FB}$ in the angular distribution, is non-zero, but small (for $\ell = e, \mu$) in the SM. 
This term can also see large enhancements in models with (pseudo-)scalar and tensor couplings of up to $F_{\rm H} \sim 0.5$. 
Recent SM predictions at low- and high-$q^2$ can be seen in Refs.~\cite{Bobeth:2007dw,Khodjamirian:2010vf,Bobeth:2011nj,Becirevic:2012fy}. 
The current experimental limits on \BRof\Bsmumu now disfavour large $C_S$ and $C_P$, and if NP is present only in tensor operators then NP contributions are expected to be in the range $|A_{\rm FB}| \lsim 5 \,\%$ and  $F_{\rm H} \lsim 0.2$. 

In addition to $A_{\rm FB}$, $F_{\rm H}$ and the differential branching fraction of the decays, it is possible to probe the universality of lepton interactions by comparing the branching fraction of decays \decay{B^{0(+)}}{K^{0(+)}\ellell} with two different lepton flavours (\eg electrons versus muons): 
\begin{align}
  R_K & = \Gamma_\mu/\Gamma_e  ~~~~~~~~(\mbox{with the same $q^2$ cuts}).
\end{align}
\noindent 
Lepton universality may be violated in extensions to the SM, such as R-parity-violating SUSY models.\footnote{
  There are hints of lepton universality violation in recent measurements of $B \to D^{(*)}\tau\nu$ by BaBar~\cite{Lees:2012xj} and Belle~\cite{Matyja:2007kt,Bozek:2010xy}.
} 
In the SM, the ratio $R_K^{\rm SM}$ is expected to be close to unity, $R_{K}^{\rm SM} = 1 + {\cal{O}}(m_\mu^2/m_B^2)$~\cite{Hiller:2003js}.  

It is also interesting to note that at high $\qsq$ the differential decay rates and \CP asymmetries of $B^{0(+)} \to K^{0(+)} \ell^+\ell^-$ and $B^{0(+)} \to K^{*0(+)} \ell^+ \ell^-$ ($\ell=e,\mu$) are correlated~\cite{Bobeth:2011nj} and exhibit the same short-distance dependence (in the SM operator basis). 
Any deviation would point to a problem for the OPE used in the high $q^{2}$ region. 

\subsubsection{Rare semileptonic $b \to d \ellell$ decays} 

\begin{figure}
\centering 
\includegraphics[scale=0.4]{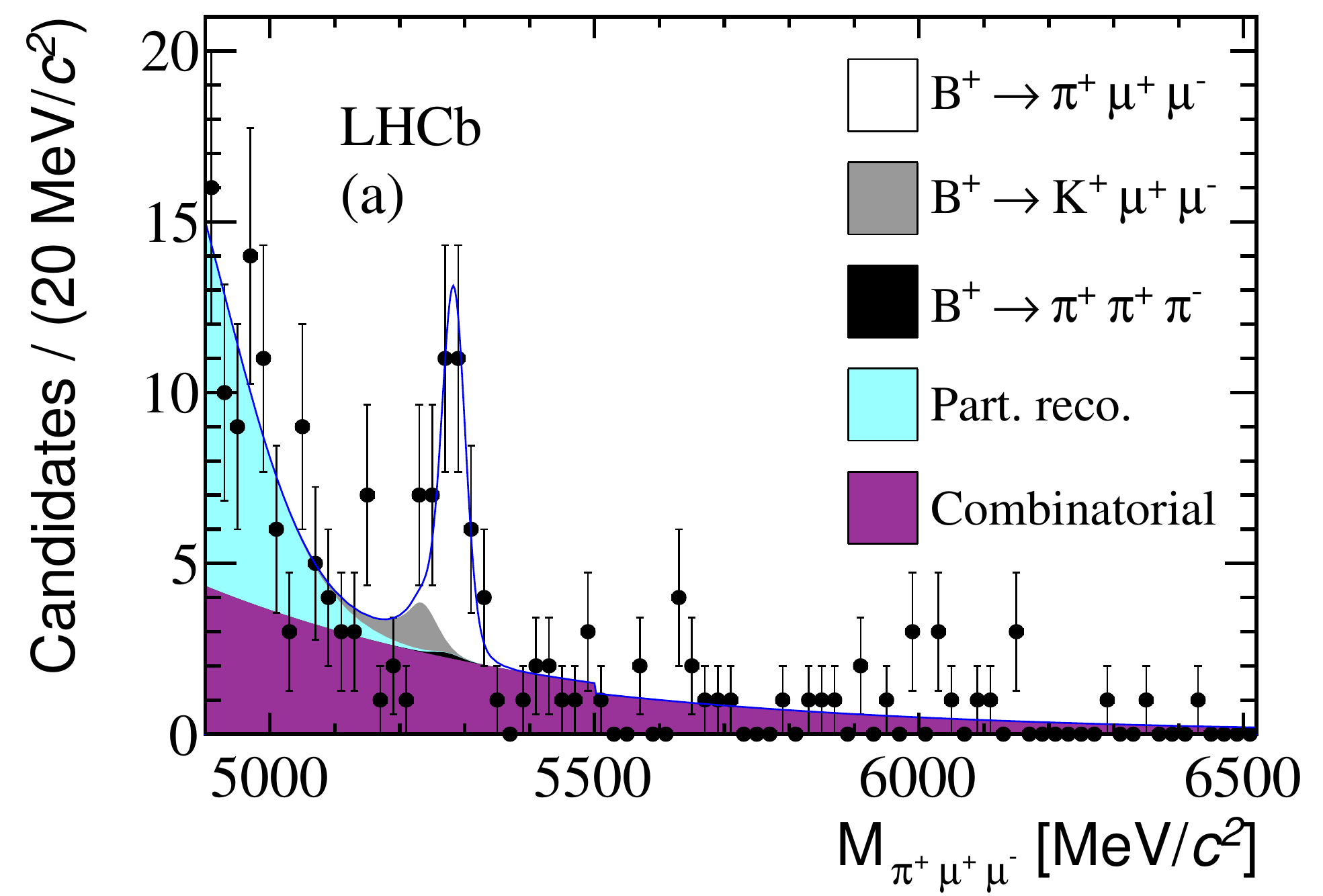}
\caption{\small
  Invariant mass of selected \decay{\Bp}{\pip\mumu} candidates in $1.0 \invfb$ of integrated luminosity~\cite{LHCb-PAPER-2012-020}. 
  In the legend, ``part. reco.'' and ``combinatorial'' refer to partially reconstructed and combinatorial backgrounds respectively. 
}
\label{fig:rare:pimumu}
\end{figure}

Rare $b\to d$ radiative decay processes, such as $B \to \rho \gamma$, have been observed at the \B factories~\cite{Taniguchi:2008ty,delAmoSanchez:2010ae}. 
In the 2011 data sample, the very rare decay \decay{\Bp}{\pip\mumu} was observed at the \lhcb experiment (see Fig.~\ref{fig:rare:pimumu}).
This is a rare $b \to d\ellell$ transition, which in the SM is suppressed by loop and CKM factors proportional to  $\left| V_{td}/V_{ts} \right|$.  
In the $1.0 \invfb$ data sample, \lhcb observes $25.3\,^{+6.7}_{-6.4}$ signal candidates corresponding to a branching fraction of
$\BF(\decay{\Bp}{\pip\mumu}) = (2.4\pm0.6\pm0.2) \times 10^{-8}$~\cite{LHCb-PAPER-2012-020}.
This measurement is in good agreement with the SM prediction, \ie\ consistent with no large NP contribution to $b \to d\ellell$ processes and with the MFV hypothesis.  

The $b \to d$ transitions can show potentially larger \CP- and isospin-violating effects than their $b \to s$ counterparts due to the different CKM hierarchy~\cite{Beneke:2004dp}. These studies would need the large statistics provided by the future \lhcb upgrade. 
A $50 \invfb$ data sample will also enable a precision measurement of the ratio of the branching fractions of \Bp meson decays to $\pip\mumu$ and $\Kp\mumu$. This ratio would enable a useful comparison of $\left| V_{td}/V_{ts} \right|$ to be made using penguin processes (with form factors from lattice QCD) and box processes (using $\Delta m_{s}/\Delta m_{d}$ and bag-parameters from lattice QCD) and provide a powerful test of MFV.

%

%

\subsubsection{Isospin asymmetry of \decay{B^{0(+)}}{K^{0(+)}\mumu} and \decay{B^{0(+)}}{K^{*0(+)}\mumu} decays}
\label{sec:rare:semi:isospin}

Analyses at hadron colliders (at LHCb and CDF) have mainly focused on
decay modes with charged tracks in the final state. 
$B$ meson decays involving $K^{0}$ mesons are experimentally much more challenging due to the long lifetimes of \KS and \KL mesons (the \KL is not reconstructable within \lhcb). 
Nevertheless, \lhcb has been able to select 60 \decay{\Bd}{K^{0}\mumu} decays, reconstructed as
\decay{\KS}{\pip\pim}, and 80 \decay{\Bp}{K^{*+}\mumu}, reconstructed
as \decay{K^{*+}}{\KS\pip}, which are comparable in size to 
the samples that are available for these modes in the full data
sets of the $B$ factories. 
The isolation of these rare decay modes enables a measurement of the isospin asymmetry of \decay{\B}{K^{(*)}\mumu} decays,
\begin{equation}
  A_{\rm I} = 
  \frac{\BF(\Bz \to K^{0} \mumu) - (\tau_{\Bz}/\tau_{\Bp})\BF(\Bp\to \Kp\mumu)}{\BF(\Bz \to K^{0} \mumu) + (\tau_{\Bz}/\tau_{\Bp})\BF(\Bp\to \Kp\mumu)}\,.
\end{equation}
At leading order, isospin asymmetries (which involve the spectator quark) are expected to be zero in the SM. 
Isospin-breaking effects are subleading in $\Lambda_{\rm QCD}/m_b$, and are difficult to estimate due to unknown power corrections. 
Nevertheless isospin-breaking effects are expected to be small and these observables may be useful in NP searches because they offer complementary information on specific Wilson coefficients~\cite{Feldmann:2002iw}.  
 
\begin{figure}
\centering
\subfigure[]{\includegraphics[scale=0.35]{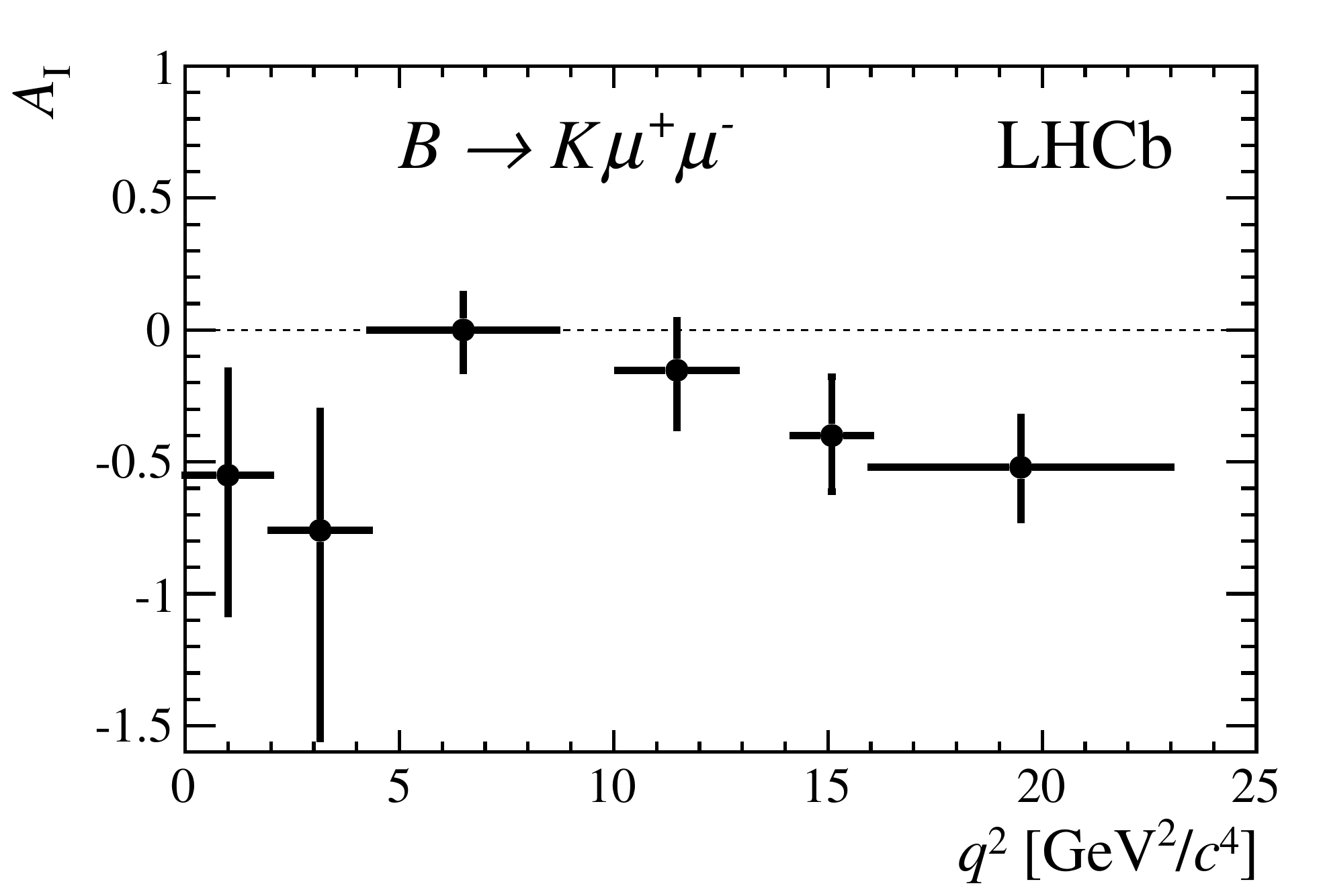}}
\subfigure[]{\includegraphics[scale=0.35]{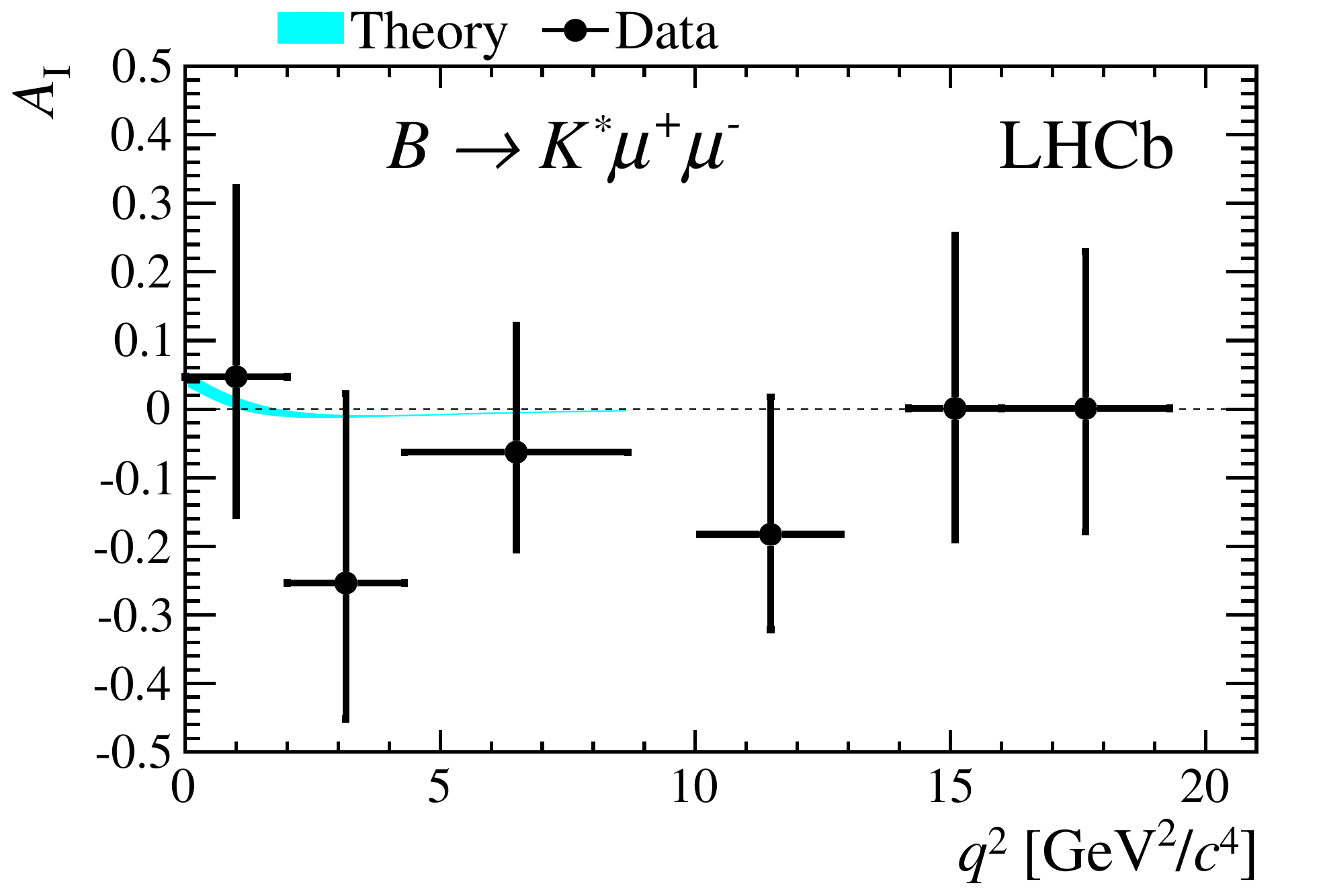}}
\caption{\small
  (a) $B \to K\mumu$ and (b) $B\to K^*\mumu$ isospin asymmetries in $1.0 \invfb$ of data collected by the LHCb collaboration in 2011~\cite{LHCb-PAPER-2012-011}.
}
\label{fig:rare:isospin}
\end{figure}

The \lhcb measurement of the $K$ and $K^{*}$ isospin asymmetries in bins of \qsq are shown in Fig.~\ref{fig:rare:isospin}.  
For the \Kstar modes $A_{\rm I}$ is compatible with the SM expectation that $A_{\rm I}^{\rm SM} \simeq 0$, but for the $\Kp/K^{0}$ modes, $A_{\rm I}$ is seen to be negative at low- and high-$q^{2}$~\cite{LHCb-PAPER-2012-011}. 
This is consistent with what has been seen at previous experiments, but is inconsistent with the na\"{i}ve expectation of $A_{\rm I}^{\rm SM} \sim 0$ at the $4\,\sigma$ level.\footnote{
  A calculation of $A_{\rm I}^{\rm SM}(B\to K \mumu)$ has recently become available~\cite{Khodjamirian:2012rm}, giving values consistent with the na\"{i}ve expectation within $1\,\%$.
}
Such a discrepancy would be hard to explain in any model that is also consistent with other experimental results.  
Improved measurements are needed to clarify the situation.

\subsection{Radiative \B decays} 
\label{sec:rare:radiative}

While the theoretical prediction of the branching ratio of the $B\to K^*\gamma$ decay is problematic due to large form factor uncertainties, the mixing-induced asymmetry\footnote{
  Note that the notation $S$ used here and in the literature for mixing-induced asymmetries is not related to the use of the notation in Sec.~\ref{sec:rare:semileptonic} for \CP-averaged properties of the angular distributions.
} 
$S_{K^*\gamma}$ provides an important constraint due to its sensitivity to the chirality-flipped magnetic Wilson coefficient $C_7^\prime$. 
At leading order it vanishes for $C_7^\prime \to 0$, so the SM prediction is tiny and experimental evidence for a large $S_{K^*\gamma}$ would be a clear indication of NP effects through right-handed currents~\cite{Atwood:1997zr,Atwood:2004jj}. 
Unfortunately it is experimentally very challenging to measure $S_{K^*\gamma}$ in a hadronic environment, requiring both flavour tagging and the ability to reconstruct the \Kstarz in the decay mode \decay{\Kstarz}{K^{0} \piz}. 
However, the channel $\Bs \to \phi\gamma$, which is much more attractive experimentally, offers the same physics opportunities, with additional sensitivity due to the non-negligible width difference in the \Bs system.
Moreover, LHCb can study several other interesting radiative $b$-hadron decays.

\subsubsection{Experimental status and outlook for rare radiative decays}
\label{sec:rare:sgamm:current}

In $1.0 \invfb$ of integrated luminosity \lhcb observes 5300 \decay{\Bd}{\Kstarz\gamma} and 690 \decay{\Bs}{\phi\gamma}~\cite{LHCb-PAPER-2012-019} candidates.
These are the largest samples of rare radiative \Bd and \Bs decays collected by a single experiment. 
The large sample of \decay{\Bd}{\Kstarz\gamma} decays has enabled \lhcb to make the world's most precise measurement of the direct \CP-asymmetry $\ACP(\Kstar\gamma) = 0.8 \pm 1.7 \pm 0.9 \,\%$, compatible with zero as expected in the SM~\cite{LHCb-PAPER-2012-019}. 

With larger data samples, it will be possible to add additional constraints on the $C_7 - C_7^\prime$ plane through measurements of $b \to s\gamma$ processes. 
These include results from time-dependent analysis of $\decay{\Bs}{\phi\gamma}$~\cite{Muheim:2008vu}, as described in detail in the \lhcb roadmap document~\cite{Adeva:2009ny}. 
Furthermore, the large $\Lb$ production cross-section will allow for measurements of the photon polarisation through the decays $\Lb \to \L^{(*)}\gamma$~\cite{Legger:2006jz,Hiller:2007ur}. 
In fact, the study of $\Lb \to \L$ transitions is quite attractive from the theoretical point of view, since the hadronic uncertainties are under good control~\cite{Wang:2008sm,Mannel:2011xg,Feldmann:2011xf}.
However, because the $\Lb$ has $J^{P} = \frac{1}{2}^{+}$ and can be polarised at production, it will be important to measure first the $\Lb$ polarisation.

\decay{B}{VP\g} decays with a photon, a vector and a pseudoscalar particle in the final state can also provide sensitivity to $C_7^\prime$~\cite{Gronau:2001ng,Gronau:2002rz,Atwood:2007qh,Kou:2010kn}. 
The decays \decay{B}{\phi K \gamma} and $B^{+} \to K_{1}(1270)^{+}\gamma$ have been previously observed at the $B$ factories~\cite{Aubert:2006he,Yang:2004as} and large samples will be available for the first time at \lhcb.  

\subsection{Leptonic \B decays}
\label{sec:rare:leptonic}

\subsubsection{\decay{\Bs}{\mumu} and \decay{\Bd}{\mumu}}

The decays \decay{B^0_{(s)}}{\mumu} are a special case amongst the electroweak penguin processes, as they are chirality-suppressed in the SM and are most sensitive to scalar and pseudoscalar operators. 
The branching fraction of \Bqmumu can be expressed as~\cite{Bobeth:2001sq,Bobeth:2001jm,Buras:2002vd,Mahmoudi:2008tp}: 

\begin{eqnarray}
\label{BRBsmumu}
{\cal B}(B^0_q \to \mu^+\mu^-) &=& \frac{G_F^2 \alpha^2}{64 \pi^3}
f_{B_q}^2 \tau_{B_q} m^3_{B_q} |V_{tb}V_{tq}^*|^2 \sqrt{1-\frac{4
    m_\mu^2}{m_{B_q}^2}} \\ 
&\times& \left\{\left(1-\frac{4 m_\mu^2}{m_{B_q}^2}\right) | C_{S}
  -C^\prime_{S} |^2 + \left | (C_{P} -C^\prime_{P}) + 2  \frac{m_\mu}{m_{B_q}}\,
    (C_{10} -C^\prime_{10}) \right |^2\right\} \;, \nonumber 
\end{eqnarray} 

\noindent where $q = s, d$.

Within the SM, $C_{S}$ and $C_{P}$ are negligibly small and the dominant contribution of $C_{10}$ is helicity suppressed. 
The coefficients $C_{i}$ are the same for \Bs and \Bd in any scenario (SM or NP) that obeys MFV. 
The large suppression of \BRof\Bdmumu with respect to \BRof\Bsmumu in MFV scenarios means that \decay{\Bs}{\mumu} is often of more interest than \Bdmumu for NP searches. 
The ratio \BRof\Bsmumu/\BRof\Bdmumu is however a very useful probe of MFV.  

The SM branching fraction depends on the exact values of the input parameters: $f_{B_{q}}$, $\tau_{B_{q}}$ and $|V_{tb}V_{tq}^*|^2$. 
The \Bs decay constant, $f_{B_s}$, constitutes the main source of uncertainty on \BRof\Bsmumu.
There has been significant progress in theoretical calculations of this quantity in recent years. 
As of the year 2009 there were two unquenched lattice QCD calculations of $f_{B_s}$, by the HPQCD~\cite{Gamiz:2009ku} and FNAL/MILC~\cite{Bernard:2009wr} collaborations, which, when averaged, gave the value $f_{B_s}=238.8\pm 9.5\mev$~\cite{Laiho:2009eu}. 
The FNAL/MILC calculation was updated in 2010~\cite{Simone:2010zz}, 
and again in 2011 to give $f_{B_s}=242\pm 9.5\mev$~\cite{Bazavov:2011aa,Neil:2011ku}. 
Also in 2011, the ETM collaboration reported a value of $f_{B_s}=232\pm 10\mev$~\cite{Dimopoulos:2011gx}.
The HPQCD collaboration presented in 2011 a result, $f_{B_s}=227\pm 10\mev$~\cite{McNeile:2011ng}, which has recently been improved upon with an independent calculation that gives $f_{B_s}=225\pm 4\mev$~\cite{Na:2012kp}. 




A weighted average of FNAL/MILC '11~\cite{Bazavov:2011aa}, HPQCD '11~\cite{McNeile:2011ng} and HPQCD '12~\cite{Na:2012kp} was presented recently~\cite{Laiho:2009eu}, giving $f_{B_s}=227.6\pm 5.0$ MeV. 
Using this value, the SM prediction for the branching ratio is~\cite{Buras:2012ts}:
\begin{equation}
\BRof\Bsmumu_{\rm SM} = (3.1 \pm 0.2) \times 10^{-9}\,.
\label{bsmumu2}
\end{equation}

\noindent
This value is taken as the nominal $\BRof\Bsmumu_{\rm SM}$.
Note that, in addition to $f_{B_s}$, other sources of uncertainty are due to the $\Bs$ lifetime, the CKM matrix element $\left| V_{ts} \right|$, the top mass $m_t$, the electroweak corrections and scale variations.
For a more detailed discussion of the SM prediction, see Ref.~\cite{Buras:2012ru}.
It is also possible to obtain predictions for $\BRof\Bsmumu_{\rm SM}$ with reduced sensitivity to the value of $f_{B_s}$ using input from either $\Delta m_s$~\cite{Buras:2010mh} or from a full CKM fit~\cite{Charles:2011va}.

Likewise for $f_{B_d}$, using the average of ETMC-11 ($f_{B_d} = 195 \pm 12$ MeV)~\cite{Dimopoulos:2011gx}, FNAL/MILC-11 ($f_{B_d}=197 \pm 9$ MeV)~\cite{Bazavov:2011aa,Neil:2011ku} and HPQCD-12 ($f_{B_d} =191 \pm 9$ MeV)~\cite{Na:2012kp} results, which gives $f_{B_d}=194 \pm 10$ MeV~\cite{Mahmoudi:2012un}, the branching ratio of $\Bdmumu$ is: 
\begin{equation}
\BRof\Bdmumu_{\rm SM} = (1.1 \pm 0.1) \times 10^{-10}\,.
\label{bdmumu}
\end{equation}

NP models, especially those with an extended Higgs sector,
can significantly enhance the \Bqmumu branching fraction even in the
presence of other existing constraints. In particular, it has been
emphasised in many works~\cite{Choudhury:1998ze,Babu:1999hn,Ellis:2005sc,Carena:2006ai,Ellis:2007ss,Mahmoudi:2007gd,Golowich:2011cx,Akeroyd:2011kd}
that the decay \Bsmumu is very sensitive to the presence of SUSY particles. 
At large $\tan\beta$ -- where $\tan\beta$ is the ratio of vacuum expectation values of the Higgs doublets\footnote{
  Note that elsewhere in this document the symbol $\beta$ is used to denote an angle of the unitarity triangle of the CKM matrix.
} -- the SUSY contribution to this process is dominated by the exchange of neutral Higgs bosons, and both $C_S$ and $C_P$ can receive large contributions from scalar exchange.  

In constrained SUSY models such as the CMSSM and NUHM1 (see Sec.~\ref{sec:rare:dependent}), predictions can be made for \BRof\Bsmumu that take into account the existing constraints from the general purpose detectors. 
These models predict~\cite{Buchmueller:2011sw}:  
\begin{equation}
  1 < \frac{\BRof\Bsmumu_{\rm CMSSM}}{\BRof\Bsmumu_{\rm SM}} < 2 \, , ~~~
  1 < \frac{\BRof\Bsmumu_{\rm NUHM1}}{\BRof\Bsmumu_{\rm SM}} < 3 \, .
\end{equation}
\noindent 
The LHCb~\cite{LHCb-PAPER-2012-007} (and CMS~\cite{Chatrchyan:2012rg}) measurements of \Bsmumu have already excluded the upper range of these predictions. 

%

Other NP models such as composite models (\eg Littlest Higgs model
with $T$-parity or Topcolour-assisted Technicolor), models with extra
dimensions (\eg Randall Sundrum models) or models with fourth
generation fermions can modify \BRof\Bsmumu~\cite{Blanke:2006eb,Blanke:2008yr,Liu:2009hv,Bauer:2009cf,Buras:2010cp,Buras:2012ts}. 
The NP contributions from these models usually arise via $(C_{10}
-C^\prime_{10})$, and they are therefore correlated with the constraints
from other $b\to s\ellell$ processes, \eg\ with $\BR(\decay{\Bp}{\Kp\mumu})$ which depends on $(C_{10}+C^\prime_{10})$. 
The term $(C_P -C^\prime_{P})$ in the branching fraction adds coherently with the SM contribution from $(C_{10} -C^\prime_{10})$, and therefore can also destructively interfere. 
In such cases, if $(C_{S} -C^\prime_{S})$ remains small, \BRof\Bsmumu could be smaller than the SM prediction. 
A measurement of \BRof\Bsmumu well below the SM prediction would be a
clear indication of NP and would be symptomatic of a model with a
large non-degeneracy in the scalar sector (where $C_P^{(\prime)}$ is enhanced but $C_{S}^{(\prime)}$ is not). 
If only $C_{10}$ is modified, these constraints currently require the 
branching ratio to be above $1.1\times10^{-10}$~\cite{Altmannshofer:2012ir}. 
In the presence of NP effects in both $C_{10}$ and $C_{10}^\prime$, even stronger suppression is possible in principle. 

At the beginning of 2012, the LHCb experiment set the world best limits on the \BRof\Bqmumu~\cite{LHCb-PAPER-2012-007}.\footnote{
  Results on \BRof\Bqmumu presented at HCP2012~\cite{LHCb-PAPER-2012-043} are not included in this discussion.
}
At 95\,\% C.L.
\begin{eqnarray}
\label{BR:bqll_results}
\BRof\Bsmumu  & < &  4.5 \times 10^{-9} \, ,\nonumber \\
\BRof\Bdmumu  & < &  1.0 \times 10^{-9} \, .\nonumber 
\end{eqnarray}
\noindent
Experimentally the measured branching fraction is the time-averaged
(TA) branching fraction, which differs from the theoretical value
because of the sizeable width difference between the heavy and light
\Bs mesons~\cite{deBruyn:2012wk,deBruyn:2012wj}.\footnote{
  This was previously observed in a different context~\cite{DescotesGenon:2011pb}.
}
In general,  
\begin{equation}
\BRof\Bsmumu_{\rm TH}
= [(1 - y_s^2)/(1 + {\cal A}_{\Delta \Gamma} y_s)] \times  \BRof\Bsmumu_{\rm TA}
\end{equation}

\noindent where ${\cal A}_{\Delta \Gamma} = +1$ in the SM and $y_s = \Delta \Gamma_s / (2 \Gamma_s) = 0.088 \pm 0.014$~\cite{LHCb-CONF-2012-002}. 
Thus the experimental measurements have to be compared to the following SM prediction for the time-averaged branching fraction:
\begin{equation}
\BRof\Bsmumu_{\rm SM, TA} =
\BRof\Bsmumu_{\rm SM, TH}/(1 - y_s) = (3.5 \pm 0.2) \times 10^{-9}\, .
\end{equation}

With $50\invfb$ of integrated luminosity, taken with an upgraded LHCb
experiment, a precision better than 10\,\% can be achieved in
\BRof\Bsmumu, and $\sim35\,\%$ on the ratio \BRof\Bsmumu/\BRof\Bdmumu. 
The dominant systematic uncertainty is likely to come from knowledge of the ratio of fragmentation fractions, $f_d/f_s$, which is currently known to a precision of 8\,\% from two independent determinations.\footnote{
  This value is valid for $B$ mesons produced from $\sqrt{s} = 7 \tev$ $pp$ collisions within the LHCb acceptance.  It will, in principle, need to be remeasured at each different LHC collision energy, and may depend on the kinematic acceptance of the detector (\ie\ on the transverse momentum and pseudorapidity of the $B$ mesons).  However, once a suitable \Bs branching fraction, such as that for $\Bs \to \Jpsi\,\phi$ or $\Bs \to \Kp\Km$, is known to good precision, normalisation can be carried out without direct need for an $f_d/f_s$ value.
}
One method~\cite{LHCb-PAPER-2011-022}\footnote{
  The results from Ref.~\cite{LHCb-PAPER-2011-022} were updated at HCP2012~\cite{LHCb-PAPER-2012-037}.
} is based on hadronic $B$ decays~\cite{Fleischer:2010ay,Fleischer:2010ca}, and relies on knowledge of the $B_{(s)} \to D_{(s)}$ form factors from lattice QCD calculations~\cite{Bailey:2012rr}.
The other~\cite{LHCb-PAPER-2011-018} uses semileptonic decays, exploiting the expected equality of the semileptonic widths~\cite{Bigi:1996si,Bigi:2011gf}.
However, the two methods have a common, and dominant, uncertainty which originates from the measurement of $\BR(\Ds\to\Kp\Km\pip)$, which in the PDG is given to $4.9\,\%$ (coming from a single measurement from CLEO~\cite{Alexander:2008aa}).
A new preliminary result from Belle has recently been presented~\cite{Wang:ICHEP} -- inclusion of this measurement in the world average will improve the uncertainty on $\BR(\Ds\to\Kp\Kp\pip)$ to $\sim 3.5\,\%$.
With the samples available with the LHCb upgrade, it will be possible to go beyond branching fraction measurements and study the effective lifetime of \Bsmumu, that provides additional sensitivity to NP~\cite{deBruyn:2012wk}.

In Sec.~\ref{sec:rare:dependent}, the NP implications of the current
measurements of \BRof\Bsmumu and the interplay with other observables,
including results from direct searches, are discussed for a selection
of specific NP models. In general, the strong experimental constraints
on \BRof\Bsmumu~\cite{LHCb-PAPER-2012-007,Chatrchyan:2012rg,Aad:2012pn,Aaltonen:2011fi} largely preclude any visible effects from scalar or pseudoscalar operators in other $b\to s\ell^+\ell^-$ decays.\footnote{
  Barring a sizeable, fortuitous cancellation among $C_{S,P}$ and $C_{S,P}^\prime$~\cite{Becirevic:2012fy}.
}  

\subsubsection{$\Bs \to \tautau$}
\label{sec:rare:tautau}

The leptonic decay $\Bs \to \tautau$ provides interesting information on the interaction of the third generation quarks and leptons. 
In many NP models, contributions to third generation quarks/leptons can be dramatically enhanced with respect to the first and second generation. 
This is true in, for example, scalar and pseudoscalar interactions in supersymmetric scenarios, for large values of $\tan\beta$.  
Interestingly, there is also an interplay between $b\to s\tautau$ processes and the lifetime difference $\Gamma_{12}^s$ in $\Bs$ mixing (see Sec.~\ref{sec:B-CPV}). 
The correlation of both processes has been discussed model-independently~\cite{Bauer:2010dga,Bobeth:2011st} and in specific scenarios, such as leptoquarks~\cite{Dighe:2007gt,Dighe:2010nj} or $Z^\prime$ models~\cite{Alok:2010ij,Kim:2010gx,Kim:2012rp}. 
There are presently no experimental limits on \decay{\Bs}{\tautau}, however the interplay with $\Gamma_{12}^s$, and the latest LHCb-measurement of $\Gamma_d/\Gamma_s$ would imply a limit of $\BR(\Bs \to \tautau) < 3\,\%$ at $90\,\%$ C.L. 
Any improvement on this limit, which might be in reach with the existing \lhcb data set, would yield strong constraints on models that couple strongly to third generation leptons.
A large enhancement in $b \to s \tautau$ could help to understand the anomaly observed by the D0 experiment in their measurement of the inclusive dimuon asymmetry~\cite{Abazov:2011yk} and could also reduce the tension that exists with other mixing observables~\cite{Bauer:2010dga,Bobeth:2011st}.  

The study of $\Bs \to \tautau$ at LHCb presents significant challenges.  
The $\tau$ leptons must be reconstructed in decays that involve at least one missing neutrino.  
Although it has been demonstrated that the decay $Z \to \tautau$ can be separated from background at LHCb, using both leptonic and hadronic decay modes~\cite{LHCb-PAPER-2012-029}, at lower energies the backgrounds from semileptonic heavy flavour decays cause the use of the leptonic decay modes to be disfavoured.
However, in the case that ``three-prong'' $\tau$ decays are used, the vertices can be reconstructed from the three hadron tracks.  
The analysis can then benefit from the excellent vertexing capability of LHCb, and, due to the finite lifetime of the $\tau$ lepton, there are in principle sufficient kinematic constraints to reconstruct the decay.  
Work is in progress to understand how effectively the different potential background sources can be suppressed, and hence how sensitive LHCb can be in this channel.

\subsection{Model-independent constraints} 
\label{sec:rare:independent}

\begin{figure}[!htb]
\includegraphics[width=\textwidth]{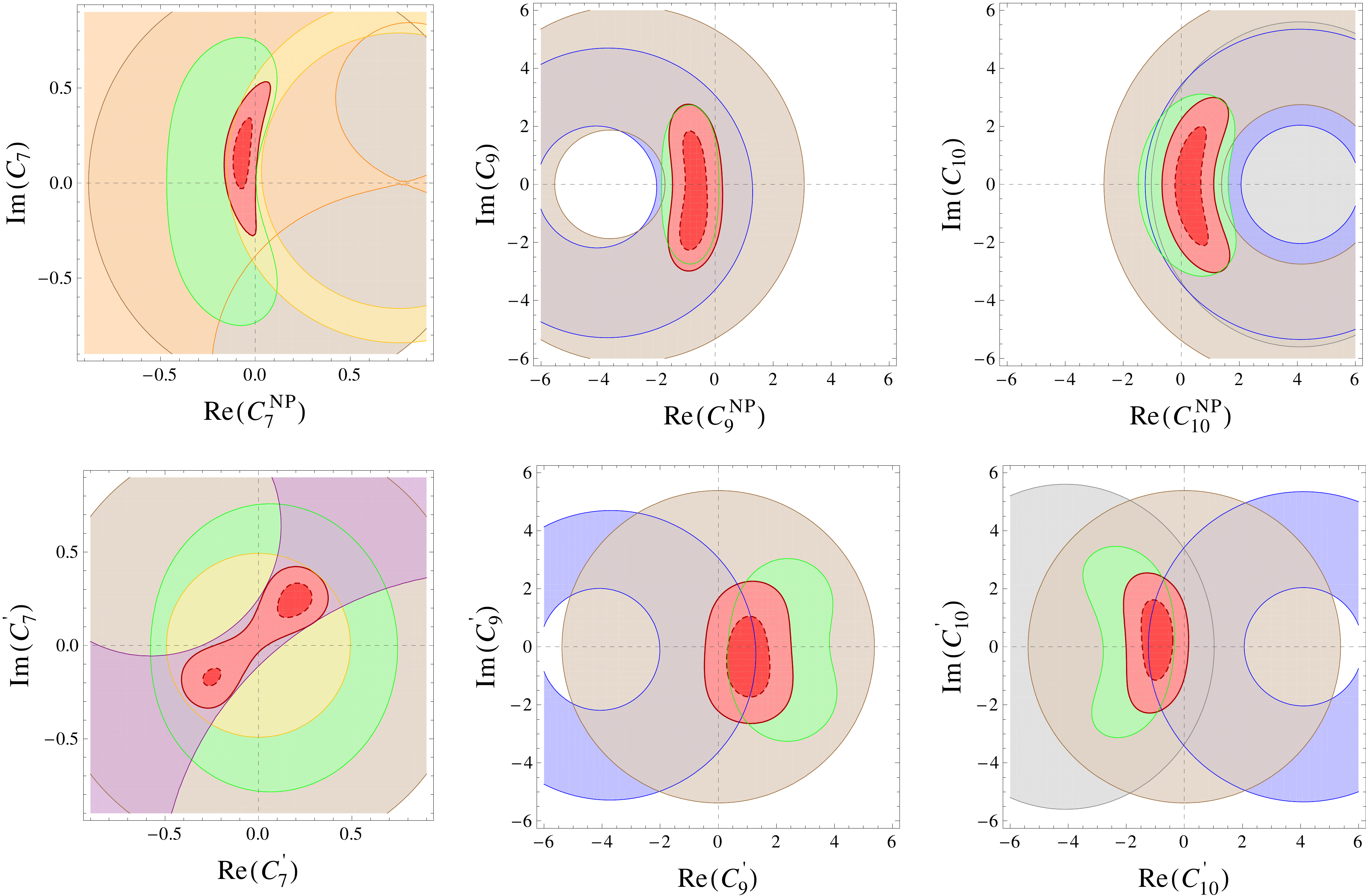}
\caption{\small
  Individual $2\,\sigma$ constraints in the complex planes of Wilson
  coefficients, coming from $B\to X_s\ell^+\ell^-$ (brown), $B\to X_s\gamma$ 
  (yellow), $A_{\CP}(b\to s\gamma)$ (orange), $B\to K^*\gamma$
  (purple), $B\to K^*\mu^+\mu^-$ (green), $B\to K\mu^+\mu^-$ (blue)
  and $\Bs\to\mu^+\mu^-$ (grey), as well as combined 1 and $2\,\sigma$
  constraints (red)~\cite{Altmannshofer:2012ir}.
} 
\label{fig:wcconstraints}
\end{figure}

Figure~\ref{fig:wcconstraints}, taken from Ref.~\cite{Altmannshofer:2012ir}, shows the current constraints on the NP contributions to the Wilson coefficients (defined in Eq.~\ref{eq:Heff}) $C_7^{(\prime)}$, $C_9^{(\prime)}$ and $C_{10}^{(\prime)}$, varying only one coefficient at a time. The experimental constraints included here are:
the branching fractions of $B \to X_s \gamma$, $B \to X_s \ell^+\ell^-$, $B \to K \mu^+\mu^-$ and $\Bs \to \mu^+\mu^-$, the mixing-induced asymmetries in $B\to K^*\gamma$ and $b\to s\gamma$ and the branching fraction and angular observables in $B \to K^* \mu^+\mu^-$. 
One can make the following observations:
\begin{itemize} 
\item At 95\,\% C.L., all Wilson coefficients are compatible with their
  SM values.  
\item For the coefficients present in the SM, \ie\ $C_7$, $C_9$ and $C_{10}$,
  the constraints on the imaginary part are looser than on the real part.  
\item For the Wilson coefficients $C_{10}^{(\prime)}$, the constraint
  on $\BR(\Bs\to\mumu)$ is starting to become competitive with the constraints 
  from the angular analysis of $B\to K^{(*)}\mu^+\mu^-$.  
\item The constraints on $C_9^\prime$ and $C_{10}^\prime$ from $B\to K\mu^+\mu^-$ and $B\to K^*\mu^+\mu^-$ are complementary and lead to a more constrained
  region, and better agreement with the SM, than with $B\to K^*\mu^+\mu^-$
  alone.  
\item A second allowed region in the $C_7$-$C_7^\prime$ plane characterised by
  large positive contributions to both coefficients, which was found previously to be allowed \eg in Refs.~\cite{Altmannshofer:2011gn,DescotesGenon:2011yn}, is
  now disfavoured at 95\,\% C.L. by the new $B\to K^*\mu^+\mu^-$ data,
  in particular the measurements of the forward-backward asymmetry from \lhcb.
\end{itemize}
The second point above can be understood from the fact that for the branching
fractions and \CP-averaged angular observables which give the strongest constraints,
only NP contributions aligned in phase with the SM can interfere with the SM
contributions. As a consequence, NP with non-standard \CP violation is in fact
constrained more weakly than NP where \CP violation stems only from the CKM phase.
This highlights the need for improved measurements of \CP asymmetries directly 
sensitive to non-standard phases.\footnote{
  LHCb has presented results on $A_{\CP}(\Bd \to \Kstarz\mumu)$ at CKM 2012~\cite{LHCb-PAPER-2012-021}.
}

Significant improvements of these constraints -- or first hints for physics beyond the SM -- can be obtained in the future by both improved measurements of the observables discussed above and by improvements on the theoretical side. 
From the theory side, there is scope for improving the estimates of the hadronic form factors from lattice calculations, which will reduce the dominant source of uncertainty on the exclusive decays. 
On the experimental side there are a large number of theoretically clean observables that can be extracted with a full angular analysis of \decay{\Bd}{\Kstarz\mumu}, as discussed in Sec.~\ref{sec:BKll:pheno}.  


\subsection{Interplay with direct searches and model-dependent constraints}
\label{sec:rare:dependent}

The search for SUSY is the main focus of NP searches in ATLAS and CMS.
Although the results so far have not revealed a positive signal, they have put strong constraints on constrained SUSY scenarios. 
The understanding of the parameters of SUSY models also depends on other measurements, such as the anomalous dipole moment of the muon, limits from direct dark matter searches, measurements of the dark matter relic density and various $B$ physics observables.
As discussed in \secref{sec:rare:leptonic}, the rare decay channels studied in LHCb, such as \Bqmumu, provide stringent tests of SUSY.  
In addition, the decays $B \to K^{(*)} \mumu$ provide many complementary observables which are sensitive to different sectors of the theory. 
In this section, the implications of the current LHCb measurements in different SUSY models are explained, both in constrained scenarios and in a more general case. 

\begin{figure}[!t]
\centering
\includegraphics[scale=0.34]{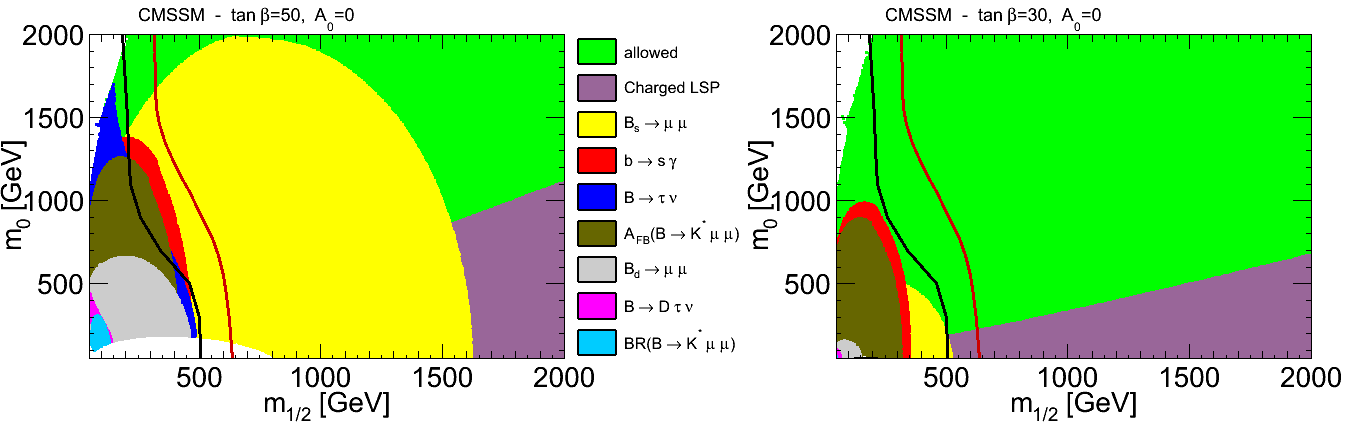}
\caption{\small
  Constraints from flavour observables in CMSSM in the plane ($m_{1/2}, m_0$) with $A_0=0$, for $\tan\beta$ = (left) 50 and (right) 30~\cite{Mahmoudi:2012uk}, using {\tt SuperIso}~\cite{Mahmoudi:2007vz,Mahmoudi:2008tp}. The black line corresponds to the CMS exclusion limit with $1.1 \invfb$ of data~\cite{Chatrchyan:2011zy} and the red line to the CMS exclusion limit with $4.4 \invfb$ of data~\cite{CMS-PAS-SUS-12-005}.
}
\label{fig:cmssm1}
\end{figure}
\begin{figure}[!t]
\centering
\includegraphics[scale=0.325]{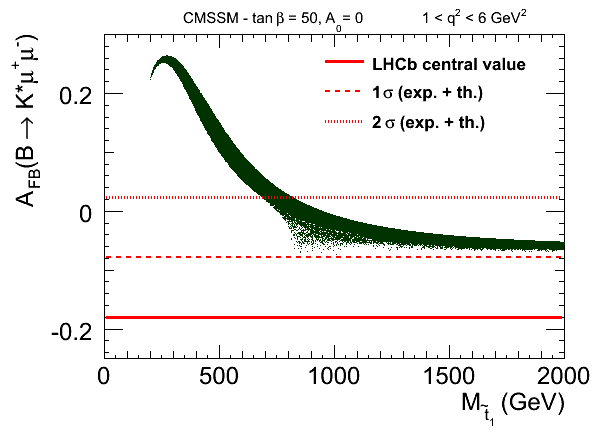}
\includegraphics[scale=0.325]{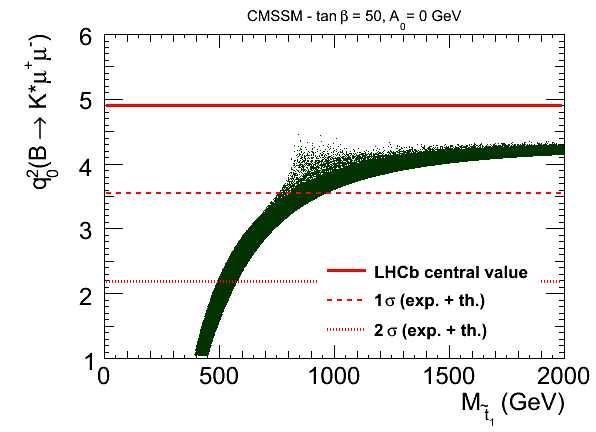}
\includegraphics[scale=0.325]{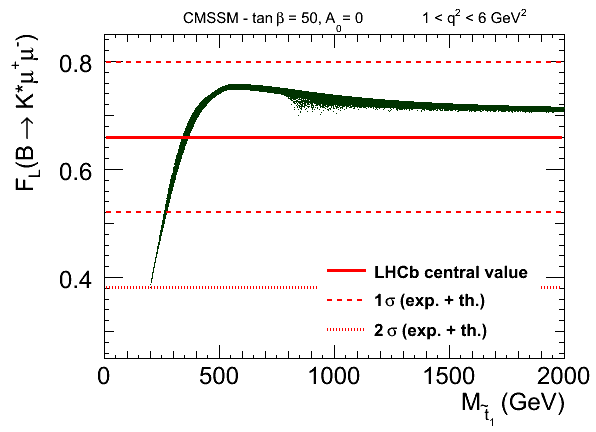}
\caption{\small
  SUSY spread of (top left) $A_{\rm FB}(B\to K^*\mu^+\mu^-)$ at low $q^2$,
  (top right) $q^2_0(B\to K^*\mu^+\mu^-)$ and 
  (bottom) $F_{\rm L}(B\to K^*\mu^+\mu^-)$ as a function of the lightest stop mass, for $A_0=0$ and $\tan\beta$ = 50~\cite{Mahmoudi:2012un}, using {\tt SuperIso}~\cite{Mahmoudi:2007vz,Mahmoudi:2008tp}.
  The solid red lines correspond to the preliminary LHCb central value with $1.0 \invfb$~\cite{LHCb-CONF-2012-008}, while the dashed and dotted lines represent the 1 and $2\,\sigma$ bounds respectively, including both theoretical and experimental errors.
} 
\label{fig:cmssm2}
\end{figure}

First consider the constrained minimal supersymmetric standard model (CMSSM) and a model with non-universal Higgs masses (NUHM1). 
The CMSSM is characterised by the set of parameters $\{m_0, m_{1/2}, A_0, \tan\beta, \mbox{sgn}(\mu)\}$ and invokes unification boundary conditions at a very high scale $m_{\rm GUT}$ where the universal mass parameters are specified. 
The NUHM1 relaxes the universality condition for the Higgs bosons which are decoupled from the other scalars, adding then one extra parameter compared to the CMSSM.  

Figure~\ref{fig:cmssm1} shows the plane ($m_{1/2}, m_0$) for large and moderate values of $\tan\beta$ in the CMSSM where, for comparison, direct search limits from CMS are superimposed. 
It can be seen that, at large $\tan\beta$, the constraints from flavour observables -- in particular \BRof\Bsmumu --  are more constraining than those from direct searches. 
As soon as one goes down to smaller values of $\tan\beta$, the flavour observables start to lose importance compared to direct searches. 
On the other hand, $B \to K^* \mu^+\mu^-$ related observables, in particular the forward-backward asymmetry, lose less sensitivity and play a complementary role. 
To see better the effect of $A_{\rm FB}(B \to K^* \mu^+\mu^-)$ at low $q^2$,\footnote{
  The effect of SUSY models on $A_{\rm FB}(B \to K^* \mu^+\mu^-)$ is discussed in Ref.~\cite{Demir:2002cj}.
} the $A_{\rm FB}$ zero-crossing point $q_0^2$ and $F_{\rm L}(B \to K^* \mu^+\mu^-)$, in \figref{fig:cmssm2} their SUSY spread is shown as a function of the lightest stop mass for $\tan\beta=50$~\cite{Mahmoudi:2012un}. 
As can be seen from the figure, small stop masses are excluded and in particular $m_{\tilde t_1} \lesssim$ 800 GeV is disfavoured by $A_{\rm FB}$ at the $2\,\sigma$ level. 

\begin{figure}[!htb]  
\begin{center}
\includegraphics[width=0.5\textwidth]{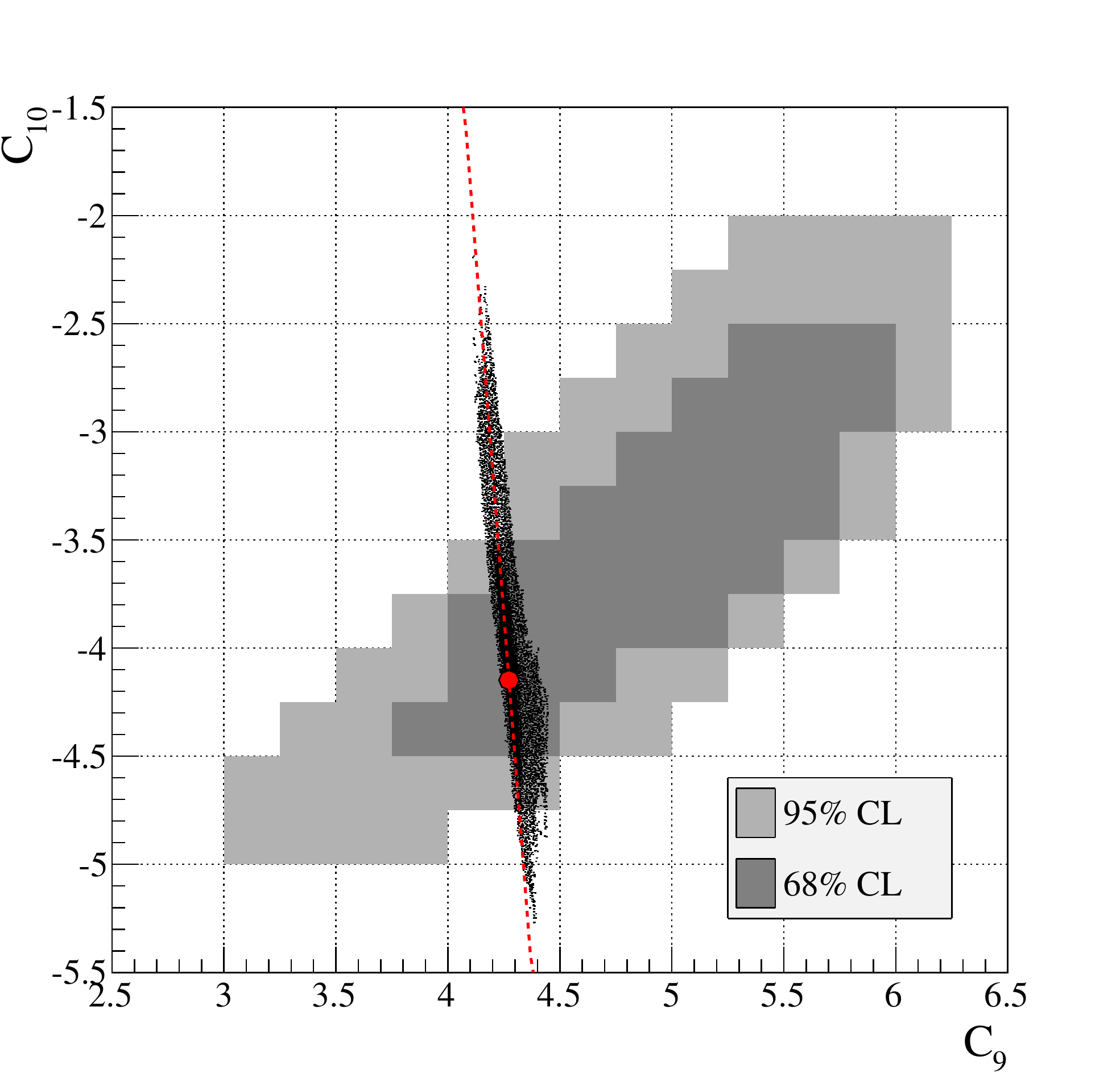} 
\caption{\small
  SUSY spread in NMFV-models~\cite{Behring:2012mv}. The light (dark) grey shaded areas are the 95\,\% (68\,\%) confidence limit (C.L.) bounds from $B \to K^{(*)} l^+ l^-$ data~\cite{Bobeth:2011nj}. The red dotted line denotes the $Z$-penguin correlation $C^{Z-\rm p}_{10}/C^{Z-\rm p}_9 =1/(4 \sin^2 \theta_W-1)$. The SM point  $(C_9^{\rm SM},C_{10}^{\rm SM})$ is marked by the red dot.
} 
\label{plot-mia-delta23uLR-c7-sm} 
\end{center}
\end{figure}

\begin{figure}[!htb]
\centering
\includegraphics[width=0.45\textwidth]{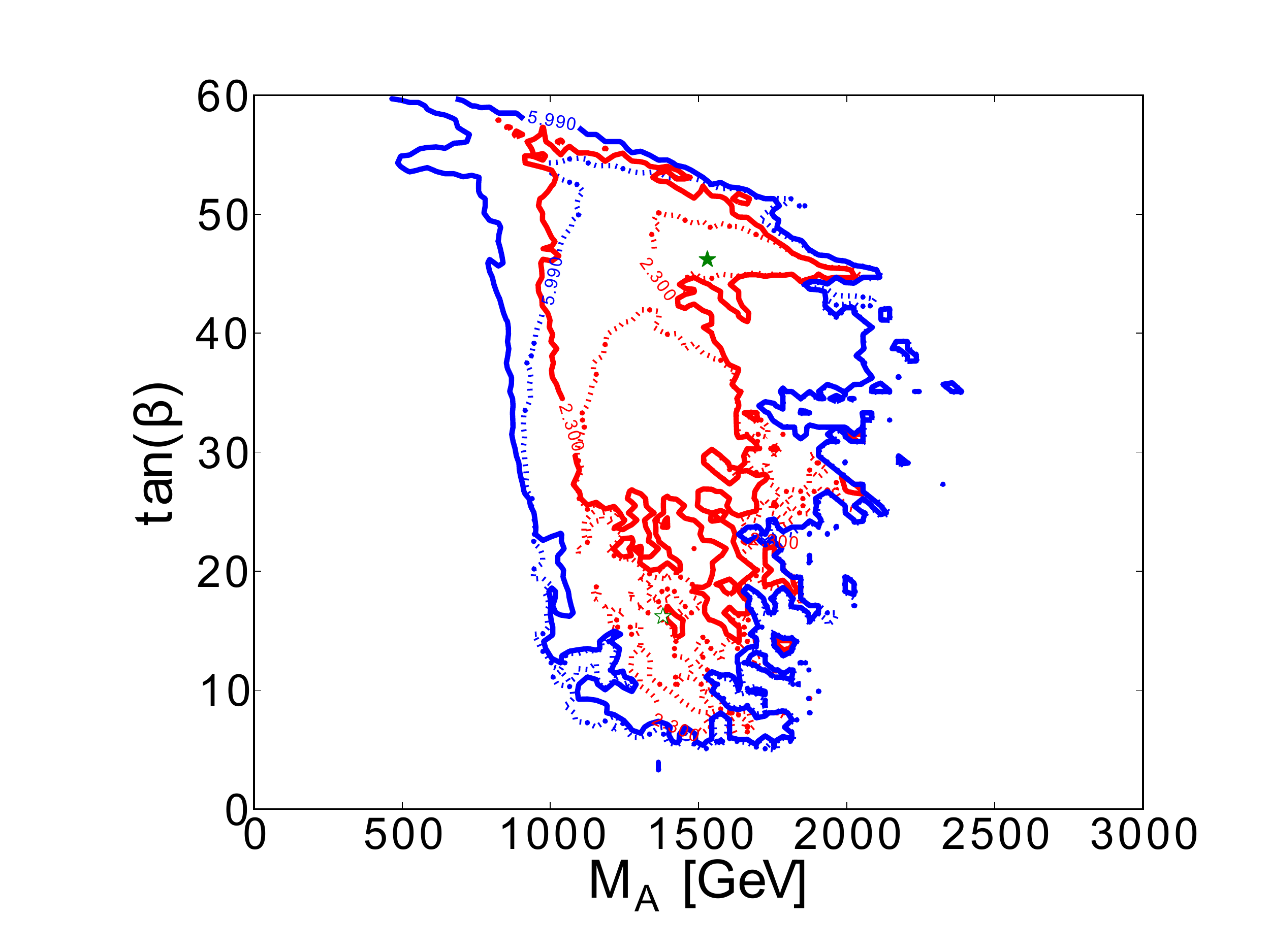} 
\includegraphics[width=0.45\textwidth]{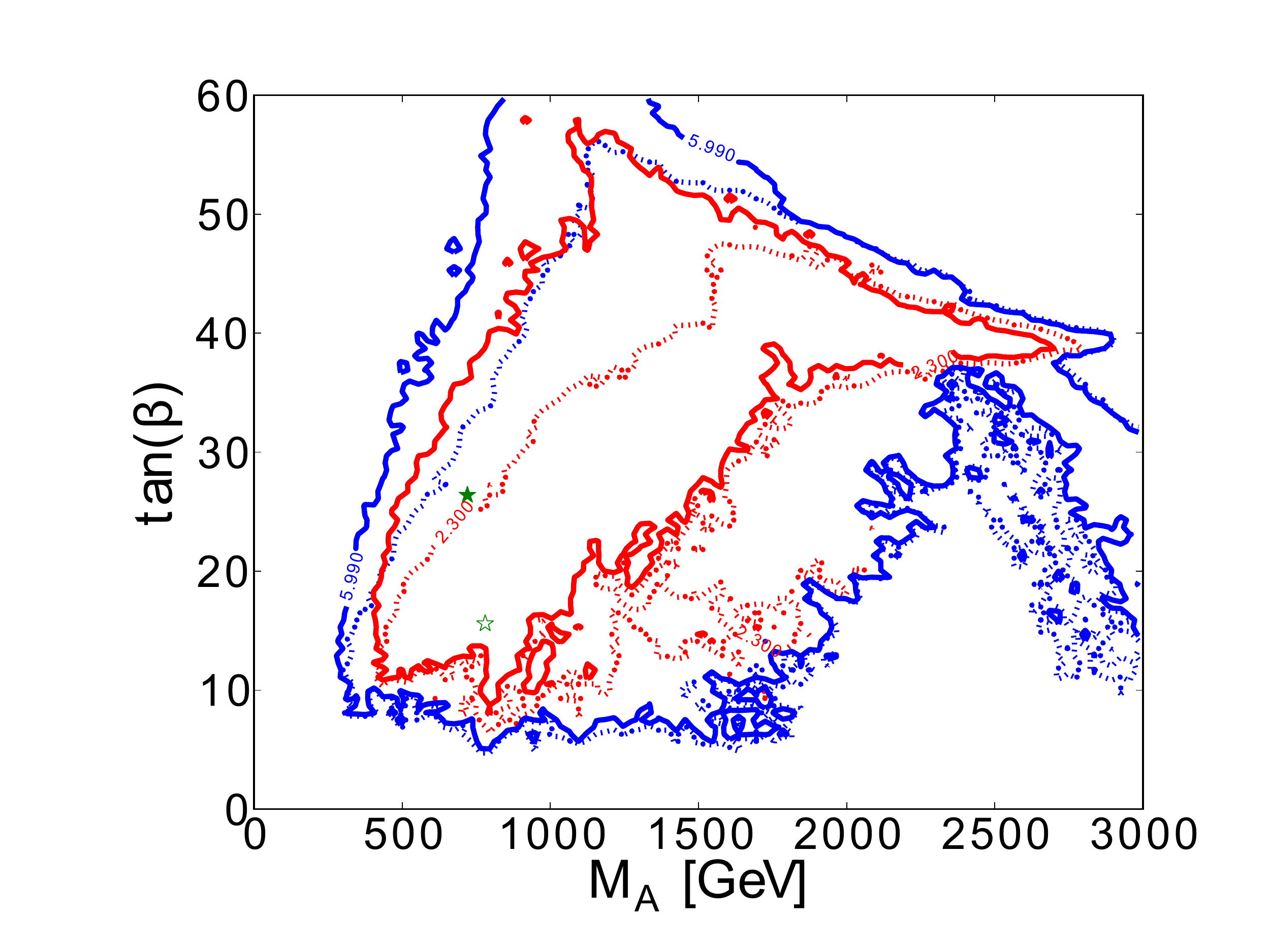} 
\caption{\small
  Impact of the latest \Bsmumu limits on the ($M_{A}$ , $\tan\beta$) plane in the (left) CMSSM and (right) NUHM1~\cite{Buchmueller:2012hv}.
  In each case, the full global fit is represented by an open green star and dashed blue and red lines for the 68 and 95\,\% C.L. contours, whilst the fits to the incomplete data sets are represented by closed stars and solid contours.
} 
\label{fig:MasterCode_1}
\end{figure}

The impact of the recent  $B \to K^{(*)} l^+ l^-$  decay data on SUSY
models beyond MFV (NMFV) with moderate $\tan \beta$ is shown in
Fig.~\ref{plot-mia-delta23uLR-c7-sm}. 
The largest effect stems from left-right mixing between top and charm super-partners. 
Due to the $Z$-penguin dominance of the SUSY-flavour contributions the constraints are most effective for the Wilson coefficient $C_{10}$ (see Sec.~\ref{sec:rare:operator}). 
SUSY effects in $C_{10}$ are reduced from about $50\,\%$ to $16\,\% \,(28\,\%)$ at $68 \,(95)\,\%$~C.L. by the recent data on the rare decay $\Bd \to \Kstarz\mumu$~\cite{Behring:2012mv}.  
The constraints are relevant to flavour models based on radiative
flavour violation (see, \eg, Ref.~\cite{Crivellin:2008mq}), and exclude
solutions to the flavour problem with flavour generation in the
up-sector and sub-TeV spectra. The flavour constraints are stronger for
lighter stops, hence there is an immediate interplay with direct
searches.  

Figure~\ref{fig:MasterCode_1} shows the ($M_{A}$, $\tan\beta$) plane from fits of the CMSSM and NUHM1 parameter space to the current data from SUSY and Higgs searches in ATLAS and CMS, as well as dark matter relic density~\cite{Buchmueller:2011sw,Buchmueller:2011ab}. 
The study in constrained MSSM scenarios is illustrative but not representative of the full MSSM. 
The strong constraints provided by the current data in the CMSSM are not necessarily reproduced in more general scenarios. 
To go beyond the constrained scenarios, consider the phenomenological MSSM (pMSSM)~\cite{Djouadi:1998di}. 
This model is the most general \CP- and R-parity-conserving MSSM, assuming MFV at the weak scale and the absence of FCNCs at tree level. 
It contains 19 free parameters: 10 sfermion masses, 3 gaugino masses, 3 trilinear couplings and 3 Higgs masses. 

\begin{figure}
\centering
\includegraphics[scale=0.32]{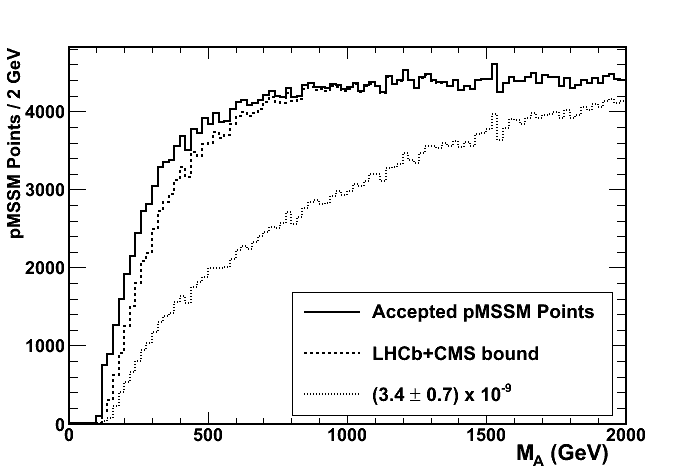}\includegraphics[scale=0.32]{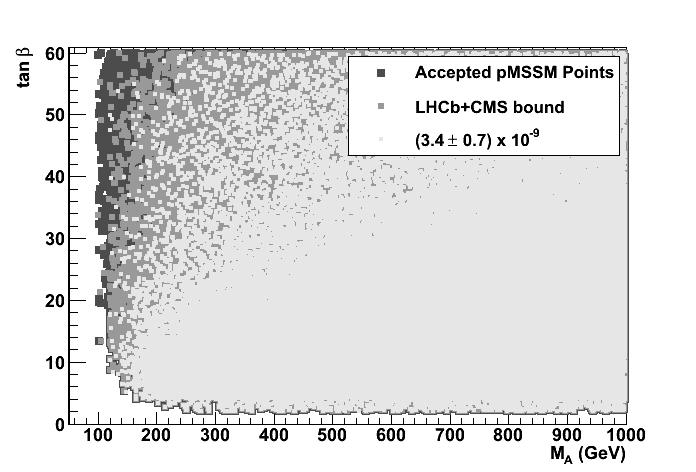}
\caption{\small
  Distribution of pMSSM points after the $\Bs \rightarrow \mu^+ \mu^-$ constraint projected on the $M_A$ (left) and ($M_A , \tan \beta$) plane (right) for all accepted pMSSM points (medium grey), points not excluded by the combination of the 2010 LHCb and CMS analyses (dark grey) and the projection for the points compatible with the measurement of the SM expected branching fractions with a 20\,\% total uncertainty (light grey)~\cite{Arbey:2011aa}.
}
\label{fig:pmssm}
\end{figure}

To study the impact of the \Bsmumu results on the pMSSM, the parameter space is scanned and for each point in the space the consistency of the model with experimental bounds is tested~\cite{Arbey:2011aa}.
The left panel of \figref{fig:pmssm} shows the density of points as a
function of $M_{A}$ before and after applying the combined 2010 LHCb
and CMS \Bsmumu limit ($1.1 \times 10^{-8}$ at 95\,\% C.L.~\cite{LHCb-CONF-2012-017}), as well as the projection for a SM--like measurement with an overall 20\,\%
theoretical and experimental uncertainty. As can be seen the density
of the allowed pMSSM points is reduced by a factor of 3, in the case
of a SM--like measurement. The right panel shows the same distribution
in the ($M_{A}$, $\tan\beta$) plane. 
Similar to the CMSSM case, the region with large $\tan\beta$ and small $M_{A}$ is most affected by the experimental constraints.

The interplay with Higgs boson searches can also be very illuminating as any viable model point has to be in agreement with all the direct and indirect limits. 
As an example, if a scalar Higgs boson is confirmed at $\sim 125 \gev$,\footnote{
  At ICHEP 2012 the observation of a new particle consistent with the SM Higgs boson was reported by ATLAS and CMS~\cite{ATLAS-Higgs,CMS-Higgs}.
} 
the MSSM scenarios in which the excess would correspond to the
heaviest \CP-even Higgs (as opposed to the lightest Higgs) are ruled
out by the \Bsmumu limit, since they would lead to a too light
pseudoscalar Higgs.  

It is clear that with more precise measurements a large part of the
supersymmetric parameter space could be disfavoured. In particular the
large $\tan \beta$ region is strongly affected by \Bsmumu as can be
seen in \figref{fig:cmssm1}. Also, a measurement of \BR(\Bsmumu) lower
than the SM prediction would rule out a large variety of
supersymmetric models. 
In addition, $B \to K^* \mu^+\mu^-$ observables play a complementary role especially for smaller $\tan \beta$ values. 
With reduced theoretical and experimental errors, the exclusion bounds in \figref{fig:cmssm2} and \figref{plot-mia-delta23uLR-c7-sm} for example would shrink leading to important consequences for SUSY parameters. 

\subsection{Rare charm decays}
\label{sec:rare:charm}

So far the focus of this chapter has been on rare \B decays, but the
charm sector also provides excellent probes for NP in the  
form of very rare decays. 
Unlike the \B decays described in the previous sections, the smallness
of the $d$, $s$ and $b$ quark masses makes the Glashow-Iliopoulos-Maiani (GIM) cancellation in loop processes very effective. 
Branching ratios governed by FCNC are hence not expected to exceed ${\cal O}(10^{-10})$ in the SM.  
These processes can then receive contributions from NP scenarios which can be several orders of magnitude larger than the SM expectation.  

\subsubsection{Search for $D^0\rightarrow \mu^+\mu^-$}

The branching fraction of the $D^0\rightarrow \mu^+\mu^-$ decay is
dominated in the SM by the long distance contributions due to the
two photon intermediate state, $D^0\rightarrow \gamma\gamma$. 
The experimental upper limit on the two photon mode can be combined with
theoretical predictions to constrain ${\cal B}(D^0\rightarrow
\mu^+\mu^-)$  in the framework of the SM: ${\cal B}(D^0\rightarrow
\mu^+\mu^-)<6\times 10^{-11} $ at 90$\,\%$ C.L.~\cite{Burdman:2001tf}.
Particular NP models where this decay is enhanced include
supersymmetric models with R-parity violation (RPV), which provides
tree-level contributions that would enhance the branching fraction. 
In such models, the branching fraction would be related to the $\Dz$--$\Dzb$ mixing parameters. 
Once the experimental constraints on the mixing parameters are taken into account, the corresponding tree-level couplings can still give rise to ${\cal B}(D^0\rightarrow \mu^+\mu^-)$ of up to ${\cal O}(10^{-9})$~\cite{Golowich:2009ii}. 

Preliminary results from a search for these rare decays have been performed by the LHCb collaboration~\cite{LHCb-CONF-2012-005}.  
The upper limit obtained with $0.9 \invfb$ of data taken in 2011 is:  
\begin{equation}
  {\cal B}(D^0\rightarrow \mu^+\mu^-) \le 1.3\, (1.1)\times 10^{-8}
  \textrm{ at 95 (90)\,\%  C.L.}\,
\end{equation}
This upper limit on the branching fraction, already an improvement of an order of magnitude on previous results, is expected to improve down to $5\times 10^{-9}$ by the end of the first data-taking phase of the LHCb experiment.

\subsubsection{Search for $D^{+}_{(s)}\rightarrow h^{+} \mu^{+}\mu^{-}$
  and $D^0\rightarrow h h^{^\prime} \mu^+\mu^-$}

The $D^{+}_{(s)}\rightarrow h^{+} \mu^{+}\mu^{-}$  decay rate is dominated by long distance contributions from tree-level $D^{+}_{(s)}\rightarrow h^{+} V$ decays, where $V$ is a light resonance ($V = \phi, \rho, \omega$). 
The long-distance contributions have an effective branching fraction (with 
\decay{V}{\mumu}) above $10^{-6}$ in the SM.  Large deviations in the
total decay rate due to NP are therefore unlikely. 
However, the regions of the dimuon mass spectrum far from these
resonances are interesting probes. Here, the SM contribution stems
only from FCNC processes, that should yield no partial branching ratio
above $10^{-11}$~\cite{Buchalla:2008jp}.  
NP contributions could enhance the branching fraction away
from the resonances by several orders of magnitude: \eg in the RPV
model mentioned above, or in models involving a fourth quark
generation~\cite{Buchalla:2008jp,Fajfer:2007dy}.  

The LHCb experiment is well-suited to search for $D^{+}_{(s)}\rightarrow h^{\mp} \mu^{+}\mu^{\pm}$ decays.  
The long distance contributions can be used to normalise the decays searched for at high and low dimuon mass: their decay rate will be measured relative to that of $D^{+}_{(s)}\rightarrow \pi^+ \phi(\mu^{+}\mu^{-})$. 
These resonant decays have a clean experimental signature and their final state only differs from the signal in the kinematic distributions, which helps to reduce the systematic uncertainties. 
The sensitivity of the LHCb experiment can be estimated by comparing
the yields of $D^{+}_{(s)}\rightarrow \pi^+ \phi(\mu^{+}\mu^{-})$
decays observed in LHCb with those obtained by the \dzero experiment, which
established the best limit on these modes so far~\cite{Abazov:2007aj}.
With an integrated luminosity corresponding to $1.0\invfb$, 
upper limits on the $\Dp$ ($\Ds$) modes are expected close to
$10^{-8}$ ($10^{-7}$) at 90\,\% C.L.

In analogy to the \B sector, there is a wealth of observables
potentially available in four-body rare decays of $D$ mesons. 
In the decays $D^0\rightarrow h h^{^\prime} \mu^+\mu^-$ (with $h^{(^\prime)}=K$ or $\pi$),   forward-backward asymmetries or asymmetries based on $T$-odd quantities could reveal NP effects~\cite{Buchalla:2008jp,Bigi:2011em,Cappiello:2012vg}.
Clearly the first challenge is to observe the decays which, depending on their branching fractions, may be possible with the 2011 data set.
However, the $50 \invfb$ collected by the upgraded LHCb detector will be necessary to exploit the full set of observables in these modes.

\subsection{Rare kaon decays}
\label{sec:rare:kaon}

The cross-section for $\KS$ production at the LHC is such that $\sim10^{12}$
$\KS\to\pi^{+}\pi^{-}$ would be reconstructed and selected in LHCb with a fully efficient trigger. 
This provides a good opportunity to search for rare $\KS$ decays in channels with high trigger efficiency, in particular \Ksmumu. 

The decay \Ksmumu is a flavour-changing neutral current that has not yet
been observed. This decay is strongly suppressed in the SM, with an expected 
branching fraction of~\cite{Ecker:1991ru,Isidori:2003ts}
\begin{equation}
\BRof\Ksmumu = (5.0 \pm 1.5) \times 10^{-12}\, ,
\end{equation}
while the current experimental upper limit is $3.2\times10^{-7}$ at 90\,\% C.L.~\cite{Jack}. 
The study of \Ksmumu has been suggested as a possible way to look for new light scalars~\cite{Ecker:1991ru}, and indeed NP contributions up to one order of magnitude above the SM expectation are allowed~\cite{Isidori:2003ts}. 
Enhancements above $10^{-10}$ are less likely. 
Bounds on \BRof\Ksmumu close to $10^{-11}$ could be useful to discriminate among NP scenarios if other modes, such as $K^{+}\to\pi^{+}\nu\bar{\nu}$, indicated a non-standard enhancement of the $s\to d ll$ transition. 
First results from LHCb, $\BRof\Ksmumu < 9 \times 10^{-9}$ at 90\,\% C.L.~\cite{LHCb-PAPER-2012-023}, have significantly better sensitivity than the existing results.
With improved triggers on low mass dimuons, LHCb could reach branching fractions of ${\cal O}(10^{-11})$ or below with the luminosity of the upgrade. 
Decays of $\KL$ mesons into charged tracks can also be reconstructed, but with much less ($\sim1\,\%$) efficiency compared to a similar decay coming from a $\KS$ meson. 
This is due to the long distance of flight of the $\KL$ state, which tends to decay outside the tracking system.

\subsection{Lepton flavour and lepton number violation}
\label{sec:rare:lepton}

The experimental observation of neutrino oscillations provided the first signature of lepton flavour violation (LFV). 
The consequent addition of mass terms for the neutrinos in the SM implies LFV also in the charged sector, but with branching fractions smaller than $10^{-40}$. 
NP could significantly enhance the rates but, despite steadily improving experimental sensitivity, charged lepton flavour violating (cLFV) processes like $\mu^-\to e^-\gamma$, $\mu$--$N \to e$--$N$, $\mu^-\to e^+e^-e^-$, $\tau^-\to\ell^-\gamma$ and $\tau^-\to\ell^+\ell^-\ell^-$ (with $\ell^-=e^-,\mu^-$) have not been observed. 
Numerous theories beyond the SM predict larger LFV effects in $\tau^-$ decays than $\mu^-$ decays, with branching fractions within experimental reach~\cite{lfvreview}. 
An observation of cLFV would thus be a clear sign for NP, while lowering the experimental upper limit will help to further constrain theories~\cite{Raidal:2008jk}.

Another approach to search for NP is via lepton number violation (LNV).  
Decays with LNV are sensitive to Majorana neutrino masses --- their discovery would answer the long-standing question of whether neutrinos are Dirac or Majorana particles.
The strongest constraints on minimal models that introduce neutrino masses come from neutrinoless double beta decay processes, but searches in heavy flavour decays provide competitive and complementary limits in models with extended neutrino sectors.

In this section, LFV and LNV decays of $\tau$ leptons and $B$ mesons with only charged tracks in the final state are discussed.

\subsubsection{Lepton flavour violation}

The neutrinoless decay $\tau^-\to\mu^+\mu^-\mu^-$ is a particularly 
sensitive mode in which to search for LFV at LHCb as the inclusive
$\tau^-$ production cross-section at the LHC is large
($\sim80\mub$, coming mainly from $\Ds$ decays\footnote{
  Calculated from the $b\bar{b}$ and $c\bar{c}$ cross-sections measured at the LHCb experiment and the inclusive branching ratios $b\ra \tau$ and $c\ra \tau$~\cite{Nakamura:2010zzi}.
}) 
and muon final states provide clean signatures in the detector. 
This decay is experimentally favoured with respect to the decays
$\tau^-\to\mu^-\gamma$ and $\tau^-\to e^+e^-e^-$ due to the considerably
better particle identification of the muons and better possibilities
for background discrimination. 
LHCb has reported preliminary results from a search for the decay $\tau^-\to\mu^+\mu^-\mu^-$ using $1.0\invfb$ of data~\cite{LHCb-CONF-2012-015}. 
The upper limit on the branching fraction was found to be
$\BF(\tau^-\to\mu^+\mu^-\mu^-)<7.8~(6.3) \times 10^{-8}$ at 95\,\%
(90\,\%) C.L, to be compared with the current best
experimental upper limit from Belle:
$\BF(\tau^-\to\mu^+\mu^-\mu^-)<2.1 \times 10^{-8}$ at 90\,\% C.L.
As the data sample increases this limit is expected to scale as the square root of the available statistics, with possible further reduction depending on improvements in the analysis.
The large integrated luminosity that will be collected by the upgraded experiment will provide sensitivity corresponding to an upper limit of a few times $10^{-9}$. 
Searches will also be conducted in modes such as $\tau^- \to \bar{p} \mu^+ \mu^-$ or $\tau^- \to  \phi \mu^-$, where the existing limits are much weaker, and low background contamination is expected in the data sample.\footnote{
  Preliminary results on $\tau^- \to \bar{p} \mu^+ \mu^-$ and $\tau^- \to p \mu^- \mu^-$ were presented at TAU 2012~\cite{LHCb-CONF-2012-027}.
}

The pseudoscalar meson decays probe transitions of the type $q\ra q^\prime
\ell \ell^\prime$ and hence are particularly sensitive to
leptoquark-models and thus provide complementarity to leptonic decay
LFV processes~\cite{Pati:1974yy,Landsberg:2004sq}. 
For the LHCb experiment, both decays from $D$ and $B$ mesons are
accessible. 
Sensitivity studies for the decays $B^0_{(s)}\ra e^- \mu^+$ and $D^0 \ra e^- \mu^+$ are ongoing. 
Present estimates indicate that LHCb will be able to match the sensitivity of the existing limits from the $B$ factories and CDF in the near future. 

\subsubsection{Lepton number violation}

In lepton number violating $B$ and $D$ meson decays a search can be made for
Majorana neutrinos with a mass of $\mathcal{O}(1\gev)$. These indirect searches
are performed by analysing the production of same sign charged leptons
in $D$ or $B$ decays such as $\Ds \ra \pim \mup\mup$ or $\Bp \ra \pim \mup\mup$~\cite{Atre:2009rg,Gorbunov:2007ak}. 
These same sign dileptonic decays can only occur via exchange of heavy Majorana neutrinos.  Resonant production may be possible if the heavy neutrino is kinematically accessible, which could put the rates of these decays within reach of the future LHCb luminosity.
Non-observation of these LNV processes, together with low energy neutrino data, would lead to better constraints for neutrino masses and mixing parameters in models with extended neutrino sectors.

Using $0.4\invfb$ of integrated luminosity from LHCb, limits have been
set on the branching fraction of $B^{+} \to
D_{(s)}^{-}\mup\mup$ decays at the level of a few times $10^{-7}$ and
on \decay{\Bp}{\pim\mup\mup} at the level of $1\times 10^{-8}$~\cite{LHCb-PAPER-2011-009, LHCb-PAPER-2011-038}. 
These branching fraction limits imply a limit on, for example, the coupling
$|V_{\mu 4}|$ between $\nu_\mu$ and a Majorana neutrino with a mass in
the range $1 < m_{N} < 4 \gevcc$ of $|V_{\mu 4}|^{2} < 5\times 10^{-5}$.  

\subsection{Search for NP in other rare decays}
\label{sec:rare:other}

Many extensions of the SM predict weakly interacting particles with
masses from a few \mev to a few \gev~\cite{Kahn:2007ru,*Dermisek:2005gg,*Bouchiat:2004sp,*Boehm:2003bt,*Gorbunov:2000cz} and there are some experimental hints for these particles from astrophysical and collider experiments~\cite{Adriani:2008zr,Chang:2008aa}. 
For example, the HyperCP collaboration has reported an excess of $\Sigma^+ \ra p \mumu$ events with dimuon invariant masses around 214\mevcc~\cite{Park:2005eka}. 
These decays are consistent with the decay $\Sigma^+ \ra p X$ with the subsequent decay $X \ra \mumu$. 
Phenomenologically, $X$ can be interpreted as a pseudoscalar or axial-vector particle with lifetimes for the pseudoscalar case estimated to be about $10^{-14} \sec$~\cite{Deshpande:2005mb,Gorbunov:2005nu,Geng:2005ra}. 
Such a particle could, for example, be interpreted as a 
pseudoscalar sgoldstino~\cite{Gorbunov:2005nu} or a light pseudoscalar Higgs boson~\cite{He:2006fr}. 

The LHCb experiment has recorded the world's largest data sample of $B$ 
and $D$ mesons which provides a unique opportunity to search for these
light particles. 
Preliminary results from a search for decays of $B^0_{(s)} \ra  \mu^+ \mu^-  \mu^+ \mu^- $ have been reported~\cite{LHCb-CONF-2012-010}. 
Such decays could be mediated by sgoldstino pair production~\cite{Demidov:2011rd}.
No excess has been found and limits of $1.3$ and $0.5 \times 10^{-8}$ at 95\,\% C.L. have been set for the \Bs and \Bd modes respectively.
The analysis can naturally be extended to $D^0 \ra \mu^+ \mu^- \mu^+ \mu^- $
decays, as well as $B^0_{(s)} \ra  V^0 \mu^+ \mu^-$ ($V^0 = K^{(*)0}, \rho^0, \phi$), where the dimuon mass spectrum can be searched for any resonant structure. 
Such an analysis has been performed by the Belle collaboration~\cite{Hyun:2010an}.    
With the larger data sample and flexible trigger of the LHCb upgrade, it will be possible to exploit several new approaches to search for exotic particles produced in decays of heavy flavoured hadrons (see, \eg Ref.~\cite{Freytsis:2009ct}).

\clearpage

\section{\boldmath \CP violation in the \B system}
\label{sec:B-CPV}

\subsection{Introduction}
 
\CP violation, \ie\ violation of the combined symmetry of charge conjugation and parity, is one of three necessary conditions to generate a baryon asymmetry in the Universe~\cite{Sakharov:1967dj}. 
Understanding the origin and mechanism of \CP violation is a key question in physics. 
In the SM, \CP violation is fully described by the CKM mechanism~\cite{Cabibbo:1963yz,*Kobayashi:1973fv}. 
While this paradigm has been successful in explaining the current experimental data, it is known to generate insufficient \CP violation to explain the observed baryon asymmetry of the Universe. Therefore, additional  sources of \CP violation are required. Many extensions of the SM naturally contain new sources of \CP violation. 
  
The \bquark hadron systems provide excellent laboratories to search for new sources of \CP violation, since new particles beyond the SM may enter loop-mediated processes such as $\bquark \to q$  FCNC  transitions with $q=s$ or $d$, leading to discrepancies between measurements of \CP asymmetries and their SM expectations. 
Two types of $\bquark \to q$ FCNC transitions are of special interest: neutral \B meson mixing ($\Delta B=2$) processes, and loop-mediated \B decay ($\Delta B=1$) processes.

The LHCb experiment exploits the large number of \bquark hadrons, including the particularly interesting \Bs mesons, produced in proton-proton collisions at the LHC to search for \CP-violating NP effects. 
Section~\ref{sec:Bmixing}  provides a review of the status and prospects in the area of searches for NP in $B^0_{(s)}$ mixing, in particular through measurements of the mixing phases $\phi_{d(s)}$ and the semileptonic asymmetries $a_{\rm sl}^{d(s)}$.  
The LHCb efforts to search for NP in hadronic $\bquark \to \squark$ penguin decays, such as $\Bs \to \phi \phi$,  are discussed in Sec.~\ref{sec:b2spenguin}. 
Section~\ref{sec:gamma} describes the LHCb programme to measure the angle $\gamma$ of the CKM unitarity triangle (UT) in decay processes described only by tree amplitudes, such as $\Bpm \to D\Kpm$, $\Bz \to D\Kstarz$ and $\Bs \to \Dsmp \Kpm$.
These measurements allow precise tests of the SM description of quark-mixing via global fits to the parameters of the CKM matrix, as well as direct comparisons with alternative determinations of $\gamma$ in decay processes involving loop diagrams, such as \BsToKK. 
At the end of  each section, a brief summary of the most promising measurements with the upgraded LHCb detector and their expected/projected sensitivities is provided. 

\subsection{$B^0_{(s)}$ mixing measurements}
\label{sec:Bmixing}

\subsubsection{$B^0_{(s)}$--$\bar{B}^0_{(s)}$ mixing observables}
\label{sec:bmixing_observables}

The effective Hamiltonian of  the $B^0_q$--$\bar{B}^0_q$ ($q=d,s$) system can be written as
\begin{equation}
\mathbf{H_q} = \left( \begin{array}{cc}
M^q_{11} & M^q_{12}  \\
M^{q*}_{12} & M^q_{22}  
\end{array} \right)
-
\frac{i}{2}\left( \begin{array}{cc}
\Gamma^q_{11} & \Gamma^q_{12}  \\
\Gamma^{q*}_{12} & \Gamma^q_{22}
\end{array}  \right) \,.
\end{equation}
where $M^q_{11}=M^q_{22}$ and $\Gamma^q_{11}=\Gamma^q_{22}$ hold under the assumption of \CPT invariance.
The off-diagonal  elements $M^q_{12}$ and $\Gamma^q_{12}$ are responsible for $B^0_q$--$\bar{B}^0_q$ mixing phenomena.
The ``dispersive'' part $M^q_{12}$ corresponds to virtual $\Delta B=2$  transitions dominated by heavy internal particles (top quarks in the SM)  while the ``absorptive'' part $\Gamma^q_{12}$ arises from on-shell transitions due to decay modes common to $B^0_q$ and $\bar{B}^0_q$ mesons. 
Diagonalising the Hamiltonian matrix leads to the two mass eigenstates $B^q_{\rm H,L}$ (${\rm H}$ and ${\rm L}$ denote  heavy and light, respectively),
with mass $M^q_{\rm H,L}$  and decay width $\Gamma^q_{\rm H,L}$, being linear combinations of flavour eigenstates with complex coefficients\footnote{
  Strictly, the coefficients $p$ and $q$ should also have subscripts $q$ to indicate that they can be different for \Bd and \Bs, but these are omitted to simplify the notation.
} 
$p$ and $q$ that satisfy $\left| p \right|^2 + \left| q \right|^2 = 1$,
\begin{equation}
|B^q_{\rm L,H}\rangle = p |B^0_q\rangle \pm q |\bar{B}^0_q \rangle\,. 
\end{equation}

The magnitudes of $M^q_{12}$ and $\Gamma^q_{12}$ and their phase difference are physical observables and can be determined from measurements of the following quantities (for more details see, \eg, Refs.~\cite{Nierste:2009wg,Lenz:2012mb}):
\begin{itemize}
\item the mass difference between the heavy and light mass eigenstates 
\begin{equation}
\Delta m_q \equiv M^q_{\rm H}- M^q_{\rm L} \approx 2|M^q_{12}|\left( 1-\frac{|\Gamma^q_{12}|^2}{8|M^q_{12}|^2}\sin^2\phi_{12}^q \right)\,,
\label{eq:dms}
\end{equation}
where $\phi_{12}^q = \arg(-M^q_{12}/\Gamma^q_{12})$ is convention-independent;

\item the decay width difference between the light and heavy mass eigenstates
\begin{equation}
\Delta \Gamma_q \equiv \Gamma^q_{\rm L}- \Gamma^q_{\rm H} \approx 2|\Gamma^q_{12}|\cos\phi_{12}^q \left(1+\frac{|\Gamma^q_{12}|^2}{8|M^q_{12}|^2}\sin^2\phi_{12}^q\right)\,;
\label{eq:dgs}
\end{equation}

\item the flavour-specific asymmetry\footnote{
  The notation $a^q_{\rm sl}$ is used to denote flavour-specific asymmetries, reflecting the fact that the measurements of these quantities use semileptonic decays.
}
\begin{equation}
  a^q_{\rm sl} \equiv \frac{\left|p/q\right|^2 - \left|q/p\right|^2}{\left|p/q\right|^2 + \left|q/p\right|^2}
  \approx \frac{|\Gamma^q_{12}|}{|M^q_{12}|}\sin\phi_{12}^q
  \approx \frac{\Delta\Gamma_q}{\Delta m_q} \tan \phi_{12}^q \,.
\end{equation}
\end{itemize}
The correction terms in Eqs.~(\ref{eq:dms}) and~(\ref{eq:dgs}) proportional to $\sin^2\phi_{12}^q$  are tiny.
In addition, the ratio of $q$ and $p$ can be written
\begin{equation}
  \left( \frac{q}{p} \right) =
  - \frac{
    \Delta m_q + \frac{i}{2}\Delta \Gamma_q
  }{
    2 (M_{12}^q - \frac{i}{2} \Gamma_{12}^q )
  } \, ,
\end{equation}
and hence in both \Bd and \Bs systems one obtains, to a good approximation, a convention-dependent expression (for an unobservable quantity) $\arg(-q/p) \approx -\arg(M_{12}^q)$.
Since $B$--$\bar{B}$ mixing is dominated by the box diagram with internal top quarks, this leads to an expression in terms of CKM matrix elements $\arg(-q/p) = 2\arg(V^*_{tb}V_{tq})$.

Further information can be obtained by measuring the phase difference between the amplitude for a direct decay to a final state $f$ and the amplitude for decay after oscillation.
In the case that the decay is dominated by $b \to c \bar{c}s$ tree amplitudes,
and where $f$ is a \CP eigenstate $f$ with eigenvalue $\eta_f$,\footnote{
  The cases for more generic final-states can be found in the literature, \eg\ Ref.~\cite{HFAG}.
}
this phase difference is denoted as
\begin{equation}
  \phi_q \equiv - \arg \left( \eta_f \frac{q}{p} \frac{\bar A_f}{A_f} \right) \, ,
\end{equation}
where $A_f$ and ${\bar A}_f$ are the decay amplitudes of $B \to f$  and ${\bar B} \to f$, respectively.
In the absence of direct \CP violation
$\bar A_f / A_f = \eta_f$.
With these approximations, the \CP-violating phases in \B mixing give the unitarity triangle angles, $\phi_d \approx 2\beta$ and $\phi_s \approx -2\beta_s$,\footnote{
  Note the conventional sign-flip between $\beta$ and $\beta_s$ ensures that both are positive in the SM.
}
where the angles are defined as~\cite{HFAG}
\begin{equation}
  \beta   \equiv \arg \left( - \frac{V_{cd}V_{cb}^*}{V_{td}V_{tb}^*} \right) \,, 
  \quad
  \beta_s \equiv \arg \left( - \frac{V_{ts}V_{tb}^*}{V_{cs}V_{cb}^*} \right) \,.
\end{equation}
Clearly, if there is NP in $M_{12}^q$ or in the decay amplitudes, the measured value of $\phi_q$ can differ from the true value of $(-)2\beta_{(s)}$.
Similarly, NP in either $M_{12}^q$ or $\Gamma_{12}^q$ can make the observed value of $a_{\rm sl}^q$ differ from its SM prediction.
Note, however, that even within the SM, there is a difference between $\phi_q$ and $\phi_{12}^q$~\cite{Lenz:2007nk}.
Nonetheless, the notations $\phi_{d(s)}$ and $\beta_{(s)}$ are usually used interchangeably.

The $\phi_s$ notation has been used in the LHCb measurements of the \CP-violating phase in \Bs mixing, using $\jpsi\,\phi$~\cite{LHCb-PAPER-2011-021,LHCb-CONF-2012-002} and $\jpsi\,f_0(980)$~\cite{LHCb-PAPER-2011-031,LHCb-PAPER-2012-006} final states.
By using the same notation for different decays, an assumption that $\arg(\bar{A}_f/A_f)$ is common for different final states is being made.  
This corresponds to an assumption that the penguin contributions to these decays are negligible.
Although this is reasonable with the current precision, as the measurements improve it will be necessary to remove such assumptions -- several methods to test the contributions of penguin amplitudes are discussed below.
These include measuring $\phi_q$ with different decay processes governed by different quark-level transitions.
Previous experiments have used the notation $2\beta^{\rm eff}$ in particular for measurements based on $b \to q\bar{q}s \ (q = u,d,s)$ transitions; for symmetry the notation $2\beta^{\rm eff}_s$ is used in corresponding cases in the $\Bs$ system, although the cancellation of the mixing and decay phases in $\Bs$ decays governed by $b \to q\bar{q}s$ amplitudes is expected to lead to a vanishing \CP violation effect (within small theoretical uncertainties).

In the SM, the mixing observables can be predicted using CKM parameters from a global fit to other observables and hadronic parameters (decay constants and bag parameters) from lattice QCD calculation.
These predictions can be compared to their direct measurements to test the SM and search for NP in neutral \B mixing.

\subsubsection{Current experimental status and outlook}
\label{sec:status_Bmixing}

The current measurements and SM predictions for the mixing observables are summarised in Table~\ref{tab:mixing_status}.

\begin{table}[!htb]
  \begin{center}
    \caption{\small
      Status of \B mixing measurements and corresponding SM predictions. 
      New results presented at ICHEP 2012 and later are not included.
      The inclusive same-sign dimuon asymmetry $A^b_{\rm SL}$ is defined below and in Ref.~\cite{Abazov:2011yk}.
    }
    \label{tab:bmixing_status}
    \resizebox{0.99\textwidth}{!}{
    \begin{tabular}{|c|c|c|c|c|}
\hline
 Observable & Measurement  & Source  & SM prediction &  References  \\
\hline
\multicolumn{5}{|c|}{\Bs system} \\
\hline
$\Delta m_s$ ($\invps$)  & $ 17.719 \pm 0.043 $  & HFAG 2012~\cite{HFAG} & $17.3 \pm 2.6 $ & \cite{Lenz:2011ti,Lenz:2006hd,Beneke:1998sy,*Ciuchini:2003ww,*Beneke:1996gn,*Beneke:2003az}\\
                         & $ 17.725 \pm 0.041 \pm 0.026  $  & LHCb ($0.34 \invfb$)~\cite{LHCb-CONF-2011-050} &                 &    \\
\hline
$\Delta \Gamma_s$ ($\invps$)  & $ 0.105 \pm 0.015 $  & HFAG 2012~\cite{HFAG}& $0.087 \pm 0.021 $ & \cite{Lenz:2011ti,Lenz:2006hd,Beneke:1998sy,*Ciuchini:2003ww,*Beneke:1996gn,*Beneke:2003az}\\
                         & $ 0.116 \pm 0.018 \pm 0.006  $  & LHCb ($1.0 \invfb$)~\cite{LHCb-CONF-2012-002} &                 &    \\
\hline
$\phi_s$  (rad) &  $-0.044 ^{+0.090}_{-0.085}$ &  HFAG 2012~\cite{HFAG}  &  $-0.036 \pm 0.002$ & \cite{Charles:2011va,Lenz:2006hd,Beneke:1998sy,*Ciuchini:2003ww,*Beneke:1996gn,*Beneke:2003az} \\
               &  $-0.002 \pm 0.083 \pm 0.027$ &  LHCb ($1.0 \invfb$)~\cite{LHCb-CONF-2012-002}  &  &  \\
\hline
$a^s_{\rm sl}$ ($ 10 ^{-4}$)   &  $-17\pm 91\,^{+14}_{-15}$ &  \dzero (no $A^b_{\rm SL}$)~\cite{Abazov:2009wg}  & $0.29^{+0.09}_{-0.08}$ & \cite{Charles:2011va,Lenz:2006hd,Beneke:1998sy,*Ciuchini:2003ww,*Beneke:1996gn,*Beneke:2003az}\\
                           &  $-105\pm 64$ &  HFAG 2012 (including $A^b_{\rm SL}$ )~\cite{HFAG}  &   & \\
\hline
\multicolumn{5}{|c|}{Admixture of \Bd and \Bs systems} \\
\hline
$A^b_{\rm SL}$ ($ 10 ^{-4}$)   &  $-78.7\pm 17.1 \pm 9.3$ &  \dzero~\cite{Abazov:2011yk}  & $-2.0 \pm 0.3 $ & \cite{Lenz:2011ti,Lenz:2006hd,Beneke:1998sy,*Ciuchini:2003ww,*Beneke:1996gn,*Beneke:2003az}\\
\hline
\multicolumn{5}{|c|}{\Bd system} \\
\hline
$\Delta m_d$ ($\invps$)  & $ 0.507 \pm 0.004 $  & HFAG 2012~\cite{HFAG}& $0.543\pm 0.091 $ & \cite{Lenz:2012mb,Lenz:2006hd,Beneke:1998sy,*Ciuchini:2003ww,*Beneke:1996gn,*Beneke:2003az}\\
\hline
$\Delta \Gamma_d / \Gamma_d$   & $ 0.015 \pm 0.018 $  & HFAG 2012~\cite{HFAG}& $0.0042 \pm 0.0008 $ & \cite{Lenz:2011ti,Lenz:2006hd,Beneke:1998sy,*Ciuchini:2003ww,*Beneke:1996gn,*Beneke:2003az}\\
\hline
$\sin2\beta$ &  $0.679 \pm 0.020$ &  HFAG 2012~\cite{HFAG}  & $0.832 \,^{+0.013}_{-0.033} $ & \cite{Charles:2011va,Lenz:2006hd,Beneke:1998sy,*Ciuchini:2003ww,*Beneke:1996gn,*Beneke:2003az}\\
\hline
$a^d_{\rm sl}$ ($ 10 ^{-4}$)   &  $-5\pm 56$ &  HFAG 2012~\cite{HFAG}  & $-6.5\,^{+1.9}_{-1.7}$ & \cite{Charles:2011va,Lenz:2006hd,Beneke:1998sy,*Ciuchini:2003ww,*Beneke:1996gn,*Beneke:2003az}\\
\hline
    \end{tabular}
  }
\label{tab:mixing_status}
  \end{center}
\end{table}

The HFAG average of the \Bs mass difference $\Delta m_s$ in Table~\ref{tab:mixing_status} is based on measurements performed at CDF~\cite{Abulencia:2006ze} and LHCb~\cite{LHCb-PAPER-2011-010, LHCb-CONF-2011-050}. 
It is dominated by the preliminary LHCb result obtained using $0.34 \invfb$ of data~\cite{LHCb-CONF-2011-050}, which is also given in Table~\ref{tab:mixing_status}. 
These are all consistent with the SM prediction. 
Improving the precision of the SM prediction is desirable to further constrain NP in $M^s_{12}$, and requires improving the accuracy of lattice QCD evaluations of the decay constant and bag parameter (see Ref.~\cite{Lenz:2012mb} and references therein).

\begin{figure}
\centering
\includegraphics[width=0.50\textwidth]{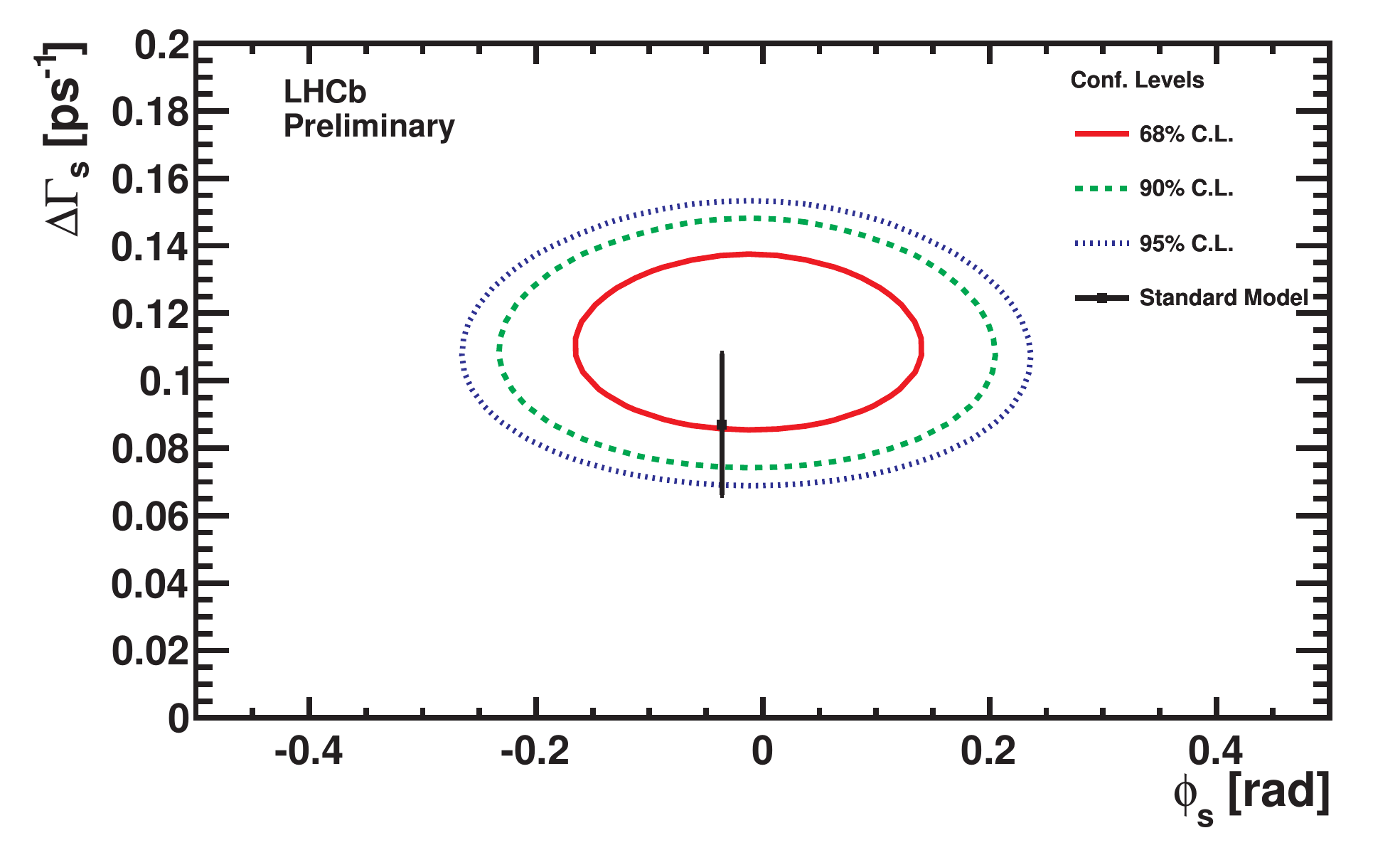}
\includegraphics[width=0.46\textwidth,bb=40 200 600 580]{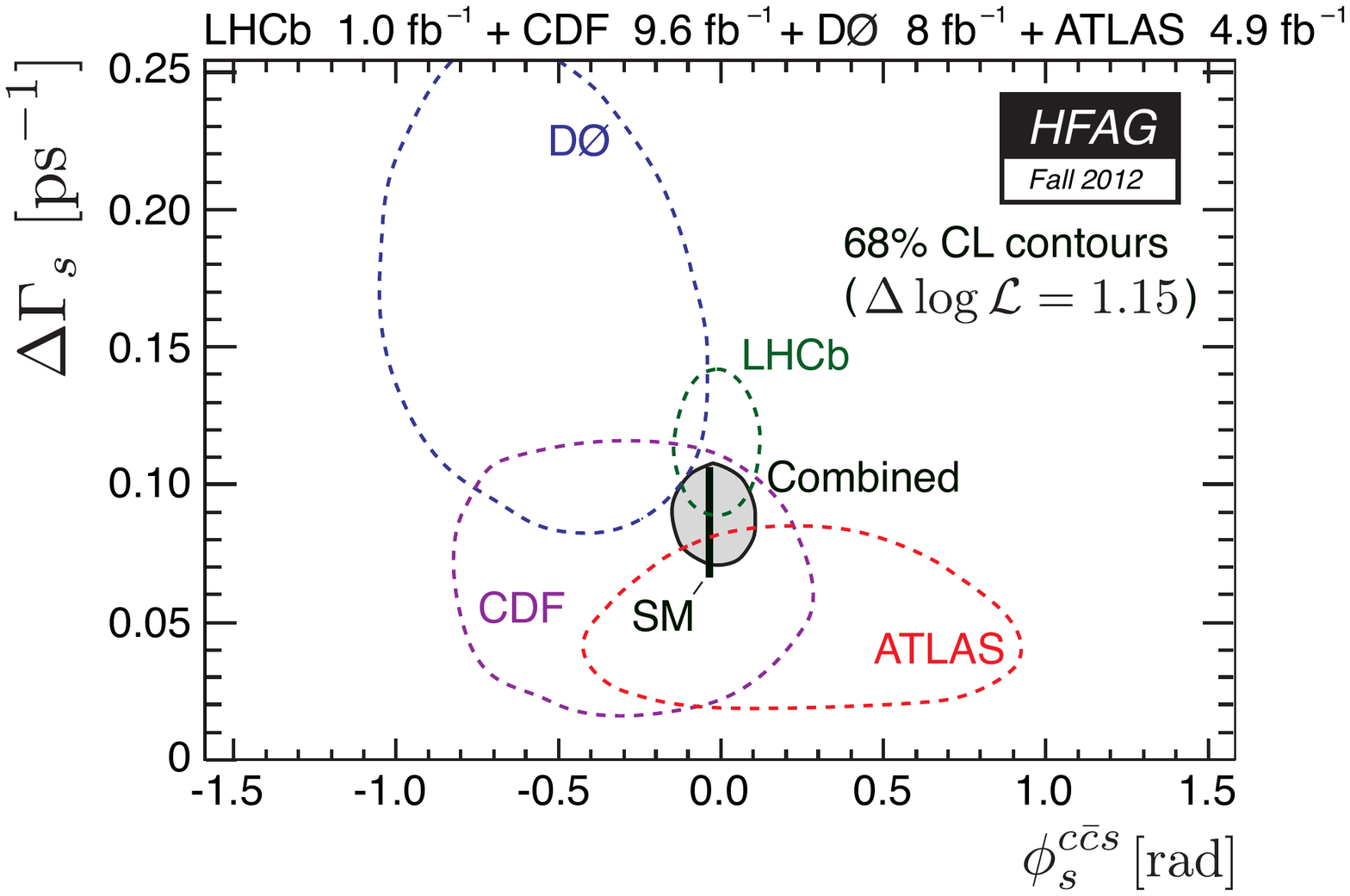}
\caption{\small
  (Left) Preliminary LHCb measurement of $\phi_s$ and $\Delta\Gamma_s$ from $\Bs\to \jpsi\,\phi$ decays using $1.0 \invfb$~\cite{LHCb-CONF-2012-002}.
  (Right) HFAG 2012 combination of $\phi_s$ and $\Delta\Gamma_s$ results, where the $1\,\sigma$ confidence region is shown for each experiment and the combined result~\cite{HFAG}.
  Note the different scales.
  \label{fig:bcpv:phis}
}
\end{figure}

The observables $\phi_s$ and $\Delta\Gamma_s$ have been determined simultaneously from $\Bs\to \jpsi\,\phi$ decays using time-dependent flavour tagged angular analyses~\cite{Dighe:1998vk,Dunietz:2000cr}. 
The first LHCb tagged analysis using $0.34 \invfb$ of data~\cite{LHCb-PAPER-2011-021} already provided a significant constraint on $\phi_s$ 
and led to the  first direct evidence for a non-zero value of $\DGs$.
LHCb  has also determined the sign of $\Delta\Gamma_s$ to be positive at $4.7\,\sigma$ confidence level~\cite{LHCb-PAPER-2011-028}
by exploiting the interference between the \KpKm S-wave and P-wave amplitudes in the $\phi (1020)$ mass region~\cite{Xie:2009fs}.
This resolved the two-fold ambiguity  in the value of $\phi_s$ for the first time. 
LHCb has made a preliminary update of the $\Bs\to \jpsi\,\phi$ analysis using the full data sample of $1.0 \invfb$ collected in 2011~\cite{LHCb-CONF-2012-002}.
The results from this analysis,
\begin{align}
 \phi_s & = -0.001 \pm 0.101 \pm 0.027 \, {\rm rad} \, , &
 \Delta \Gamma_s  & = 0.116 \pm 0.018 \pm 0.006 \invps \, ,&
\end{align}
are shown in Fig.~\ref{fig:bcpv:phis} (left), and are in good agreement with the SM expectations.
 
\begin{figure}
\centering
\includegraphics[width=0.51\textwidth]{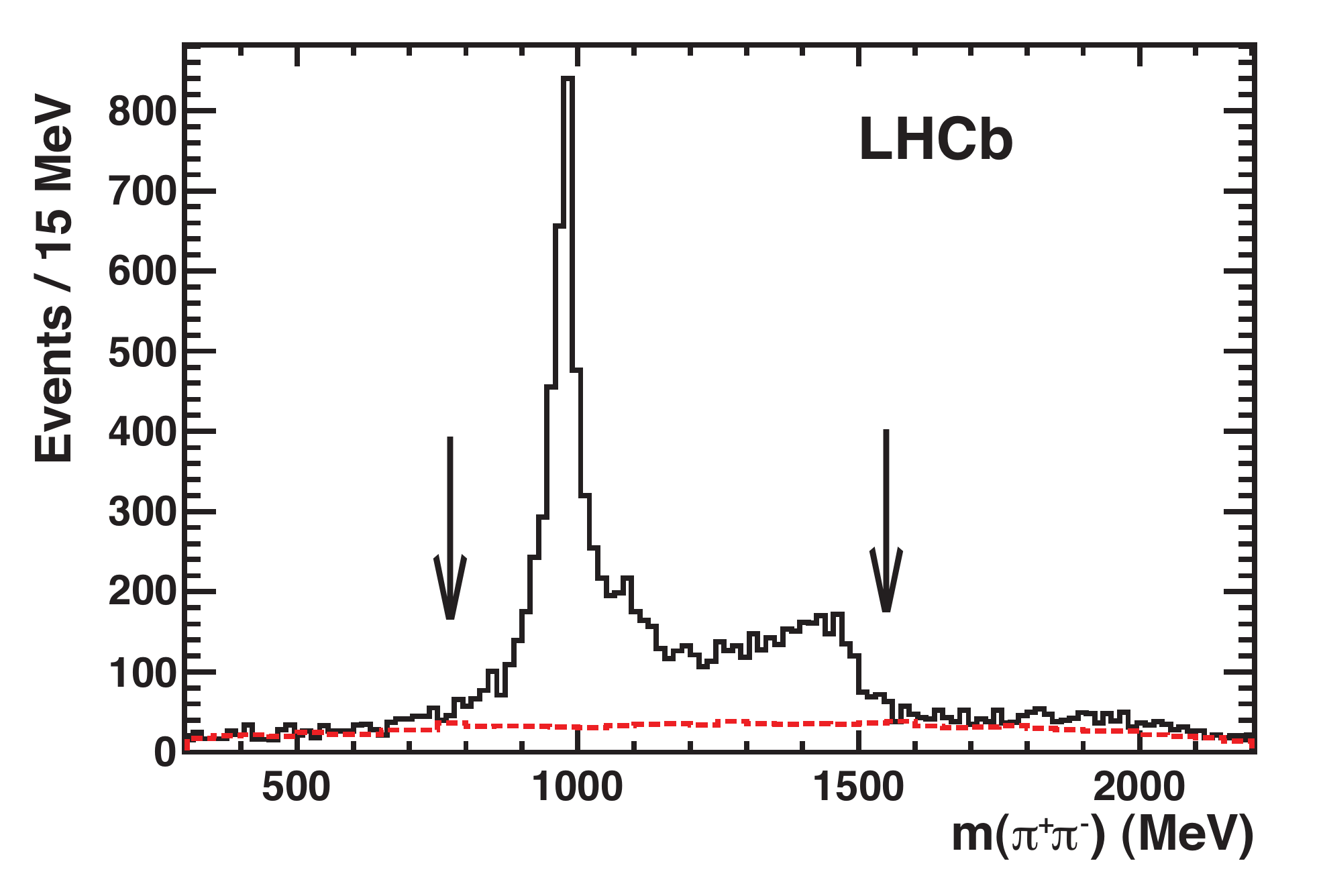}
\includegraphics[width=0.46\textwidth]{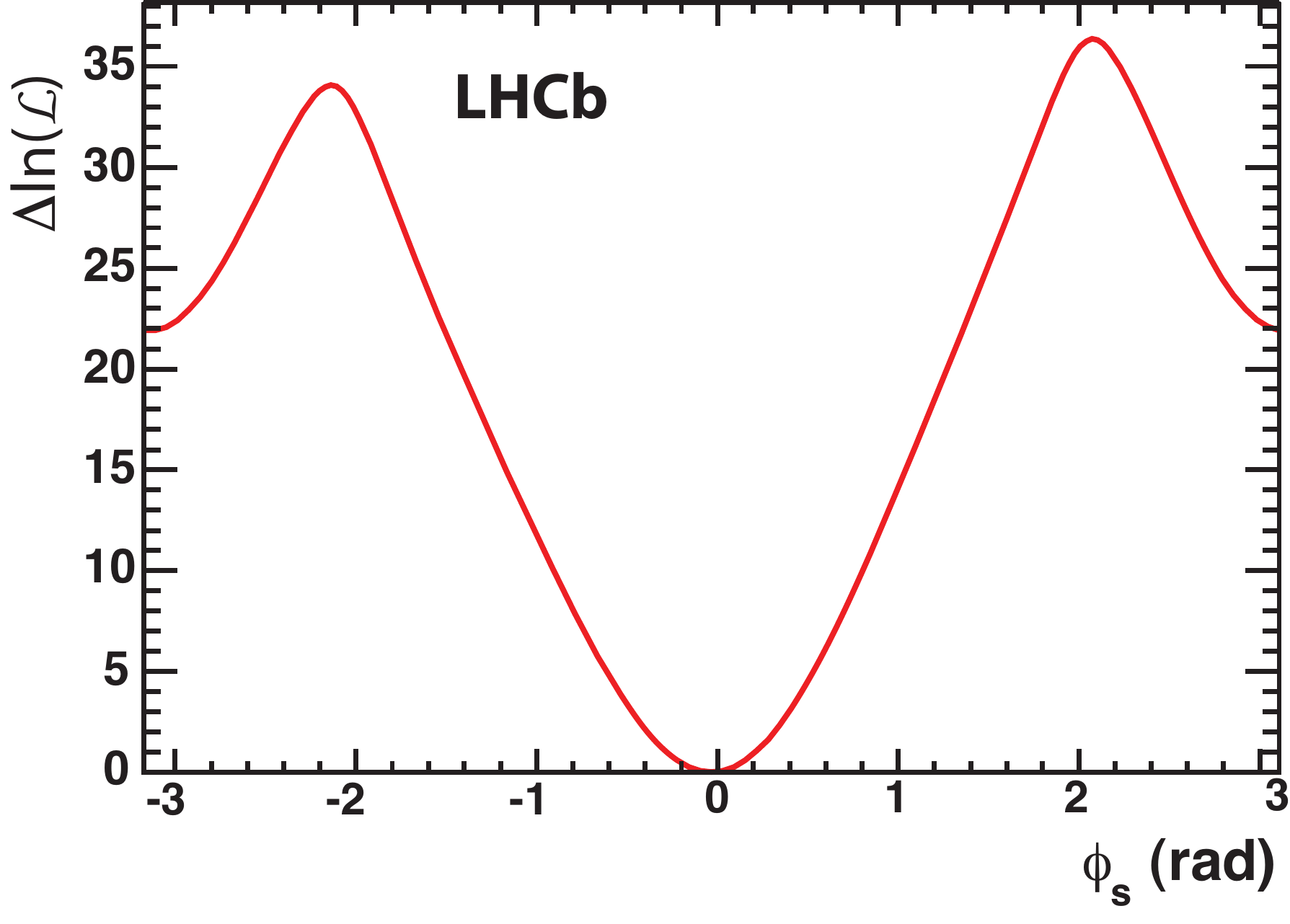}
\caption{\small
  (Left) $\pi^+\pi^-$ mass distribution of selected $\Bs\to \jpsi\,\pi^+ \pi^-$ candidates and range used for the $\phi_s$  measurement.
  (Right) log-likelihood difference as a function $\phi_s$~\cite{LHCb-PAPER-2012-006}.
  \label{fig:bcpv:jpsipipi}
}
\end{figure}

LHCb has also studied the decay $\Bs\to \jpsi\,\pi^+ \pi^-$. 
This decay process is expected to proceed dominantly via $b\to \ccbar s$ (the $s\bar{s}$ produced in the decay rescatters to $\pip\pim$ through either a resonance such as $f_0(980)$ or a nonresonant process).
Therefore, these events can be used to measure $\phi_s$. 
The $\pi^+\pi^-$ mass range 775--1550 \mev shown in Fig.~\ref{fig:bcpv:jpsipipi} (left) is used for the measurement. 
In contrast to $\Bs\to \jpsi\,\phi$, no angular analysis is needed to disentangle the \CP eigenstates, since the final state is determined to be dominantly \CP-odd in this mass range~\cite{LHCb-PAPER-2012-005}. 
On the other hand, \DGs cannot be determined in this decay channel alone.\footnote{
  The effective lifetime of $\Bs\to \jpsi\,f_0(980)$ is sensitive to \DGs and \CP violation parameters~\cite{Fleischer:2011au} and has been measured by LHCb~\cite{LHCb-PAPER-2012-017}.
}
Using as input the value of $\DGs$ obtained from $\Bs\to \jpsi\,\phi$, the measurement from the analysis of $\Bs\to \jpsi\,\pi^+ \pi^-$  with $1.0 \invfb$ is~\cite{LHCb-PAPER-2012-006} 
\begin{equation}
\phi_s = -0.019 \,^{+0.173}_{-0.174}\,^{+0.004}_{-0.003}\, {\rm rad}\,.
\end{equation}
Figure~\ref{fig:bcpv:jpsipipi} (right) shows the log-likelihood scan for the $\phi_s$ parameter for the $\Bs\to \jpsi\,\pi^+ \pi^-$ analysis.
The latest HFAG average in Table~\ref{tab:mixing_status} combines the LHCb results with the $\Bs\to \jpsi\,\phi$ analysis results from CDF using $9.6 \invfb$~\cite{CDF:2011af} and \dzero using $8.0 \invfb$~\cite{Abazov:2011ry}.
The LHCb result dominates the combination, which is in good agreement with the SM predictions, as seen in Fig.~\ref{fig:bcpv:phis} (right).\footnote{
  Results from ATLAS and CMS, presented at ICHEP2012 or later, are not included in this compilation.
}

The LHCb $\Bs\to \jpsi\,\phi$ and $\Bs\to \jpsi\,\pi^+ \pi^-$  analyses discussed above only used opposite side flavour tagging~\cite{LHCb-PAPER-2011-027,LHCb-CONF-2012-026}. 
Future updates of these analyses will gain in sensitivity by also using the same side kaon tagging information, which so far has been used in a preliminary determination of $\Delta m_s$~\cite{LHCb-CONF-2011-050,LHCb-CONF-2012-033}.
Currently, the systematic uncertainty on $\phi_s$ is dominated by imperfect knowledge of the background, angular acceptance effects and by neglecting potential contributions of direct \CP violation.
All of these uncertainties are expected to be reduced with more detailed understanding and some improvements in the analysis.
Therefore it is expected that the determination of $\phi_s$ will remain limited by statistical uncertainties, even with the data samples available after the upgrade of the LHCb detector.
In addition to $\Bs\to \jpsi\,\phi$ and $\Bs\to \jpsi\,\pi^+ \pi^-$, other $b\to \ccbar s$  decay modes of \Bs mesons, such as $\jpsi \eta$, $\jpsi \eta^{\prime}$~\cite{Fleischer:2011ib} and $\D_s^+\D_s^-$~\cite{Fleischer:2007zn} will be investigated. 
These decays have been measured at LHCb~\cite{LHCb-CONF-2012-009,LHCb-PAPER-2012-022}.

The SM prediction $\phi_s = -0.036 \pm 0.002$ rad could receive a small correction from doubly CKM-suppressed penguin contributions in the decay.
The value of this correction is not precisely known, and may depend on the decay mode.  
Moreover, NP in the $b\to \ccbar s$ decay may also affect the results. 
Although such effects are already constrained by results from $\Bu$ and $\Bd$ decays, NP in the decay amplitudes can lead to polarisation-dependent mixing-induced \CP asymmetries and triple product asymmetries in $\Bs\to \jpsi\,\phi$~\cite{Chiang:2009ev}.  
Such effects will be searched for in future analyses.

The flavour-specific asymmetries provide important complementary constraints on $\Delta B = 2$ processes.
The \dzero collaboration has performed a direct measurement of $a^s_{\rm sl}$ in semileptonic \Bs decays~\cite{Abazov:2009wg}, which is only weakly constraining.\footnote{
  An updated measurement has been presented by \dzero at ICHEP 2012~\cite{Abazov:2012zz}.
}
However, a measurement of the inclusive same-sign dimuon asymmetry provides better precision, and shows evidence of a large deviation from its SM prediction~\cite{Abazov:2011yk}.  
The inclusive measurement is sensitive to a linear combination of the flavour-specific asymmetries, $A_{\rm SL}^b = C_d\, a_{\rm sl}^d + C_s\, a_{\rm sl}^s$, where $C_q$ depend on the production fractions and mixing probabilities, and are determined to be $C_d = 0.594 \pm 0.022$, $C_s = 0.406 \pm 0.022$~\cite{Abazov:2011yk}.\footnote{
  The factors $C_d$ and $C_s$ depend in principle on the collision environment and the kinematic acceptance, though the dependence appears to be weak.
  Trigger requirements can also affect the values of these parameters.
}
As discussed in Sec.~\ref{sec:bmixing_constraint}, the D0 $A_{\rm SL}^b$ result is in tension with other $\Delta B=2$ observables.
Improved measurements of $a^s_{\rm sl}$ and $a^d_{\rm sl}$ from LHCb are needed to solve this puzzle. 

In LHCb, $a^s_{\rm sl}$ can be determined from the asymmetry between the time-integrated untagged decay rates of \Bs decays to $D_s^+\mu^- X$ and $D_s^-\mu^+ X$, with $\Dspm \to \phi \pipm$, $\phi \to \Kp\Km$ (or with the full $\Dspm \to \Kp\Km\pipm$ Dalitz plot).
Detector- and trigger-induced asymmetries can be calibrated in control channels, and the fact that data is taken with both magnet dipole polarities can be used as a handle to reduce systematic uncertainties.
The effect of \Bs production asymmetry is cancelled due to the fast oscillation, so the asymmetry in the yields of $\Dsp\mu^- X$ and $\Dsm\mup X$ decays is trivially related to $a^s_{\rm sl}$. 
A first preliminary LHCb result on $a^s_{\rm sl}$, based on $1.0 \invfb$, has been reported at ICHEP 2012, and is the most precise measurement of this quantity to date~\cite{LHCb-CONF-2012-022},
\begin{equation}
  a^s_{\rm sl} = \left( -0.24 \pm 0.54 \pm 0.33 \right)\,\%\,.
\end{equation}
It will also be possible to measure $a^d_{\rm sl}$ using $D^+ \mu^- X$ final states with $D^+ \to K^- \pi^+ \pi^+$.
In this case extra care must be taken to calibrate the difference between $K^+$ and $K^-$ detection efficiencies and an independent measurement of the \Bz production asymmetry is needed as input.
Moreover, the \CP-symmetric background from charged \B decays is significant and must be accurately subtracted.

\begin{figure}[!htb]
  \centering
  \includegraphics[width=0.44\textwidth]{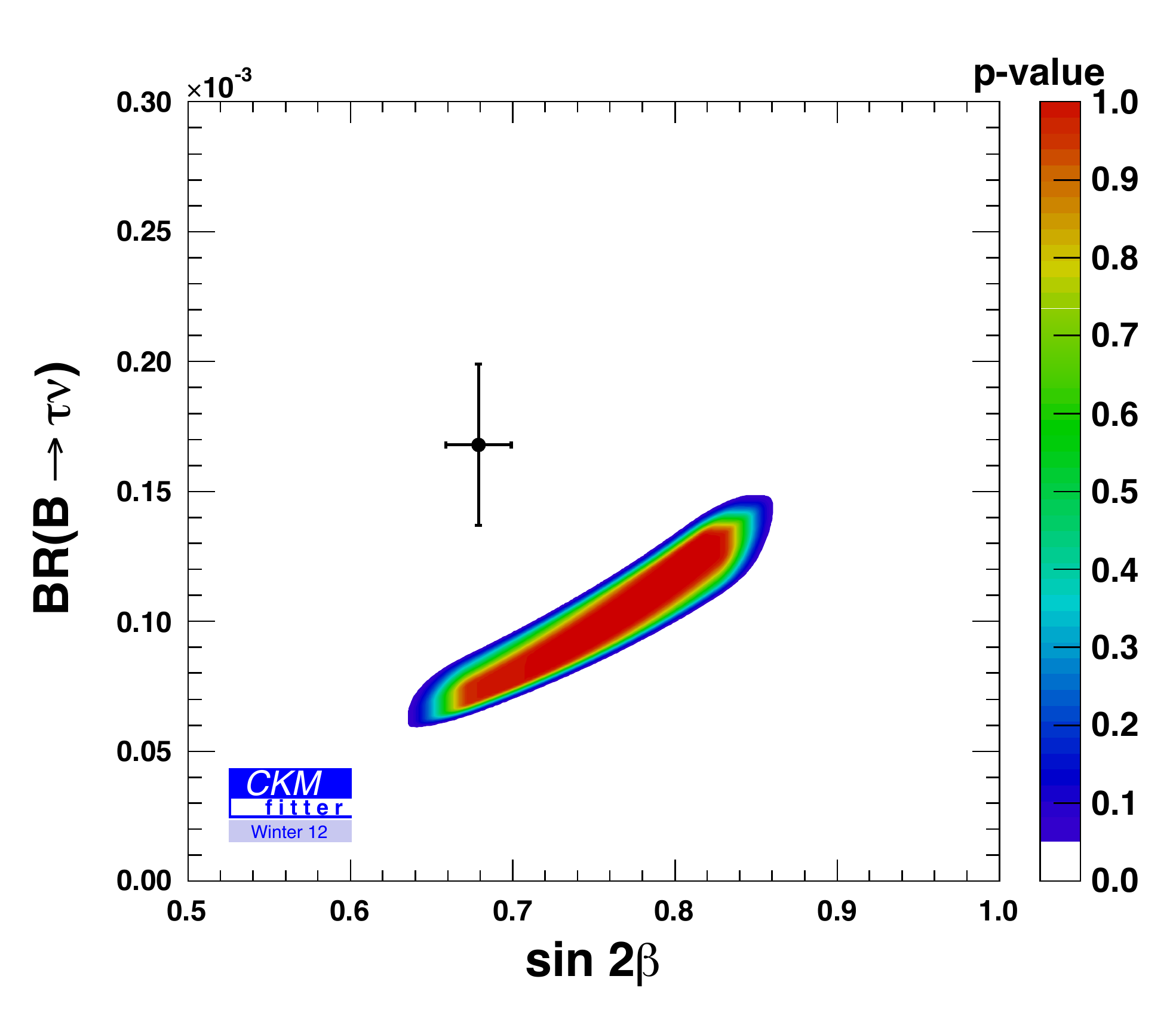}
  \caption{\small
    Comparison of direct and indirect determinations of $\sin \phi_d \equiv \sin 2\beta$ {\it vs.} $\BR(B^+ \to \tau^+ \nu)$, from Ref.~\cite{Charles:2004jd}. 
  }
  \label{fig:bcpv:bdtension}
\end{figure}

In the \Bz system, $\Delta m_d$ and $\sin \phi_d$ (\ie\ $\sin 2\beta$) have been measured precisely by the \B factories~\cite{HFAG}. 
The measurements of $\Delta\Gamma_d$ and $a^d_{\rm sl}$ are consistent with their SM predictions, but their uncertainties are at least an order of magnitude larger than those of the predictions.
Hence a large improvement in precision is needed to test the SM using these observables. 
In the \Bz sector there has been for some time a tension between the measurements of $\sin 2\beta$~\cite{HFAG} and the branching ratio $\BR(B^+ \to \tau^+ \nu )$~\cite{Hara:2010dk,Lees:2012ju}, as shown in Fig.~\ref{fig:bcpv:bdtension},\footnote{
  An updated measurement of $\BR(B^+ \to \tau^+ \nu)$ using the hadronic tag method was presented by Belle at ICHEP 2012~\cite{Adachi:2012mm}: this new result reduces, but does not completely remove, the tension in the fits.  The analyses discussed here do not include this new result.
}
and discussed in Sec.~\ref{sec:globalfit}. 
This motivates improved measurements of $\sin 2\beta$ and improved understanding of the possible effects of penguin contributions to this observable.

LHCb has already presented first results on $\Delta m_d$~\cite{LHCb-PAPER-2011-010,LHCb-PAPER-2012-032} and $\sin 2\beta$~\cite{LHCb-PAPER-2012-035}. 
The $\Delta m_d$ result is the world's most precise single measurement of this quantity, while the sensitivity on $\sin 2\beta$ will be competitive with the \B factory results using the data sample that will be collected by the end of 2012. 
LHCb can also search for enhancements in the value of $\Delta\Gamma_d$  above the tiny value expected in the SM, \eg\ by comparing the effective lifetimes of $\Bz \to \Jpsi \KS$ and $\Bz \to \Jpsi K^{*0}$~\cite{Gershon:2010wx}. 
Significantly improving the precisions of the \Bz mixing observables is an important goal of the LHCb upgrade, as will be discussed in Sec.~\ref{sec:bmixing_upgrade}.

The SM predictions of $b$-hadron lifetimes and $\Delta\Gamma_q$ are all obtained within the framework of the heavy quark expansion.
LHCb is actively working on measurements of $b$-hadron lifetimes and lifetime ratios, which will be used to test these predictions. 
The knowledge obtained from this work will allow to improve the SM predictions of $\Delta\Gamma_q$ for the purpose of searching for NP. 
Furthermore, a more precise measurement of the ratio of \Bs to \Bz lifetimes could either support or strongly constrain the existence of NP in $\Gamma^s_{12}$~\cite{Bauer:2010dga,Bobeth:2011st,Lenz:2011ti,Lenz:2012mb,Lenz:2012az}.

\subsubsection{Model independent constraints on new physics in \B mixing}
\label{sec:bmixing_constraint}
Neutral $B_q$ meson mixing is described in terms of the three parameters
$|M_{12}^q|$, $|\Gamma_{12}^q|$ and $\phi_q = \mbox{arg} (- M_{12}^q /
\Gamma_{12}^q)$ for each of the two systems $q = d, s$. 
In the context of model-independent analyses, the NP  contributions can be
parametrised in the form of two complex quantities $\Delta_q$ and ${\Lambda}_q$~\cite{Bobeth:2011st,Lenz:2011zz}
\begin{align} 
  \label{eq:NPcontribution:def}
  M_{12}^q & 
    = M_{12}^{q,{\rm SM}} \, |\Delta_q| \, e^{i \phi_q^\Delta} \,, &
  \Gamma_{12}^q &
    = \Gamma_{12}^{q,{\rm SM}} \, |{\Lambda}_q| \, e^{i \phi_q^{{\Lambda}}} \,,
\end{align}
\ie, 4 real degrees of freedom. 
The observables which depend on these parameters are the mass and decay width differences and flavour-specific \CP-asymmetries. 
They can be expressed in terms of the SM predictions and NP parameters as
\begin{align} 
  \label{eq:NP:dmdg}
  \dmq & 
    = (\dmq)_{\rm SM} \hspace{0.5mm} |\Delta_q| \,, &
  \DGq & 
    = (\DGq)_{\rm SM} \hspace{0.25mm} |{\Lambda}_q| \, 
   \frac{\cos \left (\phi_{12}^{q,{\rm SM}} + \phi_q^\Delta - \phi_q^{\Lambda} \right )} 
        {\cos \phi_{12}^{q,{\rm SM}}},
\end{align}
\begin{align}
  \label{eq:NP:afs}
  a_{\rm sl}^q & 
    = (a_{\rm sl}^q)_{\rm SM}\, \frac{|{\Lambda}_q|}{|\Delta_q|} \; 
    \frac{\sin \left (\phi_{12}^{q,{\rm SM}} + \phi_q^\Delta - \phi_q^{\Lambda} \right )}
         {\sin \phi_{12}^{q,{\rm SM}}}\,,
\end{align}
up to corrections suppressed by tiny $(\Gamma_{12}^q/M_{12}^q)^2$.
Note that the expressions of Eqs.~(\ref{eq:NP:dmdg}) and~(\ref{eq:NP:afs}) depend only on the difference $(\phi_q^\Delta - \phi_q^{\Lambda})$. 
The SM predictions of $\dmq$, $\DGq$ and $a_{\rm sl}^q$ can be found in Table~\ref{tab:mixing_status} and for $\phi^q_{12}$~\cite{Lenz:2011ti}
\begin{align}
  \phi_{12}^{d,{\rm SM}} & = (-0.075 \pm 0.024)\rad\,, &
  \phi_{12}^{s,{\rm SM}} & = (0.0038 \pm 0.0010)\rad.
\end{align}
The values of $\dmq$ have been precisely measured, giving rather strong constraints on $|\Delta_q|$ which are limited by the knowledge of hadronic matrix elements. 
The new $\DGs$ measurement of \lhcb starts to provide useful constraints.
As discussed above, the \CP-asymmetries $a_{\rm sl}^q$ are currently rather weakly constrained.

Further information can be extracted from the mixing-induced \CP-asymmetries in
$\Bz\to \jpsi \KS$ and $\Bs\to \jpsi \phi$ decays
\begin{align}
  \label{eq:NP:phi}
  \phi_d & = 2 \beta   + \phi_d^\Delta - \delta_d, &
  \phi_s & = -2 \beta_s + \phi_s^\Delta - \delta_s,
\end{align}
where $\delta_d$ and $\delta_s$ denote shifts of $\phi_d$ and $\phi_s$ induced by either SM penguin diagrams or NP contributions in the decay process.
In the SM $\phi_d$ and $\phi_s$ are related to the angles $\beta$ and $\beta_s$ of the according unitarity triangles.  
When short-distance NP contributions are introduced, $\phi_q$ depends on the phase $\phi_q^\Delta$ of $M_{12}^q$, whereas the phase $\phi_q^{\Lambda}$ of $\Gamma_{12}^q$ does not enter. 
The SM penguin pollution to $\delta_q$ is expected to be negligible for the current precision of $\phi_q$, and is discussed in detail in Sec.~\ref{sec:bmixing_penguin}.
Beyond the SM, NP can contribute to $\delta_q$ in principle in both the tree $b\to \ccbar s$ decay and the penguin process.
However, in the model-independent analysis described here, NP contributions in the $b \to \ccbar s$ decay are neglected and any observed deviation from the SM will be interpreted as effects of NP in neutral \B meson mixing.
When $\delta_q$ is neglected, Eqs.~(\ref{eq:NP:dmdg}), (\ref{eq:NP:afs}) and~(\ref{eq:NP:phi}) allow to determine the NP parameters $|\Delta_q|$, $\phi_q^\Delta$, $|\Lambda_q|$ and $\phi_q^\Lambda$.

\begin{figure}[!htb]
\centering
\includegraphics[width=0.48\textwidth]{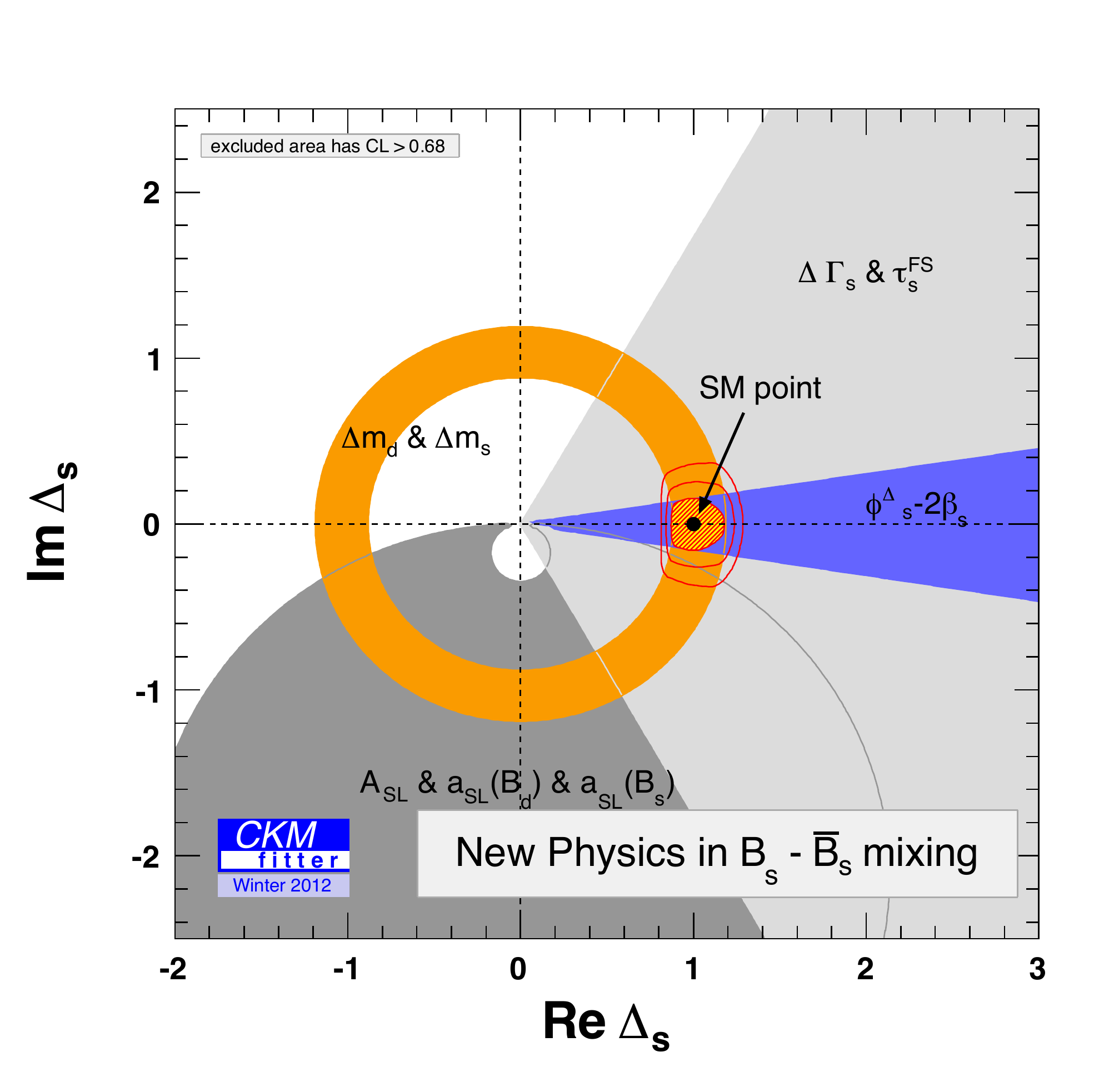}
\includegraphics[width=0.48\textwidth]{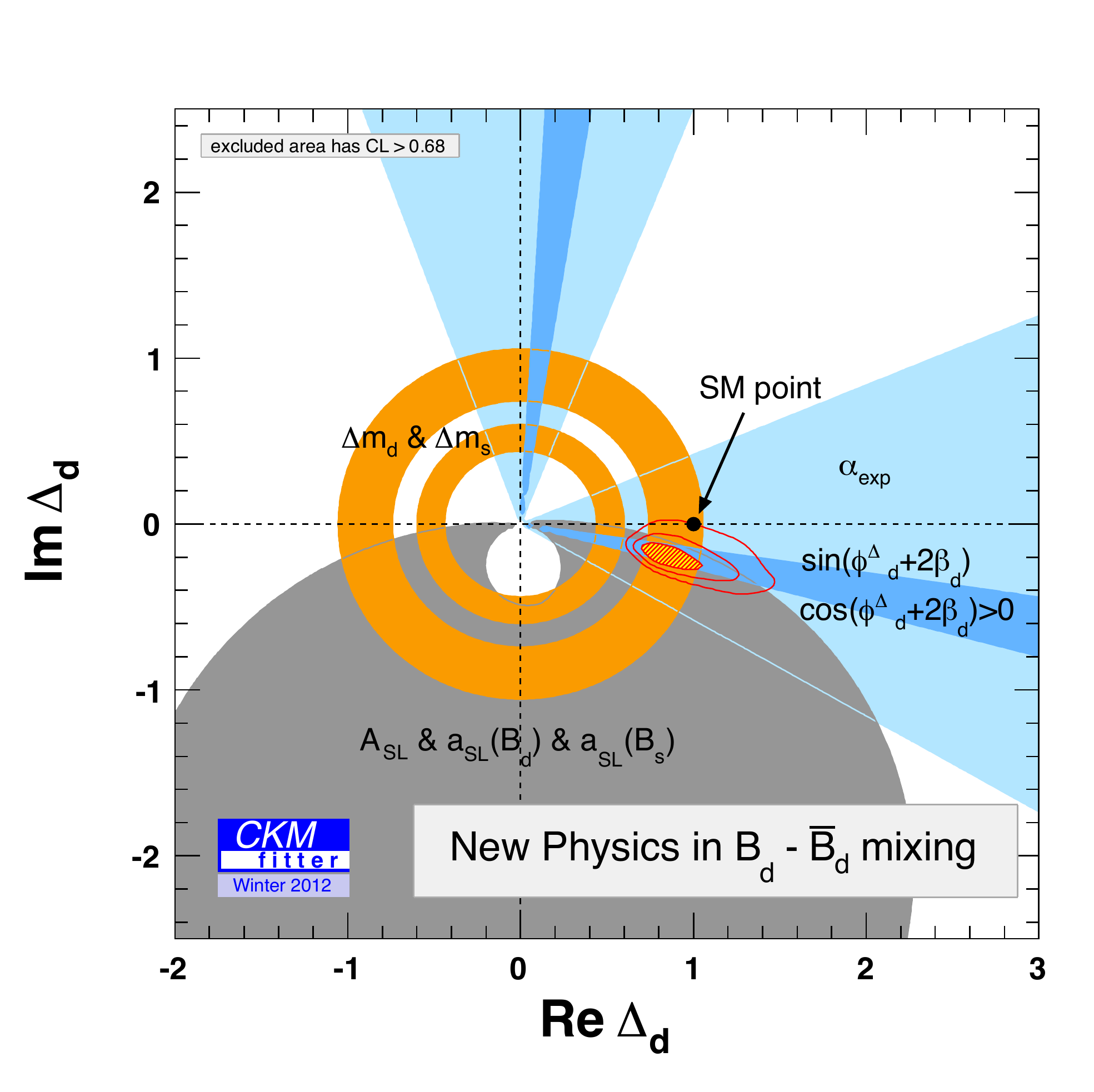}
\caption{\small
  Model-independent fit~\cite{Lenz:2012az} in the scenario that NP affects $M^q_{12}$ separately. 
  The coloured areas represent regions with C.L. $< 68.3\,\%$ for the individual constraints. 
  The red area shows the region with C.L. $< 68.3\,\%$ for the combined fit, with the two additional contours delimiting the regions with C.L. $< 95.45\,\%$ and C.L. $< 99.73\,\%$.
 \label{fig:bcpv:newphysics}
}
\end{figure}

The assumption of NP in $M_{12}^q$ only, or equivalently in $\Delta B = 2$ processes only, implies that there is no NP in $\Delta B = 1$ processes which contribute to the absorptive part $\Gamma_{12}^q$. 
Consequently, NP can only decrease $\DGq$ (since $\cos(\phi_{12}^{q,{\rm SM}})$ is maximal, see Eq.~(\ref{eq:NP:dmdg})) with respect to the SM~\cite{Grossman:1996era,Dunietz:2000cr}. 
This scenario has been studied in extensions of the CKM fit of the SM which includes $\Delta B = 2$ measurements to constrain the CKM elements $V_{tq}$~\cite{Lenz:2010gu,Lenz:2012az}, in combination with many other flavour-changing processes. 
Including LHCb measurements~\cite{LHCb-CONF-2012-002,LHCb-PAPER-2011-010}\footnote{
  But not including results shown for the first time at ICHEP 2012 or later.
}
the SM point $\Delta_d = \Delta_s = 1$ is disfavoured by $2.4\,\sigma$~\cite{Lenz:2012az} (prior to the LHCb results being available, a similar analysis gave a discrepancy of $3.6\,\sigma$ driven mainly by the anomalous dimuon asymmetry~\cite{Lenz:2010gu}).
The analysis gives $\Delta_s$ consistent with the SM, within large uncertainties, whereas the more precise data in the \Bd system hint at a deviation in $\Delta_d$ (see Fig.~\ref{fig:bcpv:newphysics}). 
Moreover, NP effects up to $30$--$40$\,\% are still allowed in both systems at the $3\,\sigma$ level. 
It should be noted, that the large deviations in the $\Bz$ sector are not only due to $A_{\rm SL}^b$, but also due to the tension between $\sin \phi_d$ and $\BR(B^+ \to \tau^+ \nu)$.

NP contributions to the absorptive part $\Gamma_{12}^q$ of $B$ mixing can enter through $\Delta B = 1$ decays $b\to q X$ with light degrees of freedom $X$ of total mass below $m_B$. 
In some particular models such contributions can arise~\cite{Dighe:2007gt,Badin:2007bv} and  interfere constructively or destructively with the SM contribution.
The recent measurements of $\DGq$ and of $A_{\rm SL}^b$ revived interest in this possibility. 
Model-independent analyses have confirmed that the $A_{\rm SL}^b$ measurement cannot be accommodated within the SM~\cite{Dobrescu:2010rh,Ligeti:2010ia}. 
A model-independent fit assuming NP in both $M_{12}^q$ and $\Gamma_{12}^q$ has been considered in the framework of an extended CKM fit~\cite{Lenz:2012az}.  
In this case, the experimental data can be accommodated, and the $\Bs$ system remains rather SM-like, but large NP contributions in the $\Bz$ system are required.
 
Model-independent analyses based on Eq.~(\ref{eq:NPcontribution:def}) are restricted to a particular set of observables, mainly those with $\Delta B=2$, since correlations with $\Delta B=1$ measurements are difficult to quantify.
Either additional assumptions on the nature of $X$ in $b\to q X$ or explicit NP models will permit better exploitation of the wealth of future experimental information. 
In fact, such analyses have found it difficult to accommodate the hypothesis of large NP in $\Gamma^q_{12}$ with current $\Delta B = 1$ measurements, therefore NP in $\Gamma^q_{12}$ seems unlikely to provide a full explanation of the measured value of $A_{\rm SL}^b$.
In the case of $X = f\bar{f}$, the $\Delta B = 1$ operators $b\to (d,s) f\bar{f}$ ($f = q$ or $\ell$) are strongly constrained~\cite{Bauer:2010dga}, with the exception of $b\to s\ccbar$ and $b\to s \tautau$. 
Currently, only a weak upper bound on $\BR(\Bp \to \Kp\tautau) \lsim 3.3 \cdot 10^{-3}$ at 90\,\% C.L.~\cite{Flood:2010zz} exists whereas other decays $\Bs\to\tautau$, $\B\to X_s\tautau$ might be indirectly constrained with additional assumptions (see also the discussion in Sec.~\ref{sec:rare:tautau}). 
As an example, the improved LHCb measurement of  $\tau_{\Bs}/\tau_{\Bd}$ allowed the derivation of a stronger bound on $\BR(\Bs \to \tautau)$.
Still, a model-independent analysis of the complete set of $b\to s \tautau$ operators does not allow for deviations larger than 35\,\% from the SM in $\Gamma_{12}^s$~\cite{Bobeth:2011st}, which is much too small to resolve the tension with $A_{\rm SL}^b$. 
 For $b\to d \tautau$ operators there exists a stronger constraint $\BR(\Bz\to \tautau) \lsim 4 \cdot 10^{-3}$ and even smaller NP effects are expected in $\Gamma_{12}^d$.  
Other proposed solutions such as the existence of new light spin-0~\cite{Bai:2010kf} or spin-$1$~\cite{Oh:2010vc} $X$ states could be seriously challenged by improved measurements of quantities, such as ratios of lifetimes, which are theoretically under good control~\cite{Lenz:2011ti}.

In summary, NP contributions to $|\Delta_q|$ are already quite constrained due to $\dmq$ measurements and theoretical progress is required in order to advance.  
Although the phases $\phi_q^\Delta$ are constrained by the recent \lhcb measurement of $\phi_s$, and \B factory measurements of $\phi_d$, there is a mild tension with the SM in model-independent fits of $\Delta B=2$ measurements~\cite{Lenz:2012az,Dobrescu:2010rh,Ligeti:2010ia,Bobeth:2011st}, especially when allowing for NP in $\Gamma_{12}^q$. 
On the other hand, NP effects in $\Gamma_{12}^q$ are expected to be limited when constraints from $\Delta B=1$ observables are taken into account.
Independent improved measurements of $a_{\rm sl}^q$ are needed in order to resolve the nature of the current discrepancies between the $\Delta B = 2$ observables with their SM expectations and other observables entering global CKM fits. 
Further, improved measurements of $\Gq$ and $\DGq$, as well as of control channels, are needed to constrain NP in $\Gamma^q_{12}$.

\subsubsection{CKM unitarity fits in SM and beyond}
\label{sec:globalfit}

This section presents the results of the unitarity triangle (UT) analysis performed by two groups: UTfit~\cite{Bona:2005vz} and CKMfitter~\cite{Charles:2004jd}.\footnote{
  Similar approaches have been developed in Ref.~\cite{Lunghi:2010gv,Eigen:2013cv}.
}
The main aim of the UT analysis is the determination of the values of the CKM parameters, by comparing experimental measurements and theoretical predictions for several observables. 
The popular Wolfenstein parametrisation allows for a transparent expansion of the CKM matrix in terms of the sine of the small Cabibbo angle, $\lambda$, with the other three parameters being $A$, $\bar{\rho}$ and $\bar{\eta}$. 
Assuming the validity of the SM, one can perform a fit to the available measurements. 
LHCb results already make important contributions to the constraints on $\gamma$ and $\Delta m_s$.
With more statistics, LHCb results are expected to impact on other CKM fit inputs, including $\alpha$ and $\sin2\beta$.
It is important to note the crucial role of lattice QCD calculations as input to the CKM fits.
For example, the parameters $f_{B_s}\sqrt{B_{B_s}}$ and $\xi$ enter the constraints on $\Delta m_s$ and $\Delta m_d/\Delta m_s$.
At the end of 2011, the precision of the calculations was at the level of $5.4\,\%$ and $2.6\,\%$, respectively~\cite{Laiho:2009eu}.
The necessary further progress to obtain the full benefit of the LHCb measurements appears to be in hand exploiting algorithmic advances as well as ever increasing computing power for the lattice calculations.

\begin{figure}[!htb]
\begin{center}
\setlength{\unitlength}{1\linewidth}
\includegraphics[width=0.475\linewidth]{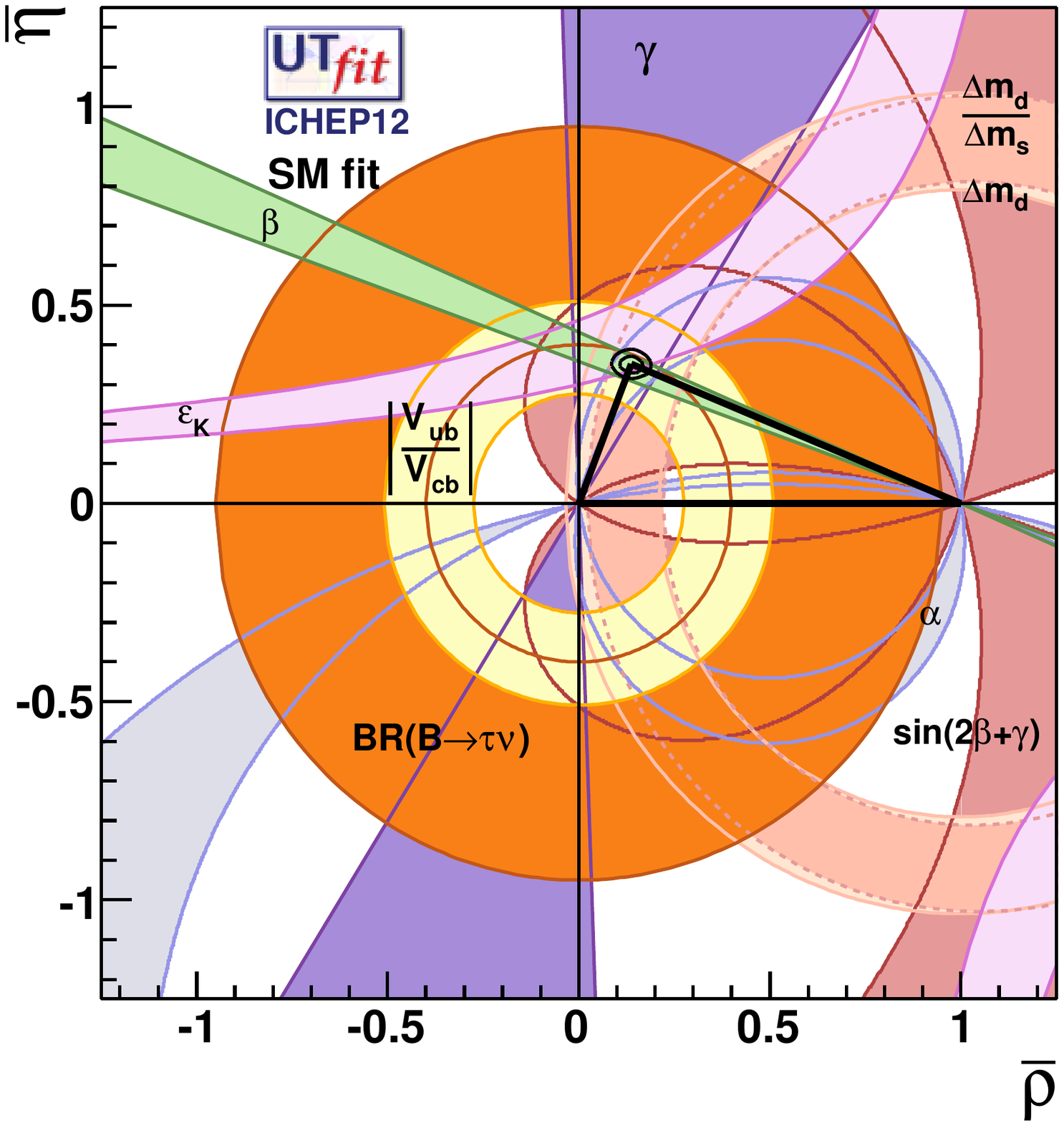}
\includegraphics[width=0.425\linewidth]{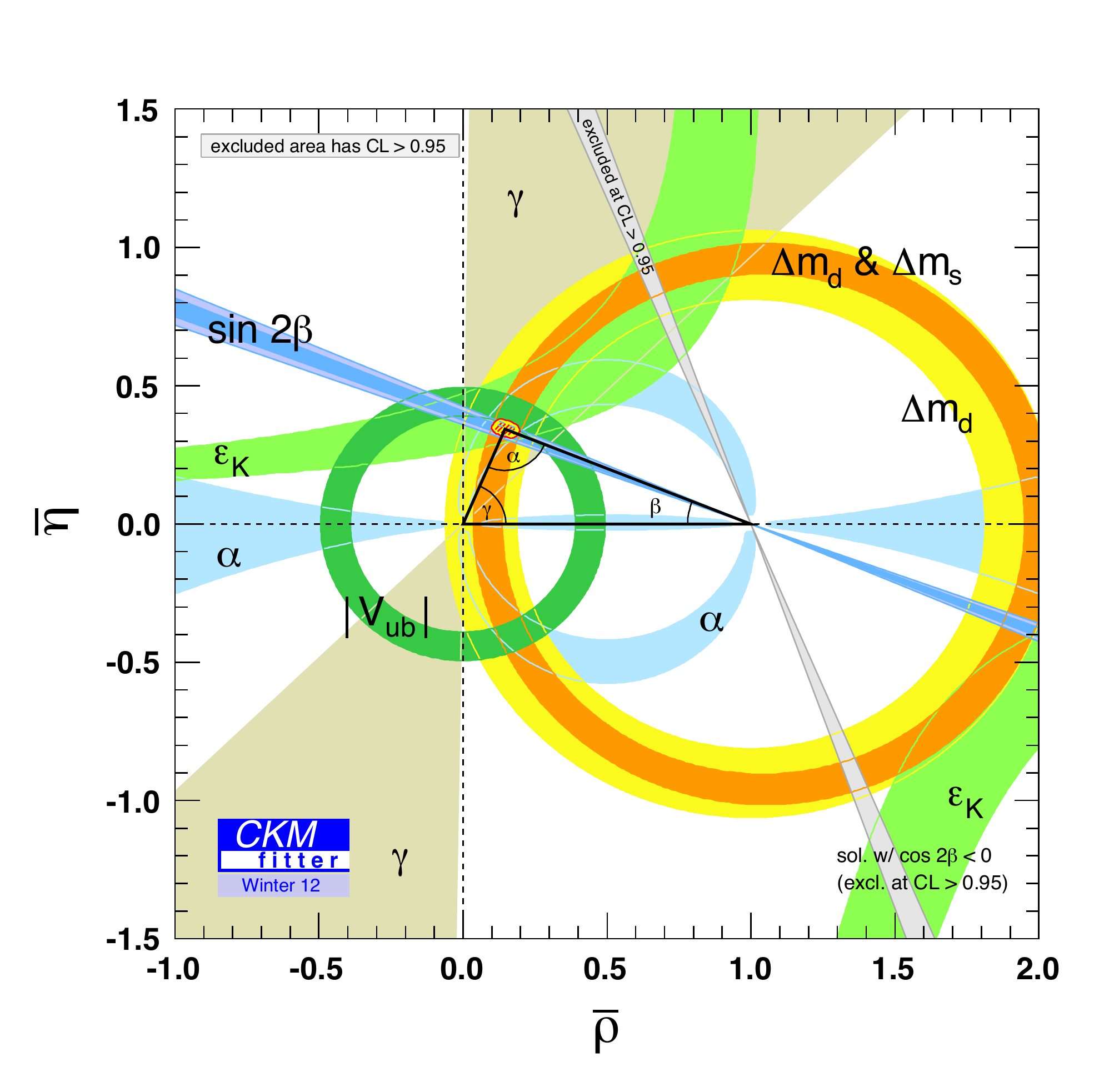}
\end{center}
\caption{\small
  Result of the UT fit within the SM: $\{\bar\rho, \bar\eta\}$ plane obtained by (left) UTfit~\cite{Bona:2005vz} and (right) CKMfitter~\cite{Charles:2004jd}. 
  The 95\,\% probability regions selected by the single constraints are also shown with various colours for the different constraints.
}
\label{fig:SMfullfit}
\end{figure}

The overall quality of the fit can be judged using the projection of the likelihoods on the $\{\bar\rho, \bar\eta\}$ plane. 
This projection is shown in Fig.~\ref{fig:SMfullfit}. 
The fit can also be made removing one of the inputs, giving a prediction for the removed parameter, which then can be compared to the experimental value.
The results of this study are presented in Table~\ref{tab:SMfullfittable}. 
Both groups find a tension between $\BR(B\to \tau\nu)$ and $\sin2\beta$, as can be seen in Fig.~\ref{fig:bcpv:bdtension}.
(As discussed in Sec.~\ref{sec:status_Bmixing} this tension will be reduced once the latest Belle result on $\BR(\Bp\to\taup\nu_\tau)$~\cite{Adachi:2012mm} is included in the fits.)
Improved measurements of $\sin2\beta$ can shed further light on this problem.

\begin{table}[!htb]
\caption{\small
  Predictions for some parameters of the SM fit and their measurements as combined by the UTfit and CKMfitter groups. 
  Note that the two groups use different input values for some parameters.
  The lines marked with (*) are not used in the full fit. 
  Details of the pull calculation can be found in Refs.~\cite{Bona:2009cj,Lenz:2010gu}.
  New results presented at ICHEP2012 and later are not included in these analyses.
}
\label{tab:SMfullfittable}
\begin{center}
\resizebox{0.99\textwidth}{!}{
\begin{tabular}{|c | c | c | c | c | c | c|}
\hline
\multirow{2}{*}{Parameter}&\multicolumn{3}{c|}{UTfit}&\multicolumn{3}{c|}{CKMfitter}\\
\cline{2-7}
				&prediction		&measurement		&pull			&prediction			&measurement 		&pull\\
\hline
$\alpha\,(^{\circ})$		&$87.5\pm3.8$		&$91.4\pm6.1$		&$+0.5\,\sigma$		& $95.9\,^{+2.2}_{-5.6}$ 			& $88.7\,^{+2.2}_{-5.9}$ 		& $-1.0\,\sigma$\\
$\sin2\beta$			&$0.809\pm0.046$	&$0.667\pm0.024$	&$-2.7\,\sigma$		& $0.820\,^{+0.024}_{-0.028}$ 		& $0.679\pm 0.020$ 		& $-2.6\,\sigma$\\
$\gamma\,(^{\circ})$		&$67.8\pm3.2$		&$75.5\pm10.5$		&$+0.7\,\sigma$		& $67.2\,^{+4.4}_{-4.6}$ 			& $66\,^{+12}_{-12}$ 		& $-0.1\,\sigma$\\
$V_{ub}\,(10^{-3})$		&$3.62\pm0.14$		&$3.82\pm0.56$		&$+0.3\,\sigma$		& $3.55\,^{+0.15}_{-0.14}$ 		& $3.92\pm 0.09\pm 0.45$ 	& $0.0\,\sigma$\\
$V_{cb}\,(10^{-3})$		&$42.26\pm0.89$		&$41\pm1$		&$-0.9\,\sigma$		& $41.3\,^{+0.28}_{-0.11}$ 		& $40.89\pm 0.38 \pm 0.59$ 	& $0.0\,\sigma$\\
$\eps_k\,(10^{-3})$		&$1.96\pm0.20$		&$2.229\pm0.010$	&$+1.3\,\sigma$		& $2.02\,^{+0.53}_{-0.52}$ 		& $2.229\pm 0.010$ 		& $0.0\,\sigma$\\
$\Delta m_s\,(\invps)$	&$18.0\pm1.3$		&$17.69\pm0.08$		&$-0.2\,\sigma$   	& $17.0\,^{+2.1}_{-1.5}$ 			& $17.731\pm 0.045$ 		& $0.0\,\sigma$ \\
$\BR(\B\to\tau\nu)\,(10^{-4})$	&$0.821\pm0.0077$	&$1.67\pm0.34$		&$+2.5\,\sigma$		& $0.733\,^{+0.121}_{-0.073}$ 		& $1.68\pm 0.31$ 		& $+2.8\,\sigma$\\
$\beta_s\,\rad$ (*)     	&$0.01876\pm0.0008$	&			&				& $0.01822\,^{+0.00082}_{-0.00080}$ 	&			&\\
$\BR(\Bs\to\mu\mu)\,(10^{-9})$ (*)	&$3.47\pm0.27$		&			&		& $3.64\,^{+0.21}_{-0.32}$		&			&\\
\hline
\end{tabular}
}
\end{center}
\end{table}

In order to estimate the origin of the tensions, the UTfit and CKMfitter groups have performed analyses including model-independent NP contributions to neutral meson mixing processes (see Refs.~\cite{Bona:2007vi,Lenz:2012az} for details). 
The NP effects are introduced through the real valued $C$ and $\phi$ parameters ($A_{\rm NP} = C e^{i\phi} A_{\rm SM}$) in case of UTfit and the complex valued $\Delta$  parameter ($A_{\rm NP} = \Delta A_{\rm SM}$) for CKMfitter. 
The parameters are added separately for the $\Bs$ and $\Bd$ sectors. 
In the absence of NP, the expected values are $C=1$, $\phi=0^{\circ}$, and $\Delta = 1$. 
For the $\Bd$ sector the fits return $C=0.94\pm0.14$ and $\phi=(-3.6\pm3.7)^{\circ}$, and $\Delta = (0.823\,^{+0.143}_{-0.095})+i (-0.199\,^{+0.062}_{-0.048})$. 
The results for both groups show some disagreement with the SM, driven by tensions in the input parameters mentioned above. 
In the $\Bs$ sector, on the other hand, the situation is much closer to the SM than before the LHCb measurements were available: $C=1.02\pm0.10$ and $\phi=(-1.1\pm2.2)^{\circ}$, and $\Delta = (0.92\,^{+0.13}_{-0.08})+i(0.00 \pm 0.10)$.

The results of the studies by both groups point to the absence of big NP effects in $\Delta B = 2$ processes.
Nevertheless there is still significant room for NP in mixing in both $\Bd$ and $\Bs$ systems.
More precise results, in particular from LHCb, can enable more careful studies.
Besides providing null tests of the SM hypothesis, improved $\phi_s$ and $a_{\rm sl}^s$ measurements are crucial to quantity effects of NP in mixing.
In addition a precise $\gamma$ determination is essential, not only for a SM global consistency test, but also to fix the apex of the UT in the extended fits.

\subsubsection{Penguin pollution in $\bquark \to \cquark \overline{\cquark} \squark$ decays }
\label{sec:bmixing_penguin}
In addition to the very clear experimental signature, precise determination of the $\Bd$ and $\Bs$ mixing phases is possible due to the fact that in the ``golden modes'', $\Bd \to \Jpsi\, \KS$ and $\Bs \to \Jpsi\, \phi$, explicit calculation of the relevant matrix elements can be avoided, once subleading doubly Cabibbo-suppressed and loop-suppressed terms are assumed to vanish~\cite{Bigi:1981qs}. 
Estimates yield corrections of the order $O(10^{-3})$ only~\cite{Boos:2004xp,Li:2006vq, Gronau:2008cc}; it is however notoriously difficult to actually calculate the relevant matrix elements, and non-perturbative enhancements cannot be excluded. 
Given the future experimental precision for these and related modes, a critical reconsideration of this assumption is mandatory. 

The main problem lies in the fact that once the assumption of negligible penguin contributions is dropped, the evaluation of hadronic matrix elements again becomes necessary, which still does not seem feasible to an acceptable precision for the decays in question. 
To avoid explicit calculation, symmetry relations can be used, exploiting either flavour SU(3) or U-spin symmetry~\cite{Fleischer:1999nz, Ciuchini:2005mg, Faller:2008gt, Faller:2008zc, Jung:2009pb, DeBruyn:2010hh, Ciuchini:2011kd}. 
Without taking into account any QCD evaluation and only using control channels to estimate the size of  the penguin amplitude, the analyses in Refs.~\cite{Ciuchini:2011kd, Faller:2008zc}  still allow a phase shift of up to a few degrees for $\phi_d$, which would correspond to a very large non-perturbative enhancement of the penguin size.
In Ref.~\cite{Faller:2008zc} a negative sign is preferred which (slightly) reduces the tension in the unitarity triangle fit shown in Fig.~\ref{fig:bcpv:bdtension}. 
The reason for the large allowed range of the shift of $\phi_d$ is due to the limited precision to which the corresponding control channels  $\Bd\to \jpsi\, \piz$ and $\Bs\to\jpsi\,\Kz$, which are Cabibbo-suppressed compared to the golden modes, are known. 
For $\phi_s$, an analogous analysis~\cite{Faller:2008gt} cannot yet constrain the penguin contribution, due to the lack of a $\B \to \jpsi\, V$ control channel data for $\Bs \to \jpsi\, \phi$. 
However, in principle the effects in the $\B \to \jpsi\, V$ modes are expected to be of the same order of magnitude as in the $\B \to \jpsi\, P$ modes. 
The control channel $\Bs \to \jpsi\, K^{*0}$ has already been observed at CDF~\cite{Aaltonen:2011sy} and LHCb~\cite{LHCb-PAPER-2012-014},
and work is ongoing to measure its decay rate, polarisations and direct \CP asymmetries.
This will enable the first direct constraint on the  shift of $\phi_s$ due to penguin contributions in the decay $\Bs \to \jpsi\,\phi$.\footnote{
  Other data-driven methods to control penguin contributions to $\Bs \to \jpsi\,\phi$ have been proposed~\cite{Bhattacharya:2012ph}.
}
For $\Bs \to \Jpsi\,f_0(980)$ there is an additional complication due to the unknown hadronic structure of the $f_0(980)$~\cite{Fleischer:2011au}.

In addition to insufficient data, there are, at present, theoretical aspects limiting the precision of this method at present, the most important of which is the violation of SU(3) symmetry.
Regarding the \Bd mixing phase, a full SU(3) analysis can be performed~\cite{Jung:2012mp} (instead of using only one control channel) to be able to model-independently include SU(3) breaking. 
The inclusion of SU(3)-breaking contributions is important: their neglect can lead to an overestimation of the subleading effects. 
Including recent data for two of the relevant modes~\cite{LHCb-PAPER-2011-024,LHCb-PAPER-2011-041}, the analysis shows that the data are at the moment actually compatible with vanishing penguin contributions, with SU(3)-breaking contributions of the order $20\,\%$.
Including the penguin contributions, an upper limit
on the shift of the mixing-induced \CP asymmetry $\Delta S = \sin \phi_d - \sin2\beta$ is derived: $|\Delta S| \lesssim 0.01$, with a negative sign for $\Delta S$ slightly preferred.\footnote{
     Note the definition of $\Delta S$ here has a sign difference to that in Ref.~\cite{Jung:2012mp}.
}
This is the most stringent limit available, despite the more general treatment of SU(3) breaking.
In this analysis still some (conservatively chosen) theoretical inputs are needed to exclude fine-tuned solutions: SU(3)-breaking effects have been restricted to at most $40\,\%$ for a few  parameters which are not well determined by the fit and also have only small influence on the \CP violation observables, and the penguin matrix elements are constrained to be at most $50\,\%$ of the leading contributions. 
Importantly, these theory inputs can be replaced by experimental measurements, namely of the \CP asymmetries in the decay $\Bs \to \Jpsi\,\KS$, the decay rate of which has already been measured at LHCb~\cite{LHCb-PAPER-2011-041} after its observation at CDF~\cite{Aaltonen:2011sy}. 
Furthermore, data from all the corresponding modes (\ie\ $B_{d,u,s} \to \Jpsi\,P$, with light pseudoscalar meson $P = \pi, \eta^{(\prime)}$ or $K$) can be used to determine the shift more precisely, \ie\ the related uncertainty is not irreducible, but can be reduced with coming data.

Turning to the second golden mode, $\Bs \to \Jpsi\, \phi$, in general, the absolute shift is not expected to be larger than in the \Bd case.
At the moment the data are not yet available to make a comparable analysis. 
While  the penguin decay mode $\Bs \to \phi \phi$  is not related by symmetry with  \BsToJPsiPhi, 
comparing their decay rates   indicates that the penguin  contributions  are small,
and  there are no huge enhancements to be expected for the penguin matrix elements in question.

Nonetheless, a quantitative analysis will ultimately be warranted here as well. 
In principle, these methods can be adapted to extract the $\Bs$ mixing phase including penguin contributions and model-independent SU(3) breaking, thereby improving the method proposed in Ref.~\cite{Faller:2008gt}. 
The corresponding partners of the golden mode $\Bs \to \Jpsi\, \phi$ are all the decays $\B_{u,d,s} \to \Jpsi\, V$, with the light vector mesons $V = K^*$, $\rho$, $\phi$ or $\omega$. 
However, the complete analysis requires results on the polarisation fractions and \CP asymmetries for each of these final states, and for some of them the experimental signature is quite challenging.
In addition, the $\phi$ meson is a superposition of octet and singlet,  therefore the ``control channels'' involving $K^*$ and $\rho$ are not as simply related as in the case with a pseudoscalar meson, but require the usage of nonet symmetry, whose precision has to be investigated in turn.

Nevertheless, significant progress can be expected. 
Several $\B \to \Jpsi\, V$ modes, including $B^0_{(s)} \to \Jpsi K^{*0}$~\cite{LHCb-PAPER-2012-014}, are being studied at LHCb.
While measurements of the modes involving $b \to d$ transitions are expected to exhibit rather large uncertainties at first, the advantage of the proposed method is the long ``lever arm'' due to the relative enhancement $\sim 1/\lambda^2$ in the control channels, so that even moderate precision will be very helpful.

\subsubsection{Future prospects with LHCb upgrade}
\label{sec:bmixing_upgrade}

Current measurements of $\phi_s$  carried out by LHCb in the $\jpsi\,\phi$  and $\jpsi\,\pi^+\pi^-$ final states show no deviation from the SM prediction within uncertainties~\cite{LHCb-PAPER-2012-006,LHCb-CONF-2012-002}, putting strong constraints on NP in \Bs mixing, as discussed in Sec.~\ref{sec:bmixing_constraint}.
Table~\ref{tab:phis} shows the current results with $1.0 \invfb$ and the projected precision for $50 \invfb$ with the upgraded detector.
A precision of $< 10 \mrad$ is expected for $50 \invfb$ with the upgraded detector. 
It is expected that even with this data sample, the main limitation will be statistical: the largest systematic uncertainties on the current measurement (background description, angular acceptance, effect of fixed physics parameters)~\cite{LHCb-CONF-2012-002} are expected to be removed with more sophisticated analyses or to scale with statistics.
Thus changes as small as a factor of two with respect to the SM should be observable with $3\,\sigma$ significance. 
This precision will make it possible either to measure a significant deviation from the SM prediction or otherwise to place severe constraints on NP scenarios.

\begin{table}[!htb]
\begin{center}
  \caption{\small
    LHCb measurements of $\phi_s$. 
    The quoted uncertainties are statistical and systematic, respectively.
  }
\begin{tabular}{cccc}
\hline
Final State & Current value (rad) with $1.0 \invfb$ & Projected uncertainty ($50 \invfb$) \\
\hline
$\jpsi\phi$ &  $-0.001\pm 0.101 \pm 0.027$  &  $0.008$ \\
$\jpsi\pi^+\pi^-$ &  $-0.019\,^{+0.173}_{-0.174}\,^{+0.004}_{-0.003} $ &  $0.014$ \\
Both & $-0.002\pm 0.083\pm 0.027$&  $0.007$\\
\hline
\end{tabular}
\end{center}\label{tab:phis}
\vspace{-4mm}
\end{table}

As discussed in Sec.~\ref{sec:bmixing_penguin}, contributions from doubly CKM-suppressed SM penguin diagrams could have a non-negligible effect on the mixing-induced \CP asymmetry and bias the  extracted value of $\phi_s$.
Naive estimates of the bias are of the order $O(10^{-3})$ only~\cite{Boos:2004xp, Li:2006vq, Gronau:2008cc}, but this must be examined with experimental data using flavour symmetries to exploit control channels.
LHCb can perform an SU(3) analysis using measurements of the decays rates and \CP asymmetries in $\Bs \to \jpsi \Kstarz$, $\Bz \to \jpsi \rho^0$ and $\Bd \to \jpsi \phi$ as control channels for $\Bs \to \Jpsi \phi$.
The necessary high precision can only be reached using the large data sample that will be collected with the upgraded LHCb detector.
The $50 \invfb$ data sample will also allow to measure $\phi_s$ in the penguin-free ($b \to c \bar{u} s / u \bar{c} s$) $\Bs \to D \phi$ decay~\cite{Fleischer:2003aj,Nandi:2011uw}.

Another important goal is a more precise determination of $\sin2\beta$ in the \Bz system, motivated by the tension between the direct and indirect determinations of $\sin2\beta$ seen by both UTfit and CKMfitter groups, as  shown in Table~\ref{tab:SMfullfittable}. 
With the upgraded detector, using the $\Bz \rightarrow \jpsi\,\KS$ final state alone, a statistical precision of $\pm 0.006$ is expected, to be compared to the current error from the $B$ factories of $\pm 0.023$~\cite{Nakamura:2010zzi}. 
Given experience with the current detector it seems feasible to control the systematic uncertainties to a similar level. 
Such precision, together with better control of the penguin pollution, will allow us to pin down any NP effects in \Bz mixing.
In addition, the penguin-free ($b \to c \bar{u} d / u \bar{c} d$) $\Bz \to D\rho^0$ channel can be used to get another handle on $\sin2\beta$~\cite{Charles:1998vf,Latham:2008zs}.


The importance of improved measurements of $\Delta\Gamma_q$ has been emphasised in Sections~\ref{sec:bmixing_observables}--\ref{sec:bmixing_constraint}.
LHCb has made a preliminary measurement of $\Delta\Gamma_s$ in $\Bs\to \jpsi\,\phi$  using a $1.0 \invfb$ data sample~\cite{LHCb-CONF-2012-002}. 
The effective lifetime of $\Bs \to \jpsi\,f_0(980)$~\cite{Fleischer:2011cw} has also been measured~\cite{LHCb-PAPER-2012-017}.
Based on this,  the statistical precision on $\Delta\Gamma_s$ with $50 \invfb$ is projected to be $\sim 0.003 \invps$. 
It is hoped that the systematic uncertainty can be controlled to the same level.

A measurement of $\Delta\Gamma_d$ is of interest as any result larger than the tiny value expected in the SM would clearly signal NP~\cite{Dighe:2007gt,Dighe:2001gc,Gershon:2010wx}. 
To determine this quantity, LHCb will compare the effective lifetimes of the two decay modes $\Bz\to \jpsi\,\KS$ with $\Bz\to \jpsi\,K^{*0}$. 
The estimated precision for $1.0 \invfb$ is $\sim 0.02 \invps$.  
With the upgraded detector and $50 \invfb$ a statistical precision of $\sim 0.002 \invps$ on $\Delta\Gamma_d$ can be achieved. 
The systematic uncertainty is under study.

The LHCb upgrade will also have sufficient statistics to make novel tests of \CPT symmetry. 
Any observation of \CPT violation indicates physics beyond the SM. 
An example of a unique test in the \Bz system uses $\Bz\to\jpsi\,K^0$ and its charge-conjugate decay, where the $K^0$ decays semileptonically~\cite{Kostelecky:1996fk,Kayser:1997rk,Du:1999wk}. 
This measurement involves looking at four separate decay paths that interfere. 
While several tests can be performed, one particular observable is the asymmetry $A_{bk}$, that can be measured without the need of flavour tagging, where
\begin{equation}
  A_{bk}=\frac{\Gamma(\Bz+\Bdb\to\jpsi[\pi^-\mu^+\nu]-\Gamma(\Bz+\Bzb\to\jpsi[\pi^+\mu^-\overline{\nu}])}{\Gamma(\Bz+\Bzb\to\jpsi[\pi^-\mu^+\nu]+\Gamma(\Bz+\Bzb\to\jpsi[\pi^+\mu^-\overline{\nu}])} \, .
\end{equation}
In terms of the \CPT violation parameter $\theta^\prime$, the kaon decay time $t_K$, the \Bz decay time $t_B$, the \Bz mass difference $\Delta m_d$ and \CP-violating phase $2\beta$, and kaon decay widths $\Gamma^K_S$ and $\Gamma^K_L$, this can be expressed
\begin{equation}
A_{bk}=\Re(\theta^\prime)\frac{2e^{-\frac{1}{2}\left(\Gamma^K_S+\Gamma^K_L\right)t_K}
\sin2\beta\left(1-\cos\Delta m_dt_B\right)}{e^{-\Gamma^K_S t_K}+e^{-\Gamma^K_L t_K}}\,.
\end{equation}
A signature of \CPT violation would be a $1-\cos\Delta m_Bt_B$ dependence of the decay rate after integrating over kaon decay times. 
Roughly 5000 such decays can be expected with the upgrade. 
It is possible to detect these decays with low background level, even with the missing neutrino, using the measured \Bz direction, the detected \jpsi four-momentum, and the kaon decay vertex.
Other methods to test \CPT symmetry (\eg\ Ref.~\cite{Kundu:2012mg}) are also under investigation.

\subsection{\CP\ violation measurements with hadronic $\bquark \to s$  penguins}
\label{sec:b2spenguin}

\subsubsection{Probes for new  physics in penguin-only $\bquark \to \squark q \bar{q}$ decays}

The presence of physics beyond the SM can be detected by looking for its contribution to $b\to s \qqbar$ ($q=s,\,d$) decays,\footnote{
  Decays mediated primarily by $b \to s u\bar{u}$ transitions are discussed in Secs.~\ref{sec:gammafromloop} and~\ref{sec:bto3h}.
} which in the SM can only proceed via FCNC loop diagrams that are dynamically suppressed.
These decays provide a rich set of observables that are rather precisely known in the SM but could potentially receive sizeable corrections from new heavy particles appearing in the loop.

\begin{itemize}

\item {\bf \boldmath Direct \CP asymmetries}. 
  In the SM $b\to s \qqbar$ decays are dominated by the penguin diagram with an internal top quark. 
  As a consequence, the direct \CP asymmetry is expected to be small.
  If there is a NP amplitude with comparable size interfering with the SM amplitude, and it has different strong and weak phases than the SM amplitude, a much larger direct \CP asymmetry can arise.

\item {\bf Polarisation and triple product asymmetries}. 
  For \B decays into two vector mesons $V_1$ and $V_2$, followed by vector to two pseudoscalar decays $V_1 \to P_1 P^{\prime}_1$ and $V_2 \to P_2 P^{\prime}_2$, there are three transversity states, labelled ``longitudinal'' ($0$), ``perpendicular'' ($\perp$) and ``parallel'' ($\parallel$).
  Measurements of the fractions of the total decay rate in each of these states, which correspond to determinations of the polarisation in the final state, provide useful information about the chiral structure of the electroweak currents, as well about non-perturbative effects such as rescattering and penguin annihilation.
  In the SM, the decay to each transversity state is dominated by a single amplitude with magnitude $|A_j|$, weak phase $\Phi_j$ and strong phase $\delta_j$.  
  The \CP-violating observables $\Im (A_{\perp}A^*_{j}-{\bar A}_{\perp}{\bar A}^*_{j})$ are then
\begin{equation}
  \Im (A_{\perp}A^*_{j}-{\bar A}_{\perp}{\bar A}^*_{j}) = 2 |A_{\perp}||A_{j}|
  \cos(\delta_{\perp}-\delta_{j})\sin(\Phi_{\perp}-\Phi_{j}),\,\, j=0,\parallel\,.
\end{equation}
The values of these observables are tiny since in the SM the weak phases are the same to a very good approximation, but $\Im (A_{\perp}A^*_{j}-{\bar A}_{\perp}{\bar A}^*_{j})$ can significantly differ from zero  if there is a sizeable \CP-violating NP contribution in the loop. 

These observables can be extracted from the differential distributions in terms of the angles $\theta_1$, $\theta_2$ and $\phi$, where $\theta_1$ ($\theta_2$) is the polar angle of $P_1$ ($P_2$) in the rest frame of $V_1$ ($V_2$) with respect to the opposite of the direction of motion of the \B meson,
and $\phi$ is the angle between the decay planes of $V_1 \to P_1P^{\prime}_1$ and $V_2 \to P_2 P^{\prime}_2$ in the rest frame of the \B meson.
The two observables can also be related to two triple product asymmetries for \CP-averaged decays\footnote{
  The triple product asymmetries in $\Bs \to \phi \phi$ and $\Bs \to \Kstarz \Kstarzb$ decay could in principle also receive contribution from non-zero mixing-induced \CP asymmetries arising from NP in \Bs mixing.
  However, this contribution is suppressed by $\DGs/\Gamma_s$ and is already highly constrained.
}
which are equal to asymmetries between the number of events with positive and negative values of $U=\sin 2\phi$ and $V={\rm sign}(\cos\theta_1 \cos\theta_2)\sin \phi$:
\begin{equation}
 \Im (A_{\perp}A^*_{\parallel}-{\bar A}_{\perp}{\bar A}^*_{\parallel}) \propto A_U = \frac{N(U>0)-N(U<0)}{N(U>0)+N(U<0)}\,,
\end{equation}
\begin{equation}
 \Im (A_{\perp}A^*_{0}-{\bar A}_{\perp}{\bar A}^*_{0}) \propto A_V = \frac{N(V>0)-N(V<0)}{N(V>0)+N(V<0)}\,.
\end{equation}
A review of this subject can be found in Ref.~\cite{Gronau:2011cf} and references therein.

\item {\bf \boldmath Mixing-induced \CP asymmetries}. 
  Mixing-induced \CP asymmetries in $b\to s \qqbar$ decays of neutral \B to \CP eigenstates are precisely predicted.
  Due to the fact that the penguin diagram with an internal top quark is expected to dominate, the values of $2\beta^{\rm eff}$ determined using $\Bd \to \phi \KS$, $\Bd \to \eta^\prime \KS$, $\Bd \to f_0(980)\KS$, \etc, are all expected to give $\approx 2\beta$ (see, \eg Refs.~\cite{Beneke:2005pu,Cheng:2005bg} and the discussion in Ref.~\cite{HFAG}).
  Similarly, the values of $2\beta_s^{\rm eff}$ determined from $\Bs \to \phi \phi$, $\Bs \to K^{*0} {\bar K}^{*0}$, \etc, are expected to vanish due to cancellation of weak phases between mixing (top box) and decay (top penguin) amplitudes. 
  Higher order corrections from subleading diagrams are expected to be small compared to the precision that can be achieved in the near-term, but further theoretical studies will be needed as the upgrade era approaches.
  NP with a flavour structure different from the SM will alter these \CP asymmetries through the decay amplitudes, even if there is no NP in \B mixing. 
  A number of quasi-two-body or three-body decay modes can be studied. 

\item {\bf Correlations between direct and mixing-induced asymmetries}.
  Penguin-only decay modes are particularly interesting as the difference between formal ``tree'' and ``penguin'' contributions boils down to a difference in the quark-flavour running in the loop of the penguins. 
  This difference, dominated by short distances, can be assessed accurately using QCD factorisation, and it can be used to correlate the branching ratio and the
\CP asymmetries of penguin-mediated modes. 
As discussed in Refs.~\cite{DescotesGenon:2006wc,DescotesGenon:2007qd,DescotesGenon:2011pb}, these observables can be correlated not only within the SM, but can also be used to extract the \Bs mixing phase even in the presence of NP affecting only this phase.

\end{itemize}

\subsubsection{Current status and outlook of LHCb measurements}
%

\begin{figure}[h]
\includegraphics[width=0.44\textwidth]{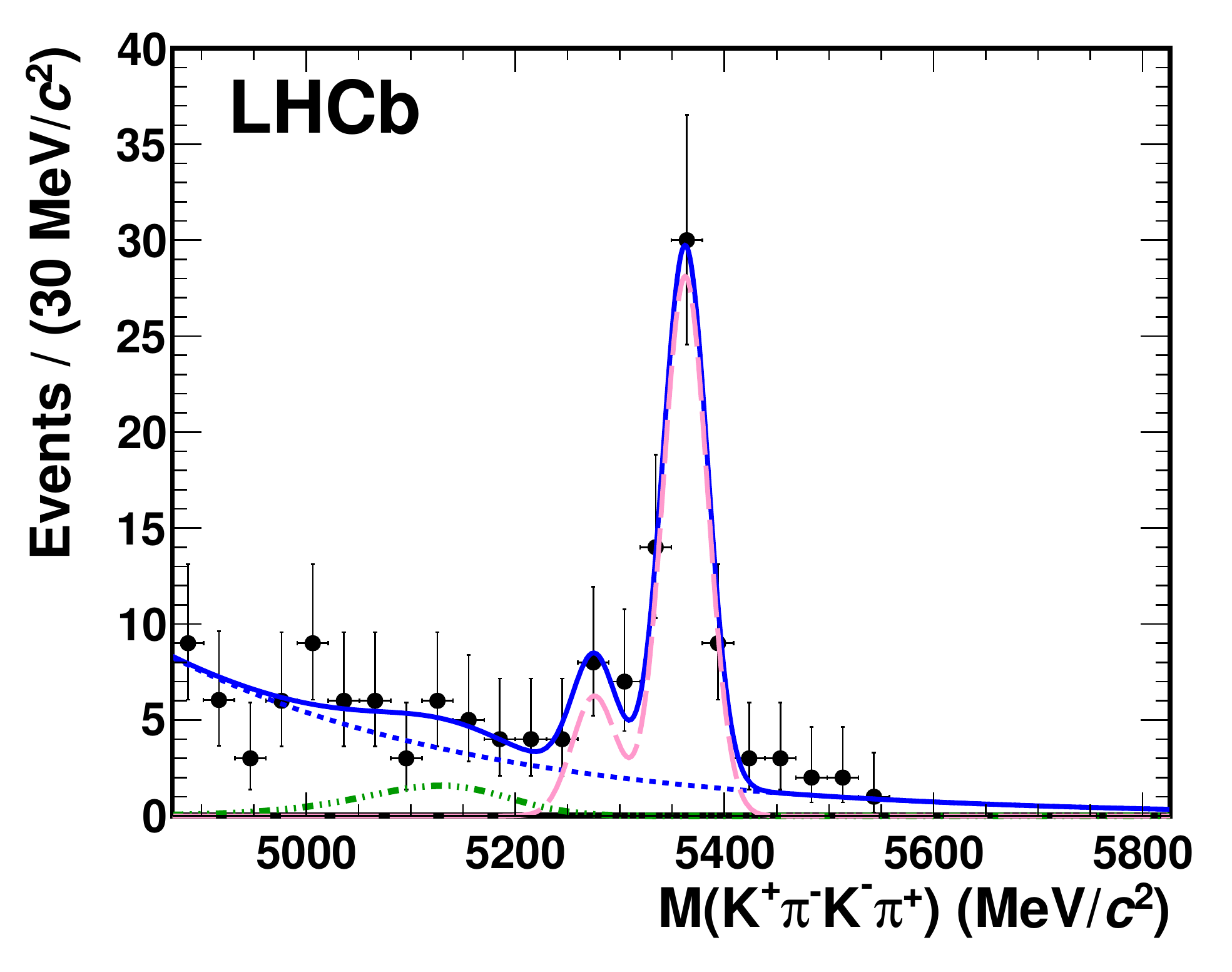}
\includegraphics[width=0.49\textwidth]{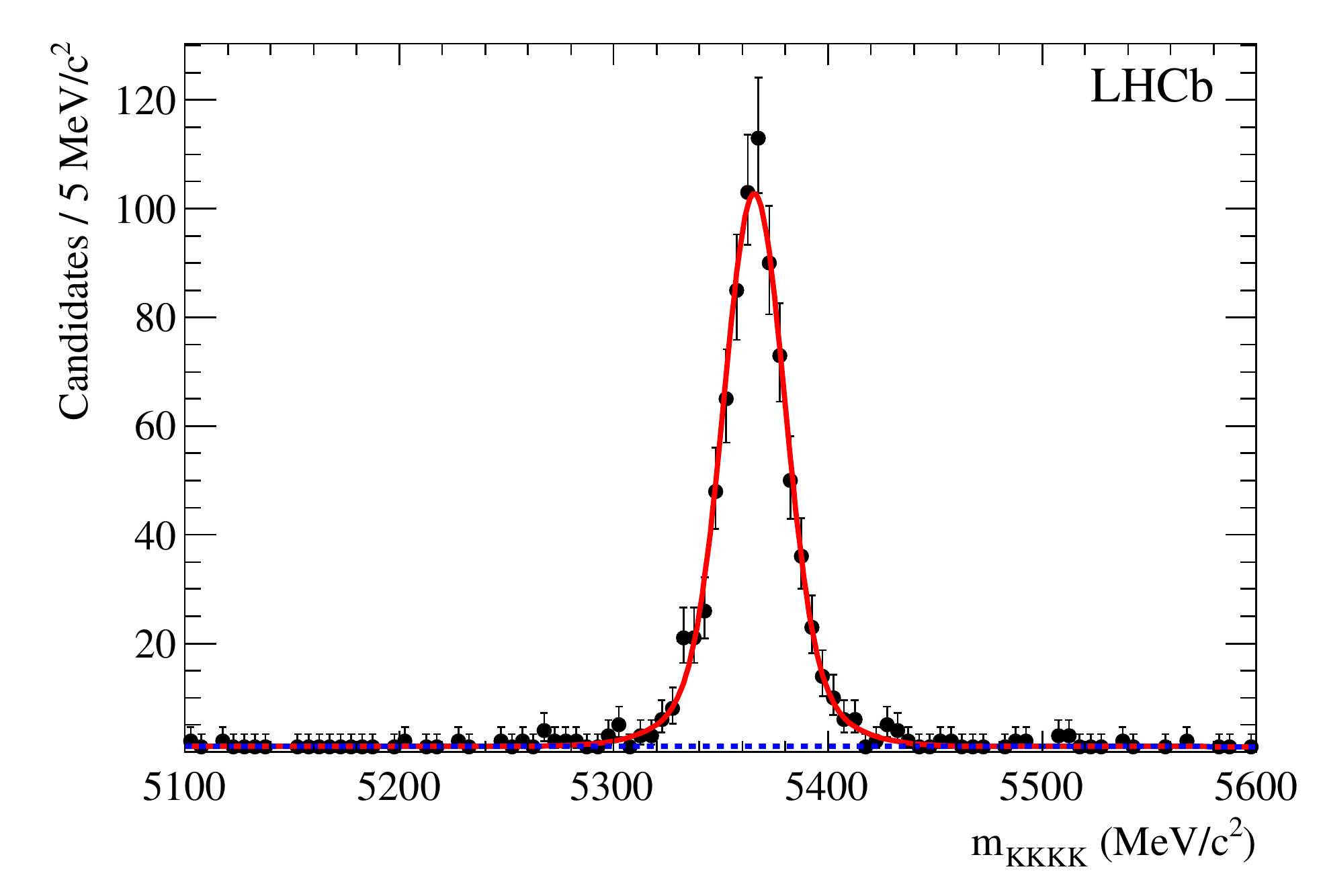}
\caption{\small 
  (Left) fit of the $K^+ \pi^- K^- \pi^+$ mass distribution for $\Bs\to\Kstarz\Kstarzb$ candidates from $35 \invpb$~\cite{LHCb-PAPER-2011-012};
  (right) fit of the $K^+ K^- K^- K^+$ mass distribution for $\Bs \to \phi\phi$ candidates from $1.0 \invfb$~\cite{LHCb-PAPER-2012-004}.}
\label{hmassplots}
\end{figure}

LHCb published the first observation and measurement of the branching ratio and polarisation amplitudes in the $\Bs \to \Kstarz \Kstarzb$ decay mode~\cite{LHCb-PAPER-2011-012} using $35~\invpb$ of data collected in 2010. 
A clean mass peak corresponding to  $50 \pm 8$ $ B^0_s \rightarrow (K^+ \pi^-)(K^+ \pi^+)$ decays is seen (Fig.~\ref{hmassplots} (left)), mostly from resonant $\Bs \to \Kstarz \Kstarzb$ decays. 
Using this signal the longitudinal polarisation amplitude is measured to be $f_{L} = 0.31 \pm 0.12({\rm stat}) \pm 0.04({\rm syst})$ and the branching ratio to be $\BR(\Bs \to \Kstarz \Kstarzb) =  (2.81 \pm 0.46({\rm stat}) \pm 0.45({\rm syst})) \times 10^{-5}$.

LHCb also published  the measurement of the polarisation amplitudes
and triple product asymmetries in $\Bs \rightarrow \phi \phi$~\cite{LHCb-PAPER-2012-004} using the  2011 data set of $1.0 \invfb$. 
In this data set $801 \pm 29$ events are observed with excellent signal-to-background ratio (see Fig.~\ref{hmassplots} (right)). 
The polarisation amplitudes are measured to be
\begin{equation}
\begin{array}{lcl}
|A_0|^2 & = & 0.365 \pm 0.022\, ({\rm stat}) \pm 0.012\, ({\rm syst})\,, \\
|A_{\perp}|^2 & = & 0.291 \pm 0.024\, ({\rm stat}) \pm 0.010\, ({\rm syst})\,, \\
|A_{\parallel}|^2 & = & 0.344 \pm 0.024\, ({\rm stat}) \pm 0.014\, ({\rm syst}) \,,
\end{array}
\end{equation}
where the sum of the square of the amplitudes is constrained to unity. 
The triple product asymmetries in this mode are measured to be
\begin{equation}
\begin{array}{lcl}
A_U & = & -0.055 \pm 0.036\,  ({\rm stat}) \pm 0.018\, ({\rm syst})\,, \\
A_V & = & \phantom{-}0.010  \pm 0.036\,  ({\rm stat}) \pm 0.018\, ({\rm syst}) \,.
\end{array}
\end{equation}
The results of this analysis are in agreement with, and more precise than, the previous measurement~\cite{Aaltonen:2011rs}, and are also consistent with the SM.
 
First measurements of \CP asymmetries in these modes from time-dependent flavour-tagged angular analyses are expected to follow.
With high statistics, it will be possible to measure polarisation-dependent direct and mixing-induced \CP asymmetries, but for the first analysis it will be more convenient to determine a single complex observable common to all polarisations (as done for $\Bs \to \jpsi\,\phi$)
\begin{equation}
  \lambda = \eta_j \frac{q}{p} \frac{\bar A_j}{A_j} 
\end{equation}
where $j$ denotes one of the three transversity states, which are also \CP eigenstates with eigenvalues $\eta_j$, and $A_j$ ($\bar A_j$) is the decay amplitude of \Bs (\Bsb) to the corresponding state.
With this approximation it will be possible to determine the magnitude $|\lambda|$ and phase $\phi^{\rm eff}_s \equiv -\arg(\lambda)$. 
The SM expectation is $|\lambda| \approx 1$ and $\phi^{\rm eff}_s \approx 0$ due to the dominance of the top-quark loop, and any observed deviation from these expectations would be a signature of NP.
Since  NP in \Bs\ mixing is already constrained by measurement of $\phi_s$ from $\Bs \to \jpsi\,\phi$, the main interest in these $b \to s$ penguin modes is to look for NP in the decay processes. 
Based on simulation studies, a sensitivity on $\phi^{\rm eff}_s$ of 0.3--0.4 radians with $1.0 \invfb$ is expected for both $\Bs \to \phi\phi$ and $\Bs \to \Kstarz \Kstarzb$.

\subsubsection{Future prospects with LHCb upgrade}

\begin{figure}[!htb]
\centering
\includegraphics[width=0.48\textwidth]{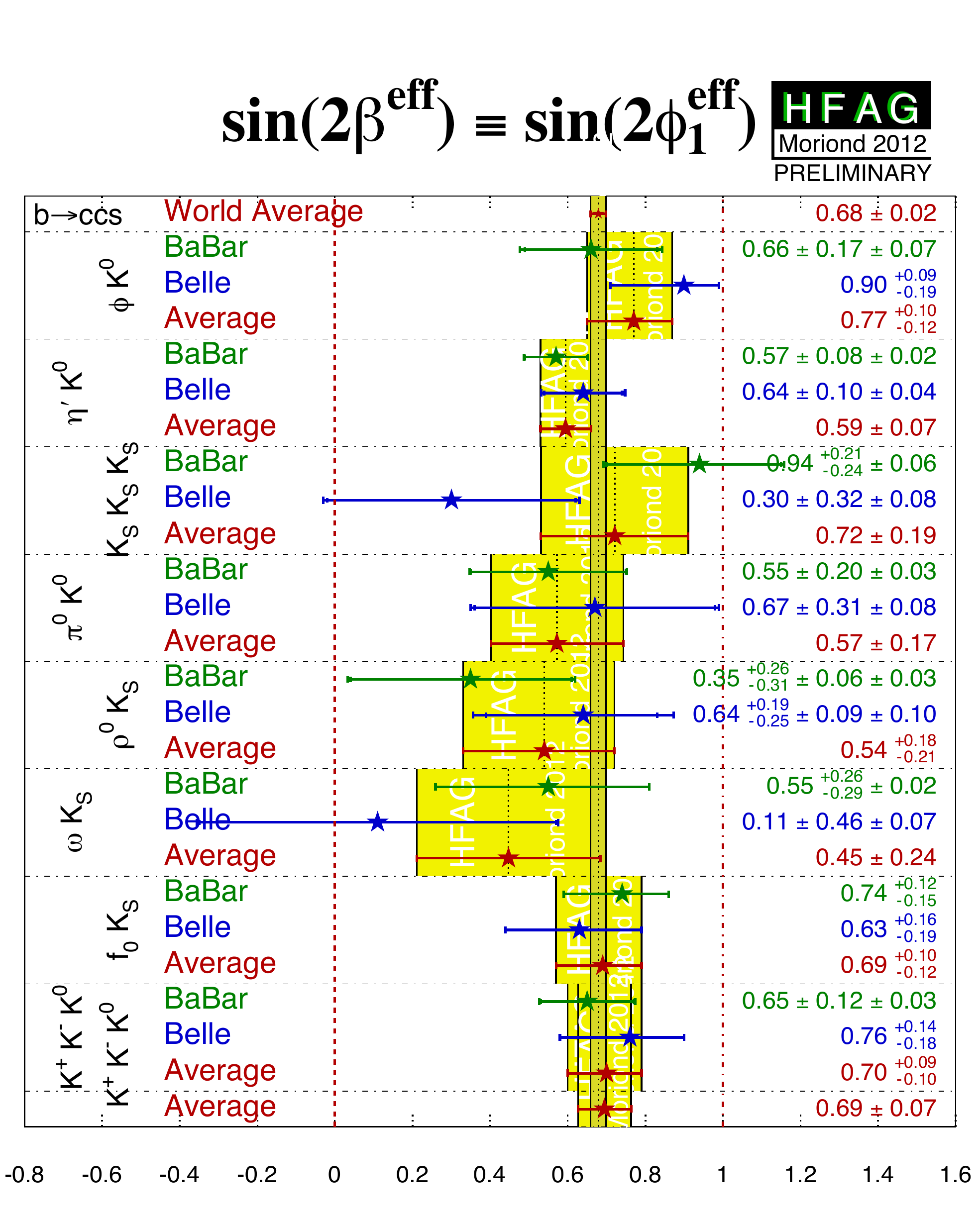}
\caption{\small
  HFAG compilation of results for $ \sin 2\beta^{\rm eff}$ in $b \to s \qqbar$ decays~\cite{HFAG}. 
}
\label{fig:b2sqq:phid}
\end{figure}

The latest results on mixing-induced \CP violation in $b \to s$ transitions show no significant deviation from the SM, as seen in Fig.~\ref{fig:b2sqq:phid}, which compares the mixing-induced \CP violation parameter $\sin 2\beta^{\rm eff}$ measured in penguin-dominated $b \to s$ decays with the value of $\sin 2\beta$ measured in the tree-dominated $b \to \ccbar s$ decays. 
In the absence of NP these observables should only differ by small amounts. 
Due to these results, large NP contributions in $b \to s \qqbar$ decays are unlikely but further tests with higher precision remain interesting. 
LHCb will be able to make competitive measurements of $\sin 2\beta^{\rm eff}$ in $\Bd \to \phi \KS$ and several other $b \to s \qqbar$ decays, but a significant improvement in precision requires the $50 \invfb$ of the LHCb upgrade.
The improved trigger efficiency in the LHCb upgrade is particularly important for these decays, which have only hadrons in the final state.
With the upgrade data sample, the statistical error of $\sin 2\beta^{\rm eff}(\Bd \to \phi \KS)$ is estimated to be roughly 0.06, which is still above the SM uncertainty of $\sim 0.02$~\cite{Bartsch:2008ps}.
 
There are several more NP probes in $b \to s \qqbar$ decays that can be exploited at LHCb and its upgrade, such as mixing-induced \CP asymmetries and triple product asymmetries in both $\Bs \to \phi \phi$ and $\Bs \to \Kstarz \Kstarzb$ decays. 
The statistical precision of $\phi_s^{\rm eff}$ with each channel is estimated to be 0.3--0.4 rad for $1.0 \invfb$.
The projected precision for $50 \invfb$ is about 0.03 rad each. 
This can be compared with the uncertainties of their SM predictions of about 0.02 rad. 
It is also possible to perform a combined analysis of $\Bs \to \Kstarz \Kstarzb$ and its U-spin related channel $\Bd \to \Kstarz \Kstarzb$, which will put strong constraint on  the subleading penguin diagrams in $\Bs \to \Kstarz \Kstarzb$, thus further reducing the theoretical uncertainty in the measurement of $\phi^{\rm eff}_s$~\cite{Fleischer:1999zi,Ciuchini:2007hx}.
The statistical precision of $A_U$ and $A_V$ is estimated to be about 0.004, compared with an upper bound of 0.02 on their possible sizes in the SM~\cite{Gronau:2011cf}. 

In summary, the LHCb upgrade will allow the exploitation of the full potential of the NP probes in $b \to s \qqbar$ decays.
Table~\ref{tab:b2sqq:upgrade:sensitivities} compares the current and projected (LHCb upgrade, \ie\ $50 \invfb$) precisions of the key observables  with the theory uncertainties of their SM predictions.

\begin{table}[!htb]
  \begin{center}
    \caption{\small
      Current and projected precisions of the key observables in $b \to s \qqbar$ decays.
    }
    \label{tab:b2sqq:upgrade:sensitivities}
    \resizebox{0.99\textwidth}{!}{
    \begin{tabular}{|c|c|c|c|}
\hline
Observable & Current   &  LHCb upgrade ($50 \invfb$)  & Theory uncertainty    \\
\hline
$A_{U,\,V}(\Bs \to \phi\phi)$ & 0.04 (LHCb $1.0 \invfb$)  & 0.004 & $0.02$~\cite{Datta:2003mj} \\
$\phi^{\rm eff}_s(\Bs \to \phi\phi)$ & -- & 0.03 & $0.02$~\cite{Bartsch:2008ps} \\
$\phi^{\rm eff}_s(\Bs \to K^{*0} \bar{K}^{*0})$ & -- & $0.03$ & $0.02$~\cite{Bartsch:2008ps} \\
$\sin 2\beta^{\rm eff}(\Bd \to \phi \KS)$ & $0.12$ (\B factories)  & 0.06 & $0.02$~\cite{Buchalla:2008jp} \\
\hline
    \end{tabular}
    }
  \end{center}
\end{table}

\subsection[Measurements of the CKM angle gamma]{Measurements of the CKM angle $\gamma$}
\label{sec:gamma}


\subsubsection{Measurements of $\gamma$ using tree-mediated decays}
\label{sec:gammafromtrees}

The CKM angle $\gamma$, defined as the phase $\gamma = \arg \left[ -V_{ud}V_{ub}^*/(V_{cd}V_{cb}^*) \right]$, is one of the angles of the unitarity triangle formed from the hermitian product of the first ($d$) and third ($b$) columns of the CKM matrix $V$. 
It is one of the least well known parameters of the quark mixing matrix.
However, since it can be determined entirely through decays of the type $B \to DK$~\footnote{ 
  By $B \to DK$ all related tree-dominated decay processes are implicitly included, including $\Bp \to D\Kp$, $\Bd \to D\Kstarz$, $\Bs \to D\phi$, $\Bs \to \Dsmp\Kmp$ and $\Bd \to D^{(*)\mp}\pipm$.
  In these specific decay processes, the notation $D$ refers to a neutral $D$ meson that is an admixture of $\Dz$ and $\Dzb$ states.
} 
that involve only tree amplitudes --- an unusual, even unique, property amongst all \CP violation parameters --- it provides a benchmark measurement.
The determination from tree level decays has essentially negligible theoretical uncertainty, at the level of $\delta\gamma/\gamma = {\cal O}(10^{-6})$, as will be shown in the next section.
This makes $\gamma$ a very appealing ``standard candle'' of the CKM sector. 
It serves as a reference point for comparison with $\gamma$ values measured from loop decays (see Sec.~\ref{sec:gammafromloop}).

Moreover, the determination of $\gamma$ is crucial to improve the precision of the global CKM fits, and resulting limits on (or evidence for) NP contributions (see Sec.~\ref{sec:globalfit}).
In particular, the measurement of $\Delta m_d$ and the oscillation phase $\sin 2\beta$ in \Bz--\Bzb mixing can be converted to a measurement of $\gamma$ (in the SM). 
This can be compared to the reference value from $B \to DK$ --- their consistency verifies that the Kobayashi-Maskawa mechanism of \CP violation is the dominant source in quark flavour-changing processes. 
Existing measurements provide tests at the level of ${\mathcal O}(10\,\%)$, but improving the precision to search for smaller effects of NP is well motivated.

Several established methods to measure $\gamma$ in tree decays exploit the $B^-\to D^{(*)}K^{(*)-}$ decays.
They are based on the interference between the $b\to u$ and $b\to c$ tree amplitudes, which arises when the neutral $D$ meson is reconstructed in a final state accessible to both $\Dz$ and $\Dzb$ decays.
The interference between the amplitudes results in observables that depend on their relative weak phase $\gamma$. 
Besides $\gamma$ they also depend on hadronic parameters, namely the ratio of magnitudes of amplitudes $r_B \equiv \left| A(b\to u) / A(b\to c)\right|$ and the relative strong phase $\delta_B$ between the two amplitudes.
These hadronic parameters depend on the $B$ decay under investigation. 
They can not be precisely calculated from theory (see, however, Ref.~\cite{Aubin:2011kv}), but can be extracted directly from data by simultaneously reconstructing several different $D$ final states.
%

The various methods differ by the $D^{(*)}$ final state that is used. 
The three main categories of $D$ decays considered so far by the $B$ factories \babar\ and Belle, and by CDF, are:
\begin{itemize}
\item \CP eigenstates (the GLW method~\cite{Gronau:1990ra,Gronau:1991dp}),
\item doubly Cabibbo-suppressed (DCS) decays (the ADS method~\cite{Atwood:1996ci,Atwood:2000ck}),
\item three-body, self-conjugate final states (the GGSZ or ``Dalitz'' method~\cite{Giri:2003ty}).
\end{itemize}
An additional category has not been possible to pursue at previous experiments due to limited event sample sizes:
\begin{itemize}
\item singly Cabibbo-suppressed (SCS) decays (the GLS method~\cite{Grossman:2002aq}).
\end{itemize}
In practise, except for the case of two-body decays, there is often no clear distinction between the different methods.  

The best sensitivity to $\gamma$ obviously comes from combining the results of all different analyses.  
This not only improves the precision on $\gamma$, but provides additional constraints on the hadronic parameters.  
It also allows one to overcome the fact that \CP-odd final states such as $\KS\piz$ are not easily accessible in LHCb's hadronic environment.
%
%

A brief review of the main ideas of the different methods follows.
The amplitudes of the $B^-\rightarrow D^0 K^-$ and $B^-\rightarrow \Dzb K^-$ processes are written as:
\begin{eqnarray}
A(B^-\rightarrow D^0 K^-) = & A_c e^{i\delta_c} \,,~~~~ & A(D^0\rightarrow f)=A_fe^{i\delta_f}\,,\\
A(B^-\rightarrow \Dzb K^-) = & A_u e^{i(\delta_u-\gamma)}\,, &A(D^0\rightarrow\bar{f})=A_{\bar{f}}e^{i\delta_{\bar{f}}}\,, \nonumber
\end{eqnarray}
where $A_c$, $A_u$, $A_f$ and $A_{\bar{f}}$ are real and positive (and \CP violation in \Dz decays has been neglected).
The subscripts $c$ and $u$ refer to the $b\rightarrow c$ and $b\rightarrow u$ transitions, respectively.
The amplitudes for the $D^0$ decay can generally include the case where the $D^0$ decays to a three-body final state.
In this case, $A_f$, $A_{\bar{f}}$, $\delta_f$ and $\delta_{\bar{f}}$ are functions of the Dalitz plot coordinates.
The amplitude of the process $B^-\to D[\rightarrow f] K^-$ can be written, neglecting \Dz--\Dzb mixing, as
\begin{equation}
  \label{eq:amp2}
  A(B^-\rightarrow D[\rightarrow f]K^-) =
  A_c A_fe^{i(\delta_c+\delta_f)}+A_uA_{\bar{f}}e^{i(\delta_u+\delta_{\bar{f}}-\gamma)}\, ,
\end{equation}
and the rate is given by
\begin{eqnarray}
\Gamma(B^-\rightarrow D[\rightarrow f]K^-)&\propto&A_c^2A_f^2 +
	A_u^2A_{\bar{f}}^2 +2A_cA_fA_uA_{\bar{f}}
	\Re(e^{i(\delta_B+\delta_D-\gamma)})\nonumber\\
	&\propto&A_c^2\left(A_f^2 +
	r_B^2A_{\bar{f}}^2 +2r_B A_f A_{\bar{f}}
	\Re(e^{i(\delta_B+\delta_D-\gamma)})\right)\,,
	\label{eq:rate2}
\end{eqnarray}
where $r_B=A_u/A_c$, $\delta_B=\delta_u-\delta_c$ and $\delta_D=\delta_{\bar{f}}-\delta_f$. 
The rate for the charge-conjugated mode (still neglecting \CP violation in \Dz decays) is obtained by exchanging $\gamma \rightarrow -\gamma$.
Taking into account CKM factors and, in the case of charged $B$ decays, colour suppression of the $b\to u$ amplitude, $r_B$ is expected to be around 0.1 for $B^-$ decays and around 0.3 for $B^0$ decays.
From Eq.~(\ref{eq:rate2}) all the relevant formulae of the GLW, ADS and GGSZ methods can be derived.

In the GLW analysis, the neutral $D$ mesons are selected in \CP eigenstates $f_{\CP\pm}$ such as $D\rightarrow K^-K^+$ ($\CP=+1$) or $D\rightarrow \KS\pi^0$ ($\CP=-1$). 
Thus $A_f/A_{\bar{f}}=1$ and $\delta_D=0,\pi$ for $\CP=\pm1$.
Eq.~(\ref{eq:rate2}) becomes:
\begin{equation}
  \label{eq:glw1}
  \Gamma(B^-\rightarrow D[\rightarrow f_{\CP\pm}]K^-)\propto A_c^2(1+r_B^2\pm 2r_B\cos(\delta_B-\gamma))\,.
\end{equation}
The $\Bm\to D\Km$ decays, where the $D$ decays to Cabibbo-favoured (CF) final states (\eg\ $D^0\to K^-\pi^+$) can be used to normalise the rates in order to construct observables that minimise the systematic uncertainties.
For those decays, to a good approximation,
\begin{equation}
  \label{eq:CA_finalstate}
  \Gamma(B^-\rightarrow D[\rightarrow K^-\pi^+]K^-) 
  = \Gamma(B^+\rightarrow D[\rightarrow K^+\pi^-]K^+) \propto A_c^2~.
\end{equation}
From Eqs.~(\ref{eq:glw1}) and~(\ref{eq:CA_finalstate}) and their \CP conjugates the usual GLW observables follow:
\begin{align}
  R_{\CP\pm} & = \frac{2[\Gamma(B^-\rightarrow D_{\CP\pm}K^-)+\Gamma(B^+\rightarrow D_{\CP\pm}K^+)]}
                       {\Gamma(B^-\rightarrow D^0 K^-)+\Gamma(B^+\rightarrow \Dzb K^+)}
                       \label{eq:glw2def} \\
  A_{\CP\pm} & =   \frac{\Gamma(B^-\rightarrow D_{\CP\pm}K^-)-\Gamma(B^+\rightarrow D_{\CP\pm}K^+)}
                       {\Gamma(B^-\rightarrow D_{\CP\pm}K^-)+\Gamma(B^+\rightarrow D_{\CP\pm}K^+)}
                       \label{eq:glw3def}\,.
\end{align}
Eqs.~(\ref{eq:glw2def}) and~(\ref{eq:glw3def}) provide a set of four observables that are connected to the three unknowns $\gamma$, $r_B$ and $\delta_B$ through
\begin{align}
  R_{\CP\pm} & = 1+r_B^2\pm2 r_B\cos\delta_B\cos\gamma \label{eq:glw2} \\
  A_{\CP\pm} & = \frac{\pm2r_B\sin\delta_B\sin\gamma}{R_{\CP\pm}} \label{eq:glw3}\,.
\end{align}
However, only three of these equations are independent since, from Eq.~(\ref{eq:glw3}), $R_{\CP+}A_{\CP+}=-R_{\CP-}A_{\CP-}$.
Analogous relations hold for $B\to D^*_{\CP}K$ and $B\to D_{\CP}K^*$ decays, with different values of the hadronic parameters characterising the $B$ decay. 
However, in the $B\to D^*_{\CP}K$ case one has to take into account a \CP flip due to the different charge conjugation quantum numbers of the \piz and the photon from the $D^*$ decay~\cite{Bondar:2004bi}: $D^*_{\CP\pm}\to D_{\CP\pm}\pi^0$, but $D^*_{\CP\pm}\to D_{\CP\mp}\gamma$. 
For analysis of $B \to D_{\CP}K^*$ the finite width of the $K^*$ resonance must be taken into account~\cite{Gronau:2002mu}.
There are related important consequences for the ADS and GGSZ analyses of $B\to D^*K$ and $B \to DK^*$ decays.

%
%

In the ADS analysis, the neutral $D$ mesons are selected in CF and DCS decays, such as $\Dz\to\Km\pip$ and $\Dz\to\pim\Kp$, respectively.
The $B$ decay rate is the result of the interference of the colour allowed $B^-\rightarrow D^0K^-$ decay followed by the DCS $\Dz\to\pim\Kp$ decay and the colour suppressed $B^-\rightarrow \Dzb K^-$ decay followed by the CF $\Dz\to\Km\pip$ decay. 
As a consequence, the interfering amplitudes are of similar magnitude and hence large interference effects can occur.
From Eq.~(\ref{eq:rate2}) one finds
\begin{equation}
  \Gamma(B^\mp\rightarrow D[\to K^\pm\pi^\mp]K^\mp) \propto r_B^2+r_D^2\pm 2r_Br_D\cos(\delta_B+\delta_D\mp\gamma)
  \label{eq:ads_rate}
\end{equation}
where both $r_D=A_f/A_{\bar{f}}=|A(\Dz\to\pim\Kp)/A(D^0\rightarrow K^-\pi^+)|$ and the phase difference $\delta_D$ are measured in charm decays. 
The value of $\delta_D$ can be determined directly using data collected from $e^+e^-$ collisions at the $\psi(3770)$ resonance, as has been done by CLEO~\cite{Asner:2008ft,Asner:2012xb}, but the most precise value comes from a global fit including charm mixing parameters.
The results provided by HFAG~\cite{HFAG} from a combination with \CP violation in charm allowed are
$r_D = 0.0575\pm0.0007$, $\delta_D = \left(202\,^{+10}_{-11}\right)^\circ$.
Defining $R_{\rm ADS}$ and $A_{\rm ADS}$ as
\begin{eqnarray}
  &&R_{\rm ADS}=
  \frac{\Gamma(B^-\rightarrow D[\to \pi^-K^+]K^-)+\Gamma(B^+\rightarrow D[\to \pi^+K^-]K^+)}
       {\Gamma(B^-\rightarrow D[\to K^-\pi^+]K^-)+\Gamma(B^+\rightarrow D[\to K^+\pi^-]K^+)} \,, \\
       &&A_{\rm ADS}=
       \frac{\Gamma(B^-\rightarrow D[\to \pi^-K^+]K^-)-\Gamma(B^+\rightarrow D[\to \pi^+K^-]K^+)}
            {\Gamma(B^-\rightarrow D[\to \pi^-K^+]K^-)+\Gamma(B^+\rightarrow D[\to \pi^+K^-]K^+)} \,,
            \label{eq:rads}
\end{eqnarray}
and using Eqs.~(\ref{eq:CA_finalstate}) and~(\ref{eq:ads_rate}) gives
\begin{eqnarray}
  &&R_{\rm ADS}=r_B^2+r_D^2+2r_B\,r_D\cos\gamma\cos(\delta_B+\delta_D)\label{eq:rads2} \,,\\
  &&A_{\rm ADS}=2r_B\,r_D\sin\gamma\sin(\delta_B+\delta_D)/R_{\rm ADS}\label{eq:aads2} \,.
\end{eqnarray}
It has been noted that for the extraction of $\gamma$ it can be more convenient to replace the pair of observables ${R_{\rm ADS}, A_{\rm ADS}}$ with a second pair, ${R_+,R_-}$, defined as:
\begin{equation}
  \label{eq:Rpm_ADS}
  R_\pm \equiv \frac{\Gamma(B^\pm\rightarrow [K^\mp\pi^\pm]_DK^\pm)}
                    {\Gamma(B^\pm\rightarrow [K^\pm\pi^\mp]_DK^\pm)} 
                    = r_B^2+r_D^2 + 2r_Br_D\cos(\delta_B+\delta_D\pm\gamma)
\end{equation}
Unlike ${R_{\rm ADS}, A_{\rm ADS}}$, the two quantities ${R_+,R_-}$ are statistically independent. 
The ADS decay chain $\Bpm\to [\pipm\Kmp]_D\Kpm$ has been observed for the first time by LHCb~\cite{LHCb-PAPER-2012-001}, confirming the evidence that had begun to accumulate in previous measurements~\cite{Belle:2011ac,delAmoSanchez:2010dz,Aaltonen:2011uu}.

%
%

In the GGSZ analysis, the neutral $D$ mesons are selected in three-body self-conjugate final states. 
The channel that has been used most to date is $D\to \KS\pi^+\pi^-$, though first results have also been presented with $D\to \KS K^+K^-$ and other channels are under consideration.
For concreteness, consider $D \to \KS\pi^+\pi^-$, with $A_fe^{i\delta_f}=f(m_-^2,m_+^2)$ and $A_{\bar{f}}e^{i\delta_{\bar{f}}}=f(m_+^2,m_-^2)$, where $m_-^2$ and $m_+^2$ are the squared masses of the $\KS\pi^-$ and $\KS\pi^+$ combinations.
The rate in Eq.~(\ref{eq:rate2}) can be re-written as:
\begin{align}
  \label{eq:dalitzrate1}
  \Gamma(B^\mp\to &D[\to \KS\pi^-\pi^+]K^\mp) \propto |f(m_\mp^2,m_\pm^2)|^2+r_B^2|f(m_\pm^2,m_\mp^2)|^2 \\
  &+ 2r_B|f(m_\mp^2,m_\pm^2)||f(m_\pm^2,m_\mp^2)|\cos(\delta_B+\delta_D(m_\mp^2,m_\pm^2)\mp\gamma)\,,\nonumber
\end{align}
where $\delta_D(m_\mp^2,m_\pm^2)$ is the strong phase difference between $f(m_\pm^2,m_\mp^2)$ and $f(m_\mp^2,m_\pm^2)$.
Due to the fact that $r_B$ is required to be positive, the direct extraction of $r_B$, $\delta_B$ and $\gamma$ can be biased. 
To avoid these biases, the ``Cartesian coordinates'' have been introduced~\cite{Aubert:2005iz}
\begin{equation}
  x_\pm=\Re[r_B e^{i(\delta_B\pm\gamma)}]\,, \quad
  y_\pm=\Im[r_B e^{i(\delta_B\pm\gamma)}]\,,
\end{equation}
allowing Eq.~(\ref{eq:dalitzrate1}) to be rewritten as
\begin{equation}
  \label{eq:dalitzrate2}
  \Gamma(B^\mp\to D[\to \KS\pi^+\pi^-]K^\mp)\propto
  |f_{\mp}|^2+r_B^2|f_{\pm}|^2+2\left[x_\mp \Re[f_{\mp}f_{\pm}^*]+y_\mp \Im[f_{\mp}f_{\pm}^*]\right]\,.
\end{equation}
Here the notation has been simplified using $f_{\pm}=f(m_\pm^2,m_\mp^2)$. 
This Dalitz plot-based method can be implemented in a model-dependent way by parametrising the amplitude as a function of the Dalitz plot of the three-body state, or in a model-independent way by dividing the Dalitz plot into bins and making use of external measurements of the $D$ decay strong phase differences within these bins~\cite{Giri:2003ty,Bondar:2005ki,Bondar:2008hh}.\footnote{
  As for $\delta_D$ in the ADS method, the strong phase differences can be determined directly from $\psi(3770) \to D\bar{D}$ data, which has been done by CLEO~\cite{Libby:2010nu}.  
  In future, it is expected that the most precise value will come from a global fit including results of time-dependent analyses of multibody charm decays.
}

%
%

\def\fdsk{\ensuremath{f}\xspace}
\def\fbdsk{\ensuremath{\bar{f}}\xspace}

\def\ldsk{\ensuremath{\lambda_{\fdsk}}\xspace}
\def\lbdsk{\ensuremath{\overline{\lambda}_{\fbdsk}}\xspace}

\def\Afdsk{\ensuremath{A_{\fdsk}}\xspace}
\def\Afbdsk{\ensuremath{A_{\fbdsk}}\xspace}
\def\Abfdsk{\ensuremath{\overline{A}_{\fdsk}}\xspace}
\def\Abfbdsk{\ensuremath{\overline{A}_{\fbdsk}}\xspace}

\def\Cfdsk{\ensuremath{C_{\fdsk}}\xspace}
\def\Cfbdsk{\ensuremath{C_{\fbdsk}}\xspace}
\def\Dfdsk{\ensuremath{D_{\fdsk}}\xspace}
\def\Dfbdsk{\ensuremath{D_{\fbdsk}}\xspace}
\def\Sfdsk{\ensuremath{S_{\fdsk}}\xspace}
\def\Sfbdsk{\ensuremath{S_{\fbdsk}}\xspace}

\def\normlamsq{\ensuremath{|\ldsk|^2}\xspace}
\def\normlambarsq{|\bar{\lambda}_{\fbdsk}|^{2}}
\def\deltag{\frac{\Delta \Gamma_{s} \, t}{2}}
\def\dMst{\Delta m_{s} \, t}
\def\poverq{\Big|\frac{p}{q}\Big|^{2}}
\def\qoverp{\Big|\frac{q}{p}\Big|^{2}}
\def\weak{\gamma -2\beta_s}

Besides the established methods based on direct \CP violation in $B\to DK$ decays, it is also possible to measure $\gamma$ using time-dependent analyses of neutral \Bd and \Bs tree decays~\cite{Dunietz:1987bv,Aleksan:1991nh,Fleischer:2003yb}.
The method still relies on the interference of $b\to u$ and $b\to c$ amplitudes, but interference is achieved through \Bd (\Bs) mixing. 
Thus one measures the sum of $\gamma$ and the mixing phase, namely $\gamma+2\beta$ and $\gamma-2\beta_s$ in the \Bd and \Bs systems, respectively. 
Since both $\sin 2\beta$ and $\beta_s$ are becoming increasingly well measured, these measurements provide sensitivity to $\gamma$.

Pioneering time-dependent measurements using the $\Bd\to D^{(*)\mp}\pi^\pm$ decays have been performed by both BaBar~\cite{Aubert:2005yf,Aubert:2006tw} and Belle~\cite{Ronga:2006hv,Bahinipati:2011yq}. 
In these decays the amplitude ratios $r_{D\pi} = |A(\Bd \to D^{(*)+}\pi^-)/A(\Bd \to D^{(*)-}\pi^+)|$ are expected to be small, $r_{D\pi} \lesssim 0.02$, limiting the sensitivity.
In the decays $\Bs \to D_s^\mp K^\pm$, however, both $b\to c$ and $b\to u$ amplitudes are of same order in the Wolfenstein parameter $\lambda$, $\mathcal{O}(\lambda^3)$, so that the interference effects are expected to be large. 
In addition, the decay width difference in the \Bs system, $\Delta\Gamma_s$, is non-zero, which adds sensitivity to the weak phase through the hyperbolic terms in the time evolution (see also Ref.~\cite{Nandi:2008rg}). 
The time-dependent decay rates of the initially produced flavour eigenstates are given by the decay equations
\begin{align}
  \frac{d\Gamma_{\Bs(\Bsb)\to \fdsk}(t)}{dt\, e^{-\Gamma_s t}} 
  & = \frac{1}{2} |\Afdsk|^2 (1+|\ldsk|^2) \nonumber\\
  & \times \left[\cosh\left(\frac{\Delta\Gamma_s t}{2}\right) - \Dfdsk \sinh\left(\frac{\Delta\Gamma_s t}{2}\right) 
    \pm \ \Cfdsk \cos\left(\Delta m_s t\right) \mp \Sfdsk \sin\left(\Delta m_s t\right) \right]\,,
  \label{eq:decay_rates}
\end{align}
where $\Gamma_s$, $\Delta\Gamma_s$, $\Delta m_s$ are the usual mixing parameters of the \Bs system and $|q/p|=1$ has been assumed.
The top (bottom) of the $\pm$ and $\mp$ signs is used when the initial particle is tagged as a $\Bs$ ($\Bsb$) meson.
In Eq.~(\ref{eq:decay_rates}), \Afdsk is the decay amplitude for a \Bs meson to decay to a final state \fdsk, and $\ldsk = (q/p)(\Abfdsk/\Afdsk)$
where \Abfdsk is the amplitude for a \Bsb to decay into \fdsk.
Similar equations hold for the charge conjugate processes
replacing \Afdsk by \Abfbdsk, \ldsk by $\lbdsk = (p/q) (\Afbdsk/\Abfbdsk)$,
and with a separate set of coefficients \Cfbdsk, \Sfbdsk and \Dfbdsk.
As each decay is dominated by a single diagram, $|\ldsk|=|\lbdsk|$.
The \CP asymmetry observables are then given by
\begin{eqnarray}
\Cfdsk  = \Cfbdsk = \frac{ 1 - \normlamsq }{ 1 + \normlamsq }\,, \quad
\Sfdsk  = \frac{ 2 \Im(\ldsk) }{ 1 + \normlamsq }\,, \quad
\Dfdsk  = \frac{ 2 \Re(\ldsk) }{ 1 + \normlamsq }\,, \nonumber \\
\Sfbdsk = \frac{ 2 \Im(\lbdsk)}{ 1 + \normlambarsq }\,, \quad
\Dfbdsk = \frac{ 2 \Re(\lbdsk)}{ 1 + \normlambarsq }\,.
\label{eq:asymm_obs}
\end{eqnarray}
The equality $\Cfdsk = \Cfbdsk $ results from $\left| q/p \right| = 1$ and $|\ldsk|=|\lbdsk|$.
The term \ldsk is connected to the weak phase by
\begin{equation}
  \ldsk = \Big( \frac{q}{p}\Big) \frac{\Abfdsk}{\Afdsk} = 
  \Big( \frac{V_{tb}^{*}V_{ts}}{V_{tb}V_{ts}^{*}} \Big) \Big( \frac{V_{ub}V_{cs}^{*}}{V_{cb}^{*}V_{us}} \Big)
  \Big| \frac{A_2}{A_1}\Big| e^{i\Delta}
  = |\ldsk|e^{i(\Delta-(\weak))}\,,
\end{equation}
where $|A_2/A_1|$ is the ratio of the hadronic amplitudes between $\Bs\to\Dsm\Kp$ and $\Bs\to\Dsp\Km$, $\Delta$ is their strong phase difference, and $\weak$ is the weak phase difference. 
An analogous relation exists for \lbdsk, $\lbdsk = |\ldsk|e^{i(\Delta+(\weak))}$.
Thus one obtains five observables from Eq.~(\ref{eq:asymm_obs}) and solves for $|\ldsk|$, $\Delta$, and $(\weak)$.

The LHCb experiment has the necessary decay time resolution, tagging power  and access to large enough signal yields to perform this time-dependent \CP measurement.\footnote{
  Preliminary results have been presented at CKM 2012~\cite{LHCb-CONF-2012-029}.
}
The signal yields can be seen from the measurement of $\BR(\Bs\to\Dsmp\Kpm)$~\cite{LHCb-PAPER-2011-022} (see Sec.~\ref{sec:gwt:expt} below).
The identification of the initial flavour of the signal \Bs candidate can be done combining both the responses of opposite-side and same-side kaon tagging algorithms, as is planned for other measurements of mixing-induced \CP-violation in $\Bs$ decays, and has already been implemented in the preliminary analysis of $\Bs \to D^-_s \pi^+$ decays~\cite{LHCb-CONF-2011-050}.


\subsubsection{Theoretical cleanliness of $\gamma$ from $B \to DK$ decays}

The answer to the question of why it is interesting to measure $\gamma$ precisely depends on the experimental precision that can be achieved. 
In the era of LHCb, the main motivation is the theoretically clean measurement of the SM CKM phase. 
The search for NP can thus be performed by comparing the extracted value of $\gamma$ to other observables, for example in the CKM fit (see Sec.~\ref{sec:globalfit}). 
However, one can also cross-check for the presence of NP in $B \to DK$ channels themselves.
One way is to test that the values of $\gamma$ determined from the many different $B \to DK$ type channels all coincide.
Another is automatically built in to the method for $\gamma$ extraction in the GGSZ analysis.
Consider the case where the decay amplitudes get modified by an extra contribution with a new strong phase $\delta_B^\prime$ and a weak phase $\gamma^\prime$. 
Then instead of the decay amplitudes in Eq.~(\ref{eq:amp2}) one finds
\begin{equation}
  A(B^\pm \to f_D K^\pm)\propto 1+r_D e^{i\delta_D}(r_B e^{i(\delta_B\pm\gamma)}+r_B^\prime e^{i(\delta_B^\prime\pm\gamma^\prime)})\,.
  \label{fBarDec}
\end{equation}
This means that for $B^+$ and $B^-$ decays the $r_B$ ratios are different
\begin{equation}
  r_{B^+}\to |r_Be^{i(\delta_B+\gamma)}+r_B^\prime e^{i(\delta_B^\prime+\gamma^\prime)}|\,,
  \quad
  r_{B^-}\to |r_Be^{i(\delta_B-\gamma)}+r_B^\prime e^{i(\delta_B^\prime-\gamma^\prime)}|\,.
\end{equation}
Discovering that $r_{B^-}\ne r_{B^+}$ would signal a \CP-violating NP contribution to the $B\to DK$ amplitude. 
One signature of NP would then be ${x_+^2+y_+^2}\ne {x_-^2+y_-^2}$,
though it is also possible that the equality could be satisfied even in the presence of NP: in this case there can be a shift in the extracted value of $\gamma$.

Existing measurements place strong constraints on tree-level NP effects, yet the possibility of discoveries in this sector in the near term is not ruled out.
In the far future, with much larger statistics, the measurement of $\gamma$ is well suited to search for high scale NP since it is theoretically very clean.
For example, NP with contributions of different chirality could give different shifts in $\gamma$, so the above test is meaningful.  

A useful question to ask is, what is the energy scale that could be probed in principle?  
To answer this, the irreducible theoretical uncertainty in the determination of $\gamma$ must be estimated. 
There are several sources that can induce a bias in the determination of $\gamma$ from $B\to DK$ decays. However, most of these can be avoided, either (i) with more statistics (for example, the Dalitz plot model uncertainty where a switch to a model-independent method is possible), or (ii) by modifying the equations used to determine $\gamma$ (an example is to correct for effects of $\Dz$--$\Dzb$ mixing~\cite{Grossman:2005rp,Bondar:2010qs}). 
The remaining, irreducible, theory uncertainties are then from the electroweak corrections.

\begin{figure}[!htb]
\begin{center}
\includegraphics[width=0.25\textwidth]{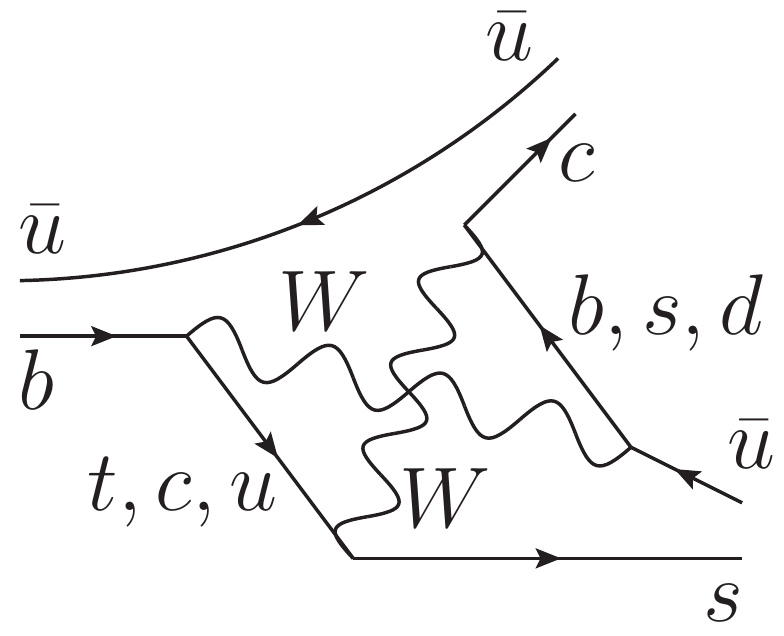}
\hspace{5mm}
\includegraphics[width=0.25\textwidth]{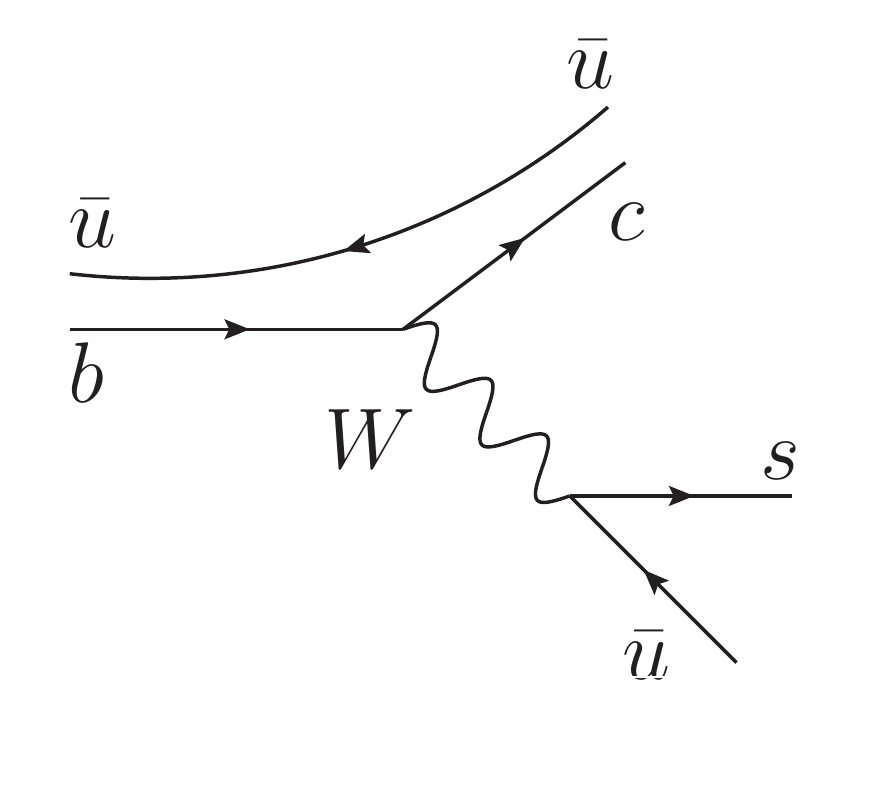}
\end{center}
\caption{\small
  A $\Bm\to\Dz\Km$ box diagram electroweak correction (left) with a different CKM structure than the leading weak decay amplitude (right).
}
\label{higherEW}
\end{figure}

\begin{table*}[!htb]
\begin{center}
\caption{\small
  Ultimate NP scales that can be probed using different observables listed in the first column. They are given by saturating the theoretical errors given respectively by 1) $\delta \gamma/\gamma=10^{-6}$, 2) optimistically assuming no error on $f_B$, so that the ultimate theoretical error is only from electroweak corrections,  3) using SM predictions in Ref.~\cite{Pirjol:2009vz},
4) optimistically assuming perturbative error estimates $\delta \beta/\beta~0.1\,\%$~\cite{Grossman:2002bu}, and 5) from bounds for $\Re\, C_1 (\Im \,C_1)$ from UTfitter~\cite{Bona:2007vi}.}
\label{table:probes}
\begin{tabular}{ccc}
\hline
Probe & $\Lambda_{\rm NP}$ for (N)MFV NP &$\Lambda_{\rm NP}$ for gen. FV NP \\ 
\hline
$\gamma$ from $B\to DK^{1)}$ & $\Lambda\sim {\mathcal O}(10^2 \tev)$   & $\Lambda\sim {\mathcal O}(10^3 \tev)$ \\
$B\to \tau \nu^{2)}$ & $\Lambda\sim {\mathcal O}(1 \tev)$   & $\Lambda\sim {\mathcal O}(30 \tev)$ \\
$b\to ss\bar d^{3)}$ & $\Lambda\sim {\mathcal O}(1 \tev)$   & $\Lambda\sim {\mathcal O}(10^3 \tev)$ \\
$\beta$ from $B\to \jpsi \KS\,^{4)}$ & $\Lambda\sim {\mathcal O}(50 \tev)$   & $\Lambda\sim {\mathcal O}(200 \tev)$ \\
$K-\bar K$ mixing{}$^{5)}$ & $\Lambda>0.4 \tev$ ($6 \tev$)   & $\Lambda>10^{3(4)} \tev$ \\
\hline
\end{tabular}
\end{center}
\end{table*}

The challenge to determine this uncertainty is that the hadronic elements can no longer be determined solely from the experiment.  
Not all electroweak corrections matter though --- the important ones are the corrections that change the CKM structure. 
For instance, vertex corrections and $Z$ exchanges do not affect $\gamma$, but corrections from box diagrams carry a different weak phase.
The dominant contribution is effectively due to $t$ and $b$ running in the loop.
For $b\to u s\bar c$ transitions there is a tree level contribution with $\sim V_{ub} V_{cs}^*$ CKM structure, while the box diagram has $\sim (V_{tb}V_{ts}^*) (V_{ub} V_{cb}^*)$. 
Since this has the same weak phase, it does not introduce a shift in $\gamma$.
For $b\to c s \bar u$ transitions, on the other hand, the tree level is $\sim V_{cb} V_{us}^*$, while the box diagram $\sim (V_{tb}V_{ts}^*) (V_{cb} V_{ub}^*)$, as illustrated in Fig.~\ref{higherEW}. 
The two contributions have {\sl different} weak phases, which means that the shift $\delta \gamma$ is non-zero.

The size of this effect is estimated by integrating over both $t$ and $b$ at the same time. 
The electroweak corrections in the effective theory are then described by a local operator whose matrix elements are easier to estimate. 
Although the Wilson coefficient of the operator contains large logarithms, $\log(m_b/m_W)$, for ${\mathcal O}(1)$ estimates, the precision obtained without resummation is sufficient.
If one resums $\log(m_b/m_W)$ then nonlocal contributions are also generated.
As a rough estimate only the local contributions need be kept.
The irreducible theory error on $\gamma$ is conservatively estimated to be $\delta \gamma/\gamma<{\mathcal O}(10^{-6})$ (most likely it is even $\delta \gamma/\gamma\lesssim {\mathcal O}(10^{-7})$).

This limit is far beyond the achievable sensitivity of any foreseeable experiment.
Nevertheless, it is interesting to consider what could be learnt in case such small deviations could be observed.
Assuming MFV one can probe $\Lambda_{\rm NP} \sim 10^2 \tev$, while assuming general flavour-violating (FV) NP one can probe $\Lambda_{\rm NP}\sim 10^3 \tev$ (where MFV and general FV NP scales are defined as in Ref.~\cite{Bona:2007vi}). 
This is by far the most precise potential probe of MFV, as shown in Table~\ref{table:probes}, due to the small theoretical uncertainty.

Since an experimental precision of $\delta \gamma/\gamma \sim 10^{-6}$ is not achievable in the near future, the NP scale reach must be adjusted for more realistic data sets.
This is easily done, since the scale $\Lambda_{\rm NP}$ probed goes as the fourth root of the yield.
With the LHCb upgrade, an uncertainty of $< 1^\circ$ on $\gamma$ can be achieved (see Sec.~\ref{sec:gammaprospects}), so that NP scales approaching $\Lambda_{\rm NP}\sim 5 (50) \tev$ can be probed for MFV (general FV) NP.

\subsubsection{Current LHCb experimental situation}
\label{sec:gwt:expt}

First results from LHCb in this area include a measurement using $\Bm \to D\Km$ with the GLW and ADS final states~\cite{LHCb-PAPER-2012-001}.\footnote{
  Results from preliminary GLW-type analyses using $\Bz \to D\Kstarz$~\cite{LHCb-CONF-2012-024} and $\Bm \to D\Km\pip\pim$~\cite{LHCb-CONF-2012-021} have been reported at ICHEP 2012.
}
A measurement of the branching ratio of $\Bs\to D^{\mp}_s \Kpm$ has also been performed~\cite{LHCb-PAPER-2011-022}.
Several other analyses, including studies of GGSZ-type final states, are in progress.\footnote{
  At CKM 2012, LHCb presented results of a model-independent GGSZ analysis of $\Bm \to D\Km$ with $D \to \KS\pip\pim$ and $D \to \KS\Kp\Km$~\cite{LHCb-PAPER-2012-027},
  preliminary results from a ADS-type analysis of $\Bm \to D\Km$ with $D \to K3\pi$~\cite{LHCb-CONF-2012-030},
  a preliminary determination of $\gamma$ from combined results using $\Bm \to D\Km$ and $\Bm \to D\pim$~\cite{LHCb-CONF-2012-032},
  and preliminary results on the time-dependent \CP violation parameters in $\Bs\to \Dsmp\Kpm$~\cite{LHCb-CONF-2012-029}.
}

These measurements all share common selection strategies.
They benefit greatly from boosted decision tree algorithms, which combine up to 20 kinematic variables to effectively suppress combinatorial backgrounds. 
Charmless backgrounds are suppressed by exploiting the large forward boost of the $D_{(s)}^+$ meson through a cut on its flight distance.

\begin{figure}[!htb]
\centering
\includegraphics[width=.9\textwidth]{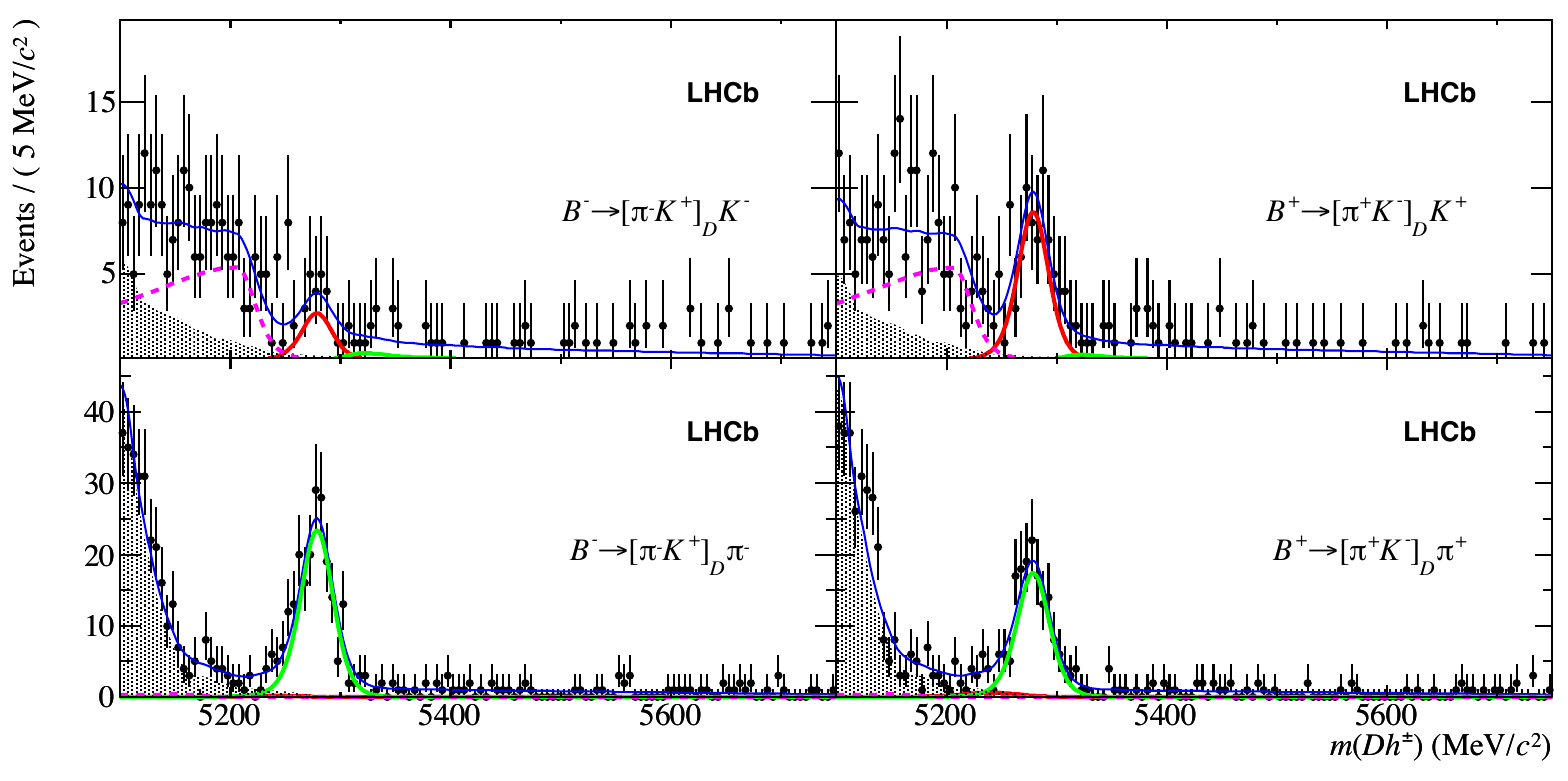}
\caption{\small
  Invariant mass distributions of selected $\Bpm\to[\pipm\Kpm]_D h^\pm$ candidate events: (left) \Bm candidates, (right) \Bp candidates~\cite{LHCb-PAPER-2012-001}. 
  In the top plots, the track directly from the $B$ vertex passes a kaon identification requirement and the \B candidates are reconstructed assigning this track the kaon mass. 
  The remaining events are placed in the bottom row and are reconstructed with a pion mass hypothesis. 
  The dark (red) curve represents the $\B\to D\Kpm$ events, the light (green) curve is $\B\to D\pipm$. 
  The shaded contribution are partially reconstructed events and the thin line shows the total fit function which also includes a linear combinatoric component.
  The broken line represents the partially reconstructed $\Bsb\to\Dz\Kp\pim$ decays where the pion is lost.
\label{fig:gwt-1}}
\end{figure}

In the GLW/ADS analysis~\cite{LHCb-PAPER-2012-001} of $1.0 \invfb$ of $\sqrt{s}=7 \tev$ data collected in 2011, the \CP eigenstates $D\to\Kp\Km$, $\pip\pim$,
and the quasi-flavour-specific $D\to\pim\Kp$ decay are used.
The \CP asymmetries defined in Eq.~(\ref{eq:rads}), and 
the ratios $R_\pm$ defined in Eq.~(\ref{eq:Rpm_ADS}), are measured for both the
$B\to DK$ signal and the abundant $B\to D\pi$ control channel.
The latter has limited sensitivity to \g but provides a large control sample from which probability density functions are shaped, and can be used to help reduce certain systematic uncertainties.
The control channel is also used to measure three ratios of partial widths
\begin{equation}
  R_{K/\pi}^f = \frac{ \Gamma(\Bm\to [f]_D\Km)+\Gamma(\Bp\to [f]_D\Kp) }{ \Gamma(\Bm\to [f]_D\pim)+\Gamma(\Bp\to [f]_D\pip) }\,,
  \label{eq:rkpi}
\end{equation}
where $f$ represents $KK$, $\pi\pi$ and the favoured $K\pi$ mode.
The signal yields are estimated by a simultaneous fit to 16 independent subsamples,
defined by the charges ($\times 2$), the $D$ final states ($\times 4$), and
the $DK$ or $D\pi$ final state ($\times 2$).
Figure~\ref{fig:gwt-1} shows the projections of the 
suppressed $\pipm\Kmp$ subsamples.
It is crucial to control the cross feed of the abundant $\Bm\to D\pim$ decays into the signal decays. 
This is achieved using the two LHCb ring-imaging Cherenkov detectors~\cite{LHCb-DP-2012-003}.
%
The systematic uncertainties are dominated by knowledge of the intrinsic asymmetry of the detector in reconstruction of positive and negative \B meson decays, and by the uncertainty on the particle identification requirements.
The results are 
\begin{align*}
R_{\CP+} &= \phantom{-} 1.007 \pm 0.038 \pm 0.012\,, \\
A_{\CP+} &= \phantom{-}  0.145 \pm 0.032 \pm 0.010\,, \\
R_-  &= \phantom{-} 0.0073 \pm 0.0023 \pm 0.0004\,, \\
R_+  &= \phantom{-} 0.0232 \pm 0.0034 \pm 0.0007\,,
\end{align*}
where the first error is statistical and the second systematic; 
$R_{\CP+}$ is computed from $R_{\CP+} \approx \langle R_{K/\pi}^{KK} , R_{K/\pi}^{\pi\pi} \rangle / R_{K/\pi}^{K\pi}$
with an additional $1\,\%$ systematic uncertainty assigned to account for the
approximation; $A_{\CP+}$ is computed as $A_{\CP+} = \langle A_{K}^{KK} , A_{K}^{\pi\pi} \rangle$.
From the $R_\pm$ one can also compute
\begin{align*}
  R_{\rm ADS}   &=       \phantom{-}  0.0152 \pm 0.0020 \pm 0.0004\,, \\
  A_{\rm ADS}   &=       -0.52 \pm 0.15 \pm 0.02\,,
\end{align*}
as $R_{\rm ADS} = (R_- + R_+) / 2$ and $A_{\rm ADS} = (R_- - R_+) / (R_- + R_+)$.
To summarise, the $\Bpm\to D\Kpm$ ADS mode is observed with \mbox{$\approx10\,\sigma$} 
statistical significance when comparing the maximum likelihood to that of the null hypothesis.
This mode displays evidence ($4.0\,\sigma$) of a large negative asymmetry, consistent with previous 
experiments~\cite{Belle:2011ac,delAmoSanchez:2010dz,Aaltonen:2011uu}.
The combined asymmetry $A_{\CP+}$ is smaller than (but compatible with) previous
measurements~\cite{delAmoSanchez:2010ji,Aaltonen:2009hz}, and is $4.5\,\sigma$ significant.
The maximum likelihood is compared with that under the null hypothesis in all 
three $DK$ final states, diluted by the non-negligible correlated systematic uncertainties. 
From this, with a total significance of $5.8\,\sigma$, direct \CP violation is observed in $\Bpm\to D\Kpm$ decays.

\newcommand{\BsDp} {\ensuremath{B^0_s \to D^-_s \pi^+}}
\newcommand{\BdDp} {\ensuremath{B^0 \to D^- \pi^+}}
\newcommand{\BsDK} {\ensuremath{B^0_s \to D^\mp_s K^\pm}}
\newcommand{\fsfdt}{\ensuremath{f_s/f_d}}
\newcommand{\BsDstarp}{\ensuremath{B^0_s \to D^{*-}_s\pi^+}}
\newcommand{\BsDstarK}{\ensuremath{B^0_s \to D^{*-}_sK^+}}
\newcommand{\BsDrho}{\ensuremath{B^0_s \to D^-_s\rho^+}}
\newcommand{\BsDkst}{\ensuremath{B^0_s \to D^-_sK^{*+}}}
\newcommand{\BsDstrho}{\ensuremath{B^0_s \to D^{*-}_s\rho^+}}
\newcommand{\BsDstkst}{\ensuremath{B^0_s \to D^{*-}_sK^{*+}}}
\newcommand{\LbDsp}{\ensuremath{\Lb \to D^{-}_s p}}
\newcommand{\LbDsstp}{\ensuremath{\Lb \to D^{*-}_s p}}

The analysis of the $\Bs\to D^{\mp}_s \Kpm$ decay mode~\cite{LHCb-PAPER-2011-022}
is based on a sample corresponding to an integrated luminosity of $0.37 \invfb$, collected in 2011 at a centre-of-mass energy of $\sqrt{s} = 7 \tev$.
This decay mode has been observed by the CDF~\cite{Aaltonen:2008ab} and Belle~\cite{Louvot:2008sc} collaborations, who measured its branching fraction with an uncertainty around 23\,\%~\cite{Nakamura:2010zzi}.
In addition to \BsDK, the channels \BdDp and \BsDp are analysed. 
They are characterised by a similar topology and therefore are good control and
normalisation channels. 
Particle identification criteria are used to separate the CF decays from the suppressed modes, and to suppress misidentified backgrounds.

\begin{figure}[!htb]
\centering
\includegraphics[width=.65\textwidth]{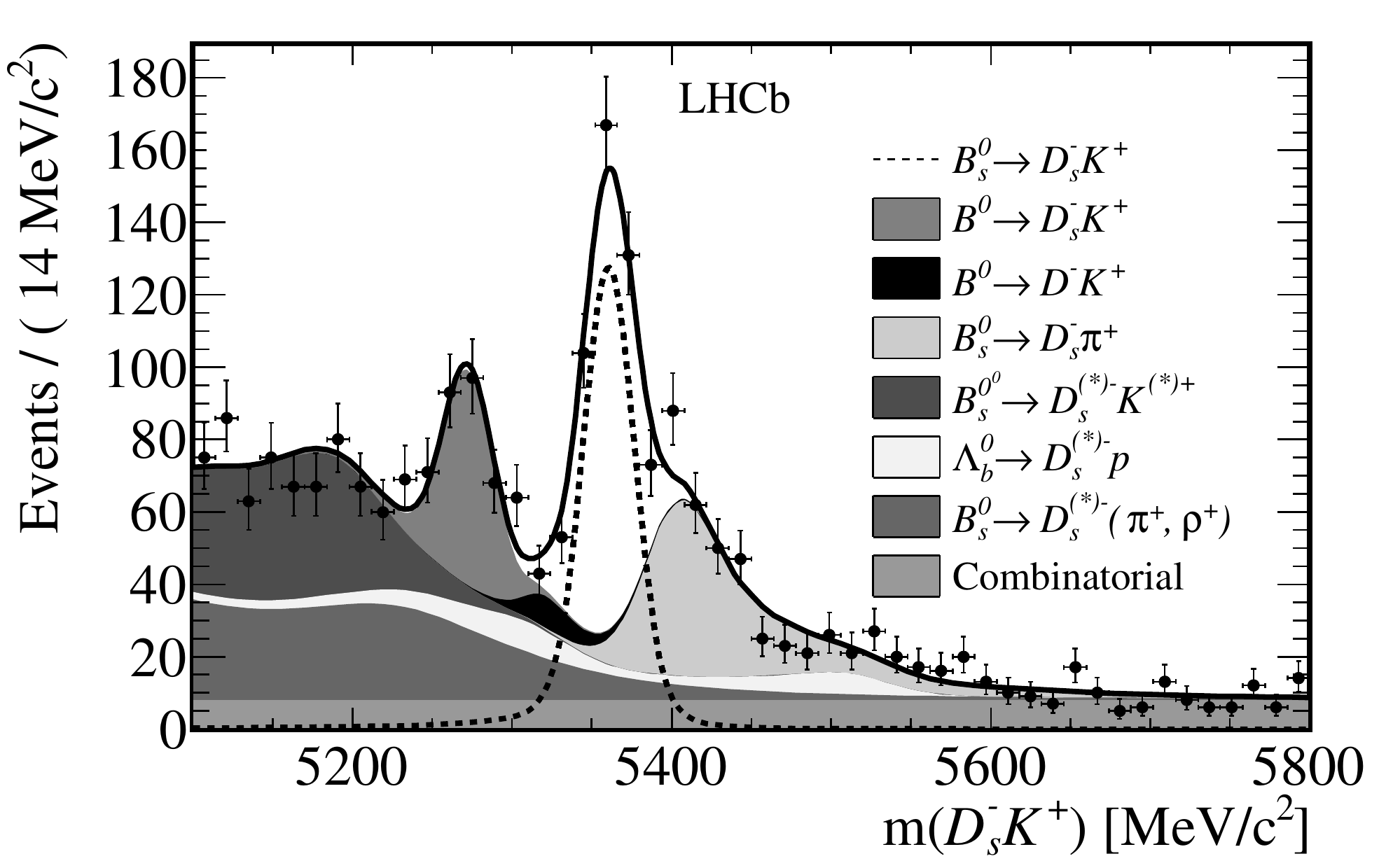}
\caption{\small 
  Mass distribution of the \BsDK~candidate events~\cite{LHCb-PAPER-2011-022}. 
  The stacked background shapes follow the same top-to-bottom order in the legend and in the plot.}
\label{fig:dsk}
\end{figure}

The signal yields are obtained from unbinned extended maximum likelihood fits to the data.
The fits include components for the combinatorial background and several sources of background from $b$ hadron decays.
The most important is the misidentified \BsDp~decay. 
Its shape is fixed from data using a reweighting procedure~\cite{LHCb-PAPER-2011-006} while the yield is left free to float. 
A similar procedure is applied to a simulated data sample to extract the shape of the $B^0 \to D^-K^+$ misidentified background. 
The fit results are shown in Fig.~\ref{fig:dsk}.

Correcting the raw signal yields for selection efficiency differences gives
\begin{equation}
  \frac{\BR(\BsDK)}{\BR(\BsDp)} = 0.0646 \pm 0.0043 \pm 0.0025 \,,
  \label{eq:Bsratio}
\end{equation}
where the first uncertainty is statistical and the second is systematic. 
Using the measured relative yield of \BdDp, the known \BdDp\ branching
fraction~\cite{Nakamura:2010zzi}, and the recent \fsfdt~measurement~\cite{LHCb-PAPER-2011-018}, the branching fractions
\begin{eqnarray}
\BR(\BsDp) &=& (2.95 \pm 0.05 \pm 0.17\,^{+0.18}_{-0.22})\times 10^{-3}\,,\\
\BR(\BsDK) &=& (1.90 \pm 0.12 \pm 0.13\,^{+0.12}_{-0.14})\times 10^{-4}
\end{eqnarray}
are obtained, where the first uncertainty is statistical, the second is the experimental systematic uncertainty, 
and the third is from the \fsfdt~measurement.
Both measurements are significantly more precise than the previous world averages~\cite{Nakamura:2010zzi}.

\subsubsection{Measurements of $\gamma$ using loop-mediated two-body $B$ decays}
\label{sec:gammafromloop}

\CP violation in $B^0_{(s)}$ decays plays a fundamental role in testing
the consistency of the CKM paradigm in the SM and in probing virtual effects of heavy new particles.

With the advent of the \B factories, the Gronau-London (GL)~\cite{Gronau:1990ka} isospin analysis of $B \to \pi \pi$ decays has been a precious source of information on the phase of the CKM matrix. 
Although the method allows a full determination of the weak phase and of the relevant hadronic parameters, it suffers from discrete ambiguities that limit its constraining power. 
It is however possible to reduce the impact of discrete ambiguities by adding information on hadronic parameters~\cite{Charles:1998qx,Bona:2007qta}. 
In particular, as noted in Refs.~\cite{Fleischer:1999pa,Fleischer:2007hj,Fleischer:2010ib}, the hadronic parameters entering the $B^{0} \to \pi^+ \pi^-$ and the $\Bs \to K^+K^-$ decays are connected by U-spin, so that experimental knowledge of $\Bs \to K^+ K^-$ can improve the extraction of the CKM phase with the GL analysis. 
Indeed, in Ref.~\cite{Bona:2007qta}, the measurement of $\BR(\Bs \to K^+ K^-)$ was used to obtain an upper bound on one of the hadronic parameters. 

LHCb has reported preliminary measurements of the time-dependent \CP asymmetries using decays to \CP eigenstates, namely $B^0\to\pi^+\pi^-$ and $B^0_s \to K^+K^-$~\cite{LHCb-CONF-2012-007}, thereby permitting the use of the U-spin strategy proposed by Fleischer (F)~\cite{Fleischer:1999pa,Fleischer:2007hj,Fleischer:2010ib} to extract the CKM phase from a combined analysis of $B^{0} \to \pi^+ \pi^-$ and the $\Bs \to K^+ K^-$ decays. 
However, as shown explicitly below, this strategy alone suffers from a sizeable dependence on the breaking of U-spin symmetry.
In Ref.~\cite{Ciuchini:2012gd}, the authors propose to perform a combined analysis of the GL modes plus $\Bs \to K^+ K^-$ to obtain an optimal determination of the CKM phase within the SM. They show that this combined strategy has a milder dependence on the magnitude of U-spin breaking, allowing for a more solid estimate of the theory error.
The experimental data used for such a determination of $\gamma$ are summarised in Table~\ref{tab:exp}.

\begin{table}[!htb]
  \centering
\caption{\small
  Experimental data on $B\to\pi\pi$ and $\Bs\to\Kp\Km$ decays.
  The correlation column refers to that between $S_f$ and $C_f$ measurements. 
  Except for the preliminary results in Ref.~\cite{LHCb-CONF-2012-007}, all other measurements have been averaged by HFAG~\cite{HFAG}.
  The \CP asymmetry of $B^+ \to \pi^+ \pi^0$ has been reported for completeness, although it has not been used in the analysis.
  New results on time-dependent \CP violation in $\Bd \to \pip\pim$ reported by Belle at CKM2012~\cite{Adachi:2013mae} are not included.
} 
\label{tab:exp}
\begin{tabular}{|c| c| c| c| c| c|}
  \hline
  Channel & $\BR \times 10^6$ & $S_f\, (\%)$ & $C_f\, (\%)$ & Corr. & Ref. \\
  \hline
  $B^{0} \to \pi^+ \pi^-$ & $5.11 \pm 0.22$ & $-65 \pm 7$ & $-38 \pm 6$ & $-0.08$ & \cite{Lees:2012kx,Ishino:2006if,Aubert:2006fha,Duh:2012ie,Bornheim:2003bv,Aaltonen:2011qt} \\
  $B^{0} \to \pi^+ \pi^-$ & -- & $-56 \pm 17 \pm 3$ & $-11 \pm 21 \pm 3$ & $0.34$ & \cite{LHCb-CONF-2012-007} \\
  $B^{0} \to \pi^0 \pi^0$ & $1.91 \pm 0.23$ & -- & $-43 \pm 24$ & -- & \cite{Lees:2012kx,Abe:2004mp,Bornheim:2003bv} \\
  $B^+ \to \pi^+ \pi^0$ & $5.48 \pm 0.35$ & -- & $-2.6 \pm 3.9$ & -- & \cite{Aubert:2007hh,Duh:2012ie,Bornheim:2003bv} \\
  $\Bs \to K^+ K^-$ & $25.4 \pm 3.7$ & $17 \pm 18 \pm 5$ & $-2 \pm 18 \pm 4$ & $0.1$ & \cite{Peng:2010ze,Aaltonen:2011qt,LHCb-CONF-2012-007}
  \\
\hline
\end{tabular}
\end{table}

The time-dependent asymmetry for a $\B$ meson decay to a \CP eigenstate $f$ can be written, with the same notation as Eqs.~(\ref{eq:decay_rates}) and~(\ref{eq:asymm_obs}),\footnote{
  In the LHCb preliminary results on $\Bd\to\pi^+ \pi^-$ and $\Bs\to K^+K^-$ decays~\cite{LHCb-CONF-2012-007} a different notation has been used: $A_f^{\rm dir} \equiv -C_f$, $A_f^{\rm mix} \equiv S_f$, $A_f^{\Delta \Gamma} \equiv -D_f$.
} as
\begin{equation}
  A_{\CP}(t)=\frac{S_{f}\sin(\Delta m t) - C_{f}\cos(\Delta m t)}{\cosh\left(\frac{\Delta\Gamma}{2}t\right) + D_{f}\sinh\left(\frac{\Delta\Gamma}{2}t\right)}\,,
\end{equation}
where $C_{f}$ and $S_{f}$ parametrise direct and mixing-induced \CP violation respectively, and the quantity $D_f$ is constrained by the consistency relation
\begin{equation}
  \left( C_f \right)^2 + \left( S_f \right)^2 + \left( D_f \right)^2 = 1\,.
  \label{eq:adeltagamma}
\end{equation}

The LHCb preliminary results on direct and mixing-induced \CP violation parameters in $\Bd\to\pi^+ \pi^-$ and $\Bs\to K^+K^-$ decays~\cite{LHCb-CONF-2012-007} are shown in Table~\ref{tab:exp}.
The measurements of $C_{\pip\pim}$ and $S_{\pip\pim}$ are compatible with those from the $B$ factories, 
whereas $C_{\Kp\Km}$ and $S_{\Kp\Km}$ are measured for the first time and are consistent with zero within the current uncertainties.

Beyond the SM, NP can affect both the $B^{0}_{(s)}$--$\bar B^{0}_{(s)}$ amplitudes and the $b \to d (s)$ penguin amplitudes. 
Taking the phase of the mixing amplitudes from other measurements, for example from $b \to c \bar c s$ decays, one can obtain a constraint on NP in $b \to s$ (or $b \to d$) penguins. 
Alternatively, assuming no NP in the penguin amplitudes, one can obtain a constraint on NP in mixing.
The analysis discussed here is based on a simplified framework~\cite{Ciuchini:2012gd}, 
using as input values $\sin2\beta = 0.679 \pm 0.024$~\cite{HFAG} and $2\beta_s =(0 \pm 5)^\circ$~\cite{LHCb-CONF-2012-002} obtained from $b\to c \bar c s$ decays.  
The optimal strategy will be to include the combined GL and Fleischer analysis in a global fit of the CKM matrix plus possible NP contributions.

The GL and Fleischer analyses were formulated with different parametrisations of the decay amplitudes. 
In order to use the constraints in a global fit
one can write\footnote{
  Note that the use here of the symbol $C$ to denote a colour-suppressed amplitude is not related to its use to denote direct \CP violation parameters in time-dependent analyses.
}
{\footnotesize
\begin{equation}
  \label{eq:ampli}     
  \begin{array}{ll}
    A(B^{0} \to \pi^+ \pi^-) = C  (e^{i \gamma} - d e^{i \theta})\,, &
    A(\bar B^{0} \to \pi^+ \pi^-) = C  (e^{-i \gamma} - d e^{i \theta})\,, \\
    A(B^{0} \to \pi^0 \pi^0) = \frac{C}{\sqrt{2}}  (T e^{i \theta_T} e^{i \gamma} + d e^{i \theta})\,, &
    A(\bar B^{0} \to \pi^0 \pi^0) = \frac{C}{\sqrt{2}}  (T e^{i \theta_T} e^{-i \gamma} + d e^{i \theta})\,,  \\
    A(B^+ \to \pi^+ \pi^0) = \frac{A(B^{0} \to \pi^+ \pi^-)}{\sqrt{2}} + A(B^{0} \to \pi^0 \pi^0)\,, &
    A(B^- \to \pi^- \pi^0) = \frac{A(\bar B^{0} \to \pi^+ \pi^-)}{\sqrt{2}} + A(\bar B^{0} \to \pi^0 \pi^0) \,,\\
  A(\Bs \to K^+ K^-) = C^\prime \frac{\lambda}{1-\lambda^2/2} (e^{i \gamma} + \frac{1-\lambda^2}{\lambda^2} d^\prime e^{i \theta^\prime})\,, &  
  A(\Bsb \to K^+ K^-) = C^\prime \frac{\lambda}{1-\lambda^2/2} (e^{-i \gamma} + \frac{1-\lambda^2}{\lambda^2} d^\prime e^{i \theta^\prime})\,,
  \end{array}
\end{equation}
}
\normalsize
\noindent
where the magnitude of $V_{ub}V_{ud}^{*}$ has been reabsorbed in $C$, and
the magnitude of $V_{cb}V_{cd}^{*}/(V_{ub}V_{ud}^{*})$ has been reabsorbed in $d$.
In the exact U-spin limit, one has $C = C^\prime$, $d = d^\prime$ and $\theta = \theta^\prime$. 
Isospin breaking in $B \to \pi \pi$ has been neglected, since its impact on the extraction of the weak phase is at the level of $1^\circ$~\cite{Gronau:1998fn,Zupan:2007fq,Gardner:2005pq,Botella:2006zi}.
The physical observables entering the analysis are
\begin{eqnarray}
  \label{eq:obs}
  && \BR(B \to f) = 
  F(B) \frac{\vert A(B \to f) \vert^2 + \vert A(\bar B \to f) \vert^2}{2}\,,  \\ 
  && C_f = 
  \frac{\vert A(B \to f) \vert^2 - \vert A(\bar B \to f) \vert^2}{\vert A(B \to f) \vert^2 + \vert A(\bar B \to f) \vert^2}\,, \qquad
  S_f = \frac{2 \mathrm{Im} \left(e^{-i \phi_M(B)} \frac{A(\bar B \to f)}{A(B \to f)} \right)}{1 + \left\vert  \frac{A(\bar B \to f)}{A(B \to f)} \right\vert^2}\,,\nonumber
\end{eqnarray}
where $\phi_M(\Bd) = 2 \beta$, $\phi_M(\Bs) = -2\beta_s$ in the SM, and $F(B^{0}) = 1$, $F(\Bp) = \tau_{B^+}/\tau_{B^{0}} = 1.08$, 
$F(\Bs) = \tau_{\Bs}/\tau_{\Bd} (m_{\Bd}^2/m_{\Bs}^2) \sqrt{(M_{\Bs}^2-4 M_{\Kp}^2)/(M_{\Bd}^2-4 M_{\pip}^2)} = 0.9112$.

\begin{figure}[!htb]
  \centering
  \includegraphics[width=.32\textwidth]{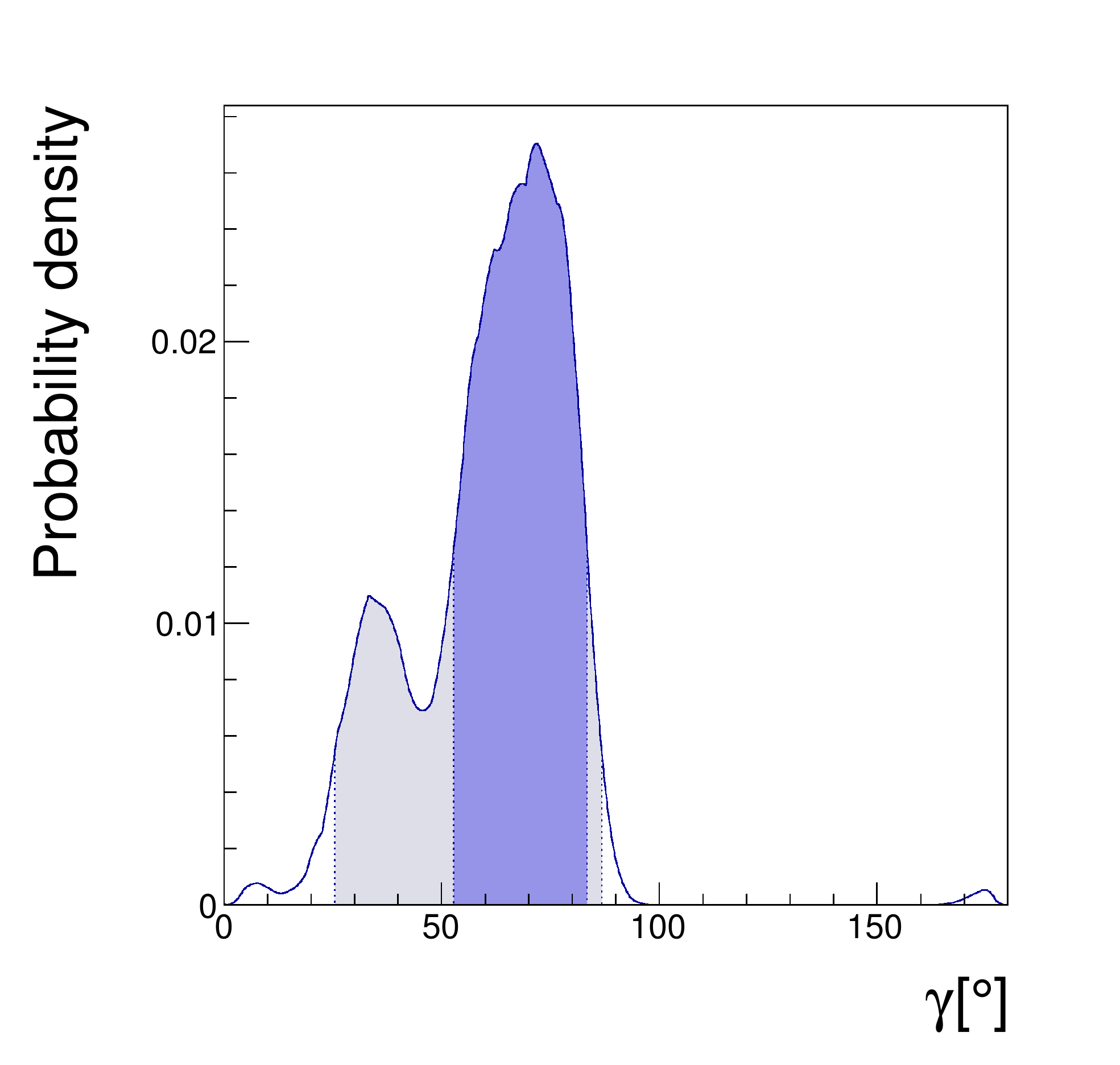}
  \includegraphics[width=.32\textwidth]{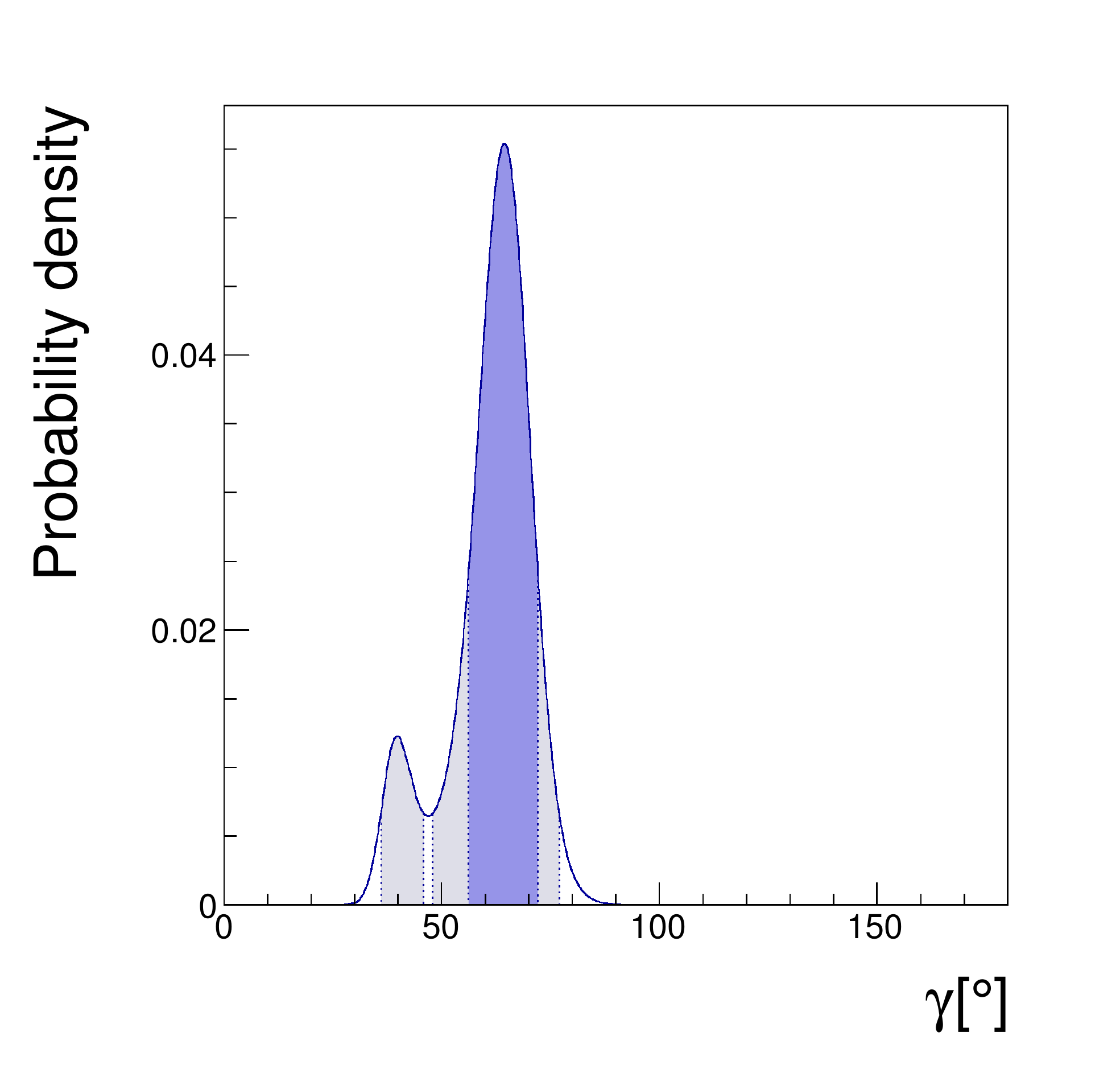}
  \includegraphics[width=.32\textwidth]{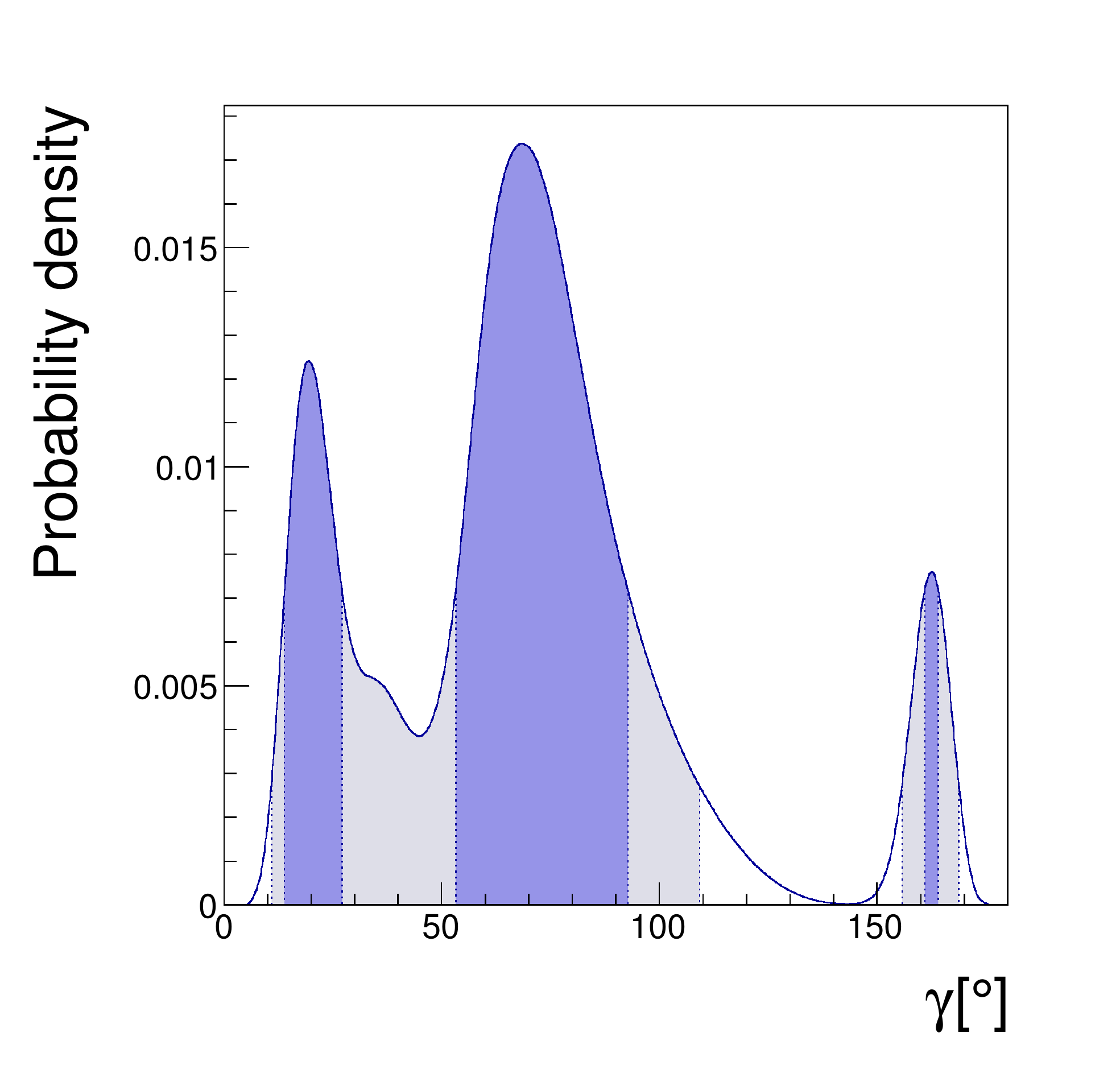}
  \caption{\small
    From left to right: PDF for $\gamma$ obtained using the
    GL method as described in the text; PDF for $\gamma$ obtained
    using the Fleischer method for $\kappa=0.1,0.5$~\cite{Ciuchini:2012gd}. 
    Here and in the following, dark (light) areas correspond to $68\,\%$ ($95\,\%$) probability regions.
    }
  \label{fig:gammaGL}
\end{figure}

In the GL approach, one extracts the probability density function (PDF) for the angle $\alpha = \pi - \beta - \gamma$ of the UT from the measurements of $\BR(B\to \pi\pi)$, $S_{\pi^+\pi^-}$, $C_{\pi^+\pi^-}$ and $C_{\pi^0\pi^0}$.
Using the unitarity of the CKM matrix, it is possible to write the $B \to \pi \pi$ decay amplitudes and observables in terms of $\alpha$ instead of $\gamma$   and $\beta$. 
However, for the purpose of connecting $B \to \pi \pi$ to $\Bs \to K K$ it is more convenient to use the parametrisation in Eq.~(\ref{eq:ampli}).
In this way, $\alpha$ (or, equivalently, $\gamma$), is determined up to discrete ambiguities, that correspond however to different values of the hadronic parameters. 
As discussed in detail in Ref.~\cite{Bona:2007qta}, the shape of the PDF obtained in a Bayesian analysis depends on the allowed range for the hadronic parameters. 
For example, using the data in Table~\ref{tab:exp}, solving for $C$ and choosing flat {\it a priori} distributions for $d\in[0,2]$, $\theta\in[-\pi,\pi]$, $T\in[0,1.5]$ and $\theta_T\in[-\pi,\pi]$  the PDF for $\gamma$ in Fig.~\ref{fig:gammaGL} is obtained, corresponding to $\gamma = (68 \pm 15)^\circ$ ($\gamma \in [25, 87]^\circ$ at $95\,\%$ probability).
Using instead the Fleischer method, one can obtain a PDF for $\gamma$ given a range for the U-spin breaking effects. 
In this method it was originally suggested to parametrise the U-spin breaking in $C^\prime/C$ using the result one would obtain in factorisation, namely
\begin{equation}
  \label{eq:rfact}
  r_\mathrm{fact} = \left\vert \frac{C^\prime}{C} \right\vert_\mathrm{fact} = 1.46 \pm 0.15\,,
\end{equation}
where the error obtained using light-cone QCD sum rule calculations~\cite{Duplancic:2008tk} has been symmetrised. 
However, this can only serve as a reference value, since there are non-factorisable contributions to $C$ and $C^\prime$ that could affect this estimate. 
In this analysis, the non-factorisable U-spin breaking is parametrised as follows
\begin{equation}
  \label{eq:su3b}
  C^\prime = r_\mathrm{fact} r_C C\,,
  \qquad
  \Re(d^\prime e^{i \theta^\prime}) = r_r \Re(d e^{i \theta}) \,,
  \qquad
  \Im(d^\prime e^{i \theta^\prime}) = r_i \Im(d e^{i \theta}) \,,
\end{equation}
with $r_C$, $r_r$ and $r_i$ uniformly distributed in the range $[1-\kappa, 1+\kappa]$.

In Fig.~\ref{fig:gammaGL} the PDF for $\gamma$ obtained with the Fleischer method for two different values of the U-spin breaking parameter $\kappa = 0.1,0.5$ is shown.
The method is very precise for small amounts of U-spin breaking ($\kappa=0.1$), 
but becomes clearly worse for $\kappa=0.5$. 
Thus, a determination of $\gamma$ from the Fleischer method alone is subject to uncertainty on the size of U-spin breaking.

\begin{figure}[!htb]
  \centering
  \includegraphics[width=.32\textwidth]{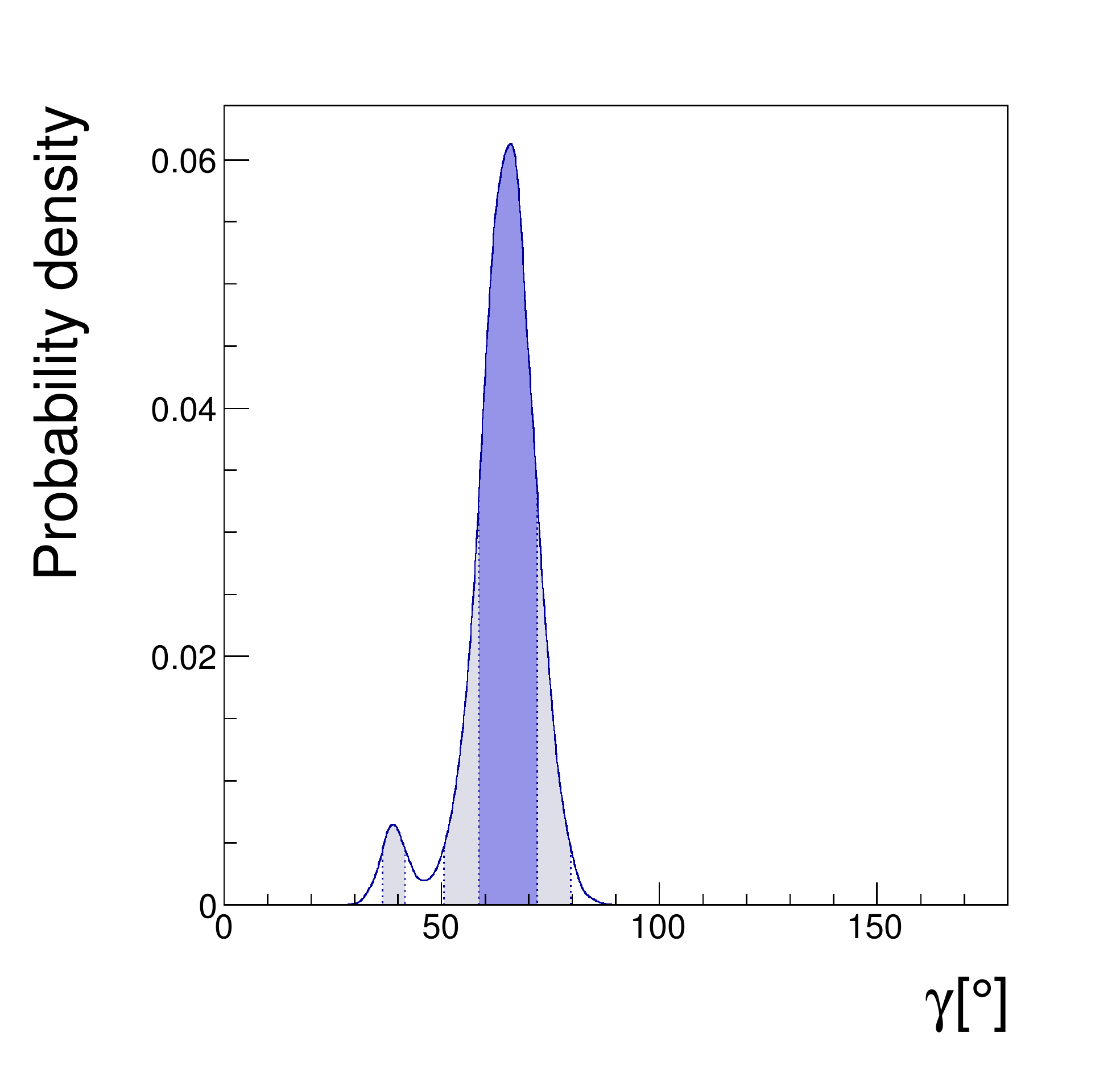}
  \includegraphics[width=.32\textwidth]{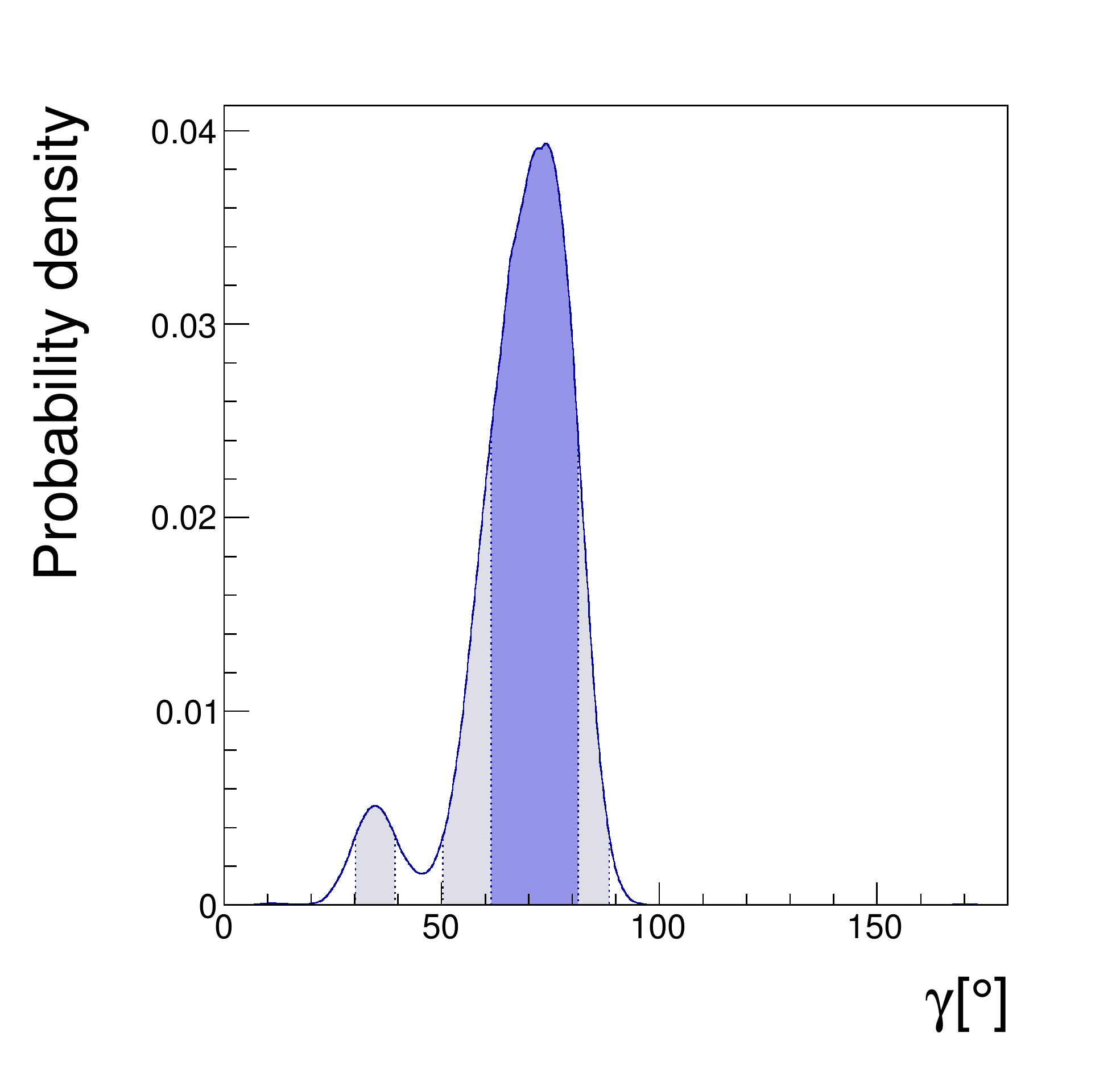}
  \includegraphics[width=.32\textwidth]{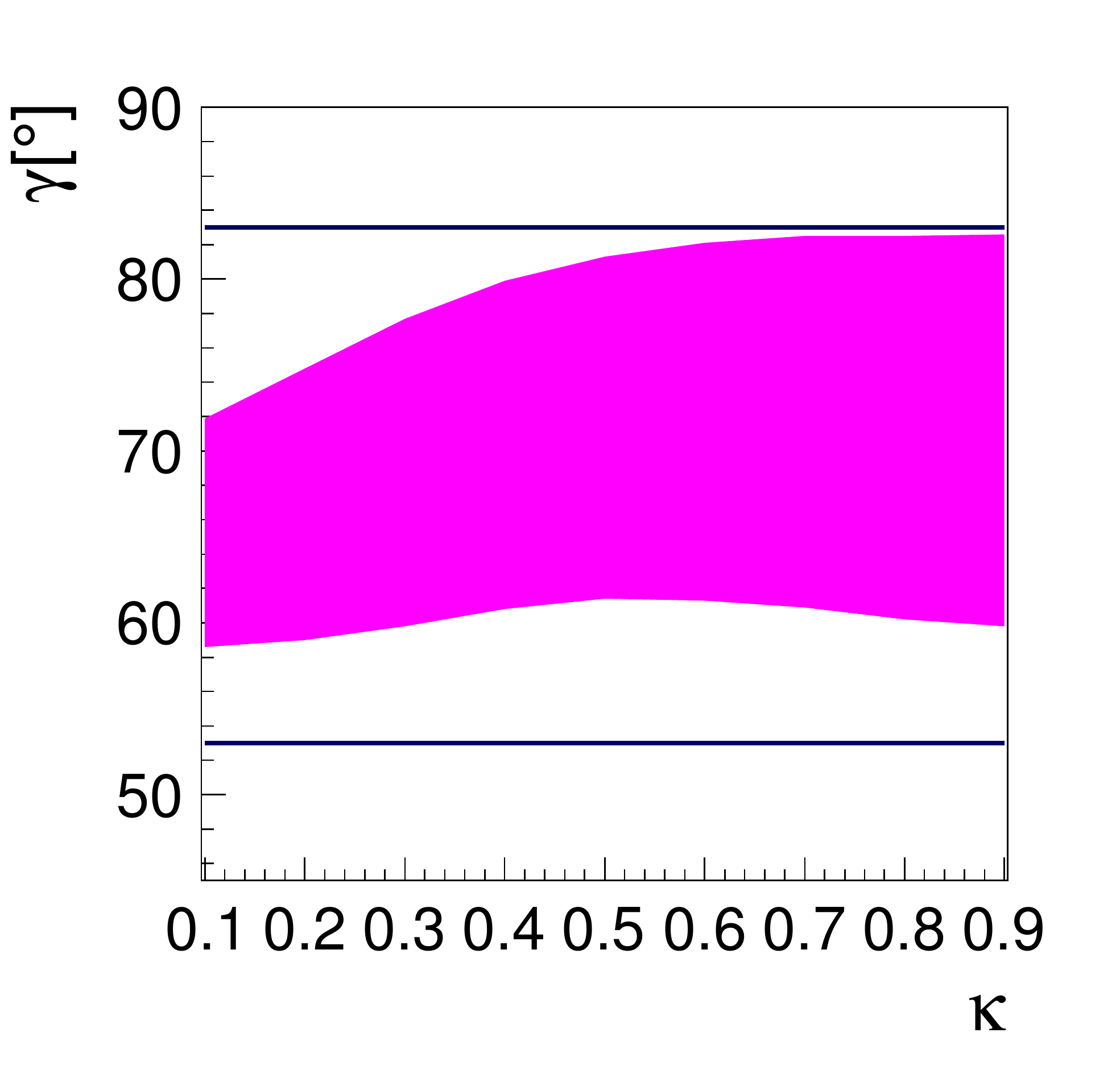}
  \caption{\small 
    From left to right: PDF for $\gamma$ obtained using the
    combined method for $\kappa = 0.1$, $0.5$; $68\,\%$
    probability region for $\gamma$ obtained using the combined method
    (filled area) or the GL method (horizontal lines) as a function of
    $\kappa$~\cite{Ciuchini:2012gd}.}
  \label{fig:gammafull}
\end{figure}

The result of the combined GL+F analysis is given in Fig.~\ref{fig:gammafull}, where the PDF for $\gamma$ for $\kappa = 0.1$ and $0.5$ is shown. 
The result of the combined analysis is much more stable against the allowed amount of U-spin breaking. 
In Fig.~\ref{fig:gammafull} the $68\,\%$ probability region for $\gamma$ obtained using the combined method as a function of $\kappa$ is also shown, and compared to the GL result. 
The combined method shows a considerable gain in precision even for very large values of $\kappa$. 

\begin{figure}[!htb]
  \centering
  \includegraphics[width=.32\textwidth]{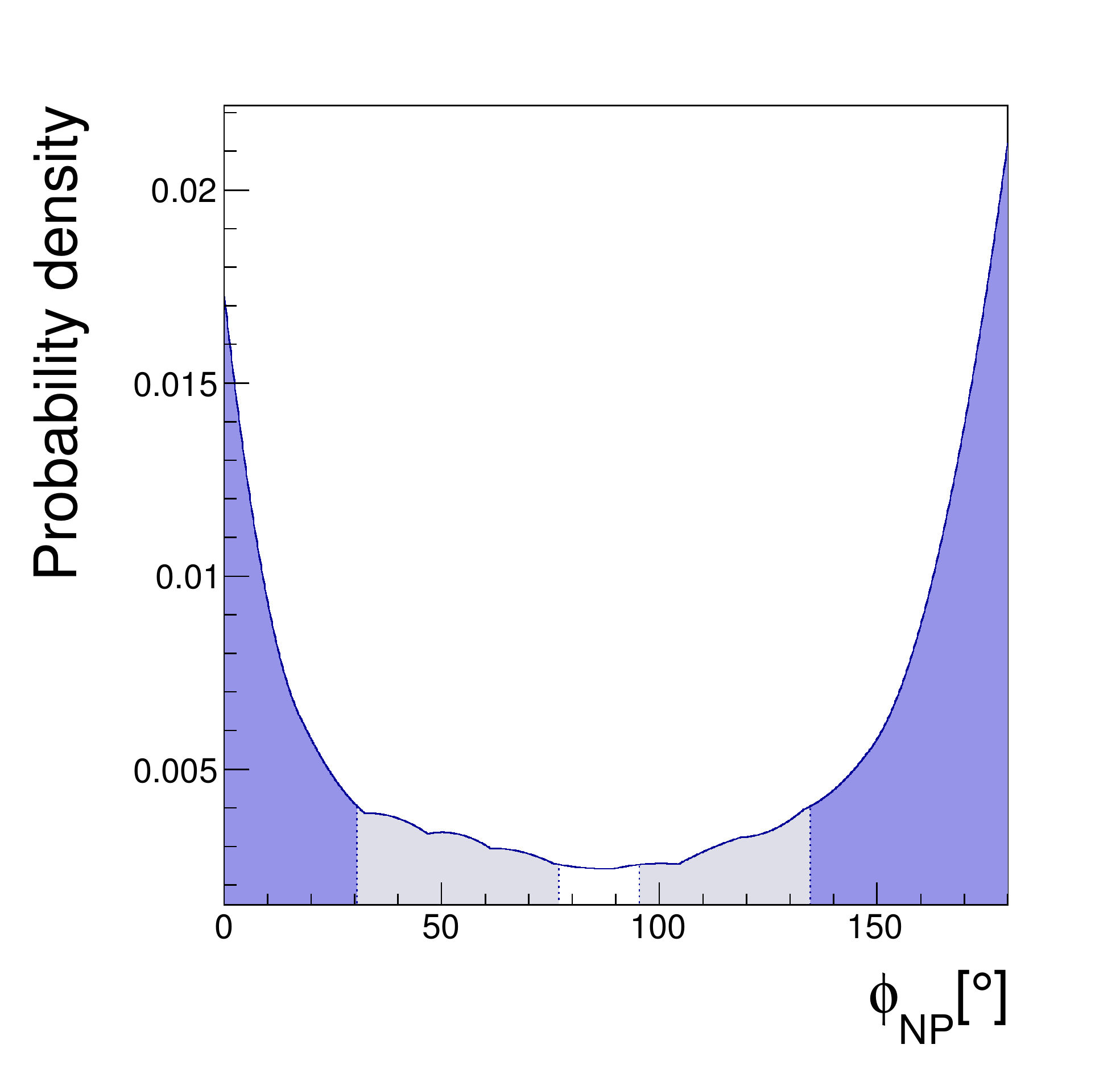}
  \includegraphics[width=.32\textwidth]{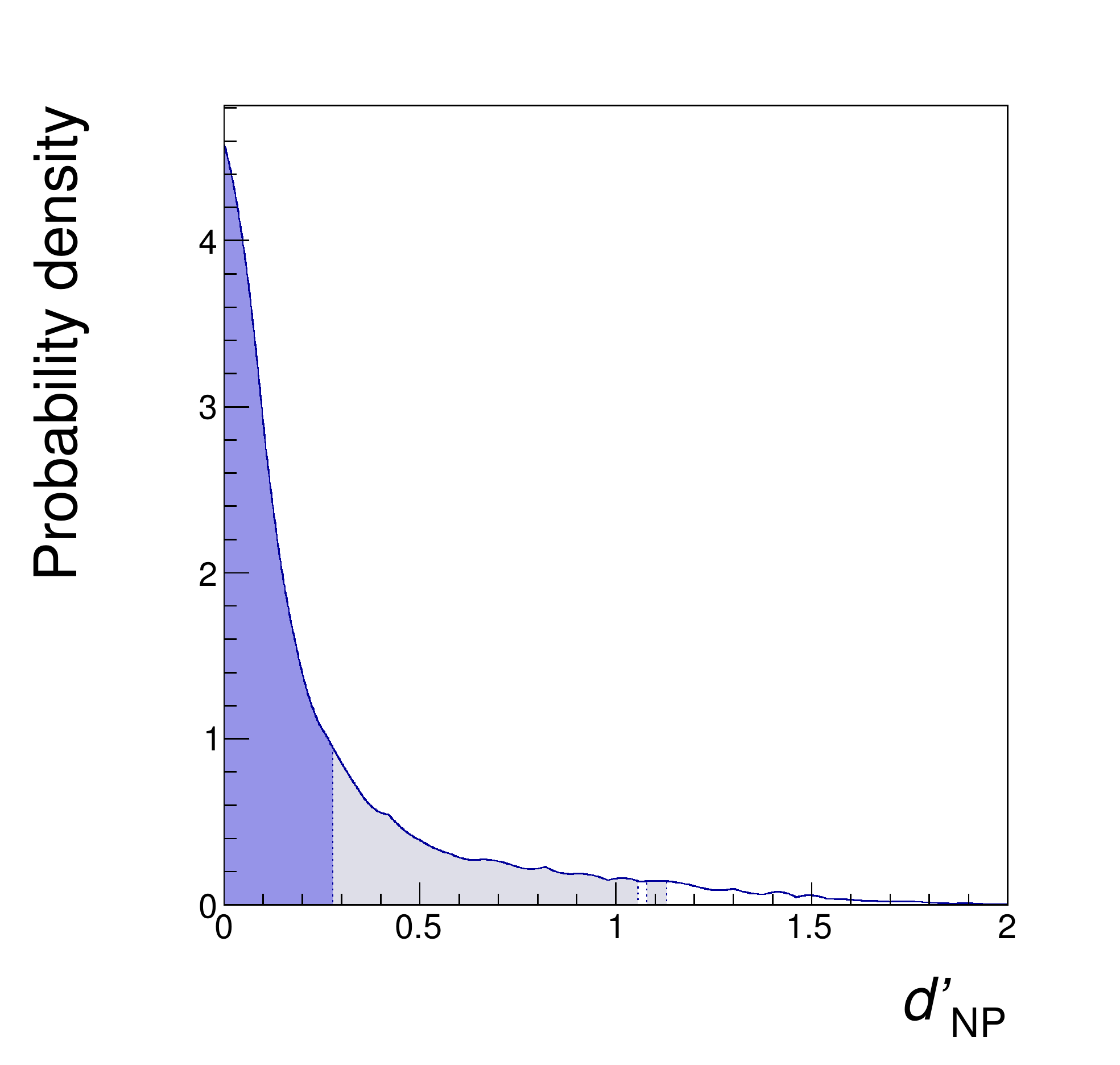}
  \includegraphics[width=.32\textwidth]{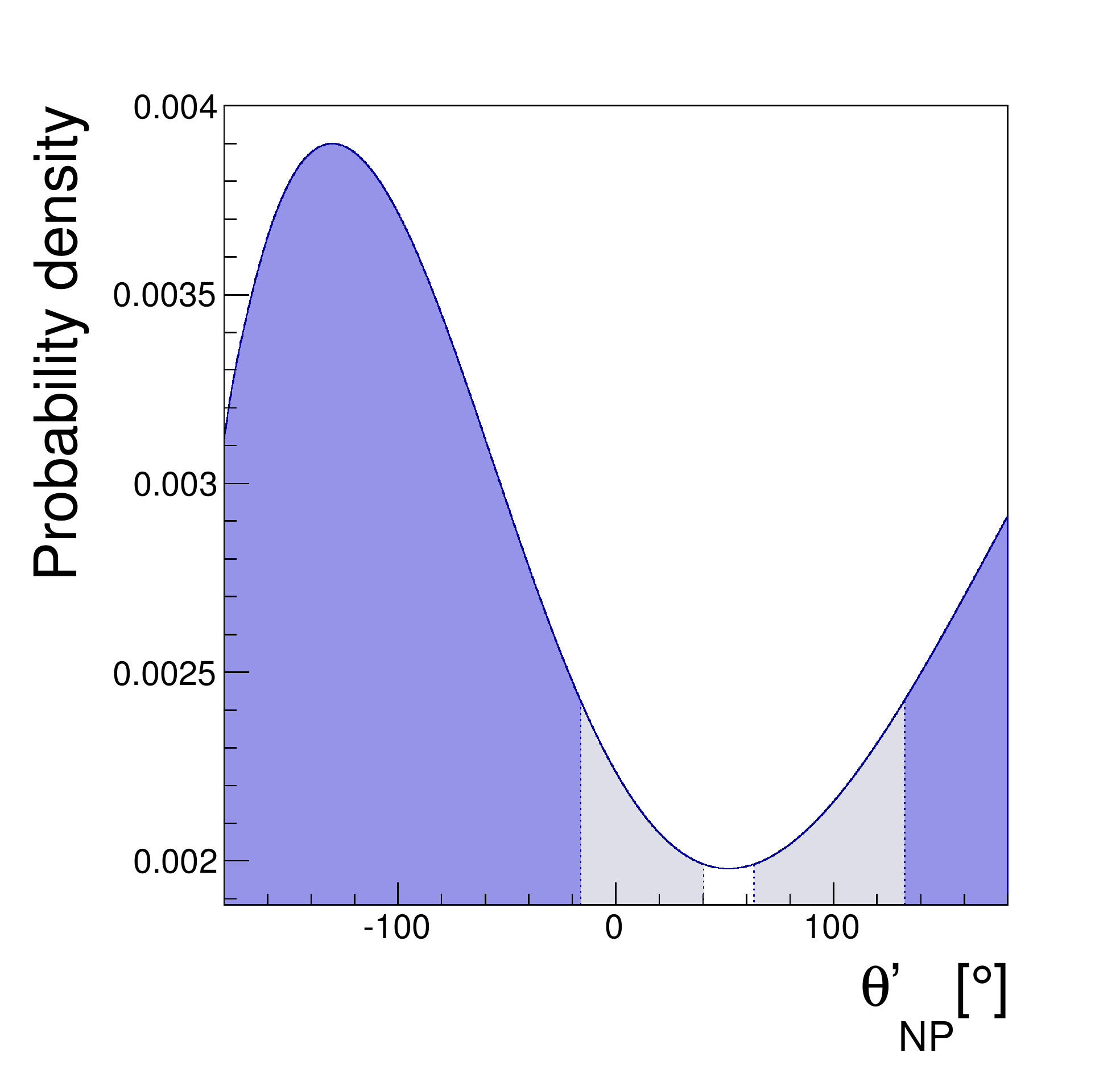}
  \caption{\small
    From left to right: 
    PDFs for $\phi_\mathrm{NP}$, $d^\prime_\mathrm{NP}$ and
    $\theta^\prime_\mathrm{NP}$ obtained using the
    combined method with $\kappa = 0.5$~\cite{Ciuchini:2012gd}.
  }
  \label{fig:np}
\end{figure}

NP could affect the determination of $\gamma$ in the combined method by giving (electroweak) penguin contributions a new \CP-violating phase. 
If one assumes that the isospin analysis of the GL channels is still valid, barring order-of-magnitude enhancements of electroweak penguins in $B \to \pi\pi$, and if one assumes for concreteness that NP enters only $b \to s$ penguins, in the framework of a global fit, one can simultaneously determine $\gamma$ and the NP contribution to $b \to s$ penguins. 
For the purpose of illustration, the value of $\gamma$ from tree-level processes, $\gamma_\mathrm{tree} = (76 \pm 9)^\circ$ is used as input~\cite{Bona:2007vi},\footnote{
  Note that the value of $\gamma$ quoted here differs from that obtained from the full CKM fit (given in Table~\ref{tab:SMfullfittable}) due to the different inputs used.
}
allowing inspection of the posterior for $\gamma$ and for the NP penguin amplitude. 
Writing
\begin{eqnarray}
  \label{eq:npampli}
  A(\Bs \to K^+ K^-) &=& C^\prime \frac{\lambda}{1-\lambda^2/2}
  \left( e^{+i \gamma} + \frac{1-\lambda^2}{\lambda^2}
  \left(
  d^\prime e^{i \theta^\prime} + e^{+i \phi_{\rm NP}} d^\prime_{\rm NP} e^{i \theta^\prime_{\rm NP}}
  \right)
  \right)\,, \\
  A(\Bsb \to K^+ K^-) &=& C^\prime \frac{\lambda}{1-\lambda^2/2}
  \left( e^{- i \gamma} + \frac{1-\lambda^2}{\lambda^2}
  \left(
  d^\prime e^{i \theta^\prime} + e^{- i \phi_{\rm NP}} d^\prime_{\rm NP} e^{i \theta^\prime_{\rm NP}}
  \right)
  \right)\,, \nonumber
\end{eqnarray}
and taking uniformly distributed $d^\prime_{\rm NP} \in [0,2]$ and $\phi_{\rm NP}, \theta^\prime_{\rm NP}\in [-\pi,\pi]$ the PDFs shown in Fig.~\ref{fig:np} are obtained for $\kappa = 0.5$. 
This yields $\gamma = (74 \pm 7)^\circ$, and a $95\,\%$ probability upper bound on $d^\prime_\mathrm{NP}$ around $1$. 
Clearly, the bound is stronger for large values of $\phi_{\rm NP}$.

Finally, $\Bs \to KK$ decays can also be used to extract $2 \beta_s$ in the SM. 
The optimal choice in this respect is represented by $\Bs \to K^{(*)0} \bar K^{(*)0}$ (with $B^{0} \to K^{(*)0} \bar K^{(*)0}$ as U-spin related control channels to constrain subleading contributions), since in this channel there is no tree
contribution proportional to $e^{i \gamma}$~\cite{Fleischer:1999zi,Ciuchini:2007hx}. 
However, the combined analysis described above, in the framework of a global SM fit, can serve for the same purpose. 
To illustrate this point, the GL+F analysis is performed, taking
as input the SM fit result $\gamma = (69.7 \pm 3.1)^\circ$~\cite{Bona:2007vi} and not using the measurement of $2 \beta_s$ from $b \to c \bar c s$ decays.
In this way, $2 \beta_s = (3 \pm 14)^\circ$ is obtained for $\kappa = 0.5$. 
The analysis can also be performed without using the measurement of $\gamma$, in this case the result is $2 \beta_s = (6 \pm 14)^\circ$. 
With improved experimental accuracy, this determination could become competitive with that from $b \to c \bar c s$ decays.
Once results of time-dependent analyses of the $B^0_{(s)} \to K^{(*)0} \bar K^{(*)0}$ channels are available these may also provide useful constraints.\footnote{
  The proposal of Ref.~\cite{Ciuchini:2007hx} has been recently critically reexamined in Ref.~\cite{Bhattacharya:2012hh}. 
  The present analysis shows no particular enhancement of the contribution proportional to $e^{i \gamma}$ in $\Bs \to K^+ K^-$, in agreement with the expectation that $\Bs \to K^{(*)0} \bar K^{(*)0}$ should be penguin-dominated to a very good accuracy.
}

To conclude, the usual GL analysis to extract $\alpha$ from $B^{0} \to \pi\pi$ can be supplemented with the inclusion of the $\Bs \to K^+ K^-$ modes, in the framework of a global CKM fit. 
The method optimises the constraining power of these decays and allows the derivation of constraints on NP contributions to penguin amplitudes or on the $\Bs$ mixing phase and illustrates these capabilities with a simplified analysis, neglecting correlations with other SM observables.

\subsubsection{Studies of \CP violation in multibody charmless $b$ hadron decays}
\label{sec:bto3h}

Multibody charmless $b$ hadron decays can be used for a variety of studies of \CP violation, including searches for NP and determination of the angle $\gamma$.
Due to the resonant structure in multibody decays, these can offer additional possibilities to search for both the existence and features of NP.
Model-independent analyses~\cite{Bediaga:2009tr,Williams:2011cd} can be performed to first establish the presence of a \CP violation effect, and then to identify the regions of the phase space in which it is most pronounced.\footnote{
  LHCb has presented preliminary results from model-independent searches for \CP violation in $\Bpm \to \pip\pim\Kpm$ and $\Bpm\to\Kp\Km\Kpm$ at ICHEP 2012~\cite{LHCb-CONF-2012-018}, and in $\Bpm \to \pip\pim\pipm$ and $\Bpm \to \Kp\Km\pipm$ at CKM 2012~\cite{LHCb-CONF-2012-028}.
}
To further establish whether any observed \CP violation can be accommodated within the Standard Model, amplitude analyses can be used to quantify the effects associated with resonant contributions to the decay.
A number of methods have been proposed to determine $\gamma$ from such processes~\cite{Ciuchini:2006kv,Ciuchini:2006st,Gronau:2006qn,Bediaga:2006jk,Gronau:2007vr,Gronau:2010dd,Gronau:2010kq,Imbeault:2010xg,Sinha:2011ky,ReyLeLorier:2011ww}, in general requiring input not only from charged $B$ decays, but also from $\Bd$ and $\Bs$ decays (to states such as $\KS h^+h^{\prime -}$ and $\pi^0 h^+h^{\prime -}$).\footnote{
  LHCb has presented preliminary branching fraction measurements of $B^0_{(s)} \to \KS h^+h^{\prime -}$ decays at ICHEP 2012~\cite{LHCb-CONF-2012-023}.
}
The potential for LHCb to study multibody charmless $\Lb$ decays adds further possibilities for novel studies of \CP violation effects.

\subsubsection{Prospects of future LHCb measurements}
\label{sec:gammaprospects}

As discussed above, the angle $\gamma$ can be determined from both tree-dominated and loop-dominated processes.
Comparisons of the values obtained provide tests of NP, and so precision measurements from both methods are needed.
Among the tree-dominated processes, in addition to the modes discussed above, any channel that involves the interference of $b\to c\bar{u}s$ and $b\to u\bar{c}s$ transitions is potentially sensitive to $\gamma$.
Many of these modes can be analysed in the upgraded phase of LHCb, including

\begin{enumerate}
\item $B^+\to DK^+\pi^-\pi^+$ where, similarly to the $B\to DK$ mode, the neutral $D$ can be reconstructed either in the two-body (ADS and GLW-like measurement) or multibody (GGSZ-like measurement) final state. 
      The observation of the CF mode in LHCb data~\cite{LHCb-PAPER-2011-040} indicates a yield only twice lower than that for the $B\to DK$ mode, which makes it competitive for the measurement of $\gamma$.\footnote{
        Preliminary results from a GLW-type analysis of this channel was presented at ICHEP 2012~\cite{LHCb-CONF-2012-021}.
      }
      However, two unknown factors affect the expected $\gamma$ sensitivity. 
      First, since this is a multibody decay, the overlap between the interfering amplitudes is in general less than 100\,\%; this is accounted for by a coherence factor between zero and unity which enters the interference term in Eqs.~(\ref{eq:glw2}), (\ref{eq:glw3}), (\ref{eq:rads2}), (\ref{eq:aads2}) as an unknown parameter. 
      Second, the value of $r_B$ can be different from that in $B\to DK$ and is as yet unmeasured, although it is expected~\cite{Gronau:2002mu} that it can be larger in this decay than in $B\to DK$.

\item $B^0\to DK^+\pi^-$. 
      Although the rate of these decays is smaller that that of $B^+\to DK^+$, both interfering amplitudes are colour-suppressed, therefore the expected value of $r_B$ is larger, $r_B\simeq 0.3$. 
      As a result, the sensitivity to $\gamma$ should be similar to that in the $B\to DK$ modes.\footnote{ 
        Preliminary results from a GLW-type analysis of $\Bd \to D\Kstarz$ were presented at ICHEP 2012~\cite{LHCb-CONF-2012-024}.
      }
      Depending on the content of $B^0\to D^0K^+\pi^-$ and $B^0\to \overline{D}{}^0K^+\pi^-$ amplitudes, the optimal strategy may involve Dalitz plot analysis of the $B^0$ decay~\cite{Gershon:2008pe,Gershon:2009qc}.  
      In this case, control of amplitude model uncertainty will become essential for a precision measurement; it can be eliminated by studying the decays $B^0\to DK^+\pi^-$ with $D\to \KS\pi^+\pi^-$~\cite{Gershon:2009qr}.

\item $\Bs\to D\phi$. This mode is not self-tagging, but sensitivity to $\gamma$ can be obtained from untagged time-integrated measurements using several different neutral $D$ decay modes~\cite{Gronau:2007bh,LHCb-PUB-2010-005}.   The first evidence for the three-body decay $\Bs \to \Dzb \Kp\Km$ has just been reported by LHCb~\cite{LHCb-PAPER-2012-018}, and investigation of its resonant structure is in progress.

\item $\Bc\to DD_s^+$. $\Bc$ production in $pp$ collisions is significantly suppressed, however, in this mode the magnitude of \CP violation is expected to  be ${\cal O}(100\,\%)$: the two interfering amplitudes are of the same magnitude because the $b\to u\bar{c}s$ amplitude is colour allowed, while the $b\to c\bar{u}s$ amplitude is colour suppressed~\cite{Masetti:1992in,Fleischer:2000pp,Giri:2001be,Giri:2006cw}.

\item $\Lb\to D\L$ and $\Lb\to DpK^-$. Measurement of $\gamma$ from analysis of the $\Lb\to D\L$ decay mode was proposed in Ref.~\cite{Giri:2001ju}. This method allows one to measure $\gamma$ in a model-independent way by comparing the $S$- and $P$-wave amplitudes. However, this mode is problematic to reconstruct at LHCb because of the poorly defined $\Lb$ vertex (both particles from its decay are long-lived) and low efficiency of $\L$ reconstruction. Alternatively, one can consider a similar measurement with the decay $\Lb\to DpK^-$. 
A preliminary observation of this mode in early LHCb data has been reported~\cite{LHCb-CONF-2011-036}.
\end{enumerate}

Table~\ref{tab:gamma_tree_sens} shows the expected sensitivity to $\gamma$ from tree level decays in the upgrade scenario.
The LHCb upgrade is the only proposed experiment which will be able to reach sub-degree precision on $\gamma$.

\begin{table}[!htb]
  \begin{center}
  \caption{\small
    Estimated precision of $\gamma$ measurements with $50\invfb$
    for various charmed $B$ decay modes.}
  \label{tab:gamma_tree_sens}
  \begin{tabular}{lc}
    \hline
    Decay mode                                         & $\gamma$ sensitivity \\
    \hline
    $B\to DK$ with $D\to hh^\prime$, $D\to K\pi\pi\pi$       & $1.3^{\circ}$    \\
    $B\to DK$ with $D\to \KS\pi\pi$                  & $1.9^{\circ}$    \\    $B\to DK$ with $D\to 4\pi$                         & $1.7^{\circ}$    \\
    $B^0\to DK\pi$ with $D\to hh^\prime$, $D\to \KS\pi\pi$ & $1.5^{\circ}$    \\
    $B\to DK\pi\pi$ with $D\to hh^\prime$                    & $\sim 3^{\circ}$ \\
    Time-dependent $\Bs\to \Dsmp \Kpm$                      & $2.0^{\circ}$    \\
    \hline
    Combined                                           & $\sim 0.9^{\circ}$ \\
    \hline
  \end{tabular}
  \end{center}
\end{table}

Measurement of $\gamma$ and $2\beta_s$ by means  of the \CP-violating observables from loop-mediated decays $\Bd \to \pi^+ \pi^-$ and  $\Bs \to K^+ K^-$ was discussed in Sec.~\ref{sec:gammafromloop}. 
Extrapolating the current sensitivity on $C$ and $S$ to the upgrade scenario, when $50 \invfb$ of integrated luminosity will be collected,  LHCb will be able to reach a statistical sensitivity $\sigma_{\rm stat}(C) \approx \sigma_{\rm stat}(S) \simeq 0.008$ in both $\Bd \to \pi^+ \pi^-$ and $\Bs \to K^+ K^-$. 
This corresponds to a precision on $\gamma$ of $1.4^{\circ}$, and on $2\beta_s$ of $0.01$ rad, assuming perfect U-spin symmetry.

\clearpage

\section{\boldmath Mixing and \CP violation in the charm sector}
\label{sec:charm}

\subsection{Introduction}
\label{sec:charm:intro}

The study of \D mesons offers a unique opportunity to access
up-type quarks in flavour-changing neutral current (FCNC) processes.
It probes scenarios where up-type quarks play a special role,
such as supersymmetric models with alignment~\cite{Nir:1993mx,Leurer:1993gy}.
It offers complementary constraints on possible NP contributions to those arising from the measurements of FCNC processes of down-type quarks (\B or \kaon mesons). 

The neutral \D system is the latest and last system of neutral mesons where mixing between particles and anti-particles has been established.
The mixing rate is consistent with, but at the upper end of, SM expectations~\cite{Falk:2004wg} and constrains many NP models~\cite{Golowich:2007ka}.
More precise \Dz--\Dzb mixing measurements will provide even stronger constraints.  
However, the focus has been shifting to \CP violation observables, which provide cleaner tests of the SM~\cite{Okun:1975di,Pais:1975qs,Bigi:1986vr}.
First evidence for direct \CP violation in the charm sector has been reported by the \lhcb collaboration in the study of the difference of the time-integrated asymmetries of $\Dz\to\Kp\Km$ and $\Dz\to\pip\pim$ decay rates through the parameter \dacp~\cite{LHCb-PAPER-2011-023}. 
No evidence of indirect \CP violation has yet been found.
As discussed in detail below, these results on \CP violation in the charm sector appear marginally compatible with the SM but contributions from NP are not excluded.

The mass eigenstates of neutral \PD mesons, $|D_{1,2}\rangle$, with
masses $m_{1,2}$ and widths $\Gamma_{1,2}$ can be written as linear
combinations of the flavour eigenstates
$|D_{1,2}\rangle=p|\Dz\rangle\pm{}q|\Dzb\rangle$, with complex
coefficients $p$ and $q$ which satisfy $|p|^2+|q|^2=1$.  The average
mass and width are defined as $m\equiv(m_1+m_2)/2$ and
$\Gamma\equiv(\Gamma_1+\Gamma_2)/2$.  The \PD mixing parameters are
defined using the mass and width difference as
$x_D\equiv(m_2-m_1)/\Gamma$ and
$y_D\equiv(\Gamma_2-\Gamma_1)/2\Gamma$. The phase convention of $p$ and
$q$ is chosen such that $\CP|\Dz\rangle=-|\Dzb\rangle$.
First evidence for mixing of neutral \Dz mesons was discovered in 2007 by \belle and \babar~\cite{Aubert:2007wf,Staric:2007dt} and is now well established~\cite{HFAG}: 
the no-mixing hypothesis is excluded at more than  $10\,\sigma$ for the world average ($x_D=0.63\,^{+0.19}_{-0.20}\,\%$, $y_D=0.75\pm0.12\,\%$).\footnote{
  At HCP 2012, LHCb presented the first observation of charm mixing from a single measurement~\cite{LHCb-PAPER-2012-038}.
}

It is convenient to group hadronic charm decays into three categories.
The CF decays, such as $\Dz \to \Km\pip$, are mediated by tree amplitudes, and therefore no direct \CP violation effects are expected.
The same is true for DCS decays, such as $\Dz \to \Kp\pim$, even though these are much more rare.
The SCS decays, on the other hand, can also have contributions from penguin amplitudes, and therefore direct \CP violation is possible, even though the penguin contributions are expected to be small.
Within this classification, it should be noted that some decays to final states containing \KS mesons, \eg $\Dz \to \KS \rho^0$, have both CF and DCS contributions which can interfere~\cite{Bigi:1994aw}.  Within the SM, however, direct \CP violation effects are still expected to be negligible in these decays. 

\lhcb is ideally placed to carry out a wide physics programme in the
charm sector, thanks to the high production rate of open charm: with a 
cross-section of $6.10\pm0.93$~mb~\cite{LHCb-CONF-2010-013,LHCb-PAPER-2012-041}, one tenth of LHC interactions produce charm hadrons.  
Its ring-imaging Cherenkov detectors provide excellent separation between pions, kaons and protons in the momentum range between 2 and 100 \GeVc, and additional detectors also provide clean identification of muons and electrons.
This allows high purity samples to be obtained both for hadronic and muonic decays.  
The large boost of the \D hadrons produced at \lhcb is beneficial for time-dependent studies.  
\lhcb has the potential to improve the precision on all the key observables in the charm sector in the next years.

In the remainder of this section the key observables in the charm sector are described, and the current status and near term prospects of the measurements at LHCb are reviewed.  
A discussion of the implications of the first LHCb charm physics results follows, motivating improved measurements and studies of additional channels.
The potential of the LHCb upgrade to make the precise measurements needed to challenge the theory is then described.

\subsubsection{Key observables}
\label{sec:charm:intro:observables}

Currently the most precise individual measurements of mixing parameters are those of the relative effective lifetime difference between \Dz and \Dzb decays to \CP eigenstates ($\hat{\Gamma}$ and $\hat{\bar{\Gamma}}$) and flavour specific final states ($\Gamma$), \ycp, which is defined as
\begin{equation}
\ycp=\frac{\hat{\Gamma} + \hat{\bar{\Gamma}}}{2\Gamma} -1 \approx
\eta_{\CP}\left[\left( 1 -\frac{1}{8}A_m^2\right)y_D\cos\phi
  -\frac{1}{2}(A_m)x_D\sin\phi\right],
\end{equation}
where terms below ${\cal O}(10^{-4})$ have been ignored~\cite{Gersabeck:2011xj}, $\eta_{\CP}$ is the \CP eigenvalue of the final state, $\phi$ is the \CP-violating relative phase between $q/p$ and $\bar{A}_f/A_f$ where \AfAfbar~~are  the decay amplitudes, and $A_m$ represents a \CP violation contribution from mixing ($|q/p|^{\pm 2}\approx1\pm A_m$).\footnote{
  $A_m$ can be determined from asymmetries in semileptonic charm decays, with the assumption of vanishing direct \CP violation.
}  
In the limit of \CP conservation \ycp is equal to the mixing parameter $y_D$.  
The resulting world average value for \ycp is $0.87\pm0.16\,\%$~\cite{Neri:Charm2012}\footnote{
  New results presented by Belle at ICHEP 2012~\cite{Peng:ICHEP} are not included in this average.
} and is consistent with the value of $y_D$ within the current accuracy.

The \CP-violating observable \agamma quantifies the difference in decay
rates of \Dz and \Dzb to a \CP eigenstate and is defined as
\begin{equation}
\agamma=\frac{\hat{\Gamma} - \hat{\bar{\Gamma}}}{\hat{\Gamma} +
  \hat{\bar{\Gamma}}} \approx
\eta_{\CP}\left[\frac{1}{2}(A_m+A_d)y_D\cos\phi-x_D\sin\phi\right],
  \label{eq:agamma}
\end{equation}
where terms below ${\cal O}(10^{-4})$ have again been ignored~\cite{Gersabeck:2011xj} and both mixing and direct \CP contributions are assumed to be small. 
The parameter $A_d$ describes the contribution from direct \CP violation  ($|\bar{A}_f/A_f|^{\pm 2}\approx1\pm A_d$).
The current world average of \agamma is $0.02 \pm 0.16\,\%$~\cite{HFAG}, consistent with the hypothesis of no \CP violation.
Due to the smallness of $x_D$ and $y_D$, \agamma provides essentially the same information as a full time-dependent \CP violation analysis of $\Dz \to \Kp\Km$ decays.

An alternative way to search for \CP violation in charm mixing is with a time-dependent Dalitz plot analysis of \Dz and \Dzb decays to $\KS\pip\pim$ or $\KS\Kp\Km$.
Such analyses have been carried out at the $B$ factories~\cite{Abe:2007rd,delAmoSanchez:2010xz}. 
Also in these cases no \CP violation was observed.

In time-integrated analyses the measured rate asymmetry is 
\begin{equation}
\label{eqn:acp}
\ACPcharm\equiv\frac{\Gamma(\Dz\to f)-\Gamma(\Dzb\to f)}{\Gamma(\Dz\to
  f)+\Gamma(\Dzb\to f)}\approx a_{\CP}^{\rm dir}-\agamma\frac{\langle t
  \rangle}{\tau},
\end{equation}
where the direct \CP asymmetry contribution is defined as
\begin{equation}
a_{\CP}^{\rm dir}\equiv\frac{|A_f|^2-|\bar{A}_f|^2}{|A_f|^2+|\bar{A}_f|^2}\approx-\frac{1}{2}A_d
\end{equation}
and $\langle t \rangle$ denotes the average decay time of the observed
candidates.

A powerful way to reduce experimental systematic uncertainties is to measure the difference in time-integrated asymmetries in related final states.  For the two-body final states $\Kp\Km$ and $\pip\pim$, this difference is given by
\begin{eqnarray}
  \dacp & \equiv & \ACP(\Kp\Km)-\ACP(\pip\pim)\nonumber \\ 
  & \approx & \dacpdir
  \left( 1+y_D\cos\phi\frac{\overline{\langle t \rangle}}{\tau} \right) +
  \left(a_{\CP}^{\rm ind}+\overline{a_{\CP}^{\rm dir}}y_D\cos\phi \right)
  \frac{\Delta\langle t \rangle}{\tau}
\end{eqnarray}
where the \CP-violating phase $\phi$ is assumed to be universal~\cite{Grossman:2006jg}, 
$\Delta a \equiv a(\Kp\Km)-a(\pip\pim)$, $\overline{a} \equiv (a(\Kp\Km)+a(\pip\pim))/2$ and the indirect \CP asymmetry parameter is defined as $a_{\CP}^{\rm ind}=-(A_m/2)y_D\cos\phi+x_D\sin\phi$.  
The ratio $\Delta \langle t \rangle / \tau$ is equal to zero
for the lifetime-unbiased \PB factory measurements~\cite{Aubert:2007if,Staric:2008rx} and is $0.098\pm0.003$ for \lhcb~\cite{LHCb-PAPER-2011-023}
and $0.25\pm0.04$ for \cdf~\cite{Aaltonen:2011se}, therefore \dacp is largely a measure of direct \CP violation.

The current most accurate measurements of \dacp are from the \lhcb and \cdf collaborations and are
$(-0.82\pm0.21\pm0.11)\,\%$~\cite{LHCb-PAPER-2011-023} and
$(-0.62\pm0.21\pm0.10)\,\%$~\cite{Aaltonen:2012qw}, respectively.\footnote{
  At ICHEP 2012, Belle also presented new results on \dacp~\cite{Ko:ICHEP}, that are consistent with, but less precise than, those from \lhcb and CDF.
}
These results show first evidence of \CP violation in the charm sector:
the world average is consistent with no \CP violation at only 
0.006\,\% C.L.~\cite{HFAG}.


\subsubsection{Status and near-term future of \lhcb measurements}
\label{sec:charm:intro:lhcb}

LHCb has a broad programme of charm physics, including searches for
rare charm decays (see Sec.~\ref{sec:rare}), spectroscopy and measurements of
production cross-sections and asymmetries (see Sec.~\ref{sec:other}). 
In this section only studies of mixing and \CP violation are discussed. 
For reviews of the formalism, the reader is referred to Refs.~\cite{Bianco:2003vb,Kagan:2009gb,Gersabeck:2011xj} and the references therein, and for an overview of NP implications to Ref.~\cite{Grossman:2006jg}.

Mixing and indirect \CP violation occur only in neutral
mesons. These are probed in a number of different decay modes,
predominantly---but not exclusively---time-dependent ratio
measurements. In most cases, the same analysis yields measurements
of both mixing and \CP violation parameters, so these are considered together. 
By contrast, direct \CP violation may occur in decays of both
neutral and charged hadrons, and the primary sensitivity to it comes
from time-integrated measurements---though it may affect
certain time-dependent asymmetries as well, as discussed in Section~\ref{sec:charm:requirements}.

Several classes of mixing and indirect \CP violation measurements are possible at LHCb, particularly:
\begin{itemize}
  \item Measurements of the ratios of the effective $D^0$ lifetimes in decays to
    quasi-flavour-specific states (\eg $\Dz \to K^- \pi^+$) and \CP
    eigenstates $f_{\CP}$ (\eg $\Dz \to K^- K^+$). These yield $y_{\CP}$.
    Comparing the lifetime of $\Dz \to f_{\CP}$ and $\Dzb \to f_{\CP}$
    yields the \CP violation parameter $A_{\Gamma}$.
  \item Measurements of the time-dependence of the ratio
    of wrong-sign to right-sign hadronic decays (\eg
    $D^0 \to K^+ \pi^-$ {\it vs.} $D^0 \to K^- \pi^+$).
    The ratio depends on $y_D^{\prime} t$ and $(x_D^{\prime 2} + y_D^{\prime 2}) t^2$ (see, \eg, Ref.~\cite{Bianco:2003vb}), where
    \begin{eqnarray*}
      x_D^{\prime} = x_D \cos \delta + y_D \sin \delta \, , \\
      y_D^{\prime} = y_D \cos \delta - x_D \sin \delta \, ,
    \end{eqnarray*}
    and $\delta$ is the mode-dependent strong phase
    between the CF and DCS amplitudes. 
    Note that $(x_D^{\prime 2} + y_D^{\prime 2}) = x_D^2 + y_D^2 \equiv r_M$.
    The mixing parameters can be measured independently
    for \Dz and \Dzb to constrain indirect \CP violation, and the
    overall asymmetry in wrong-sign decay rates for
    \Dz and \Dzb gives the direct \CP violation parameter $A_d$.
  \item Time-dependent Dalitz plot fits to self-conjugate
    final states (\eg $D^0 \to \KS \pi^- \pi^+$).
    These combine features of the two methods above, along
    with simultaneous extraction of the strong phases relative to \CP eigenstate final states.
    Consequently they yield measurements of $x_D$ and $y_D$ directly.
    Likewise, the indirect \CP violation parameters $|q/p|$ and
    $\phi$ may be extracted, along with the asymmetry in phase
    and magnitude of each contributing amplitude
    (in a model-dependent analysis).
  \item Measurements of the ratio of time-integrated rates
    of wrong-sign to right-sign semileptonic decays
    (\eg $\Dz \to \Dzb \to K^+ l^- \bar{\nu}_l$ {\it vs.} $\Dz \to K^- l^+ \nu_l$).
    These yield $r_M$ and $A_m$.
\end{itemize}
Within \lhcb, analyses are planned or in progress for each of these methods.
A measurement of $y_{\CP}$ and $A_{\Gamma}$ from the 2010 data sample has been published~\cite{LHCb-PAPER-2011-032}. 
In addition, a preliminary result on the time-integrated wrong-sign rate in $\Dz \to K \pi$ from the 2010 sample is available~\cite{LHCb-CONF-2011-029}.\footnote{
  Results of charm mixing parameters in wrong-sign $\Dz \to \Kp\pim$ decays have been presented at HCP 2012~\cite{LHCb-PAPER-2012-038}.
}
A summary of what can be achieved with the 2010--2012 prompt charm samples is given in Table~\ref{tab:charm:currentStatus:indirect}. 
Note that the observables are generally related to several physics parameters, such that the combined constraints are much more powerful than individual measurements.
After analysing $2.5 \invfb$ of data, the mixing parameters $x_D$ and $y_D$ are expected to be determined at the level of $\mathcal{O}(10^{-4})$, and $A_{\Gamma}$ to be measured with a similar uncertainty. 
This will represent a significant improvement in precision compared to the current world averages, which have uncertainties
$\sigma_{x_D} = 0.19\,\%$,
$\sigma_{y_D} = 0.12\,\%$, and 
$\sigma_{A_{\Gamma}} = 0.23\,\%$.

\begin{table}[!htb]
  \caption{\small
    Projected statistical uncertainties with 1.0 and $2.5 \invfb$ of LHCb data. 
    Yields are extrapolated based on samples used in analyses of 2011 data; 
    sensitivities are projected from these yields assuming $1/\sqrt{N}$ scaling based on reported yields by LHCb, and using published input from \babar, \belle, and \cdf.
    The projected \CP-violation sensitivities may vary depending on the true values of the mixing parameters.
  }
  \label{tab:charm:currentStatus:indirect}
  \begin{center}
    \begin{tabular}{lccc}
      \hline
      Sample & Observable & Sensitivity ($1.0 \invfb$) & Sensitivity ($2.5 \invfb$) \\ \hline
      Tagged $KK$ & $y_{\CP}$ & $5 \times 10^{-4}$ & $4 \times 10^{-4}$ \\
      Tagged $\pi\pi$ & $y_{\CP}$ & $10 \times 10^{-4}$ & $7 \times 10^{-4}$ \\
      Tagged $KK$ & $A_{\Gamma}$ & $5 \times 10^{-4}$ & $4 \times 10^{-4}$ \\
      Tagged $\pi\pi$ & $A_{\Gamma}$ & $10 \times 10^{-4}$ & $7 \times 10^{-4}$ \\
      Tagged WS/RS $K\pi$ & $x_D^{\prime 2}$ & $10 \times 10^{-5}$ & $5 \times 10^{-5}$ \\
      Tagged WS/RS $K\pi$ & $y_D^{\prime}$ & $20 \times 10^{-4}$ & $10 \times 10^{-4}$ \\
      Tagged $\KS \pi \pi$ & $x_D$ & $5 \times 10^{-3}$ & $3 \times 10^{-3}$ \\
      Tagged $\KS \pi \pi$ & $y_D$ & $3 \times 10^{-3}$ & $2 \times 10^{-3}$ \\
      Tagged $\KS \pi \pi$ & $|q/p|$ & 0.5 & 0.3 \\
      Tagged $\KS \pi \pi$ & $\phi$ & $25^{\circ}$ & $15^{\circ}$ \\ \hline
    \end{tabular}
  \end{center}
\end{table}

For direct \CP violation, control of systematic uncertainties
associated with production and efficiency asymmetries is essential. 
To date, two techniques have been used to mitigate these effects:
\begin{itemize}
\item Measurement of differences in asymmetry between two related final
  states, such that systematic effects largely cancel---for example,
  $\ACP(D^0 \to K^-K^+) - \ACP(D^0 \to \pi^-\pi^+)$~\cite{LHCb-PAPER-2011-023}. 
  This is simplest with two-body or quasi-two-body decays. 
  This is discussed in more detail in
  Sec.~\ref{sec:charm:intro:dacp}.
\item Searching for asymmetries in the distributions of multi-body decays, such that differences in overall normalisation can be neglected and effects
  related to lab-frame kinematics are largely washed out --- for example,
  in the Dalitz plot distribution of $D^+ \to K^- K^+ \pi^+$~\cite{LHCb-PAPER-2011-017}.
\end{itemize}

In the longer term, the goal is to extract the \CP asymmetries for $\Dz \to K^+K^-$ and $\Dz \to \pi^+\pi^-$ separately, along with those for other decay modes. 
To achieve this, it will be necessary to determine the production and detector efficiencies from data. 
Progress has been made in this area, notably in the $D_s^+$ production asymmetry measurement~\cite{LHCb-PAPER-2012-009}, which involves determination of the pion reconstruction efficiency from $\Dstarp \to \Dz \pi^+, \Dz \to K^- \pi^- \pi^+ \pi^+$ decays in which one of the \Dz daughter pions is not used in the reconstruction.\footnote{
  The pion reconstruction efficiency asymmetry has also been used in the determination of the $\Dp$ production asymmetry~\cite{LHCb-PAPER-2012-026}.
} 
The detector asymmetries need to be determined as functions of the relevant variables, and similarly, the production asymmetries can vary as functions of transverse momentum and pseudorapidity.
Understanding these systematic effects with the level of precision and granularity needed for \CP asymmetry measurements is difficult and it cannot be assumed that these challenges will be solved in a short time scale.
Moreover, production asymmetries can be determined only with the assumption of vanishing \CP asymmetry in a particular (usually CF) control mode.
Therefore ultimately the resulting measurements of \CP asymmetries for individual decay modes are essentially \dacp measurements relative to CF decays.

A summary of analyses that are in progress or planned with the
2011--2012 data is given below:
\begin{description}
  \item $D^0 \to K^- K^+, \, \pi^- \pi^+$:
    Updates to the $0.6 \invfb$ \dacp analysis~\cite{LHCb-PAPER-2011-023}
    are in progress, using both prompt charm and charm from
    semileptonic $B$ decays (see Sec.~\ref{sec:charm:intro:dacp}).
  \item $D_{(s)}^+ \to \KS h^+, \, \phi h^+$:
    A \dacp-style analysis is possible by comparing
    asymmetries in a CF control mode (\eg $D^+ \to \KS \pi^+$)
    and the associated SCS mode (\eg $D^+ \to \phi \pi^+$),
    taking advantage of the inherent symmetry of the
    $\KS \to \pi^- \pi^+$ and $\phi \to K^- K^+$ decays.\footnote{
      A small difference in kinematic distributions can occur in
      $\phi \to K^- K^+$ due to crossing resonances.
    } 
    The different kinematic distributions of the tracks (requiring binning or reweighting) and the \CP asymmetry in the $\KS$ decay need to be taken into account.
  \item $D^+ \to \pi^+ \pi^- \pi^+, \, K^+ K^- \pi^+$:
    A search for \CP violation in $D^+ \to K^+ K^- \pi^+$ with the
    model-independent (so-called ``Miranda'') technique~\cite{Bediaga:2009tr}
    was published with the 2010 data sample~\cite{LHCb-PAPER-2011-017}, 
    comprising $0.04 \invfb$. 
    With such small data samples, detector effects are negligible. 
    However, from studies of control modes such as $D_s^+ \to K^- K^+ \pi^+$ it is found that this is no longer the case with $1.0 \invfb$ of data or more, 
    so an update will require careful control of systematic effects. 
    The $\pi^+ \pi^- \pi^+$ final state should be more tractable, since the
    $\pi^{\pm}$ interaction asymmetry does not depend strongly on momentum.
  \item $D^0 \to \pi^- \pi^+ \pi^- \pi^+, \, K^- K^+ \pi^- \pi^+$:
    Previous publications have focused mainly on $T$-odd moments~\cite{delAmoSanchez:2010xj}, but there is further information in the distribution of final-state particles.
    A Miranda-style binned analysis or a comparable unbinned method~\cite{Williams:2011cd} can be used.\footnote{
      Preliminary results on the $D^0 \to \pi^- \pi^+ \pi^- \pi^+$ decay were presented at ICHEP 2012~\cite{LHCb-CONF-2012-019}.
    }
  \item Baryonic decays:
    LHCb will collect large samples of charmed baryons, enabling novel searches for \CP-violation effects~\cite{Bigi:2012ev}. 
    Triggering presents a challenge, but trigger lines for several $\Lc$ decay modes of the form $\L h^+$ or $p h^- h^{\prime +}$ are already incorporated, allowing large samples to be recorded. 
    In addition to the  considerations outlined above for $D$ meson decays, the large proton-antiproton interaction asymmetry and the possibility of polarisation in the initial state must be taken into account.
\end{description}

\subsubsection{Experimental aspects of $\dacp$ and related measurements}
\label{sec:charm:intro:dacp}
The raw asymmetry measured for \Dstarp-tagged \Dz decays to a final
state $f$ is defined as:
\begin{equation}
  {\cal A}_{\mathrm{raw}}(f) \equiv \frac{
    N(D^{*+} \to D^0( f)\pi_s^+) \, - \, N(D^{*-} \to \Dzb (f)\pi_s^-)
  } {
    N(D^{*+} \to D^0( f)\pi_s^+) \, + \, N(D^{*-} \to \Dzb (f)\pi_s^-)
  },
\label{def:astarrawdef}
\end{equation}
where $N(X)$ refers to the number of reconstructed events of decay $X$
after background subtraction. This raw asymmetry arises from several sources:
the \Dstarp production asymmetry ${\cal A}_{\rm P}$,
the asymmetry in selecting the tagging slow pion ${\cal A}_{\rm D}(\pi^+_\mathrm{s})$,
the asymmetry in selecting the \Dz decay into the final state ${\cal A}_{\rm D}(f)$,
and the \CP asymmetry in the decay $\ACP(f)$.

Consider the general case of a measured rate $n_{\pm}$, an efficiency (or other correction) $\varepsilon_{\pm}$,
and the corrected rate $N_{\pm}$, where the subscript refers to \Dz or \Dzb.
Then:
\begin{equation}
  \frac{N_+}{N_-} = \frac{ n_+ / \varepsilon_+ }{ n_- / \varepsilon_- }
  = \frac{n_+}{n_-} \frac{\varepsilon_-}{\varepsilon_+} \,  .
  \label{eq:charm:intro:dacp:ratio}
\end{equation}
Defining a generic asymmetry $A_x$ as
\begin{displaymath}
  A_x \equiv \frac{x_+ - x_-}{x_+ + x_-} \,,
\end{displaymath}
gives the identity
\begin{displaymath}
  \frac{x_+}{x_-} = \frac{1 + A_x}{1 - A_x} \,.
\end{displaymath}
Then applying this to Eq.~(\ref{eq:charm:intro:dacp:ratio}),
\begin{equation}
  \frac{1+A_n}{1-A_n} = \left( \frac{1 + A_N}{1 - A_N} \right) \left( \frac{1+A_\varepsilon}{1-A_\varepsilon} \right) \, .
  \label{eq:charm:intro:dacp:uglyratios}
\end{equation}
Applying the Taylor series expansion
to Eq.~(\ref{eq:charm:intro:dacp:uglyratios}), gives
\begin{displaymath}
  \left( 1 + 2A_n + 2 A_n^2 + ... \right) = 
  \left( 1 + 2A_N + 2 A_N^2 + ... \right) 
  \left( 1 + 2A_\varepsilon + 2 A_\varepsilon^2 + ... \right) \,,
\end{displaymath}
and thus
\begin{equation}
  A_n = A_N + A_\varepsilon + \mbox{(terms of order $A^2$)} \, .
\end{equation}
Generalising this to include multiple asymmetries, the formula used in the published analysis~\cite{LHCb-PAPER-2011-023} is obtained
\begin{equation}
  {\cal A}_{\mathrm{raw}}(f) = \ACP(f) + {\cal A}_{\rm P} + A_{\rm D}(\pi^+_\mathrm{s}) + {\cal A}_{\rm D}(f) \, ,
\end{equation}
which is correct up to terms of second order in the asymmetries.
In practise, for $\Dz \to h^+h^-$, the asymmetries are
${\cal A}_{\rm P} \sim 1\,\%$, ${\cal A}_{\rm D}(\pi^+_\mathrm{s}) \sim 1$--$2\,\%$, and ${\cal A}_{\rm D}(f) = 0$ by construction.
Thus, the second-order correction is $\mathcal{O}(10^{-4})$.\footnote{
  Note that the LHCb dipole magnet creates regions of parameter space with large ${\cal A}_{\rm D}(\pi^+_\mathrm{s})$, particularly at the left and right edges of the acceptance. 
  These regions are excluded with fiducial cuts.
}
Further, ${\cal A}_{\rm D}(\pi^+_\mathrm{s})$ and ${\cal A}_{\rm P}$ are the same for $f=K^+ K^-$ and $f=\pi^+\pi^-$ (leaving aside differences in kinematic distribution, considered below) and so many terms cancel in the difference:\footnote{
  Note in particular that if $\ACP(K^+K^-) = \ACP(\pi^+\pi^-) = 0$, the approximation becomes exact at all orders.
}
\begin{displaymath}
  \dacp = {\cal A}_{\rm raw}(K^+K^-) - {\cal A}_{\rm raw}(\pi^+\pi^-)
  \approx \ACP(K^+K^-) - \ACP(\pi^+\pi^-) \, .
\end{displaymath}
At the present level of precision, with a statistical uncertainty of around 0.2\,\%, this approximation is perfectly adequate. 
However, when more data is accumulated --- and certainly after the upgrade --- it will be necessary to change the analysis to take second-order terms into account. 
This can be done using the ratio formulation of Eq.~(\ref{eq:charm:intro:dacp:ratio}), \ie\
\begin{eqnarray*}
  & 
  \frac{N_{KK,+}}{N_{KK,-}} = \left( \frac{n_{KK,+}}{n_{KK,-}} \right) \left( \frac{\varepsilon_-}{\varepsilon_+} \right)\,,
  \hspace{5mm}
  \frac{N_{\pi\pi,+}}{N_{\pi\pi,-}} = \left( \frac{n_{\pi\pi,+}}{n_{\pi\pi,-}} \right) \left( \frac{\varepsilon_-}{\varepsilon_+} \right) & \\
  & 
  \Rightarrow
  \frac{ N_{KK,+} / N_{KK,-} }{ N_{\pi\pi,+} / N_{\pi\pi,+} } =
  \frac{ n_{KK,+} / n_{KK,-} }{ n_{\pi\pi,+} / n_{\pi\pi,-} }
  & 
\end{eqnarray*}

The nuisance asymmetries ${\cal A}_{\rm P}$ and ${\cal A}_{\rm D}(\pi^+_\mathrm{s})$ cancel between the $K^+K^-$ and $\pi^+\pi^-$ final states because these are properties of the \Dstarp and of the tagging slow pion, respectively, which do not depend on the decay of the \Dz meson. 
However, an artificial correlation between these asymmetries and the decay mode can arise if the asymmetry varies as a function of some variable\footnote{
  The discussion is framed in terms of kinematic variables, since there
  are clear mechanisms that could cause problems there, but the same logic
  can be applied to magnet polarity, trigger conditions, \etc
} (\eg\ the momentum of the \Dstarp) {\em and} the reconstructed
distributions in this variable are different for the $K^+K^-$ and $\pi^+\pi^-$
final states (\eg\ due to detector acceptance of the daughter tracks). 
In such a scenario, the two modes would populate regions with different raw asymmetries and so the nuisance asymmetries would not cancel fully. 
Two techniques have been used to address this:
\begin{itemize}
  \item the data can be partitioned into smaller kinematic regions such that
    within each region the raw asymmetries are constant and/or the
    $K^+K^-$ and $\pi^+\pi^-$ kinematic distributions are equal;
  \item the data can be reweighted such that the $K^+K^-$ and $\pi^+\pi^-$
    kinematic distributions are equalised.
\end{itemize}
The first approach was used in the published LHCb result, and the second
in the CDF result~\cite{Aaltonen:2011se}.

There is another way in which the formalism could be broken: through the
presence of peaking backgrounds which
  (a) fake the signal,
  (b) occur at different levels for the $K^+K^-$ and $\pi^+\pi^-$ final states, {\em and}
  (c) have a different raw asymmetry from the signal.
The signal extraction procedure used in the published LHCb analysis is a fit to
the mass difference from threshold $\delta m \equiv m((h^+h^-)_{\Dz}\pi^+_\mathrm{s})) - m(h^+h^-) - m(\pi^+)$.
This is vulnerable to a class of background in which a real \Dstarp decay occurs
and the correct slow pion is found but the \Dz decay is partly misreconstructed, \eg $\Dz \to K^- \pi^+ \pi^0$ misidentified as $\Dz \to K^- K^+$. 
This typically creates a background which peaks in $\delta m$ but is broadly distributed in $m(h^+h^-)$. 
Only cases which lie within the narrow $m(h^+h^-)$ signal window will survive. This is more common for the $K^+ K^-$ final state than for $\pi^+ \pi^-$: the energy of a missing particle can be made up by misidentifying a pion as a kaon, but apart from $\Dz \to \pi^- e^+ \nu_e$ there is little that can fake the kinematics of $\Dz \to \pi^+ \pi^-$. 
In practise, the charged hadron identification at LHCb suppresses these background greatly, and their raw asymmetries are not expected to be very different from the signal. 
In the published LHCb analysis, the impact of these backgrounds on the asymmetry was estimated by measuring their size and asymmetry in the $h^+h^-$ mass sidebands and computing the effect of such a background on the signal with a toy Monte Carlo study. 
The alternative approach would be to use a full 2D fit to $m(h^+h^-)$ and $\delta m$, which would distinguish this class of peaking background from the signal by its $m(h^+h^-)$ distribution.

The three issues discussed above --- terms entering at second order in the asymmetries, non-cancellation due to kinematic correlations, and peaking backgrounds --- are particular to this analysis and will require some changes to the procedure as larger data samples become available. 
In addition, there are more generic systematic uncertainties associated with the fit procedure and with the handling of events with more than one candidate.
These are summarised in Table~\ref{tab:charm:dacp:intro:systematics}.

\begin{table}
  \caption{\small
    Summary of absolute systematic uncertainties for \dacp.
  }
  \begin{center}
    \begin{tabular}{lc}
      Source & Uncertainty \\ 
      \hline
      Fiducial requirement & 0.01\,\% \\
      Peaking background asymmetry & 0.04\,\% \\
      Fit procedure & 0.08\,\% \\
      Multiple candidates & 0.06\,\% \\
      Kinematic binning & 0.02\,\% \\
      \hline
      Total & 0.11\,\% \\
    \end{tabular}
  \end{center}
  \label{tab:charm:dacp:intro:systematics}
\end{table}

\subsection{Theory status of mixing and indirect \CP violation}
\subsubsection{Theoretical predictions for $\DG_{\PD}$, $\dm_{\PD}$ and indirect \CP violation in the Standard Model}
\label{sec:Alexey}

As discussed in Sec.~\ref{sec:charm:intro}, mixing of charmed mesons provides outstanding opportunities to search for physics beyond the SM. 
New flavour-violating interactions at some high-energy scale may,
together with the SM interactions, mix the flavour eigenstates 
giving mixing parameters that differ from their SM expectations.
It is known experimentally that \Dz--\Dzb mixing proceeds extremely slowly, 
which in the SM is usually attributed to the absence of super-heavy
quarks. 

Both SM and NP contributions to mass and width differences can be summarised as
\begin{equation} \label{XandY}
\begin{split}
  x_{\PD} &= \frac{1}{2M_{\PD}\Gamma_{\PD}}\, {\rm Re}\, 
    \left[ 2\bra{\Dzb} {H}^{|\Delta C|=2}\,\ket{\Dz} +
    \bra{\Dzb} \,i\! \int\! {\rm d}^4 x\, T \Big\{
    {\cal H}^{|\Delta C|=1}_w (x)\, {\cal H}^{|\Delta C|=1}_w(0) \Big\}
    \ket{\Dz} \right],
  \\
  y_{\PD} &= \frac{1}{2 M_{\PD}\Gamma_{\PD}}\, 
        {\rm Im}\, \bra{\Dzb}
        \,i\! \int\! {\rm d}^4 x\, T \Big\{
        {\cal H}^{|\Delta C|=1}_w (x)\, {\cal H}^{|\Delta C|=1}_w(0) \Big\}
        \ket{\Dz}.
\end{split}
\end{equation}
These formulae serve as the initial point of calculations of the mass and 
lifetime differences.
They include contributions from local (at charm mass scale) $\Delta C = 2$
interactions generated by the $b$-quark~\cite{Golowich:2005pt,Bobrowski:2010xg,Georgi:1992as,Ohl:1992sr,Bigi:2000wn} or NP particles and from SM-dominated time-ordered products of two $\Delta C = 1$ interaction Hamiltonians (see, however, Ref.~\cite{Golowich:2006gq}). 

A simple examination of Eq.~(\ref{XandY}) reveals that the local $\Delta C = 2$
interactions only affect $x_D$, thus
one can conclude that it is more likely that $x_{\PD}$ receives large NP contributions.
Hence, it was believed that an experimental observation of
$x_{\PD} \gg y_{\PD}$ would unambiguously reveal NP contributions to charm mixing.
This simple signal for NP was found to not be realised in nature, but it is interesting that the reverse relation, $x_{\PD} < y_{\PD}$  with $y_{\PD}$ expected to be determined by the SM processes, might nevertheless significantly affect the sensitivity to NP of experimental analyses of \PD mixing~\cite{Bergmann:2000id}.
Also, it is important to point out that, contrary to the calculations of the SM contribution to mixing, the contributions of NP models can be calculated relatively unambiguously~\cite{Golowich:2007ka,Gedalia:2009kh,Isidori:2010kg}.

The calculation of the SM contribution to the mixing amplitudes is rather sophisticated. 
In the SM $x_{\PD}$ and $y_{\PD}$ are generated only at second order in flavour SU(3)$_{\rm f}$ breaking, 
\begin{equation}
  x_{\PD}\,,\, y_{\PD} \sim \sin^2\theta_C \times [{\rm SU}(3)_{\rm f} \mbox{ breaking}]^2\,,
\end{equation}
where $\theta_C$ is the Cabibbo angle.
Therefore, predicting the SM values of $x_{\PD}$ and $y_{\PD}$ depends crucially on estimating the size of SU(3)$_{\rm f}$ breaking~\cite{Falk:2001hx,Falk:2004wg}.

There are currently two approaches, neither of which give very reliable results because $m_c$ is in some sense intermediate between heavy and light.
The ``inclusive'' approach is based on the OPE.
In the $m_c \gg \Lambda_{\rm QCD}$ limit, where $\Lambda_{\rm QCD}$ is a scale characteristic of the strong interactions, $\dm_{\PD}$ and $\DG_{\PD}$ can be expanded in terms of matrix elements of local operators~\cite{Bobrowski:2010xg,Georgi:1992as,Ohl:1992sr,Bigi:2000wn}.
Such calculations typically yield $x_{\PD}, y_{\PD} < 10^{-3}$.  
The use of the OPE relies on local quark-hadron duality (see, for example, Ref.~\cite{Chibisov:1996wf}), 
and on $\Lambda_{\rm QCD}/E_{\rm released}$ (with $E_{\rm released} \sim m_c$) being small enough to allow a truncation of the series.
Moreover, a careful reorganisation of the OPE series is needed, as terms with
smaller powers of $m_s$ are numerically more important despite being more
suppressed by powers of $1/m_c$~\cite{Bobrowski:2010xg,Georgi:1992as,Ohl:1992sr,Bigi:2000wn}.
The numerically dominant contribution is composed of over twenty unknown
matrix elements of dimension-12 operators, which are very hard to estimate.
As a possible improvement of this approach, it would be important 
to perform lattice calculations of those matrix elements, as well as make
perturbative QCD (pQCD) corrections to Wilson coefficients of those operators.

The ``exclusive'' approach sums over intermediate hadronic states, which may
be modelled or fit to experimental data~\cite{Donoghue:1985hh,Wolfenstein:1985ft,Colangelo:1990hj,Kaeding:1995zx,Anselm:1979fa,Cheng:2010rv}.
Since there are cancellations between states within a given SU(3)$_{\rm f}$
multiplet, one needs to know the contribution of each state with high 
precision.
However, the $\PD$ meson is not light enough that its decays are dominated by a few final states.
In the absence of sufficiently precise data on many decay rates and on strong
phases, one is forced to use some assumptions.
While most studies find $x_{\PD}, y_{\PD} < 10^{-3}$,
Refs.~\cite{Donoghue:1985hh,Wolfenstein:1985ft,Colangelo:1990hj,Kaeding:1995zx,Anselm:1979fa,Cheng:2010rv}
obtain $x_{\PD}$ and $y_{\PD}$ at the $10^{-2}$ level by arguing that
SU(3)$_{\rm f}$ violation is of order unity.
Particular care should be taken if experimental data are used to estimate
the mixing parameters, as the large cancellations expected in the calculation
make the final result sensitive to uncertainties in the experimental inputs.
It was shown that phase space effects alone provide enough SU(3)$_{\rm f}$ 
violation to induce $x_{\PD}, y_{\PD}\sim10^{-2}$~\cite{Falk:2001hx}.
Large effects in $y_D$ appear for decays close to threshold, where
an analytic expansion in SU(3)$_{\rm f}$ violation is no longer possible;
a dispersion relation can then be used to show that $x_{\PD}$ would receive
contributions of similar order of magnitude.
The dispersion calculation suffers from uncertainties associated with unknown
(off-shell) $q^2$-dependences of non-leptonic transition amplitudes and thus
cannot be regarded as a precision calculation, although it provides a realistic estimate of $x_{\PD}$.
As a possible improvement of this approach, an estimate of SU(3)$_{\rm f}$ breaking
in matrix elements should be performed.
In addition, a calculation with $\Vubbare \neq 0$ should also be done, which 
is important to understand the size of \CP violation in charm mixing.

Based on the above discussion, it can be seen that it is difficult to find a
clear indication of physics beyond the SM in \Dz--\Dzb mixing measurements alone.
However, an observation of large \CP violation in charm mixing would be a robust signal of NP.

\CP violation in $\PD$ decays and mixing can be searched for by a variety of methods.
Most of the techniques that are sensitive to \CP violation make use of the decay asymmetry $\ACPcharm(f)$~\cite{Grossman:2006jg,Kagan:2009gb}.
For instance, time-dependent decay widths for $\PD \to \kaon \pion$ are 
sensitive to \CP violation in mixing.
In particular, a combined analysis of $\PD \to \kaon \pion$ and $\PD \to \kaon\kaon$ can yield interesting constraints on \CP-violating parameters $y_{\CP}$ and $A_\Gamma$, as discussed in Sec.~\ref{sec:charm:intro:observables}.

With the \Dz--\Dzb transition amplitudes defined as follows:
\begin{equation}
  \bra{\Dz} {\cal H} \ket{\Dzb} = M_{12}-\frac i2\Gamma_{12}\,, \qquad
  \bra{\Dzb} {\cal H} \ket{\Dz} = M_{12}^*-\frac i2\Gamma_{12}^*\,,
\end{equation}
then in the limit where direct \CP violation is neglected, one can measure~\cite{Grossman:2006jg,Kagan:2009gb} four quantities, $x_{\PD}$,
$y_{\PD}$, $A_m$, and $\phi$, which are described by three physical variables,\footnote{
  Among various possible phase definitions, only $\phi_{12}$, the relative phase between $M_{12}$ and $\Gamma_{12}$, is convention-independent and so has physical consequences.
}
\begin{equation}\label{theo_params}
  x_{12} =  \frac{2 |M_{12}|}{\Gamma}\,, \
  y_{12} =  \frac{|\Gamma_{12}|}{\Gamma}\,, \
  \phi_{12} = \arg(M_{12}/\Gamma_{12})\,.
\end{equation}
This implies that there is a model-independent relation among experimental quantities~\cite{Kagan:2009gb,Grossman:2009mn},
\begin{equation}
  \frac{x_{\PD}}{y_{\PD}} = -\frac{1}{2} \frac{A_m}{\tan\phi}\,.
\end{equation}



\subsubsection{New physics in indirect \CP violation}
\label{sec:Oram}

Indirect \CP violation in charm mixing and decays is a unique probe for NP, since within the SM the relevant processes are described by the physics of the first two generations to an excellent approximation.
Hence, observation of \CP violation in \Dz--\Dzb mixing at a level higher than $\mathcal{O}(10^{-3})$ (which is the SM contribution) would constitute an unambiguous signal of NP.

The commonly used theoretical parameters $x_{12}$ and $\phi_{12}$ defined in Eq.~(\ref{theo_params}) can be expressed in terms of $x_D$, $y_D$ and $|q/p|$ as:
\begin{equation}\label{xphicp}
\begin{split}
  x_{12}^2&=x_D^2\frac{(1+|q/p|^2)^2}{4|q/p|^2}+
  y_D^2\frac{(1-|q/p|^2)^2}{4|q/p|^2}\,,\\
  \sin^2\phi_{12}&=\frac{(x_D^2+y_D^2)^2(1-|q/p|^4)^2}
  {16x_D^2y_D^2|q/p|^4+(x_D^2+y_D^2)^2(1-|q/p|^4)^2}\,.
\end{split}
\end{equation}

The latest fit\footnote{Not including results presented at ICHEP 2012 or later.} yields the following ranges~\cite{HFAG}
\begin{equation}
  x_D \in [0.24,0.99]\,\% \,, \qquad y_D \in [0.51,0.98] \,\% \,, \qquad
  \left|q/p\right| \in [0.59,1.26] \,,
\end{equation}
all at 95\,\% C.L.
The fit also provides 95\,\% C.L. ranges also for the theoretical parameters from Eq.~(\ref{theo_params}):
\begin{equation} \label{exp_x12}
  x_{12} \in [0.25,0.99]\,\% \,, \qquad y_{12} \in [0.51,0.98] \,\%
  \,, \qquad \phi_{12} \in [-8.4^\circ,24.6^\circ] \,.
\end{equation}
It should be noted that the experimental precision on the \CP violation parameters is more than two orders of magnitude away from their SM predictions.

\begin{figure}[!htb]
\centering
\includegraphics[width=0.6\textwidth]{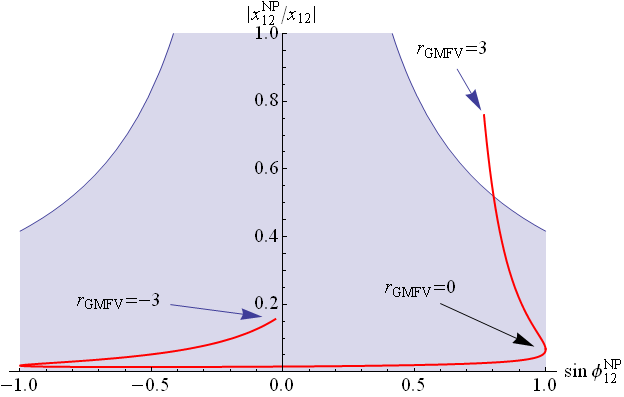}
\caption[NP allowed region in \Dz--\Dzb system $x_{12}^{\rm NP}/x_{12}-\sin\phi_{12}^{\rm NP}$]{\small
  Allowed region (shaded) in the $x_{12}^{\rm NP}/x_{12}-\sin\phi_{12}^{\rm NP}$ plane. 
  The red line corresponds to a GMFV prediction (see text for details) with
$r_{\rm GMFV} \in [-3,3]$.
  \label{fig:x12_phi12}
}
\end{figure}

It is reasonable to assume that there are no accidental strong
cancellations between the SM and the NP contributions to $M_{12}\,$.
Useful bounds can thus be obtained by taking the NP contribution to saturate
the upper limits in Eq.~(\ref{exp_x12}).
The resulting constraints are presented in the
$x_{12}^{\rm NP}/x_{12}-\phi_{12}^{\rm NP}$ plane in Fig.~\ref{fig:x12_phi12}.
One can also translate the data into model-independent bounds on four-quark
operators, as performed \eg\ in Refs.~\cite{Gedalia:2009kh, Isidori:2010kg}.

The generic NP analysis can also be applied to models with MFV, where new contributions to FCNCs originate only from the Yukawa matrices $Y_{u,d}\,$.
The relevant basis is then the up mass basis, where $Y_u$ is diagonal, so that flavour violation comes from powers of $Y_d Y_d^\dagger\,$.
The leading contribution is to the operator $(u_L^\alpha \gamma_\mu c_L^\alpha)^2$ ($\alpha$ is a colour index), and it is given in terms of its Wilson coefficient $C_1$ by
\begin{equation} \label{zmfv}
  C_1\propto [y_s^2(V_{cs}^*V_{us})+(1+r_{\rm GMFV})\times
  y_b^2(V_{cb}^*V_{ub})]^2.
\end{equation}
Here $r_{\rm GMFV}$ parameterises the effect of resummation of higher powers of the Yukawa matrices when these are important, namely in general MFV (GMFV) models~\cite{Kagan:2009bn}.

The contribution to $x_{12}$ in the linear MFV case ($r_{\rm GMFV}=0$) is orders of magnitude below the current experimental sensitivity, assuming $\mathcal{O}(1)$ proportionality coefficient in Eq.~(\ref{zmfv}). 
Yet in the context of GMFV with two Higgs doublets and large $\tan \beta$, such that $y_b \sim 1$, observable signals can be obtained, as shown in Fig.~\ref{fig:x12_phi12} for $r_{\rm GMFV}$ in the range $\left[-3,3\right]$. 
Note that strictly speaking $r_{\rm GMFV}$ (and thus the resulting signal) is not bounded, but higher absolute values than those considered here are much less likely in realistic models.
Indeed in the current example $r_{\rm GMFV} \gtrsim2$ is excluded, as shown in the figure.

The available data on \Dz--\Dzb mixing can also be used to constrain the parameter space of specific theories, such as SUSY and warped extra dimensions (WED)~\cite{Randall:1999ee}. 
This has been done \eg in Refs.~\cite{Gedalia:2009kh, Isidori:2010kg} or Refs.~\cite{Blum:2009sk,Gedalia:2012pi} where the interplay between the constraints from the \kaon and \PD systems is presented.
Here the influence of improving the current bounds is demonstrated.

\begin{figure}[!htb]
\label{fig:susy_ddbar_cpv}
\centering
\includegraphics[width=0.48\textwidth]{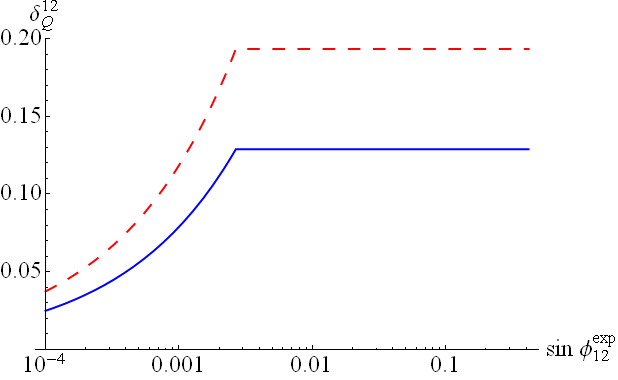}
\hspace{0.02\textwidth}
\includegraphics[width=0.48\textwidth]{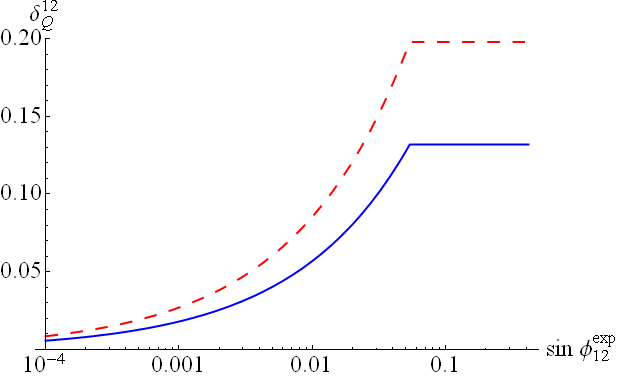}
\caption[Bound on squark mass degeneracy from \Dz--\Dzb \CP violation]{
  \small
  Bound on the squark mass degeneracy $\delta^{12}_{Q}$, defined in Eq.~(\ref{sqaurk_degen}), as a function of the experimental constraint on \CP violation in $\Dz$--$\Dzb$ mixing, parametrised by $\sin \phi_{12}^{\rm exp}$.
  The alignment angle from the down sector is $\lambda_{\rm C}^5$ (left panel)
  or $\lambda_{\rm C}^3$ (right panel).
  The solid blue line in each panel is for
  $\bar m_{\tilde Q}=m_{\tilde g}=1 \tev$ and the dashed red line is for
  $\bar m_{\tilde Q}=m_{\tilde g}=1.5 \tev$.}
\end{figure}

Within a SUSY framework, one can focus on the first two generations of the left-handed squark mass-squared matrix, $\tilde m_Q^2\,$, as the source of flavour violation.
As an additional assumption, the framework can be aligned with the down sector, where the constraints are generically stronger.
As in realistic alignment models (see \eg\ Refs.~\cite{Nir:1993mx,Nir:2002ah}),
the off-diagonal element of $\tilde m_Q^2$ in the down mass basis
(which induces $s \leftrightarrow d$ FCNCs) is taken to be small but not zero,
with comparable real and imaginary parts.
For concreteness, values of either $\lambda_{\rm C}^5$ or $\lambda_{\rm C}^3$ (with $\lambda_{\rm C}$ being the Cabibbo angle) are examined, where in both cases the dominant bounds still arise from \Dz--\Dzb mixing and not from the $K$ system~\cite{Gedalia:2012pi}.
The constrained parameter is the squark mass degeneracy, defined by
\begin{equation} \label{sqaurk_degen}
  \delta^{12}_{Q} \equiv \frac{m_{\tilde Q_2}-m_{\tilde
  Q_1}}{m_{\tilde Q_2}+m_{\tilde Q_1}}\,.
\end{equation}
In order to analyze the effect of improving the experimental
constraints on indirect \CP violation in charm (assuming that no such violation is actually observed), 
for simplicity the bound on $x_{12}$ is kept fixed as in Eq.~(\ref{exp_x12}), while that on $\phi_{12}$ is varied.
This is shown in Fig.~\ref{fig:susy_ddbar_cpv} for the two alignment angles
mentioned above and for two points in the SUSY parameter space
$\bar m_{\tilde Q}=m_{\tilde g}=1$ and $1.5$~\tev, where $\bar
m_{\tilde Q}$ is the average squark mass and $m_{\tilde g}$ is
the gluino mass.
The right edge of each of the four lines in
the plots marks the current situation, where the dominant
constraint is from $\Delta m_D\,$.
It is evident that after a certain level of improvement, the bound from
\CP violation becomes the important one, and this happens more quickly for a
weaker alignment model ($\lambda_{\rm C}^3$) than for
$\lambda_{\rm C}^5$ alignment. The reason is the larger phase
in the former case.

To conclude, the experimental search for indirect \CP violation
in charm is one of the most promising channels for discovering
NP or obtaining strong constraints.
This is not negated by the large hadronic uncertainties in the $\PD$ system,
because of the very small SM short distance contribution to \CP
violation in \Dz--\Dzb mixing.

\subsection{The status of calculations of \dacp in the Standard Model}
\label{sec:Alex}

As discussed above, the LHCb collaboration has measured a surprisingly large time-integrated \CP asymmetry difference~\cite{LHCb-PAPER-2011-023},
\begin{equation} \label{eq:dacp:lhcb}
  \dacp \equiv \ACPcharm(\DzToKK) - \ACPcharm(\DzTopipi)  
        = (-0.82 \pm 0.21 \pm 0.11)\,\%,
\end{equation} 
which has recently been supported by a result from the CDF collaboration~\cite{Aaltonen:2012qw}.\footnote{
  New results presented at ICHEP 2012, including a new result from Belle on \dacp~\cite{Ko:ICHEP}, are not included in the averages discussed here.
}
Inclusion of the \babar and \belle measurements of the individual
$\Km\Kp$ and $\pim\pip$ time-integrated \CP
asymmetries~\cite{Aubert:2007if,Staric:2008rx} and the \babar, \belle, and \lhcb measurements of the indirect \CP asymmetry
$\agamma$~\cite{Aubert:2007en,Staric:2007dt,LHCb-PAPER-2011-032} yields the world average for the
direct \CP asymmetry difference~\cite{HFAG} 
\begin{equation} \label{DeltaACP}
  \dacpdir \equiv \ACPdir(\DzToKK) - \ACPdir(\DzTopipi) = (-0.67 \pm 0.16)\,\%.
\end{equation}
The naive penguin-to-tree amplitude ratio is 
$\mathcal O([\Vcbbare \Vubbare / \Vcsbare \Vusbare] \as /\pi) \sim 10^{-4}$,
yielding $\dacpdir < 0.1\,\%$.
This has led to extensive speculation in the literature that the measurement
of \dacpdir is a signal for NP.
This is a particularly exciting possibility, given that reasonable NP 
models can be constructed in which all related flavour-changing neutral current
(FCNC) constraints, \eg, from \Dz--\Dzb mixing, are satisfied.
A summary of NP interpretations is given in Sec.~\ref{sec:Jernej}.
First, a discussion of \dacpdir in the SM is given.

The naive expectation for the SM penguin-to-tree ratio is based on estimates of
the ``short-distance" penguins with $b$-quarks in the loops.
In fact, there is consensus that a SM explanation for \dacpdir would have to
proceed via dynamical enhancement of the long-distance ``penguin contraction"
contributions to the penguin amplitudes, \ie, penguins with \squark and
\dquark quarks inside the ``loops".
Research addressing the direct \CP asymmetry in the SM has largely fallen into
one of two categories:
(i) flavour SU(3)$_{\rm f}$ or U-spin fits to the $\PD$ decay rates, to check that an
enhanced penguin amplitude can be accommodated~\cite{Golden:1989qx,Pirtskhalava:2011va,Bhattacharya:2012ah,Feldmann:2012js,Brod:2012ud,Cheng:2012xb,Hiller:2012xm}
(this, by itself, would not mean that \dacpdir is due to SM dynamics);
(ii) rough estimates of the magnitudes of certain contributions to the
long-distance penguin contractions~\cite{Buccella:1994nf,Brod:2012ud,Li:2012cf,Franco:2012ck}, to check if, in fact, it is reasonable that SM dynamics could yield the enhanced penguin amplitudes returned by the SU(3)$_{\rm f}$ or U-spin fits.

The results obtained using the flavour symmetry decompositions can be summarised as follows.
An SU(3)$_{\rm f}$ analysis of the \decay{\PD}{\PP\PP} decay amplitudes that
incorporates \CP violation effects was first carried out about 20 years ago~\cite{Golden:1989qx,Wolfenstein:1985ft,PhysRevLett.75.2460}.
Already in this study the possibility of large direct \CP asymmetries was
anticipated, \eg, as large as the percent level assuming that the penguins
receive a large enhancement akin to the $\Delta I = 1/2$ rule in kaon decays.
An updated analysis, working to first order in SU(3)$_{\rm f}$ breaking, has been presented~\cite{Pirtskhalava:2011va}, making use of branching ratio measurements for the \decay{\PD}{\kaon\pion, \pion\pion} and \decay{\Dz}{\Km\Kp, \Kzb\Peta} decay modes.
The authors concluded that \dacpdir can be easily reconciled with the measured branching ratios.
This was also the conclusion of a study based on a diagrammatic SU(3)$_{\rm f}$ amplitude decomposition~\cite{Bhattacharya:2012ah}, which considered a larger set of \decay{\PD}{\PP\PP} decay modes.
Again, this is only a statement about the possibility of accommodating the required amplitudes in the flavour decomposition,
not about their realisation via long distance QCD dynamics.
Both studies observe that a SM explanation of \dacpdir could be combined with
precise measurements of the individual asymmetries 
$\ACPdir(\DzToKK)$ and $\ACPdir(\DzTopipi)$ to obtain predictions for
$\ACPdir(\decay{\Dz}{\piz\piz})$.
The conclusion, based on current data, is that percent level asymmetries for
the latter could be realised.  
Ref.~\cite{Bhattacharya:2012ah} also discusses implications for
$\ACPdir(\decay{\Dp}{\Kp\Kzb})$.

Studies employing U-spin symmetry~\cite{Feldmann:2012js,Brod:2012ud} necessarily focus on amplitude fits to the smaller set of decay modes
\decay{\Dz}{\Km\pip, \pim\Kp, \pim\pip, \Km\Kp}, as the \Dz is a U-spin
singlet, while the four final states and the operators mediating these decays
in the SM $\Delta C=1$ effective Hamiltonian each consist of a U-spin triplet
and a singlet.
Working to first order in U-spin breaking, the four decay amplitudes can be written as
\begin{equation}\label{U-spin-decomp}
\begin{split}
  \amplitude{\DzbToKpi} &= \Vcsbare \Vudbare^* \left( T -\tfrac{1}{2} \delta T
\right)
    ,~~\amplitude{\DzbTopiK} = \Vcdbare \Vusbare^* \left(T +\tfrac{1}{2} \delta T  \right) \\ 
  \amplitude{\decay{\Dzb}{\pip\pim, \,\Kp\Km}} &= \mp \tfrac{1}{2}\big(\Vcsbare \Vusbare^* - \Vcdbare \Vudbare^*\big) \left( T  \pm \delta S  \right) 
        - \Vcbbare \Vubbare^* \left( P \mp \tfrac{1}{2} \delta P \right),
\end{split}
\end{equation}
where the U-spin triplet ``tree" amplitude $T$ and the singlet ``penguin"
amplitude $P$ arise at $0^{\mathrm{th}}$ order in U-spin breaking,
and $\delta T$, $\delta S$ and $\delta P$ are the first order U-spin breaking
corrections, which transform in turn as a triplet, singlet, and singlet under
U-spin.
The singlet amplitude $\delta S $ accounts for the large rate difference
$\Gamma(\DzToKK) / \Gamma(\DzTopipi) = 2.8$ (after accounting for phase space).
A ratio $\delta S / T  \sim 0.5 $ is found in Refs.~\cite{Feldmann:2012js,Brod:2012ud}, and in the SU(3)$_{\rm f}$ study of Ref.~\cite{Pirtskhalava:2011va} which effectively contains the above U-spin decomposition.
Realisation of Eq.~(\ref{DeltaACP}) requires $|P / T | \sim 3$, for ${\cal O}(1)$ strong phases and $\ACPdir(\DzToKK) \sim -\ACPdir(\DzTopipi)$, where the last
relation becomes an equality in the U-spin limit.
This amounts to an order of magnitude enhancement of the penguin amplitude
beyond the naive estimate.

The \CP-averaged experimental ``sum-rule'' relation, 
\begin{equation}\label{sumrule-exp}
  \Sigma_{\text{sum-rule}} = { 
        \left|\amplitude{\DzToKK}\,/\,\Vcsbare \Vusbare \right| 
        +  \left|\amplitude{\DzTopipi}\,/\,\Vcdbare \Vudbare \right|
     \over \left| \amplitude{\DzTopiK}\, /\, \Vcdbare \Vusbare \right|
        + \left| \amplitude{\DzToKpi}\, /\, \Vcsbare \Vudbare \right| } -1
  = (4.0\pm 1.6)\,\%\,,
\end{equation}
together with the observation of small ($\approx 15\,\%$) U-spin breaking in $\amplitude{\DzTopiK}$ {\it vs.} $\amplitude{\DzToKpi}$, can be interpreted as suggesting that U-spin is a good symmetry in these decays~\cite{Brod:2012ud}.
Other authors take the large difference between $\Gamma(\DzToKK)$ and $\Gamma(\DzTopipi)$ or $\delta S / T \sim 0.5 $ as evidence for large U-spin breaking in SCS decays.
In Ref.~\cite{Brod:2012ud}, rather than interpreting the amount of U-spin breaking implied by $\delta S$ by comparing it to $T$, as in other works, $\delta S$ is compared to $P$.  
It is observed that whereas \dacpdir implies that $P$ must be dominated by the sum of the long distance \squark- and \dquark- quark penguin contractions, nominal U-spin breaking would imply that $\delta S$ must be dominated by their difference.
A consistent picture emerges in which 
direct \CP asymmetries of order a few per mille are not surprising
given the size of $\Gamma(\DzToKK) / \Gamma(\DzTopipi)$.
However, as always in the flavour decomposition approach, accommodation need not translate to realisation by QCD dynamics.
One consequence of this picture is that $\ACPdir(D^0 \to \Ks \Ks)$ could be as large as  $\sim 0.6\,\%$ for ${\cal O}(1)$ strong phases.

Finally, the estimates for the long-distance penguin contractions~\cite{Brod:2011re,Franco:2012ck} are reviewed to see if the required enhancement can be realised. 
Ref.~\cite{Brod:2011re} employs the one-gluon exchange approximation.
The essential ingredients are:  (i) $1/N_c$ counting; 
(ii)  $\PD$ branching ratio data which shows that certain formally $1/m_c$
power-suppressed amplitudes are of same order as their leading $(1/m_c)^0$
counterparts; 
(iii) translation of this breakdown of the $1/m_c$ expansion to the penguin
contraction amplitudes, in the approximation of a hard gluon exchange;
(iv) use of a partonic quantity as a rough estimator of the hadronic
interactions, \eg, final state interactions, underlying the penguin contraction
``loops".
This results in a rough estimate for \dacpdir at the few per mille level.
The authors of Ref.~\cite{Brod:2011re} thus conclude that a SM explanation is
plausible, given that their estimate suffers from large uncertainties.  
In Ref.~\cite{Franco:2012ck} the penguin contractions are estimated using isospin
and information from $\pion\pion$ scattering and unitarity.  
A fit of the \CP-conserving contributions from the \CP-averaged branching ratios provides information on the isospin amplitudes
and the underlying renormalisation group invariant amplitude contributions.
Allowing for three coupled channel contributions to $\pion\pion, \kaon\kaon$ scattering 
the authors conclude that the observed asymmetries are marginally compatible with the SM.   

To summarise, flavour SU(3) or U-spin fits to the \decay{\PD}{\PP\PP} data can accommodate the enhanced penguin amplitudes required to reproduce \dacpdir.
There is consensus that in this case $\ACPdir(\decay{\Dz}{\piz\piz})$ could lie at the percent level, while $\ACPdir(\decay{\Dp}{\Kp\Kzb})$ could certainly lie at the few per mille level.
Under the assumption of nominal SU(3)$_{\rm f}$ breaking in \decay{\PD}{\PP\PP} decays, the enhancement of the long-distance penguin contractions required to realise \dacpdir is not surprising, given the large difference between the \DzToKK and \DzTopipi decay rates.
It would of course be of interest to extend the above \CP violation studies to the SCS \decay{\PD}{\PV\PP} and \decay{\PD}{\PV\PV} decay modes.
Finally, among the works which have attempted to estimate directly the magnitudes of the long distance penguin contractions, there is no consensus on whether they can be enhanced by an order of magnitude beyond the naive penguin amplitude estimates, as would be required in order to explain \dacpdir.
Ultimately this question will have to be answered directly via lattice studies.

In the following section, future prospects are discussed.
In subsequent sections, several definitive \CP-violating signals for NP in SCS $\PD$ decays will be discussed.


\subsection{\dacp in the light of physics beyond the Standard Model}
\subsubsection{General considerations}
\label{sec:Jernej}

Potential NP contributions to \dacp can be parametrised in terms of an
effective Hamiltonian valid below the $W$ and top mass scales
\begin{equation}\label{eq:HNP}
  \mathcal H^{\rm eff-NP}_{|\Delta C|=1} = \frac{G_F}{\sqrt 2} \sum_i C_i^{\rm NP(\prime)} \mathcal Q_i^{(\prime)}\,,
\end{equation}
where the relevant operators $\mathcal Q_i^{(\prime)}$ are defined in Ref.~\cite{Isidori:2011qw}.
Introducing the ratios $R^{{\rm NP},i}_{\kaon,\pion} $ as the relevant NP
hadronic amplitudes 
(matrix elements $\bra{\Km\Kp, \pim\pip} { \mathcal Q^{(\prime)}_i } \ket{\PD}$)
normalised to the leading \CP-conserving SM contributions and writing
$C_i^{\rm NP} = v_{\rm EW}^2 / \Lambda_{\rm NP}^2$, the relevant NP scale $\Lambda_{\rm NP}$ is given by~\cite{Isidori:2011qw}
\begin{equation}
  \frac{(10~{\tev})^2}{\Lambda_{\rm NP}^2} = \frac{(0.61\pm 0.17)-0.12 \, {\rm Im}(\Delta R^{\rm SM})}{{\rm Im}(\Delta R^{{\rm NP},i})}\,,
\end{equation}
where  $\Delta R^i = R^i_{\kaon} + R^i_{\pion}$ and $R_{\kaon,\pion}^{\rm SM}$
parametrise the unknown hadronic amplitude ratios associated with the \CP-violating SM contributions.
Comparing this estimate to the much higher effective scales probed by \CP-violating observables in $\PD$ mixing and also in the kaon sector, one first needs to verify if such large contributions can still be allowed by other flavour constraints.
Within the effective theory approach, this can be estimated via so-called ``weak mixing'' of the effective operators.
In particular, time-ordered correlators of
$\mathcal H^{\rm eff-NP}_{|\Delta C|=1}$ with the SM effective weak Hamiltonian
can, at the one weak-loop order, induce important contributions to \CP
violation in both $\PD$ meson mixing and kaon decays ($\epsilon^\prime/\epsilon$).
On the other hand, analogous correlators quadratic in
$\mathcal H^{\rm eff-NP}_{|\Delta C|=1}$ turn out to be either chirally
suppressed and thus negligible, or yield quadratically divergent contributions,
which are thus highly sensitive to particular UV completions of the effective
theory~\cite{Isidori:2011qw}. 

\subsubsection{Universality of \CP violation in flavour-changing decay processes}

The strongest bounds can be derived for a particular class of operators, which
transform non-trivially only under the SU(3)$_Q$ subgroup of the global SM
quark flavour symmetry $\mathcal G_F = {\rm SU}(3)_Q \times {\rm SU}(3)_U \times {\rm SU}(3)_D$,
respected by the SM gauge interactions.
In particular one can prove that their \CP-violating contributions to
$\Delta F=1$ processes (here $F$ generically represents a flavour quantum number) have to be approximately universal between the up and down sectors~\cite{Gedalia:2012pi}.
Within the SM one can identify two unique sources of SU(3)$_Q$ breaking given
by $\mathcal{A}_u \equiv (Y_u Y_u^\dagger)_{\slashed{\mathrm{tr}}}$ and $\mathcal{A}_d \equiv (Y_d Y_d^\dagger)_{\slashed{\mathrm{tr}}}$, where $Y_q$ are the Yukawa matrices and $\slashed{\mathrm{tr}}$ denotes the traceless part.
Then in the two generation limit, one can construct a single source of \CP
violation, given by $J\equiv i [\mathcal{A}_u, \mathcal{A}_d]$~\cite{Gedalia:2010zs,Gedalia:2010mf}.
The crucial observation is that $J$ is invariant under $SO(2)$ rotations
between the $\mathcal{A}_u$ and $\mathcal{A}_d$ eigenbases.
Introducing now SU(2)$_Q$ breaking NP effective operator contributions  of the
form
$\mathcal  Q_L = \Big[(X_L)^{ij}\, \overline Q_i \gamma^\mu Q_j \Big] L_\mu$,
where $Q_i$ stands for the left-handed quark doublets, $i$ and $j$ are generation indices, $X_L$ is a traceless Hermitian flavour matrix and $L_\mu$ denotes a flavour singlet current.
It follows that the \CP-violating contributions have to be proportional to $J$ and thus invariant under flavour rotations.
The universality of \CP violation induced by $\mathcal Q_L$ can be expressed
explicitly as~\cite{Gedalia:2012pi}
\begin{equation}\label{Uni2gen}
  {\rm Im}(X^u_L)_{12} = {\rm Im}(X^d_L)_{12}\propto {\rm Tr} \left( X_L \cdot
J\right)\, .
\end{equation}
The above identity holds to a very good approximation even in the
three-generation framework.
In the SM, large values of $Y_{\bquark,\tquark}$ induce a SU(3)/SU(2) flavour
symmetry breaking pattern~\cite{Kagan:2009bn} which allows one to decompose
$X_L$ under the residual SU(2) in a well defined way.
Finally, residual SM SU(2)$_Q$ breaking is necessarily suppressed by small mass
ratios $m_{\cquark,\squark}/m_{\tquark,\bquark}$, and small CKM mixing angles.
The most relevant implication of Eq.~(\ref{Uni2gen}) is that it predicts a
direct correspondence between SU(3)$_Q$ breaking NP contributions to
\dacp and $\epsilon^\prime/\epsilon$~\cite{Gedalia:2012pi}.
It follows immediately that stringent limits on possible NP contributions to
the latter require SU(3)$_Q$ breaking contributions to the former to be below
the per mille level (for $\Delta R^{{\rm NP},i}=\mathcal O (1)$).
As a corollary, one can show that within NP scenarios which only break
SU(3)$_Q$, existing stringent experimental bounds on new contributions to
\CP-violating rare semileptonic kaon decays
\decay{\KL}{\ensuremath{\piz (\neu\neub, \ellp\ellm)}} put robust constraints
on \CP asymmetries of corresponding rare charm decays
\decay{\PD}{\ensuremath{\pion (\neu\neub, \ellp\ellm)}}.
In particular, the SU(3)$_Q$-violating contribution to the \CP asymmetry in $D \to \pion \epem$ has been shown to be less than 2\,\%~\cite{Gedalia:2012pi}.

The viability of the remaining 4-quark operators in
$\mathcal H^{\rm eff-NP}_{|\Delta C|=1}$ as explanations of the experimental
\dacp value depends crucially on their flavour and chiral structure (a full
list can be found in Ref.~\cite{Isidori:2011qw}).
In particular, operators involving purely right-handed quarks are unconstrained
in the effective theory analysis but may be subject to severe constraints from
their UV sensitive contributions to $\PD$ mixing observables.
On the other hand, QED and QCD dipole operators are at present only weakly
constrained by nuclear electric dipole moments (EDMs) and thus present the best candidates to address the \dacp puzzle~\cite{Isidori:2011qw}.

Finally, note that it was shown that the impact of universality of \CP within the alignment framework is to limit the amount of \CP violation in \Dz--\Dzb mixing to below $\sim 20\,\%$, which is interestingly near the current bound. 
The expected progress in this measurement with the LHCb detector is therefore going to start probing this framework.

\subsubsection{Explanations of \dacp within NP models}

Since the announcement of the LHCb result, several prospective explanations of \dacp within various NP frameworks have appeared.
In the following the implications within some of the well-motivated NP models are discussed. 

In the MSSM, the right size of the QCD dipole operator contributions can be generated with non-zero left-right up-type squark mixing contributions $(\delta^u_{12})_{LR}$~\cite{Grossman:2006jg,Giudice:2012qq,Hiller:2012wf}.
Such effects in \dacp can be parametrised as~\cite{Giudice:2012qq}
\begin{equation}
  \label{eq:dacp:mia}
  | \dacpother{\ensuremath{\rm SUSY}}| \approx 0.6\,\% \left( \frac{|{\rm Im}(\delta_{12}^u)_{LR}|}{10^{-3}} \right) \left( \frac{\tev}{\tilde m} \right)\,,
\end{equation}
where $\tilde m$ denotes a common squark and gluino mass scale.
At the same time dangerous contributions to $\PD$ mixing observables are
chirally suppressed.
It turns out however that even the apparently small $(\delta^u_{12})_{LR}$
value required implies a highly nontrivial flavour structure of the UV theory;
in particular, large trilinear ($A$) terms and sizeable mixing among the first
two generation squarks ($\theta_{12}$) are required~\cite{Giudice:2012qq}. 
\begin{equation}
  \label{eq:dacp:susy}
  {\rm Im}(\delta^u_{12})_{LR} \approx 
  \frac{{\rm Im}(A) \theta_{12} m_c}{\tilde m} \approx
  \left(\frac{{\rm Im}(A)}{3}\right) \left(\frac{\theta_{12}}{0.3}\right) \left(\frac{\tev}{\tilde m}\right) 0.5 \times 10^{-3}\,.
\end{equation}

Similarly, WED models that explain the quark spectrum through flavour anarchy~\cite{Randall:1999ee,Goldberger:1999uk,Huber:2000ie,Gherghetta:2000qt} can naturally give rise to QCD dipole contributions affecting \dacp as~\cite{Delaunay:2012cz}
\begin{equation}
  | \dacpother{\ensuremath{\rm WED}} | \approx 0.6\,\% \left( \frac{Y_5}{6} \right)^2 \left( \frac{3 \tev}{m_{\rm KK}} \right)^2\,,
\end{equation}
where $m_{\rm KK}$ is the Kaluza-Klein (KK) scale and $Y_5$ is the five-dimensional Yukawa coupling in appropriate units. 
Reproducing the experimental value of \dacp requires near-maximal 5D Yukawa
coupling, close to its perturbative bound~\cite{Agashe:2008uz,Csaki:2009wc}
of $4\pi/\sqrt{N_{\rm KK}} \simeq 7$ for $N_{\rm KK} = 3$ perturbative KK states.
In turn, this helps to suppress unrealistic tree-level contributions to \CP
violation in \Dz--\Dzb mixing~\cite{Isidori:2010kg,Gedalia:2009kh}.
This scenario can also be interpreted within the framework of partial compositeness in four dimensions, but generic composite models typically require smaller Yukawa couplings to explain $\dacp$ and consequently predict sizeable contributions to \CP violation in $\Delta F=2$ processes~\cite{KerenZur:2012fr}.  

On the other hand, in the SM extension with a fourth family of chiral fermions
\dacp can be affected by $3\times 3$ CKM non-unitarity and $b^\prime$ penguin
operators
\begin{equation}
  | \dacpother{\ensuremath{\rm 4th\, gen}} | \propto {\rm Im} \left( \frac{\lambda_{b^\prime}}{\lambda_d - \lambda_s} \right)\,.
\end{equation}
However, due to the existing stringent constraints on the new \CP-violating
phases entering $\lambda_{b^\prime}$~\cite{Bobrowski:2010xg,Nandi:2010zx}, only
moderate effects comparable to the SM estimates are allowed~\cite{Feldmann:2012js}.

Finally, it is possible to relate $\dacp$ to the anomalously large forward-backward asymmetry in the $t\bar{t}$ system measured at the Tevatron~\cite{cdfAFBnew} through a minimal model. 
Among the single-scalar-mediated mechanisms that can explain the top data, only the $t$-channel exchange of a colour-singlet weak doublet, with a very special flavour structure, is consistent with the total and differential $t\bar{t}$ cross-section, flavour constraints and electroweak precision measurements~\cite{Blum:2011fa}. 
The required flavour structure implies that the scalar {\it unavoidably} contributes at tree level to $\dacp$~\cite{Hochberg:2011ru}.  
The relevant electroweak parameters are either directly measured, or fixed by the top-related data, implying that, for a plausible range of the hadronic parameters, the scalar-mediated contribution is of the right size.


\subsubsection{Shedding light on direct \CP violation via \decay{\PD}{\PV\Pgamma} decays}
\label{sec:Gino}

The theoretical interpretation of \dacp is puzzling: it is above its naive
estimate in the SM and it could well be a signal of NP, but it
is not large enough to rule out a possible SM explanation. 
It is then important to identify possible future experimental tests able to
distinguish standard {\it vs.} non-standard explanations of \dacp.
Among the NP explanations of \dacp, the most interesting ones are those based
on a new \CP-violating phase in the $\Delta C=1$ chromomagnetic  operator. 
A general prediction of this class of models, that could be used to test this
hypothesis from data, is enhanced direct \CP violation (DCPV) in radiative decay modes~\cite{Isidori:2012yx}.

\begin{enumerate}
\item 
The first key observation to estimate DCPV asymmetries in
radiative decay modes is the strong link between the $\Delta C=1$
chromomagnetic operator
($Q_{8} \sim  \bar u_L \sigma_{\mu\nu} T^a g_s G_a^{\mu\nu} c_R$) and the
$\Delta C=1$ electromagnetic-dipole operator 
($Q_{7} \sim \bar u_L \sigma_{\mu\nu}   Q_u  e F^{\mu\nu} c_R $).
In most explicit new-physics models the short-distance Wilson coefficients  of
these two operators ($C_{7,8}$) are expected to be similar.
Moreover, even assuming that only a non-vanishing $C_{8}$ is generated at
some high scale, the mixing of the two operators  from strong interactions
implies $C_{7,8}$ of comparable size at the charm scale.
Thus if \dacp is dominated by NP contributions generated by $Q_{8}$, it can be inferred that 
$ |\mathrm{Im}[C^{\rm NP}_7(m_c)] | \approx  |\mathrm{Im}[C^{\rm NP}_8(m_c)] | = (0.2 \textendash{} 0.8) \times 10^{-2} $.

\item
The second important ingredient   is the observation 
that in the Cabibbo-suppressed \decay{\PD}{\PV\Pgamma} decays, 
where $\PV$ is a light vector meson with \uubar valence quarks 
($\PV = \Prho^0,\Pomega$), $Q_7$ has a sizeable hadronic matrix element.
More explicitly, the short-distance contribution induced by $Q_7$, relative to the total (long-distance) amplitude,  
is substantially larger with respect to the corresponding relative weight of
$Q_8$ in \decay{\PD}{\PP^+\PP^-} decays.
Estimating the SM long-distance contributions from data, and evaluating the
short-distance \CP-violating contributions under the hypothesis that
\dacp is dominated by (dipole-type) NP, leads to the following estimate for 
the maximal direct \CP asymmetries in the \decay{\PD}{(\Prho, \Pomega) \Pgamma}
modes~\cite{Isidori:2012yx}:
\begin{equation} \label{eq:maxCPrho} 
  |\ACPdir(\decay{\PD}{(\Prho, \Pomega) \Pgamma})|^{\rm max} = 
        0.04 \left|\frac{{\rm Im}[C_7(m_c)]} {0.4\times 10^{-2}} \right| 
        \times \left[\frac{10^{-5}}{\BF(\decay{\PD}{(\Prho, \Pomega) \Pgamma})}\right]^{1/2}  \lsim 10\,\%\,. \quad
\end{equation}
The case of the $\phi$ resonance, or better the $\Kp\Km\Pgamma$ final state
with $M_{\PK\PK}$ close to the $\Pphi$ peak, is more involved since the 
matrix element  of  $Q_7$ vanishes in the large $m_c$ limit for a pure
\ssbar state. 
However, a non-negligible \CP asymmetry can be expected also in this case
since:
1) the matrix element of  $Q_7$ is not expected to be identically zero because
of sizeable $\mathcal{O}(\Lambda_{\rm QCD}/m_c)$ corrections; 
2) nonresonant contributions due to (off-shell)  $\Prho$ and $\Pomega$
exchange can also contribute to the $\Kp\Km\Pgamma$ final state. 
Taking into account these effects, the following estimates for the maximal
direct \CP asymmetries are obtained~\cite{Isidori:2012yx}:
\begin{equation} \label{eq:maxCPphi}
\begin{split}
  |\ACPdir(\decay{\PD}{\Kp\Km\Pgamma})|^{\rm max}  &\approx  2\,\%\,, \quad   2m_K <  \sqrt{s} < 1.05~\gev~, \\
  |\ACPdir(\decay{\PD}{\Kp\Km\Pgamma})|^{\rm max}  &\approx  6\,\%\,,  \quad  1.05~\gev <  \sqrt{s} < 1.20~\gev~. 
\end{split}
\end{equation}
In the first bin, close to the $\Pphi$ peak, the leading contribution is due
to the $\Pphi$-exchange amplitude.
The contribution due to the nonresonant amplitudes becomes more significant further from the $\Pphi$ peak, where the \CP asymmetry can become larger.  

\item
In order to establish the significance of these results, two important issues
have to be clarified:
1) the size of the \CP asymmetries within the SM,
2) the role of the strong phases. 

As far as the SM contribution is concerned, it can first be noticed that short-distance contributions generated by the operator $Q_7$ are safely negligible.
Using the result in Ref.~\cite{Greub:1996wn}, asymmetries are found to be below the $0.1\,\%$ level.
The dominant SM contribution is expected from the leading non-leptonic
four-quark operators, for which the general arguments discussed in Ref.~\cite{Isidori:2011qw} can be applied.
The \CP asymmetries can be decomposed as
$|\ACPother{\SM}(f)|  \approx 2   \xi ~\mathrm{Im} (R_{f}^{\SM}) \approx 0.13\,\% \times \mathrm{Im} (R_{f}^{\SM})$,
where $\xi \equiv \left| \Vcbbare\Vubbare / \Vcsbare\Vusbare \right| \approx 0.0007$
and $R_{f}^{\SM}$ is a ratio of suppressed over leading hadronic amplitudes, 
naturally expected to be smaller than one.
This decomposition holds both for $f = \pion\pion,\kaon\kaon$ and for $f = \PV\Pgamma$ channels.
The SM model explanations  of $\dacpother{}$ require
$R_{\pion\pion,\kaon\kaon}^{\SM}\sim3$.
While this possibility cannot be excluded from first principles, 
a further enhancement of one order of magnitude in the \decay{\PD}{\PV\Pgamma}
mode is beyond any reasonable explanation in QCD. 
As a result, an observation of $|\ACPdir(\decay{\PD}{\PV\Pgamma})| \gsim 3\,\%$
would be a clear signal of physics beyond the SM, and a clean indication of 
new \CP-violating dynamics associated to dipole operators. 
\end{enumerate}

Having clarified that large values of $|\ACPdir(\decay{\PD}{\PV\Pgamma})|$
would be a clear footprint of non-standard dipole operators, 
it can be asked if potential tight limits on $|\ACPdir(\decay{\PD}{\PV\Pgamma})|$ could exclude this non-standard framework.
Unfortunately, uncertainty on the strong phases does not allow this conclusion to be drawn.
Indeed the maximal values for the DCPV asymmetries presented 
above are obtained in the limit of maximal constructive interference 
of the various strong phases involved.
In principle, this problem could be overcome via time-dependent studies of
\decay{\PD(\Db)}{\PV\Pgamma} decays or using photon polarisation, 
accessible via lepton pair conversion in
\decay{\PD}{\PV(\decay{\ensuremath{\Pgamma^\ast}}{\ellp\ellm})};
however, these types of measurements are certainly more challenging from the 
experimental point of view.


\subsubsection{Testing for \CP-violating new physics in the $\Delta I = 3/2$ amplitudes}
\label{sec:Jure}

It is possible, at least in principle, to distinguish between NP and the SM as the origin of \dacp. 
If \dacp is due to a chromomagnetic operator, \ie\ due to $\Delta I=1/2$
contributions, one can measure \CP violation in radiative $\PD$ decays, as
explained in the previous section.
Examples of NP models that can be tested in this way are, \eg, flavour-violating supersymmetric squark-gluino loops that mediate the \decay{\cquark}{\uquark\Pg} transition~\cite{Grossman:2006jg,Giudice:2012qq,Hiller:2012wf}.
On the other hand, if \dacp is due to $\Delta I=3/2$ NP one can use isospin
symmetry to write sum rules for direct \CP asymmetries in $\PD$
decays~\cite{Grossman:2012eb}.
If the sum rules are violated, then NP would be found.
An example of a NP model that can be tested in this way is an addition of a
single new scalar field with nontrivial flavour couplings~\cite{Hochberg:2011ru}.

The basic idea behind the $\Delta I=3/2$ NP tests~\cite{Grossman:2012eb,Atwood:2012ac}
is that in the SM the \CP violation in SCS $\PD$ decays arises from penguin amplitudes which are $\Delta I=1/2$ transitions.
On the other hand, $\Delta I=3/2$ amplitudes are \CP-conserving in the SM. 
Moreover, there are no $\Delta I=5/2$ terms in the SM short-distance effective Hamiltonian, and though such contributions can be generated by electromagnetic rescattering (as has been discussed in the context of $B \to \pi\pi$ decays~\cite{Gardner:2001gc,Gronau:2005pq}) they would also be \CP conserving.
Observing any \CP violation effects in $\Delta I=3/2$ amplitudes would therefore be a clear signal of NP.

In the derivation of the sum rules it is important to pay attention to the
potentially important effects of isospin breaking.
Isospin symmetry is broken at ${\mathcal O}(10^{-2})$, which is also the size of the interesting \CP asymmetries.
There are two qualitatively different sources of isospin breaking: due to electromagnetic interactions, \uquark and \dquark quark masses, which are all \CP-conserving effects, and due to electroweak penguin operators that are a \CP-violating source of isospin breaking.
The \CP-conserving isospin breaking is easy to cancel in the sum rules.
As long as the \CP-conserving amplitudes completely cancel in the sum rules,
which is the case in Ref.~\cite{Grossman:2012eb}, the isospin breaking will only enter suppressed by the small \CP violation amplitude and is therefore negligible.
The electroweak penguin operators, on the other hand, are suppressed by
$\alpha/\as \sim {\mathcal O}(10^{-2})$ compared to the leading \CP-violating but isospin conserving penguin contractions of the $Q_{1,2}$ operators, and can thus also be safely neglected.

Among the SCS decays, the \decay{\PD}{\pion\pion}, \decay{\PD}{\Prho\pion},
\decay{\PD}{\Prho\Prho}, \decay{\PD}{\Kb\kaon\pion}, and \decay{\Ds}{\Kstar\pion} modes carry enough information to construct tests of $\Delta I=3/2$ NP. 
%
The sum rules for \decay{\PD}{\pion\pion} decays have the nice feature that the charged decay \decay{\Dp}{\pip\piz} is purely $\Delta I=3/2$.
In the SM therefore 
\begin{equation}
  \ACPdir(\decay{\Dp}{\pip\piz}) = 0\,.
\end{equation}
If this \CP asymmetry is measured to be nonzero, it would be a clear signal of $\Delta I=3/2$ NP.
However, if it is found experimentally to be very small, it is still possible that this is only because the strong phase between the SM and NP amplitudes is accidentally small. 

This possibility can be checked with more data if time-dependent \decay{\PD(t)}{\pim\pip} and \decay{\PD(t)}{\piz\piz} measurements become available,\footnote{
  Time-dependent \decay{\PD(t)}{\piz\piz} measurements could in principle be feasible using photon conversions~\cite{Ishino:2007pt}.
} or if there is additional information on relative phases from a charm factory running on the $\Ppsi(3770)$.
The strategy amounts to measuring the weak phase of the $\Delta I=3/2$ amplitude \ampsub{3} via generalised triangle constructions that also take isospin breaking into account~\cite{Grossman:2012eb}.
If
\begin{equation}\label{triangle-rel}
  \frac{1}{\sqrt2} \ampsub{\pim\pip} + \ampsub{\piz\piz} 
   - \frac{1}{\sqrt2} \ampbarsub{\pip\pim} - \ampbarsub{\piz\piz} =
  3\big(\ampsub{3} - \ampbarsub{3} \big)
\end{equation}
is found to be nonzero, this would mean there is \CP-violating NP in the $\Delta I=3/2$ amplitude. 

The above results apply also to \decay{\PD}{\Prho\Prho} decays, but for each polarisation amplitude separately.
The corrections due to finite $\Prho$ width can be controlled experimentally
in the same way as in \decay{\PB}{\Prho\Prho}~decays~\cite{Falk:2003uq}.
As long as the polarisations of the $\Prho$ resonances are measured (or if the
longitudinal decay modes dominate, as is the case in \decay{\PB}{\Prho\Prho}
decays), the search for $\Delta I=3/2$ NP could be easier experimentally in
\decay{\PD}{\Prho\Prho} decays since there are more charged tracks in the final
state.
The most promising observable where polarisation measurement is not needed is $\ACPcharm(\decay{\Dp}{\Prho^+\Prho^0})$, which if found nonzero (after the correction for the effect of finite $\Prho$ decay widths) would signal $\Delta I=3/2$ NP.

Another experimentally favourable probe is the isospin analysis of the \decay{\Dz}{\pip\pim\piz} Dalitz plot in terms of \decay{\PD}{\Prho\pion} decays~\cite{Gaspero:2008rs}. 
There are two combinations of measured amplitudes that are proportional to $\Delta I=3/2$ amplitudes
\begin{equation}\begin{split}
  \ampsub{\Prho^+\piz} + \ampsub{\Prho^0\pip} &=3 \sqrt{2} \ampsub{3}\,,\\
  \ampsub{\Prho^+\pim} + 2 \ampsub{\Prho^0\piz} + \ampsub{\Prho^-\pip} &= 6 \ampsub{3}\,.
\end{split}\end{equation} 
A measurement of the second sum can be obtained from the \decay{\Dz}{\pip\pim\piz} Dalitz plot.
If the related \CP asymmetry 
\begin{equation}
  |\ampsub{\Prho^+\pim} + 2 \ampsub{\Prho^0\piz} + \ampsub{\Prho^-\pip}|^2 -
  |\ampbarsub{\Prho^-\pip} + 2 \ampbarsub{\Prho^0\piz} + \ampbarsub{\Prho^+\pim}|^2 = 36\big(| \ampsub{3}|^2-|\ampbarsub{3}|^2\big),\label{Arhopi-sumrule}
\end{equation}
is found to be nonzero, this would mean that the $\Delta I=3/2$ NP contribution
is nonzero.
If it is found to vanish, however, it could be due to the strong phase difference being vanishingly small. 

A definitive answer can be provided by another test that is directly sensitive
to the weak phase of $\ampsub{3}$.
This test is possible if the time-dependent \decay{\PD(t)}{\pip\pim\piz}
Dalitz plot is measured.
In this case the relative phases between the \decay{\Dz}{\Prho\pion} and
\decay{\Dzb}{\Prho\pion} amplitudes can be obtained (alternatively one could
use time integrated entangled decays of $\Ppsi(3770)$ at the charm factory). 
The presence of a weak phase in $\ampsub{3}$ can then be determined from the
following sum-rule
\begin{equation}\label{a3rhopi}
  \big(\ampsub{\Prho^+\pim} + \ampsub{\Prho^-\pip} + 2 \ampsub{\Prho^0\piz}\big)- \big(\ampbarsub{\Prho^-\pim} + \ampbarsub{\Prho^+\pim} + 2 \ampbarsub{\Prho^0\piz}\big)=6 \big(\ampsub{3} - \ampbarsub{3}\big)\,.
\end{equation}
A non-vanishing result for Eq.~(\ref{a3rhopi}) would provide a definitive proof
for $\Delta I=3/2$ NP.
A similar sum rule for the \CP asymmetries rather than the amplitudes was
given in Eq.~(\ref{Arhopi-sumrule}).
In that case the time-integrated Dalitz plot suffices to determine the sum
rule inputs.

The sum rules involving \decay{\PD}{\kaon^{(*)}\Kb^{(*)}\pion} decays are somewhat more complex because there are at least three particles in the final state.
Nevertheless, it is possible to construct purely $\Delta I=3/2$ matrix elements from appropriate sums of decay amplitudes, and these can in principle be determined from amplitude analyses of the multibody final states.
It is also possible to search for \CP violation in $\Delta I=3/2$ amplitudes
using \decay{\Dsp}{\Kstar\pion} decays.
The sum 
\begin{equation}
  \sqrt{2} \amplitude{\decay{\Dsp}{\piz\Kstarp}} + \amplitude{\decay{\Dsp}{\pip\Kstarz}}=3 \ampsub{3}\,,
\end{equation}
is $\Delta I=3/2$ and can be measured from the common Dalitz plot for
\decay{\Dsp}{\KS\pip\piz} decay.
Direct \CP violation in this sum, \ie, 
\begin{equation}
\begin{split}\label{DCPVDs}
  &|\sqrt{2} \amplitude{\decay{\Dsp}{\piz\Kstarp}} + \amplitude{\decay{\Dsp}{\pip\Kstarz}}|^2 \, - \\
  &|\sqrt{2} \amplitude{\decay{\Dsm}{\piz\Kstarm}} + \amplitude{\decay{\Dsm}{\pim\Kstarzb}}|^2 \ne 0\,,
\end{split}
\end{equation}
would necessarily be due to $\Delta I=3/2$ NP contributions.
Additional information on the absolute value of
$|\amplitude{\decay{\Dsp}{\pip\Kstarz}}|$ can be obtained from the
\decay{\Dsp}{\pip\Kp\pim} three-body decay. 
Analogous tests using \decay{\Ds}{\Prho\Kstar} decays also exist.


\subsection{Potential for lattice computations of direct \CP violation and
mixing in the $\Dz$--$\Dzb$ system}
\label{sec:Stephen}

In searches for NP using charmed mesons, it is obviously crucial to determine accurately the size of SM contributions. 
In the next few paragraphs the prospects for such a determination in the future using the methods of lattice QCD are discussed.

Lattice QCD provides a first-principles method for determining
the strong-interaction contributions to weak decay and mixing
processes. It has developed into a precision tool, allowing
determinations of the light hadron spectrum, decay constants,
and matrix elements such as $B_K$ and $B_B$ with percent-level
accuracy. For reviews and collections of recent results,
see Refs.~\cite{Laiho:2009eu,Colangelo:2010et}.
The results provide confirmation that QCD indeed describes
the strong interactions in the non-perturbative regime, 
as well as providing predictions that play an important role in searching for new physics by looking for inconsistencies in unitarity triangle analyses.

Results with high precision are, however, only available for
processes involving single hadrons and a single insertion of
a weak operator. For the \Dz system, the ``high-precision''
quantities are thus the matrix elements describing the short-distance
parts of \Dz--\Dzb mixing and the matrix elements
of four-fermion operators arising after integrating out NP.
The methodology for such calculations is in place (and has been applied
successfully to the $\PK$ and $\PB$ meson systems),
and results are expected to be forthcoming in the next one to two years.

More challenging, and of course more interesting, are calculations
of the decay amplitudes to $\pion\pion$ and $\kaon\Kb$.
For kaon physics, this is the present frontier of lattice calculations.
One must deal with two technical challenges: (i) the fact that
one necessarily works in finite volume so the states are not
asymptotic two-particle states and (ii) the need to calculate
Wick contractions (such as the penguin-type contractions) which
involve gluonic intermediate states in some channels.
The former challenge has been solved in principle by the
work of L\"uscher~\cite{Luscher:1986pf,Luscher:1990ux}
and Lellouch and L\"uscher~\cite{Lellouch:2000pv} 
for the $K\to\pi\pi$ case, 
while advances in lattice algorithms and computational
power have allowed the numerical aspects of both challenges to
be overcome. There are now well controlled results for
the \decay{\PK}{\ensuremath{(\pion\pion)_{I=2}}} amplitude~\cite{Blum:2011ng}
and preliminary results for the \decay{\PK}{\ensuremath{(\pion\pion)_{I=0}}}
amplitude~\cite{Blum:2011pu}.
It is likely that results to $\sim 10\,\%$ accuracy for
all amplitudes will be available in a few years.
Note that, once a lattice calculation is feasible,
it will be of roughly equal difficulty to obtain
results for the \CP-conserving and \CP-violating parts. 

To extend these results to the charm case, one must
face a further challenge. This is that, even when one has
fixed the strong-interaction quantum numbers of a final state,
say to $I=S=0$, the strong interactions necessarily bring in
multiple final states when $E=m_D$. For example, $\pion\pion$ and
$\kaon\Kb$ states mix with $\Peta\Peta$, $4\pion$, $6\pion$, \etc
The finite-volume states that are used by lattice QCD are 
inevitably mixtures of all these possibilities, and one
must learn how, in principle and in practise, to disentangle
these states so as to obtain the desired matrix element.
Recently, in Ref.~\cite{Hansen:2012tf}, a first step towards 
developing a complete method has been taken, in which the
problem has been solved in principle for any number of two-particle
channels, assuming that the scattering is dominantly S-wave.
This is encouraging, and it may be that this method will allow semi-quantitative results for the amplitudes of interest to be obtained. 
Turning this method into practise is expected to take three to five years due to a number of numerical challenges (in particular the need to calculate several energy levels with good accuracy).
It is also expected to be possible to generalise the methodology to include four particle states; several groups are actively working on the theoretical issues. 
It is unclear at this stage, however, what time scale one should assign to this endeavour. 

Finally, the possibility of calculating long-distance contributions to \Dz--\Dzb mixing using lattice methods should be considered. 
Here the challenge is that there are two insertions of the weak Hamiltonian, with many allowed states propagating between them.
Some progress has been made recently on the corresponding problem for kaons~\cite{Christ:2010zz,Yu:2011gk} but the \Dz system is much more challenging.
The main problem is that, as for the decay amplitudes, there are many strong-interaction channels with $E< m_D$.
Further theoretical work is needed to develop a practical method.


\subsection{Interplay of \dacp with non-flavour observables}
\subsubsection{Direct \CP violation in charm and hadronic electric dipole moments}
\label{sec:Paride}

Models in which the primary source of flavour violation is linked to the
breaking of chiral symmetry (left-right flavour mixing) are natural candidates
to explain direct \CP violation in SCS $\PD$ meson decays, via enhanced $\Delta C=1$ chromomagnetic operators.
Interestingly, the chromomagnetic operator generates contributions to
\Dz--\Dzb mixing and $\epsilon^\prime /\epsilon$ that are always suppressed by
at least the square of the charm Yukawa couplings, thus naturally explaining
why they have remained undetected.

On the other hand, the dominant constraints are posed by the neutron and nuclear EDMs, which are expected to be close to their experimental bounds.
This result is fairly robust because the Feynman diagram contributing to quark
EDMs has essentially the same structure as that contributing to the chromomagnetic operator.

In the following the connection between \dacpdir and hadronic EDMs in concrete NP scenarios is discussed, following the analyses of Refs.~\cite{Giudice:2012qq,Hiller:2012wf}.

\vspace{1ex}
{\noindent \bf Supersymmetry}
\label{sect:SUSY}

\noindent
The leading SUSY contribution to \dacpdir stems from loops involving
up-squarks and gluinos and off-diagonal terms in the squark squared-mass
matrix in the left-right up sector, the so-called
$(\delta^{u}_{12})_{LR}$ mass-insertion.
As can be seen from Eqs.~(\ref{eq:dacp:mia})--(\ref{eq:dacp:susy}) 
and taking into account the large uncertainties involved in the evaluation of the matrix element, it can be concluded that a supersymmetric theory with left-right up-squark mixing can potentially explain the LHCb result.

Among the hadronic EDMs, the best constraints come from mercury and neutron EDMs.
Their current experimental bounds are
$|d_n| < 2.9 \times 10^{-26}~e\,\rm{cm}~(90\,\%~\rm{C.L.})$
and $|d_{\rm Hg}| < 3.1 \times 10^{-29}~e\,\rm{cm}~(95\,\%~\rm{C.L.})$.
In the mass-insertion approximation one can find
\begin{equation}
  |d_n| \approx 3\times 10^{-26}
        \left(\frac{\left|\mathrm{Im} \left( \delta^{u}_{11}\right)_{LR} \right|}{10^{-6}}\right)
        \left(\frac{{\tev}}{{\tilde m}}\right)~e\,\rm{cm}~.
\label{eq:LR12dec:dn}
\end{equation}%
and therefore it has to be seen whether a concrete SUSY scenario can naturally
account for the required level of suppression
$|\mathrm{Im}(\delta^{u}_{11})_{LR}| \lesssim 10^{-6}$.

\vspace{1ex}
{\noindent \bf \boldmath Generalised trilinear terms}

\noindent
While scenarios in which flavour violation is restricted to the trilinear terms can be envisaged, it is natural to generalise the structure of Eq.~(\ref{eq:dacp:susy}) to all squarks and take
\begin{equation}
  ( \delta^{q}_{ij})_{LR}  \sim   \frac{ A \, \theta^q_{ij} \, m_{q_j} }{\tilde m}\,,~~~~~q=\uquark,\dquark\,,
\label{eq:LRgen}
\end{equation}%
where $\theta^q_{ij}$ are generic mixing angles.
This pattern can be obtained when the matrices of the up and down trilinear
coupling constants follow the same hierarchical pattern as the corresponding
Yukawa matrices but they do not respect exact proportionality.

It is found that $\theta^q_{ij}$ can all be of order unity not only in the up,
but also in the down sector, thanks to the smallness of the down-type quark
masses entering $(\delta^{d}_{ij})_{LR}$.
The only experimental bounds in tension with this scenario are those on $|\theta^{u,d}_{11}|$ coming from the neutron EDM.

\vspace{1ex}
{\noindent \bf Split families}

\noindent 
The severe suppression of $\left(\delta^{u}_{21}\right)^{\rm eff}_{RL}$
stemming from the charm mass can be partially avoided in a framework with
split families, where the first two generations of squarks are substantially
heavier than $\tilde t_{1,2}$ and $\tilde b_{L}$, the only squarks required to
be close to the electroweak scale by naturalness arguments.
In this case the effective couplings relevant to \dacpother{\rm SUSY} can be decomposed as follows
\begin{equation}
  \left(\delta^{u}_{12}\right)^{\rm eff}_{RL} =
        \left(\delta^{u}_{13}\right)_{RR}\left(\delta^{u}_{33}\right)_{RL}
        \left(\delta^{u}_{32}\right)_{LL}~,  \qquad 
  \left(\delta^{u}_{12}\right)^{\rm eff}_{LR} =
        \left(\delta^{u}_{13}\right)_{LL}\left(\delta^{u}_{33}\right)_{RL}
        \left(\delta^{u}_{32}\right)_{RR}~.
  \label{eq:12LReffSUSY}
\end{equation}%
Notice that this scenario takes advantage of the large
$(\delta^{u}_{33})_{LR}\sim A m_{t}/\tilde m$ which is assumed to be of order one.
The following two options can be considered to explain the LHCb results:
\begin{equation}
\begin{split}
  \left(\delta^{u}_{32}\right)_{LL}=O(\lambda^2),\quad \left(\delta^{u}_{13}\right)_{RR} =  O(\lambda^2) \quad 
        & \to \quad \left(\delta^{u}_{12}\right)^{\rm eff}_{RL} = O(\lambda^4) = O( 10^{-3})~, \\
  \left(\delta^{u}_{13}\right)_{LL}=O(\lambda^3),\quad \left(\delta^{u}_{32}\right)_{RR} =  O(\lambda) \quad 
        &\to \quad \left(\delta^{u}_{12}\right)^{\rm eff}_{LR} = O(\lambda^4) = O( 10^{-3})~.
 \label{pufpuf}
\end{split}
\end{equation}

Gluino-squark loops yield an EDM ($d_u$) and a chromo-EDM ($d_u^c$) for the up
quark proportional to
$d^{(c)}_u \sim \mathrm{Im}\left[(\delta^{u}_{13})_{LL}(\delta^{u}_{31})_{RR}\right]$
and it turns out that
\begin{equation}
  \left| \dacpother{\ensuremath{\rm SUSY}} \right| \approx
  10^{-3} \times \left| \frac{d_n}{3 \times 10^{-26}} \right|
  \left| \frac{{\rm Im}\left(\delta^{u}_{32}\right)_{RR} }{0.2} \right|
  \left| \frac{10^{-3}}{ {\rm Im}\left(\delta^{u}_{31}\right)_{RR} }\right|~.
  \label{eq:acp_vs_dn}
\end{equation}%
In conclusion, the EDM bounds require a strong hierarchical structure in the
off-diagonal terms of the $RR$ up-squark mass matrix, as happens in models 
predicting $(\delta^{u}_{ij})_{RR} \sim (m_{u_i}/m_{u_j})/|V_{ij}|$.

\vspace{1ex}
{\noindent \bf \boldmath Supersymmetric flavour models}

\noindent
In models where the flavour structure of the soft breaking terms is dictated by
an approximate flavour symmetry, $(\delta^u_{LR})_{12}$ is generically flavour-suppressed by $(m_c\left|V_{us}\right|/\tilde{m})$, which is of order a few times $10^{-4}$. 
There is however additional dependence on the ratio between flavour-diagonal parameters, $A/\tilde{m}$, and on unknown coefficients of order one, that can provide enhancement by a small factor. 
In most such models, the selection rules that set the flavour structure of the soft breaking terms relate $(\delta^u_{LR})_{12}$ to $(\delta^d_{LR})_{12}$ and to $(\delta^{u,d}_{LR})_{11}$, which are bounded from above by, respectively, $\epsilon^\prime/\epsilon$ and EDM constraints.
Since both $\epsilon^\prime/\epsilon$ and EDMs suffer from hadronic uncertainties, small enhancements due to the flavour-diagonal supersymmetric parameters cannot be ruled out. 
It is thus possible to accommodate $\dacp \sim 0.006$ in supersymmetric models that are non-minimally flavour violating, but -- barring hadronic enhancements in charm decays -- it takes a fortuitous accident to lift the supersymmetric contribution above the permille level~\cite{Hiller:2012wf}.

\vspace{1ex}
{\noindent \bf \boldmath New-physics scenarios with $Z$-mediated FCNC}
\label{sect:Zmediated}

\noindent
Effective FCNC couplings of the $Z$ boson to SM quarks can appear in the SM
with non-sequential generations of quarks, models with an extra $U(1)$ symmetry
or models with extra vector-like doublets and singlets. The effective FCNC
Lagrangian can be written as
\begin{equation}
  \mathcal{L}^{Z-{\rm FCNC}}_{\rm eff} =
        -\frac{g}{2 \cos \theta_W} {\bar q}_i\gamma^\mu
        \left[ (g^{Z}_{L})_{ij} \, P_L + (g^{Z}_{R})_{ij} \, P_R \right] q_j ~ Z_\mu + \text{ h.c.}\,,
  \label{eq:eff_lagr_Z}
\end{equation}%
The  chromomagnetic operator is generated at the one-loop level, with leading
contribution from $Z$--top exchange diagrams leading to
\begin{equation}
  \left| \dacpother{Z-{\rm FCNC}} \right| \approx 
  0.6\,\%~\left| \frac{\mathrm{Im}\left[(g^{Z}_{L})^*_{ut}(g^{Z}_{R})_{ct}\right] }{2\times 10^{-4}}\right|~.
  \label{eq:acpZNP}
\end{equation}%
The presence of new \CP-violating phases in the couplings $(g^{Z}_{L,R})_{ij}$
are also expected to generate hadronic EDMs. In particular, one can find
\begin{equation}
  |d_n| \approx 3  \times 10^{-26}~
        \left| \frac{ {\rm Im} \left[ (g^{Z}_{L})^*_{ut} (g^{Z}_{R})_{ut} \right] }{ 2\times 10^{-7}} \right|
        ~{e\,\rm{cm}}\,,
  \label{eq:dZNP}
\end{equation}
and therefore $\Delta {\cal A}^{Z-{\rm FCNC}}_{\CP}= {\cal O}(10^{-2})$ only, 
provided ${\rm Im}(g^{Z}_{R})_{ut}/{\rm Im}(g^{Z}_{R})_{ct}\lesssim 10^{-3}$.

In NP scenarios with $Z$-mediated FCNCs, the most interesting FCNC processes
in the top sector are $t\to cZ$ and $t\to uZ$, which arise at the tree level.
In particular, 
\begin{equation}
  \BR(t \to cZ) \approx 0.7\times 10^{-2}
        \left| \frac{ (g^{Z}_{R})_{tc} }{10^{-1}}\right|^2\,,
\end{equation}%
which is within the reach of the LHC for the values of $(g^{Z}_{R})_{tc}$
relevant to $\Delta {\cal A}^{Z-{\rm FCNC}}_{\CP}$.

\vspace{1ex}
{\noindent \bf New-physics scenarios with scalar-mediated FCNC}
\label{sect:Hmediated}

\noindent
Finally, it is instructive to analyse a new-physics framework with effective FCNC couplings to SM quarks of a scalar particle $h$. 
The effective Lagrangian reads
\begin{equation}
  \mathcal{L}^{h-{\rm FCNC}}_{\rm eff} = -{\bar q}_i
        \left[ (g^{h}_{L})_{ij} \, P_L + (g^{h}_{R})_{ij} \, P_R \right] q_j ~ h
        + \text{ h.c.}
  \label{eq:eff_lagr_h}
\end{equation}
Also in this case the  chromomagnetic operator is generated at the one-loop
level, with a leading contribution from $h$--top exchange diagrams.
This leads to
\begin{equation}
  \left| \dacpother{\rm FCNC} \right| \approx 0. 6\,\%
        \left| \frac{ \mathrm{Im}
        \left[(g^{h}_{L})^*_{ut} (g^{h}_{R})_{tc}\right]  }{2 \times 10^{-4}} \right|~.
  \label{eq:acphNP}
\end{equation}
As in all the other frameworks, the most severe constraints are posed by the
hadronic EDMs
\begin{equation}
  |d_n| \approx 3 \times 10^{-26}~\left|
        \frac{  \mathrm{Im}\left[(g^{h}_{L})^*_{ut} (g^{h}_{R})_{tu}\right]  }{2\times 10^{-7}} \right|
  ~{e\,\rm{cm}}\,.
  \label{eq:dHNP}
\end{equation}

With scalar-mediated FCNCs, the potentially most interesting signals are the
rare top decays $t\to ch$ or $\tquark \to \uquark h$, if kinematically allowed.
In particular, 
\begin{equation}
  \BF(\tquark \to qh) \approx 0.4\times 10^{-2}
        \left| \frac{(g^{h}_{R})^{tq}}{10^{-1}}\right|^2\,,
  \label{eq:tqh}
\end{equation}%
which could be within the reach of the LHC.


\subsubsection{Interplay of collider physics and a new physics origin for \dacp}
 \label{sec:Cedric}

The first evidence for direct \CP violation in SCS $\PD$ decays may have interesting implications for NP searches around the \tev scale at the LHC.
The NP contribution to \dacpdir can be fully parametrised by a complete set of
$\Delta C=1$ effective operators at the charm scale.
As shown by the authors of Ref.~\cite{Isidori:2011qw} only a few of these
operators can accommodate the LHCb result without conflicting with present
bounds from \Dz--\Dzb mixing and $\epsilon^\prime / \epsilon$.
In particular four-fermion operators of the form
$\mathcal{O}^q=(\bar{u}_R\gamma^\mu c_R)(\bar{q}_R\gamma_\mu q_R)$ with
$q=\uquark,\dquark,\squark$ are promising since they do not lead to flavour
violation in the down-type quark sector.
The corresponding Wilson coefficients are defined as $1/\Lambda_q^2$.
Assuming the SM expectation for \dacpdir is largely subdominant,
the LHCb measurement suggests a scale of $\Lambda_q\simeq 15\tev$~\cite{Isidori:2011qw}.

There is an immediate interplay between charm decay and flavour (and \CP)
conserving observables at much higher energies provided $\mathcal{O}^q$ arises
from a heavy NP state exchanged in the $s$-channel.
Under this mild assumption $\mathcal{O}^q$ factorises as the product of two
quark currents and the same NP induces \Dz--\Dzb mixing and quark
compositeness through the $(\bar u_R \gamma_\mu c_R)^2$ and
$(\bar q_R \gamma_\mu q_R)^2$ operators, respectively.
Denoting their respective Wilson coefficients by $\Lambda_{\bar u c}$ and
$\Lambda_{\bar q q}$, the relation
$\Lambda_q=\sqrt{\Lambda_{\bar u c}\Lambda_{\bar q q}}$ is predicted.
The \Dz--\Dzb mixing bound on NP implies
$\Lambda_{\bar u c}\gtrsim 1200\,$\tev~\cite{Isidori:2010kg}.
Combining this stringent $\Delta C=2$ bound with the $\Delta C=1$ scale
suggested by \dacpdir thus generically requires
$\Lambda_{\bar q q}\lesssim 200\,$\gev, which is a rather low compositeness scale
for the light quark flavours. 

Quark compositeness can be probed at the LHC through dijet searches.
Actually for the up or the down quark the low scale suggested by
\dacpdir is already excluded by the Tevatron~\cite{CDF:9609,Abazov:2009ac}.
On the other hand dijet searches are less sensitive to contact interactions
involving only the strange quark since the latter, being a sea quark, has a
suppressed parton distribution function in the proton.
The authors of Ref.~\cite{DaRold:Futureaa} showed that a first estimation at
the partonic level of the extra dijet production from a
$(\bar s_R \gamma_\mu s_R)^2$ operator with a scale of
$\Lambda_{\bar s s}\sim 200\,$\gev is marginally consistent, given the
$\mathcal{O}(1)$ uncertainty of the problem, with the bounds from the
ATLAS and CMS experiments~\cite{ATLAS-CONF-2012-038,Chatrchyan:2012bf}. 

One concludes that an $\mathcal{O}^s$ operator induced by a $s$-channel
exchanged NP can accommodate the \dacpdir measurement without conflicting with
$\Delta C=2$, $\epsilon^\prime/\epsilon$ and dijet searches.
Furthermore such a NP scenario makes several generic predictions both for
charm and high-\pt physics:
1) most of the \CP asymmetry is predicted to be in the \Kp\Km channel,
2) \CP violation in \Dz--\Dzb mixing should be observed in the near future, and
3) an excess of dijets at the LHC is expected at a level which should be
visible in the 2012 data.

\subsection{Future potential of \lhcb measurements}
\subsubsection{Requirements on experimental precision}
\label{sec:charm:requirements}

The ultimate goal of mixing and \CP violation measurements in the charm sector is to reach the precision of the SM predictions (or better).
In some cases this requires measurements in several decay modes in order to distinguish enhanced contributions of higher order SM diagrams from effects caused by new particles.

Indirect \CP violation is constrained by the observable \agamma (see Eq.~(\ref{eq:agamma})).
The \CP-violating parameters in this observable are multiplied by the mixing parameters $x_D$ and $y_D$, respectively.
Hence, the relative precision on the \CP-violating parameters is limited by the relative precision of the mixing parameters.
Therefore, aiming at a relative precision below $10\,\%$ and taking into account the current mixing parameter world averages, the target precision would be $2$--$3\times10^{-4}$.
Indirect \CP violation is expected in the SM at the order of $10^{-4}$, and therefore the direct \CP violation parameter contributing to \agamma has to be measured to a precision of $10^{-3}$ in order to distinguish the two types of \CP violation in \agamma.

Direct \CP violation is not expected to be as large as the current world average of \dacp in most other decay modes.
However, a few large \CP violation signatures are expected in various models, as discussed in the previous sections.
Estimations based on flavour-SU(3) and U-spin symmetry lead to expectations of $\ACPdir(\decay{\Dp}{\Kp\Kzb})\gtrsim0.1\%$  and $\ACPdir(D^0 \to \KS\KS) \sim 0.6\%$.
Considerations assuming universality of $\Delta F=1$ transitions lead to a limit of $\ACPdir(\decay{\PD}{\Ppi\ep\en})\lesssim2\%$.
Enhanced electromagnetic dipole operators can lead to $\ACPdir(\decay{\PD}{V\gamma})$ of a few \%, equivalent to the influence of chromomagnetic dipole operators on \dacp.
Additional information can be obtained from time-dependent studies of \decay{\PD}{V\gamma} decays or from angular analyses of \decay{\PD}{Vl^+l^-} decays.

Analyses of $\Delta I=3/2$ transitions involve asymmetry measurements of several related decay modes.
Examples are the decays \decay{\PD}{\Ppi\Ppi}, \decay{\PD}{\Prho\Ppi}, \decay{\PD}{\Prho\Prho}, \decay{\PD}{\overline{\PK}\PK\Ppi}, and \decay{\Ds}{\Kstar\Ppi}.
The number of final state particles in these decays varies from two to six (counting the pions from \KS decays) and many of these modes contain neutral pions in their final state.
The precision for modes involving neutral pions or photons will be limited by the ability of the calorimeter to identify these particles in the dense hadronic environment.
An upgraded calorimeter with smaller Moli\`{e}re radius would greatly extend the physics reach in this area.\footnote{
   Such an upgrade to the calorimeter system is not in the baseline plan for the LHCb upgrade~\cite{CERN-LHCC-2011-001,CERN-LHCC-2012-007}.
}

In general, a precision of $5\times10^{−4}$ or better for asymmetry differences as well as individual asymmetries is needed for measurements of other SCS charm decays. 
While measurements of time-integrated raw asymmetries at this level should be well within reach, the challenge lies in the control of production and detection asymmetries in order to extract the physics asymmetries of individual decay modes.
This can be achieved by assuming that there is no significant \CP violation in CF decay modes.

\subsubsection{Prospects of future \lhcb measurements}

Numbers of events in various channels are projected directly from the numbers
reconstructed in the 2011 data set, in most cases.
This involves assumptions that the prompt charm cross-section will increase by a factor of 1.8 when doubling the centre-of-mass energy from $\sqrt{s} = 7 \tev$ to $\sqrt{s} = 14 \tev$, that the integrated luminosity will increase from $1\invfb$ to $50\invfb$, and that the trigger efficiency for charm will increase by a factor of 2 as the current hardware trigger requirement is effectively removed (or substantially relaxed).
Additionally, a factor of 3.5 times greater efficiency in channels with \decay{\KS}{\pim\pip} daughters is predicted based on progress made in the trigger software between 2011 and 2012.
This primarily results from reconstructing candidates which decay downstream of the vertex detector.
The results of this exercise are summarised in Table~\ref{tab:D0NumbersTable}
for \Dz decays and in Table~\ref{tab:DplusNumbersTable}
for \Dp and \Dsp decays.

\begin{table}[!htb]
\begin{center}
\label{tab:D0NumbersTable}
\caption{\small
  Numbers of \Dz and \decay{\Dstarp}{\Dz\pip} signal events observed in the 2011 data in a variety of channels and those projected for $50\invfb$.
  These channels can be used for mixing studies, for indirect \CP violation studies, and for direct \CP violation studies.
  As discussed in the text, the numbers of events in any one channel can vary from one analysis to another, depending on the level of cleanliness required. 
  Hence, all numbers should be understood to have an inherent variation of a factor of 2.
  To control systematic uncertainties with the very high level of precision that will be required by the upgrade, it may be  necessary to sacrifice some of the statistics.
}
\begin{tabular}{lrr}
Mode & 2011 yield  & $50\invfb$ yield \\
  & ($10^3$ events) & ($10^6$ events)    \\ \hline
Untagged \decay{\Dz}{\Km\pip} &  $ 230 \, 000 $  & $ 40 \, 000 $ \\
\decay{\Dstarp}{\Dz\pip}; \decay{\Dz}{\Km\pip} & $  40\, 000 $  & $ 7 \, 000 $ \\
\decay{\Dstarp}{\Dz\pip}; \decay{\Dz}{\Kp\pim} & 130 &  20 \\
\decay{\Dz}{\Km\Kp} & $ 25 \, 000 $  & $ 4\, 600 $\\
\decay{\Dz}{\pim\pip} & $ 6 \, 500 $ & $ 1\, 200 $ \\
\decay{\Dstarp}{\Dz\pip}; \decay{\Dz}{\Km\Kp} & $ 4 \, 300 $ & 775 \\
\decay{\Dstarp}{\Dz\pip}; \decay{\Dz}{\pim\pip} & $ 1 \, 100 $ & 200 \\
\decay{\Dstarp}{\Dz\pip}; \decay{\Dz}{\KS\pim\pip} & $ 300 $ & 180 \\
\decay{\Dstarp}{\Dz\pip}; \decay{\Dz}{\KS\Km\Kp}  & $ 45 $ & 30 \\
\decay{\Dstarp}{\Dz\pip}; \decay{\Dz}{\Km\pip\pim\pip}  & $ 7 \, 800 $ & $ 1 \, 400 $ \\
\decay{\Dstarp}{\Dz\pip}; \decay{\Dz}{\Km\Kp\pim\pip} & $ 120 $ & 20 \\
\decay{\Dstarp}{\Dz\pip}; \decay{\Dz}{\pim\pip\pim\pip} & $ 470 $ & 85 \\
\decay{\Dstarp}{\Dz\pip}; \decay{\Dz}{\Km\mup\PX}  & -- & $ 4 \, 000 $\\
\decay{\Dstarp}{\Dz\pip}; \decay{\Dz}{\Kp\mun\PX}  & -- & $ 0.1  $\\

\end{tabular}
\end{center}
\end{table}

\begin{table}[!htb]
\begin{center}
\label{tab:DplusNumbersTable}
\caption{\small
  Numbers of \Dp and \Dsp signal events observed in the 2011 data in a variety of channels and those projected for $50\invfb$.
  These channels can be used for direct \CP violation studies.
  As discussed in the text, the numbers of events in any one channel can vary from one analysis to another, depending on the level of cleanliness required. 
  To control systematic uncertainties with the very high level of precision that will be required by the upgrade, it may be necessary to sacrifice some of the statistics.
}
\begin{tabular}{lrr}
Mode & 2011 yield  & $50 \invfb$ yield \\
  & ($10^3$ events) & ($10^6$ events)    \\ \hline
\decay{\Dp}{\Km\pip\pip} &  $ 60 \, 000 $ & $ 11 \, 000 $ \\
\decay{\Dp}{\Kp\pip\pim} &  $ 200 $ & $ 40 $ \\
\decay{\Dp}{\Km\Kp\pip}  & $ 6 \, 500 $  & $ 1 \, 200 $ \\
\decay{\Dp}{\Pphi\pip}   & $ 2 \, 800 $ & 500 \\
\decay{\Dp}{\pim\pip\pip}& $ 3 \, 200 $ & 575 \\
\decay{\Dp}{\KS\pip}  & $ 1 \, 500 $ & $1\,000$ \\
\decay{\Dp}{\KS\Kp}  & 525 & 330 \\
\decay{\Dp}{\Km\Kp\Kp} & 60 & 10 \\
 & & \\
\decay{\Dsp}{\Km\Kp\pip} & $ 8 \, 900 $ &  $ 1 \, 600 $ \\
\decay{\Dsp}{\Pphi\pip}, (\decay{\Pphi}{\Km\Kp}) & $ 5 \, 350 $ & $1\,000$ \\
\decay{\Dsp}{\pim\pip\pip} & $ 2 \, 000 $ & 360 \\
\decay{\Dsp}{\Km\pip\pip}  &  & \\
\decay{\Dsp}{\pim\Kp\pip}  & 555 & 100 \\
\decay{\Dsp}{\Km\Kp\Kp}  & 50 & 10 \\
\decay{\Dsp}{\KS\Kp} & 410 & 260 \\
\decay{\Dsp}{\KS\pip}  & 33 & 20 \\ 
\end{tabular}
\end{center}
\end{table}

Estimating the physics reach with the projected data sets requires a number of assumptions.
The statistical precision generally improves as $1 / \sqrt{N}$.
Estimating the systematic error, and therefore ultimate physics reach, is more of an art.
It is often the case that data can be used to control systematic uncertainties at the level of the statistical error, but the extent to which this will be possible cannot be reliably predicted.
In some cases controlling systematic uncertainties will require sacrificing some of the statistics to work with cleaner signals or with signals which populate only parts of the detector where the performance is very well understood.
Estimates of sensitivity to \CP violation in mixing generally depend on
the values of the mixing parameters  --  the larger the number of
mixed events, the larger the effective statistics contributing
to the corresponding \CP violation measurement.

\begin{table}[!htb]
\label{tab:ProjectedD0Measurements}
\begin{center}
\caption{\small
  Estimated statistical uncertainties for mixing and \CP violation measurements which can be made with the projected samples for $50\invfb$ described in Table~\ref{tab:D0NumbersTable}.
}
\begin{tabular}{lcr}
Sample & Parameter(s)  & Precision \\ \hline
WS/RS $\kaon\pion$  &  $ ( x_D^{\prime 2}, y_D^{\prime} ) $ &
  $ {\cal O} [ (  10^{-5} ,  10^{-4} ) ] $ \\
WS/RS $\kaon\Pmu\Pnu $ & $ r_M  $ & $ {\cal O} (  5 \times 10^{-7} ) $ \\
WS/RS $\kaon\Pmu\Pnu $ & $ | p/q|_D $ &$ {\cal O} (  1\% ) $  \\
\decay{\Dstarp}{\Dz\pip}; \decay{\Dz}{\Km\Kp, \pim\pip} &
  \dacp & $  0.015\% $ \\
\decay{\Dstarp}{\Dz\pip}; \decay{\Dz}{\Km\Kp} &
  \ACPcharm & $  0.010\% $ \\
\decay{\Dstarp}{\Dz\pip}; \decay{\Dz}{\pim\pip} &
  \ACPcharm & $  0.015\% $ \\
\decay{\Dstarp}{\Dz\pip}; \decay{\Dz}{\KS\pim\pip} &
  $  (x_D,y_D) $ & $ (  0.015\%, 0.010\% ) $ \\
\decay{\Dstarp}{\Dz\pip}; \decay{\Dz}{\Km\Kp, (\pim\pip)} &
  \ycp  & $ 0.004\% \, (0.008\%) $ \\
\decay{\Dstarp}{\Dz\pip}; \decay{\Dz}{\Km\Kp, (\pim\pip)} &
  \agamma & $ 0.004\% \, (0.008\%) $ \\
\decay{\Dstarp}{\Dz\pip}; \decay{\Dz}{\Km\Kp\pim\pip} &
  $ {\cal A}_{\rm T} $ & $    2.5 \times 10^{-4}  $ \\
\end{tabular}
\end{center}
\end{table}

\begin{table}[!htb]
\label{tab:ProjectedDplusMeasurements}
\begin{center}
\caption{\small
  Estimated statistical uncertainties for \CP violation measurements which can be made with the projected \Dp samples for $50\invfb$ described in Table~\ref{tab:DplusNumbersTable}.
}
\resizebox{\textwidth}{!}{
\begin{tabular}{lcr}
Sample & Parameter(s)  & Precision \\ \hline
\decay{\Dp}{\KS\Kp}  & phase-space integrated \CP violation & $ 10^{-4} $ \\
\decay{\Dp}{\Km\Kp\pip}  & phase-space integrated \CP violation & $ 5 \times 10^{-5} $ \\
\decay{\Dp}{\pim\pip\pip} & phase-space integrated \CP violation & $ 8 \times 10^{-5} $ \\
\decay{\Dp}{\Km\Kp\pip}  & \CP violation in phases, amplitude model & $ (0.01 - 0.10)^{\circ} $ \\
\decay{\Dp}{\Km\Kp\pip}  & \CP violation in fraction differences, amplitude model & $ (0.01 -0.10) \% $ \\
\decay{\Dp}{\pim\pip\pip} & \CP violation in phases, amplitude model & $ (0.01 - 0.10)^{\circ} $ \\
\decay{\Dp}{\pim\pip\pip} & \CP violation in fraction differences, amplitude model & $ (0.01 -0.10) \% $ \\
\decay{\Dp}{\Km\Kp\pip} & \CP violation in phases, model-independent & $ (0.01 - 0.10)^{\circ} $ \\
\decay{\Dp}{\Km\Kp\pip}  & \CP violation in fraction differences, model-independent & $ (0.01 -0.10) \% $ \\
\decay{\Dp}{\pim\pip\pip} & \CP violation in phases, model-independent & $ (0.01 - 0.10)^{\circ} $ \\
\decay{\Dp}{\pim\pip\pip} & \CP violation in fraction differences, model-independent & $ (0.01 -0.10) \% $ \\
\end{tabular}
}
\end{center}
\end{table}

The estimated statistical precisions for parameters of mixing and \CP violation in the \Dz system are presented in Table~\ref{tab:ProjectedD0Measurements}.
The precision for measuring $(x_D^{\prime 2}, \, y_D^\prime)$
using the time-dependence of the wrong-sign (WS) to right-sign (RS)
$\kaon\pion$ rate comes from extrapolating the BaBar~\cite{Aubert:2007wf}
and Belle~\cite{Zhang:2006dp} sensitivities.\footnote{
    The LHCb measurements of charm mixing parameters from wrong-sign $K\pi$ decays~\cite{LHCb-PAPER-2012-038} are consistent with the estimated sensitivities.
}
The precision for measuring $ r_M $ using the ratio of WS to RS $\kaon\Pmu\Pnu$
events assumes the central value to be $ 2.5 \times 10^{-5} $.
The S/B ratio is assumed to be 30 times better than reported
by BaBar~\cite{Aubert:2004bn} for their similar $\kaon\electron\Pnu$ analysis.
Background can be reduced by a factor of 10 using LHCb's excellent
vertex resolution to remove candidates with decay time less than twice 
the \Dz lifetime -- a requirement which only modestly reduces the WS signal as its decay time distribution has the form $dN/dt \, \propto \, t^2 \, e^{-\Gamma t}$.
In addition, the excellent vertex resolution and the decay time
requirement allow the neutrino momentum, and hence the
$\Dstarp-\Dz$ mass difference to be measured with better resolution
than was possible in the $\ep\en$ experiments.
BaBar demonstrated that using a doubly-tagged sample of semileptonic decay
candidates provides the same mixing sensitivity as the more
traditional singly-tagged sample~\cite{Aubert:2007aa}.
By combining singly- and doubly-tagged samples, it should be 
possible to effectively double the statistics.

The projected sensitivities for the two-body direct \CP violation measurements
are relatively solid:
the 2011 \dacp measurements provide benchmark samples with 
full analysis cuts including fiducial cuts necessary to control
systematic uncertainties for measuring \dacp.
The systematic errors for the separate $\ACPcharm(\Km\Kp)$
and $\ACPcharm(\pim\pip)$  measurements
will be more challenging and may require sacrificing statistical precision.
The projections for measuring \ycp and \agamma
using $\Km\Kp$ and $\pim\pip$ should also be robust as the same samples will be used for these analyses as for the \ACPcharm measurements.

The projected precision for measuring $ (x_D,y_D) $ from
\decay{\Dz}{\KS\pim\pip}  comes from
scaling the Belle~\cite{Abe:2007rd} and BaBar~\cite{Aubert:2008zh}
sensitivities.
The statistical precisions could be even better as LHCb's
prompt sample will be enhanced at higher decay times
where the mixing effects are larger. 
By contrast, \Dz mesons from semileptonic \PB decays should be unbiased in this variable, providing a useful sample at lower decay times.

The estimated statistical precisions for DCPV in \Dp measurements are presented in Table~\ref{tab:ProjectedDplusMeasurements}.
The estimates for the phase-space integrated \CP violation rates are scaled by $1/\sqrt{N}$ and are then increased by a factor of two to allow for using tighter cuts to control systematic uncertainties.
The estimates for measuring \CP violation in the magnitudes and phases of 
quasi-two-body amplitudes contributing to three-body final states
come from scaling the BaBar sensitivities for time-integrated
\CP violation in \decay{\Dz}{\pim\pip\piz}
and \decay{\Dz}{\Km\Kp\piz} by $1/\sqrt{N}$.
The angular moments of the cosine of the helicity angle of the \PD decay
products reflect the spin and mass structure of the intermediate resonant
and nonresonant amplitudes with no explicit model dependence.
The difference between the angular moment distributions observed
in \Dz and \Dzb decays
provides sensitivity to \CP violation in the magnitudes (or fractions) and phases of 
amplitudes about equal to that of model-dependent fits.
The angular moment differences are robust, in the sense that they are model-independent, but they are less specific compared to the results from model-dependent analyses: they indicate only the spins and mass ranges where particle and antiparticle amplitudes differ, but do not identify a specific \CP-violating
intermediate state or how much it varies.
The sensitivity to \CP violation in any contributing amplitude depends on how much
it contributes to the three-body decay, and also on the other amplitudes with which it interferes.
For this reason, ranges of sensitivity are indicated rather than single values.
No sensitivities for \CP violation measurements in three-body \Ds decay channels are estimated explicitly.
They can be estimated roughly by extrapolating from the numbers for \Dp decays by scaling by $1/\sqrt{N}$.
These estimates should be degraded slightly as the lifetime of the \Dp is about twice that of the \Ds meson, making it easier to select clean \Dp samples.

\subsection{Conclusion}

\lhcb has proven its capability of performing high-precision charm physics measurements.
The experiment is ideally suited for \CP violation searches and for measurements of decay-time-dependent processes such as mixing.

Finding evidence for a non-zero value of \dacp has raised the question of whether or not this may be interpreted as the first hint of physics beyond the SM at the LHC.
Within the SM the central value can only be explained by significantly enhanced penguin amplitudes.
This enhancement is conceivable when estimating flavour SU(3) or U-spin breaking effects from fits to $\PD\to PP$ data.
However, attempts at estimating the long distance penguin contractions directly have not yielded conclusive results to explain the enhancement.

Lattice QCD has the potential of assessing the penguin enhancement directly.
However, several challenges arise which make these calculations impossible at the moment.
Following promising results on $\PK\to\pi\pi$ decays, additional challenges arise in the charm sector as $\pi\pi$ and $\PK\PK$ states mix with $\eta\eta$, $4\pi$, $6\pi$ and other states.
Possible methods have been proposed and results may be expected in three to five years time.

General considerations on the possibility of interpreting \dacp in models beyond the SM have led to the conclusion that an enhanced chromomagnetic dipole operator is required.
These operators can be accommodated in minimal supersymmetric models with non-zero left-right up-type squark mixing contributions or, similarly, in warped extra dimensional models.
Tests of these interpretations beyond the SM are needed.
One promising group of channels are radiative charm decays where the link between the chromomagnetic and the electromagnetic dipole operator leads to predictions of enhanced \CP asymmetries of several percent.
These can be measured to sufficient precision at the \lhcb upgrade.

Another complementary test is to search for contributions beyond the SM in $\Delta I=3/2$ amplitudes.
This class of amplitudes leads to several isospin relations which can be tested in a range of decay modes, \eg $\PD\to\pi\pi$, $\PD\to\rho\pi$, $\PD\to\PK\bar{\PK}$, \etc
Several of these measurements, such as the Dalitz plot analysis of the decay $\Dz\to\pip\pim\piz$, can be performed at \lhcb.

Beyond charm physics, the chromomagnetic dipole operators would affect the neutron and nuclear EDMs, which are expected to be close to the current experimental bound.
Similarly, rare FCNC top decays are expected to be enhanced, if kinematically allowed.
Furthermore, quark compositeness can be related to the \dacp measurement and tested in dijet searches.
Current results favour the NP contribution to be located in the $\Dz\to\Km\Kp$ decay as the strange quark compositeness scale is less well constrained.
Measurements of the individual asymmetries of sufficient precision will be possible at the \lhcb upgrade.

The charm mixing parameters have not yet been precisely calculated in the SM.
An inclusive approach based on an operator product expansion relies on the expansion scale being small enough to allow convergence and furthermore involves the calculation of a large number of unknown matrix elements.
An exclusive approach sums over intermediate hadronic states and requires very precise branching ratio determinations of these final states which are currently not available.
Contrary to the SM, contributions beyond the SM can be calculated reliably.
With the SM contribution to indirect \CP violation being $<10^{-4}$, the \lhcb upgrade is ideally suited to cover the parameter space available for enhanced asymmetries beyond the SM.
Measurements in several complementary modes will permit the extraction of the underlying theory parameters with high precision.

The \lhcb upgrade will allow to constrain \CP asymmetries and mixing observables to a level of precision which, in most of the key modes, cannot be matched by any other experiment foreseen on a similar timescale.
This level of precision should permit us not only to discover \CP violation in charm decays but also to unambiguously understand its origin.

\clearpage

\section{The LHCb upgrade as a general purpose detector in the forward region}
\label{sec:other}

The previous sections have focussed on flavour physics observables that are sensitive to physics beyond the SM.
However, LHCb has excellent potential in a range of other important topics. 
As discussed in this section, the detector upgrade will further enhance the capability of LHCb in these areas, so that it can be considered as a general purpose detector in the forward region.
LHCb may also be able to make a unique contribution to the field of heavy ion physics, by studying soft QCD and heavy flavour production in $pA$ collisions. The first $pA$ run of the LHC will clarify soon the potential of LHCb in this field.

\subsection{Quarkonia and multi-parton scattering}

\label{subsec:onia}
The mechanism of heavy quarkonium production is a long-standing problem in QCD.
An effective field theory, non-relativistic QCD (NRQCD),
provides the foundation for much of the current theoretical work. 
According to NRQCD, the production of heavy quarkonium 
factorizes into two steps:  a heavy quark-antiquark pair is first created 
perturbatively at short distances and subsequently evolves
non-perturbatively into quarkonium at long distances. 
The NRQCD calculations depend on the colour-singlet (CS) and colour-octet (CO) matrix elements, 
which account for the probability of a heavy quark-antiquark pair 
in a particular colour state to evolve into heavy quarkonium. The CS model~\cite{Kartvelishvili:1978id,Baier:1981uk}, which provides
a leading-order (LO) description of quarkonia production, was first used 
to describe experimental data. However, it underestimates the~observed
cross-section for single \jpsi~production at high $p_{\mathrm{T}}$ at the 
\tevatron~\cite{Abe:1992ww}. To~resolve this
discrepancy the CO mechanism was introduced~\cite{Braaten:1994vv}. 
The corresponding matrix elements were determined from the
large-$p_{\mathrm{T}}$ data as the CO cross-section falls more slowly
than that for CS. More recent higher-order calculations~\cite{Campbell:2007ws,Gong:2008sn,Artoisenet:2008fc,Lansberg:2008gk} close the gap between the CS predictions and the experimental
data~\cite{Brambilla:2010cs} reducing the need for large CO contributions.

\begin{figure}[!htb]
\begin{center}
\includegraphics[height=8cm]{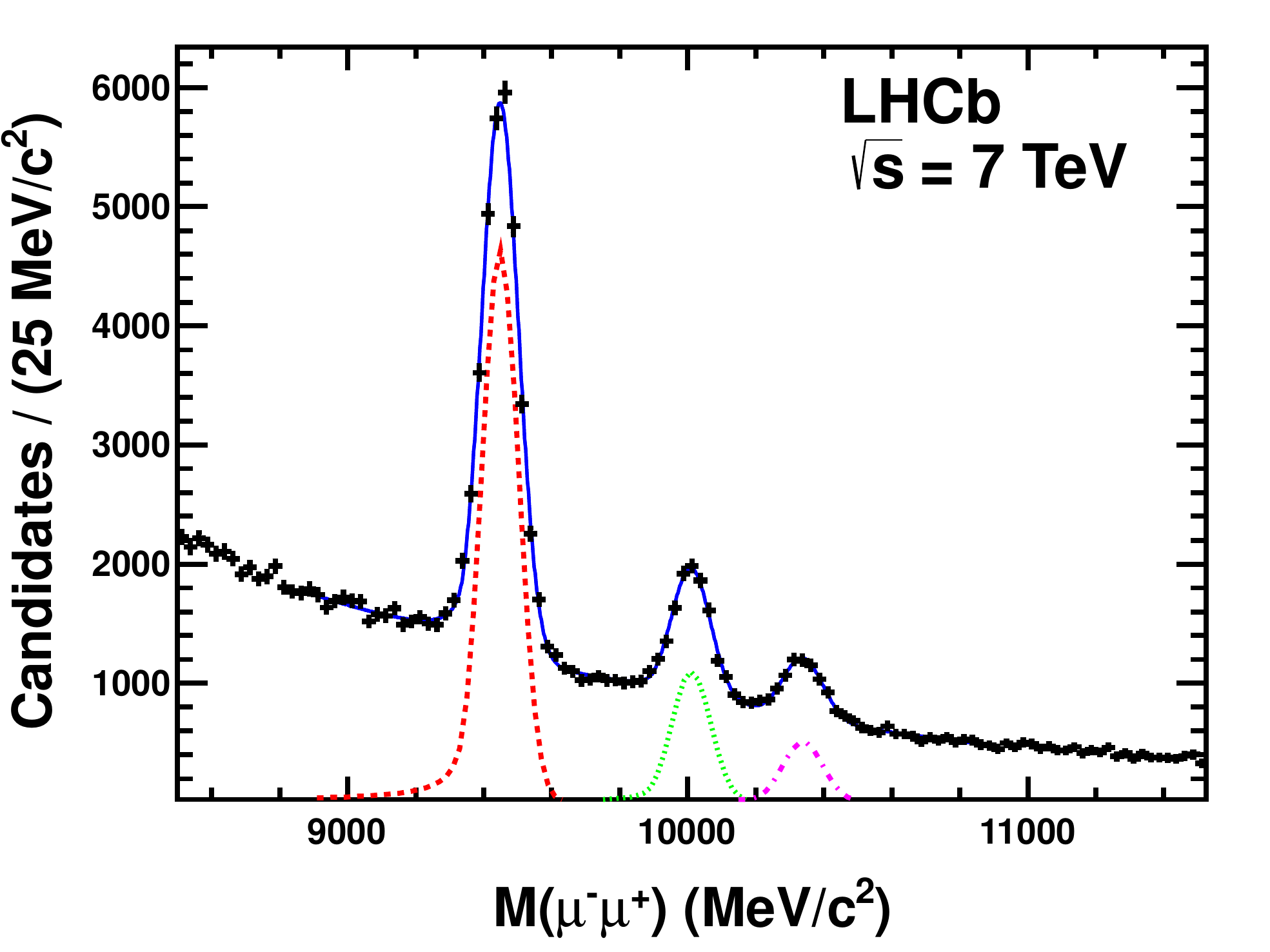}
\end{center}
\caption{\small
  Invariant mass distribution of selected $\Upsilon$ candidates from $25 \invpb$ of data collected in 2010~\cite{LHCb-PAPER-2011-036}. 
  The \OneS, \TwoS and \ThreeS states are clearly resolved. 
  The results of a maximum likelihood fit are superimposed.
}
\label{upsmass}
\end{figure}

Traditionally, quarkonia production studies at hadron colliders have focussed on the study of \jpsi, \psitwos and \nS decays to dimuon or dielectron pairs~\cite{Brambilla:2010cs}. 
The LHCb programme so far has followed this pattern with measurements of many cross-sections already published~\cite{LHCb-PAPER-2011-003, LHCb-PAPER-2011-019, LHCb-PAPER-2011-030,LHCb-PAPER-2011-036, LHCb-PAPER-2011-045}. 
As an example of the quality of the data, Fig.~\ref{upsmass} shows the $\Upsilon$ mass distribution.
By the time of the upgrade in 2018, data samples corresponding to several \invfb will have been collected at $\sqs = 7, 8$ and $14 \tev$ and the results will be dominated by systematic uncertainties. 
Therefore, new probes of quarkonia production will be pursued. 
Two possibilities are detailed here: multiple quarkonia production and quarkonia production via hadronic decay modes. 
These studies will profit from the higher integrated luminosity and improved trigger. 
These modes provide clear signals in the detector and will be relatively uneffected by the increased pile-up. 

As the cross-sections for charmonium production at the LHC are large~\cite{LHCb-PAPER-2011-003,LHCb-PAPER-2011-019,LHCb-PAPER-2011-030,LHCb-PAPER-2011-045}, the question of multiple production of these states in a single proton-proton collision naturally arises. 
Studies of double hidden charm and hidden and associated open charm production have been proposed as probes of
the quarkonium production mechanism~\cite{Brodsky:2009cf}. In proton-proton collisions contributions from other mechanisms, such
as double parton scattering (DPS)~\cite{Kom:2011bd, Baranov:2011ch, Novoselov:2011ff} or the
intrinsic charm content of the proton~\cite{Brodsky:1980pb}, are possible. 
First studies of both processes have been carried out with the current LHCb data; 
more details can be found in Refs.~\cite{LHCb-PAPER-2011-013,LHCb-PAPER-2012-004}.  

LO colour singlet calculations for the $gg \to \jpsi{}\jpsi$~process in perturbative QCD exist and give results consistent with the data~\cite{Kartvelishvili:1984ur,Humpert:1983yj,Berezhnoy:2011xy}.
In the LHCb fiducial region ($2<y_{\jpsi}<4.5$, $p^{\mathrm{T}}_{\jpsi}<10~\mathrm{GeV}/c$, where $y_{\jpsi}$ and $p^{\mathrm{T}}_{\jpsi}$ represent the rapidity and transverse momentum of the $\jpsi$, respectively) these calculations predict the $\jpsi{}\jpsi$~production cross-section to be $4.1\pm1.2~\mathrm{nb}$~\cite{Berezhnoy:2011xy} in agreement with the measured value of $5.1\pm1.0~\mathrm{nb}$~\cite{LHCb-PAPER-2011-013}. 
Similar calculations exist for the case of double \OneS production. 
For the case of $\jpsi$ plus \OneS production no
leading order diagrams contribute and hence the rate is expected to be
suppressed in Single Parton Scattering (SPS). 
This leads to an ``unnatural'' ordering of the cross-section values: $\sigma^{\jpsi \jpsi}_{gg} > \sigma^{\OneS \OneS}_{gg} > \sigma^{\OneS \jpsi}_{gg}$.

The DPS contributions to all these double onia production modes can be estimated, 
neglecting partonic correlations in the proton,
as the product of the measured cross-sections of the sub-processes involved divided by an effective 
cross-section~\cite{Kom:2011bd, Baranov:2011ch, Novoselov:2011ff,Luszczak:2011zp}. The value of the latter is determined from 
multi-jet events at the \tevatron to be $\sigma^{\mathrm{DPS}}_{\mathrm{eff}}= 14.5\pm1.7\,^{+1.7}_{-2.3}~\mathrm{mb}$~\cite{Abe:1997xk}. At $\sqs = 7~\tev$ the
contribution from this source to the total cross-section is similar in
size to the LO contribution from SPS. 
For DPS the ordering of the cross-section values is: $\sigma^{\jpsi \jpsi}_{\rm DPS} >  \sigma^{\OneS \jpsi}_{\rm DPS}  >\sigma^{\OneS \OneS}_{\rm DPS}$.

The expected cross-sections for a few double quarkonia processes, together with their yields, are summarized in Table~\ref{doubleprodpred}. 
Measurements of the cross-sections and properties in these modes will allow the two contributions to be disentangled.
\begin{table}[!htb]
\begin{center}
\caption{\small
  Expected cross-sections in the LHCb acceptance and 
  yields for double quarkonia production with $50 \invfb$ at $\sqs = 14 \tev$.}
\begin{tabular}{lcccc} 
Mode & $\sigma_{gg}$ [\nb] &Yield [SPS]  & $\sigma_{\rm DPS}$ [\nb] & Yield [DPS]    \\ \hline
\jpsi \jpsi  & 7.2 & 270\,000 & 11 & 430\,000 \\
\jpsi \psitwos & 3.2 & 14\,000 & 4.0 & 19\,000  \\
\psitwos \psitwos & 0.4  & 180  & 0.6 & 300  \\
\jpsi \chiczero & - & - & 4.3 & 200   \\
\jpsi \chicone & - & - & 6.6 & 14\,000  \\
\jpsi \chictwo & - & - & 8.6  & 11\,000 \\
\jpsi \OneS & 0.0036 &  360  & 0.27 & 20\,000  \\
\jpsi \TwoS & 0.0011 &  90 & 0.07 & 5300  \\
\jpsi \ThreeS & 0.0005 &  50 & 0.035 & 2000  \\
\OneS \OneS & 0.014 &  1100 & 0.0027 & 200  \\
\end{tabular}
\label{doubleprodpred}
\end{center}
\end{table}

As well as probing the production mechanism these studies are sensitive to a potential 
first observation of tetraquark states~\cite{Berezhnoy:2011xy} and of 
$\chib$ and $\etab$ states decaying in the double $\jpsi$ mode.  Based on the cross-sections and branching ratios given in
Ref.~\cite{PhysRevD.80.094008}, 500 (1500) fully reconstructed \chibzero (\chibtwo)
are expected with the upgraded detector and these decays 
will be visible at LHCb.
In the case of the $\eta_b$ state, several estimates exist, based on values of the branching ratio  $\eta_b\to\jpsi\jpsi$ ranging from $10^{-6}$ to $10^{-8}$~\cite{jia}, corresponding to yields of 0.02 to 5 events.  

The upgraded detector is expected to have excellent hadron
identification capabilities both offline and at the trigger level. As discussed in Ref.~\cite{Barsuk:2012ic}, 
this allows charmonium studies to be performed in hadronic decay modes. A particularly convenient mode is the $p\overline{p}$ final 
state. This is accessible for the \jpsi, \etac, \chicj, \hc and \psitwos mesons. Extrapolating from studies with the current detector large 
inclusive samples of these decays will be collected. For example around 0.5 million $\etac \rightarrow p \overline{p}$ will be collected.  

Hadronic decays of heavy bottomonium have received less attention in
the literature~\cite{jia}. 
The high mass implies a large phase space for many decay modes, but consequently the branching ratio for each individual mode is reduced. 
In Ref.~\cite{jia} it is estimated that the 
$\eta_b \rightarrow D^{*} \overline{D}$ branching fraction is $10^{-5}$ and the 
$\eta_b \rightarrow D \overline{D} \pi$ rate may be a factor of ten higher. 
Though no specific studies have been performed, based on the studies
of double open charm production given in Ref.~\cite{LHCb-PAPER-2012-004} it is plausible
that an $\eta_b$ signal will be detected in this mode with the upgraded detector.

\subsection{Exotic meson spectroscopy}
\label{subsec:exotic-spectro}

The spectroscopy of bound states formed by heavy quark-antiquark 
pairs ($c$ or $b$ quarks), has been extensively studied from both theoretical and experimental points of view since the discovery of the $\jpsi$ state in 1974~\cite{Aubert:1974js,Augustin:1974xw} and the discovery of the $\OneS$ state in 1977~\cite{Herb:1977ek}.
Until recently, all experimentally observed charmonium ($c\bar{c}$) and bottomonium ($b\bar{b}$) states matched well with expectations. 

However, in 2003, a new and unexpected charmonium state was observed by the
\belle experiment ~\cite{x3872} and then confirmed independently 
by the \babar~\cite{xbabar}, CDF~\cite{xcdf} and D0~\cite{Abazov:2004kp} experiments.
This new particle, referred to as the $X(3872)$, was observed in $B\to X(3872) K$ decays, in the decay mode $X(3872)\rightarrow \jpsi \pi^+ \pi^-$ and has a 
mass indistinguishable (within uncertainties) from the  $D^{*0} \Dzb$
threshold~\cite{Brambilla:2010cs}. 
Several of the $X(3872)$ parameters are unknown (such as its spin) 
or have large uncertainties, but this state does not match any  
predicted charmonium state~\cite{Brambilla:2010cs}. The discovery of
the $X(3872)$ has led to a resurgence of interest in exotic spectroscopy and
subsequently many new states have been claimed. For example:   the $Y$ family, 
$Y(4260), Y(4320)$ and $Y(4660)$, of spin parity $1^{-}$, or the puzzling 
charged $Z$ family, $Z(4050)^+, Z(4250)^+$ and $Z(4430)^+$, so far 
observed only by the \belle experiment~\cite{Choi:2007wga,Mizuk:2008me,Mizuk:2009da}, and not confirmed by \babar~\cite{Aubert:2008aa,Lees:2011ik}. 
The nature of these states has drawn much theoretical attention and many models
have been proposed.   
One possible explanation is that they are bound molecular states of open charm mesons~\cite{Swanson:2006st}.  
Another is that these are tetraquarks~\cite{Drenska:2009cd} states formed of four quarks (\eg $c,\bar{c}$, one light quark and one light anti-quark). 
Other interpretations have been postulated such as quark-gluon hybrid~\cite{Drenska:2009cd} or hadrocharmonium models~\cite{Voloshin:2007dx}, but experimental data are not yet able to conclude definitely. 
For reviews, see Refs.~\cite{Swanson:2006st,Voloshin:2007dx,Godfrey:2008nc,Nielsen:2009uh,Brambilla:2010cs,Drenska:2010kg,Eidelman:2012vu}.

The bottomonium system should exhibit similar exotic states to the charmonium case. 
The \belle experiment recently reported the observation of exotic bottomonium charged particles $Z_b(10610)^+$ and $Z_b(10650)^+$ in the decays $Z_b\rightarrow \nS \pi^+$ and $Z_b \rightarrow h_b(nP) \pi^+$~\cite{Belle:2011aa}.
Evidence for a neutral isopartner has also been reported~\cite{Adachi:2012im}.\footnote{
  At ICHEP 2012, Belle reported observations of the $Z_b$ states decaying to $B\bar{B}^{(*)}$~\cite{Adachi:2012cx}.
}
These states appear similar to, but narrower than, the $Z(4430)^+$ observed in the charmonium case.
In addition, neutral states analogous to the $X(3872)$ and the $Y$ states are expected in the bottomonium system.

Studies of the $X(3872)$ have already been performed with the current
detector~\cite{LHCb-PAPER-2011-034}.  
The $50 \invfb$ of integrated luminosity collected with the upgraded detector will contain over one million $X(3872)\rightarrow \jpsi\, \pi \pi$ candidates, by far the largest sample ever collected and allow study of this meson with high precision.
A significant fraction of the $X(3872)$ sample will originate from the decays of \B mesons (the remainder being promptly produced) allowing the quantum numbers and other properties to be determined. 
With such a large sample the missing $^3D_2$ state of the charmonium
system~\cite{PhysRevD.69.094019} will be also be observed and studied
with high precision. 

Another study being pursued with the current detector is to clarify the status of the $Z(4430)^+$ state. 
If confirmed, the $Z(4430)^+$ will be copiously produced at $\sqrt{s}=14~\tev$ and the larger data set will allow detailed study of its properties in different $B$ decay modes, thus setting the basis for all future searches for exotic charged states.

Similar to the charmonium-like states, exotic bottomonium states will
mainly be searched for in the  $\nS \pi^+\pi^-$ channel, with $\nS \rightarrow \mu^+\mu^-$. 
The excellent resolution observed in the $\nS$ analysis~\cite{LHCb-PAPER-2011-036}
allows efficient separation of the three states, which is crucial in 
searching for exotic bottomonium states in these channels.
 
All these studies, and searches for other exotica such as pentaquarks
will profit from the increased integrated luminosity.

\subsection{Precision measurements of $b$- and $c$-hadron properties}
\label{subsec:hadrons}

A major focus of activity with the current LHCb detector is the study of the properties of beauty and charm hadrons. 
This is a wide ranging field including studies of properties such as mass and lifetime, observation of excited $b$ hadrons and the measurements of branching ratios. 
These studies provide important input to pQCD models. 
Three topics are considered here: $b$ decays to charmonia, $B^+_c$, and $b$-baryon decays.

One important field being studied with the current detector is exclusive $b$ decays to charmonia. 
Studies of these modes are important to improve understanding of the shape of the momentum spectrum of $\jpsi$ produced in $b$ hadron decays, as measured by the \B factories~\cite{Balest:1994jf,Aubert:2002hc}. 
To explain the observed excess at low momentum, new contributions to the total $b \rightarrow \jpsi X$ rate are needed. 
Several sources have been proposed in the literature:
intrinsic charm~\cite{changhou}, baryonium formation~\cite{Brodsky:1997yr} and as yet unobserved exotic states~\cite{PhysRevD.83.114029}. 
One of the first proposed explanations for the excess was a contribution 
from an intrinsic charm component to the $b$-hadron wave-function~\cite{changhou}. 
This would lead to an enhancement of $b$-hadron decays to $\jpsi$ in
association with open charm. The \B-factories have set limits on such
decays at the level of $10^{-5}$~\cite{Nakamura:2010zzi}, 
which considerably restricts, but does not exclude, contributions from 
intrinsic charm models. 
The branching ratios of these decays have been estimated in pQCD~\cite{eilam}. 
In the case of $B^0 \rightarrow \jpsi D^0$ the branching ratio has been estimated to be $7 \times 10^{-7}$. 
If this value is correct, several hundred fully reconstructed events will be collected  with the upgraded detector. 
Similar decay modes are possible for $\Bs$ and $\Bc$ mesons though no limits (or predictions) exist.

Another possibility to explain the shape of the $\jpsi$ spectrum is contributions from exotic strange baryonia formed in decays such as $B^+ \rightarrow \jpsi \Lbar^0 p$. 
This decay has been observed by \babar~\cite{Aubert:2003ww}, with a branching ratio of $(1.18 \pm 0.31) \times 10^{-5}$. 
The related decay $\Bd \rightarrow \jpsi p \overline{p}$ is unobserved, with an upper limit on the branching ratio of $8.3 \times 10^{-7}$ at 90\,\% confidence level~\cite{Xie:2005tf}. 
At present, these decays are experimentally challenging due to the
low Q-values involved. The larger data samples available at the time of the 
upgrade, together with improved proton identification at low momentum,  
may lead to their observation.

Compared to the case of $B^0$ and $B^+$, the $\Bs$ sector is less well explored both experimentally and theoretically. 
Decays such as $\Bs \rightarrow \jpsi K^{*0} \overline{K}^{*0}$ and $\Bs \rightarrow \jpsi \phi \rho$ should be observable with the present detector. 
With the upgraded apparatus, the decay modes $\Bs \rightarrow \jpsi \KS \KS$ and $\Bs \rightarrow \jpsi \phi \phi$ will also become accessible. 
The latter channel is interesting as the low Q-value will allow a precision determination of the $\Bs$ mass.  

\begin{figure}[!htb]
\begin{center}
\includegraphics[height=12cm]{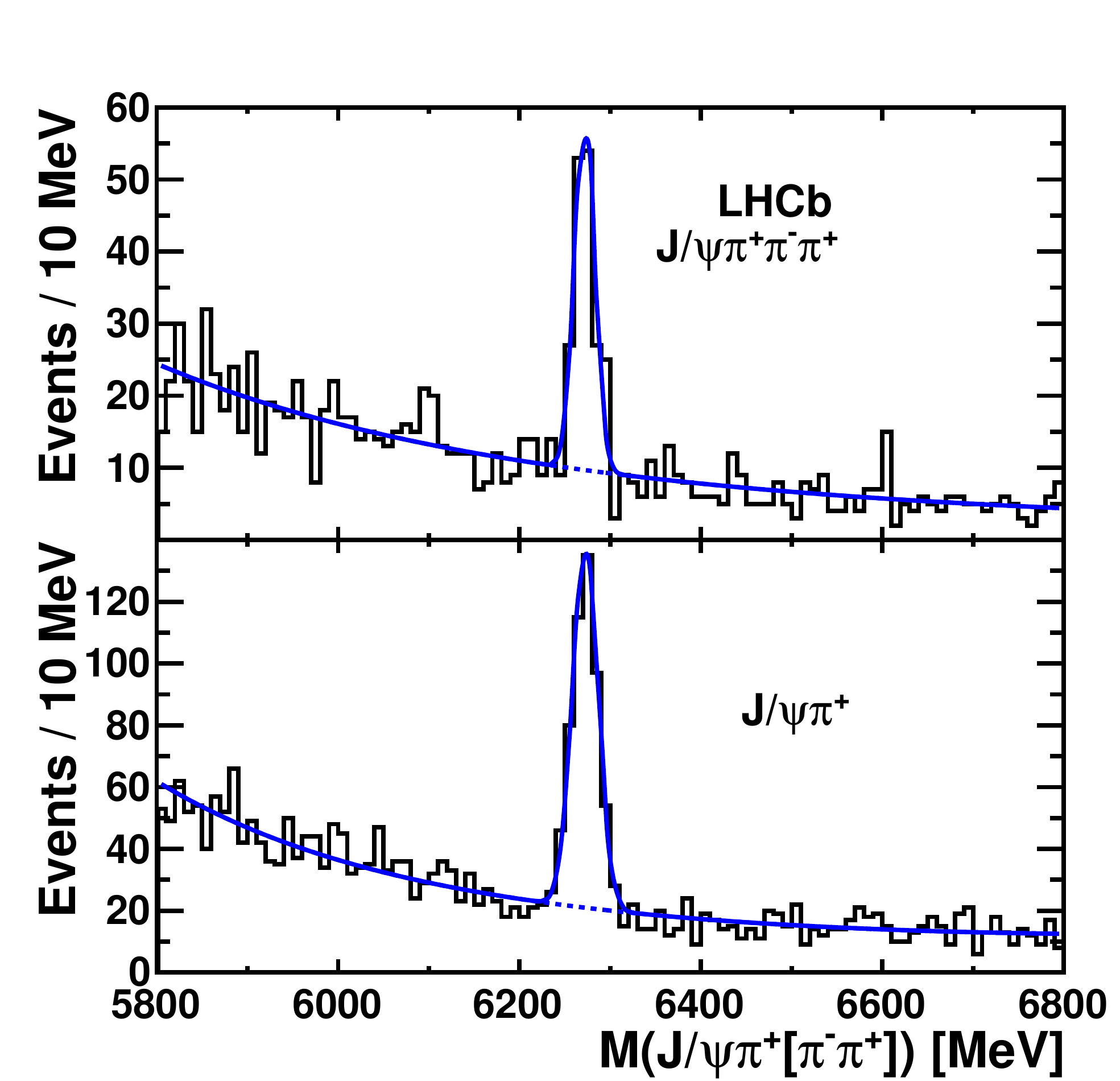}
\end{center}
\caption{\small
  Invariant mass distribution of (top) $\Bc \to \jpsi 3\pi^+$ and (bottom) $\Bc \rightarrow \jpsi \pi^+$ candidates using $0.8 \invfb$ of integrated luminosity collected in 2011~\cite{LHCb-PAPER-2011-044}. 
  The results of maximum likelihood fits are superimposed.
}
\label{bcmass}
\end{figure}

As  the lowest bound state of two heavy quarks $\overline{b}$ and $c$, 
the $B^+_c$ meson forms a unique flavoured, weakly decaying quarkonium system. 
Studies of the properties of $B^+_c$ mesons such as the mass, lifetime and two-body non-leptonic decay modes are being performed with the current detector. 
As an example, Fig.~\ref{bcmass} shows the signals observed for $\Bc \rightarrow \jpsi \pi^+$ and $\Bc \rightarrow \jpsi 3\pi^+$. 
The large data set collected with the upgraded detector will allow these studies
to be pursued with higher precision together with first studies of \CP and triple-product asymmetries in the $\Bc$ system. 
In Table~\ref{bcyields} the expected yields of selected decay modes
are estimated extrapolating from the yields of 
$\Bc \rightarrow \jpsi \pi^+$ and $\Bc \rightarrow \jpsi 3\pi^+$ 
observed with the current detector.
\begin{table}[!htb]
\begin{center}
\caption{\small
  Branching ratios and expected yields for selected $\Bc$ decays to final states containing a $\jpsi$ or $\psitwos$ meson. 
  The branching ratios for the $\jpsi$ modes are taken from Ref.~\cite{lhcreport}, with the additional constraint of the ratio of the $\Bc \rightarrow \jpsi 3\pi^+$ to $\Bc \rightarrow \jpsi \pi^+$ reported in Ref.~\cite{LHCb-PAPER-2011-044}. 
  The $\psi(2S)$ mode branching ratios are estimated assuming that they are 0.5 of the $\jpsi$ values, as observed in many modes (see for example Ref.~\cite{LHCb-PAPER-2012-010}). 
  Only dimuon modes are considered for the $\jpsi$ and $\psi(2S)$, and only the $K^+K^{-}\pi^{+}$ ($K^+\pi^{-}\pi^{+}$) modes are considered for the $D^+_s$ ($D^+$) modes. 
The $\Bc \rightarrow K^+ K^{*0}$ numbers are taken from Ref.~\cite{Gao:2010zzc}.}
\begin{tabular}{lcc} 
Mode & Branching ratio  &  Expected yield [$50 \invfb$] \\ \hline
$ \Bc \rightarrow \jpsi \pi^+$ & $2 \times 10^{-3}$  & 52\,000  \\
$ \Bc \rightarrow \jpsi 3 \pi^+$ & $5 \times 10^{-3}$    & 17\,000\\
$ \Bc \rightarrow \jpsi K^+$ & $(1\textendash{}2) \times 10^{-4}$  & 3000--4000  \\
$ \Bc \rightarrow \jpsi K_{1}^{+}$ & $3 \times 10^{-5}$  & 1000 \\
$ \Bc \rightarrow \psitwos \pi^+$ & $ 1 \times 10^{-3}$ & 3000\\
$ \Bc \rightarrow \psitwos 3 \pi^+$ & $2.5 \times 10^{-3}$ & 1000 \\
$ \Bc \rightarrow \jpsi D^+_s$ & $(2\textendash{}3) \times 10^{-3}$ & 1400--1900 \\
$ \Bc \rightarrow \jpsi D^+$ & $(5\textendash{}13) \times 10^{-4}$  & 8--100 \\
$ \Bc \rightarrow K^+ K^{*0}$ & $10^{-6}$  & 500 \\
\end{tabular}
\label{bcyields}
\end{center}
\end{table}
As well as studies of the branching ratios and searches for NP, these modes will allow precision measurements of the $\Bc$ mass and lifetime to be made. Based on ongoing studies with the current detector, a statistical precision of $0.1~\mevcc$ on the mass will be achieved. 
The uncertainty on the mass will most likely be dominated by systematic errors related to the momentum scale. 
Precision of $10^{-4}$ on this variable would translate to an uncertainty of $0.3 \mevcc$ on the mass. 
Measurements of the $\Bc$ lifetime using the $\jpsi \pi^+$ decay are ongoing. 
Extrapolating these results to $50 \invfb$, a statistical precision of 0.004~\ps will be achieved.

The large $\Bc$ data set will open possibilities for many other studies.
Decay modes of the $\Bc$ meson to a $\Bs$ or $\Bd$ meson together with a pion or kaon will also be accessible. 
Studies of the $\Bc \rightarrow \Bs \pi^+$ decay have been started with the data collected in 2011 where a handful of events are expected. 
As discussed in Ref.~\cite{lhcreport}, semileptonic $\Bc$ decays to $\Bs$ can be used to provide a clean tagged decay source for \CP violation studies. 
Finally, signals of the currently unexplored excited $B^+_c$ meson states are expected to be observed~\cite{excitedbc,Godfrey:2004ya,Kiselev:1994rc,Dowdall:2012ab}. 
As discussed in Ref.~\cite{Gao:2010zzc} observation of the $B_c^{*+}$ decay is extremely challenging due to the soft photon produced in the decay to to the ground state. 
The prospects for observation of the first P-wave multiplet decays decaying radiatively to the ground state are more promising. 

\begin{figure}[!htb]
\begin{center}
\includegraphics[height=8cm]{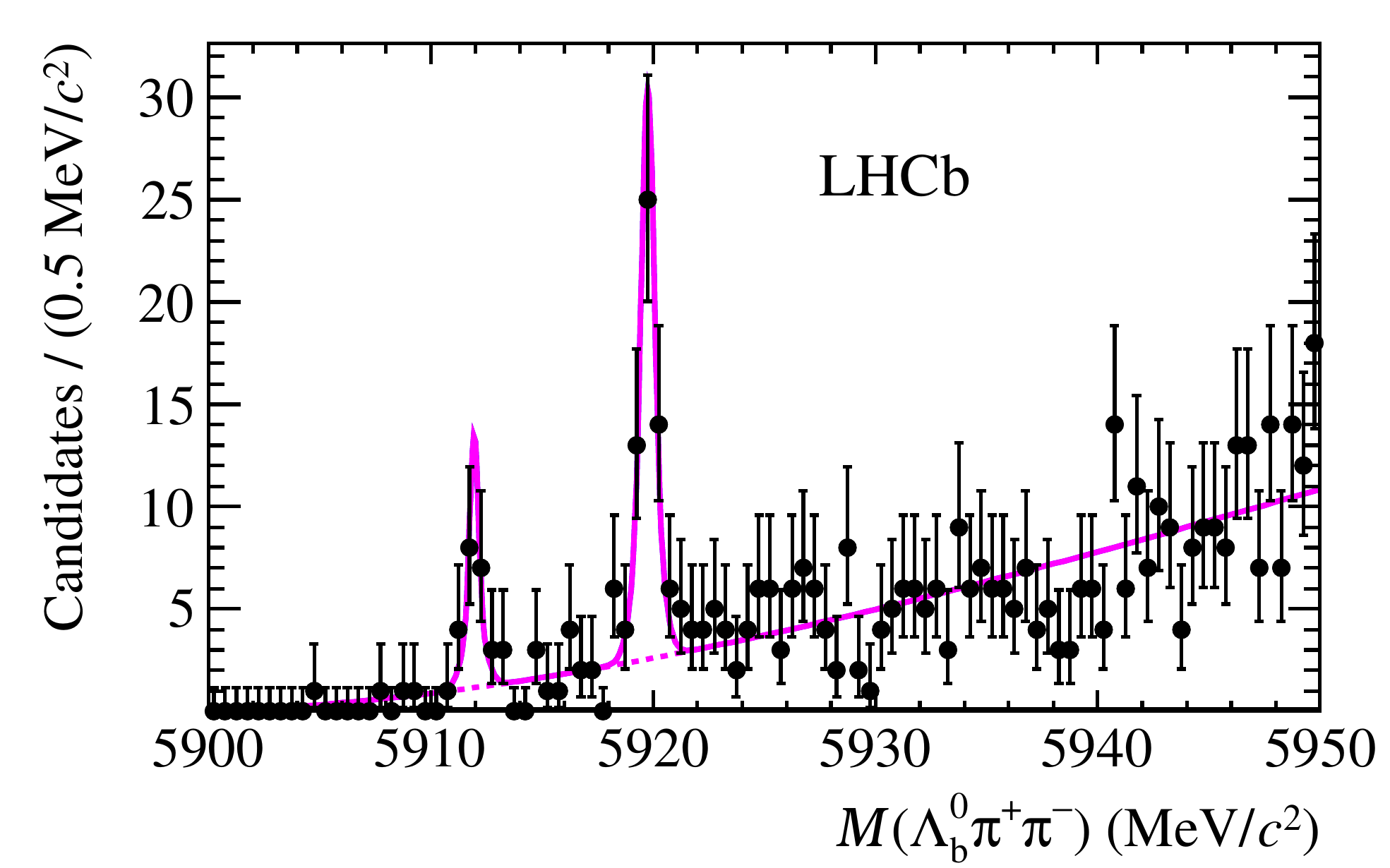}
\end{center}
\caption{\small
  Invariant mass spectrum of $\L_b^0 \pi^{+} \pi^{-}$~\cite{LHCb-PAPER-2012-012}. 
  The points with error bars are the data, the solid line is the result of a fit to this distribution, and the dashed line is the fitted background contribution.
}
\label{lambdabstar}
\end{figure}

Large samples of $b$ baryons decaying to final states containing charmonia will also be collected. 
Precision measurements of the properties of the already known states will be possible.
For example, extrapolating the preliminary studies with $0.3 \invfb$ discussed in Ref.~\cite{LHCb-CONF-2011-060}, 10\,000 $\Xi_{b} \rightarrow \jpsi \Xi$ and 2000 $\Omega_{b} \rightarrow \jpsi \Omega$ events will be collected. 
This will allow the $\Xi_b$ ($\Omega_b$) mass to be measured to a precision of 0.1~\mevcc (0.5~\mevcc). 
Precise $b$-baryon lifetime measurements, that will allow tests of the heavy quark expansion~\cite{Uraltsev:1996ta,Voloshin:1999pz,Bigi:2011gf}, should also be possible.
Studies of excited $b$ baryons, for example determination of the quantum numbers of the $\L_b^*$ baryons that have recently been observed by LHCb (Fig.~\ref{lambdabstar})~\cite{LHCb-PAPER-2012-012}, will also be made.

Baryonic states containing two heavy quarks will also be observable.
The lightest of these, the $\Xi_{cc}$ isodoublet, have an estimated cross-section of $\mathcal{O}(10^2) \nb$~\cite{Chang:2006xp,Zhang:2011hi} and so should be visible with $5 \invfb$ collected with the current detector. 
However, the statistics may be marginal for follow-on analyses: measurements of the lifetime and ratios of branching fractions, searches for excited states, and so forth. 
They will certainly be insufficient for angular analyses aimed at confirming
the quark model predictions for the spin-parity of these states.
These studies will require the statistics and improved triggering of the
LHCb upgrade. 
Heavier states such as the $\Omega_{cc}$, $\Xi_{bc}$, and $\Xi_{bb}$ have still smaller production cross-sections~\cite{Zhang:2011hi}. 
First studies towards $\Xi_{bc}$ detection are in progress.
These indicate that at best a handful of events can be expected in $5 \invfb$, but that this state should be observable with the upgrade.

\subsection{Measurements with electroweak gauge bosons}
\label{subsec:ew}

Two of the most important quantities in the LHC electroweak physics programme are the sine of the effective electroweak mixing angle for leptons, $\sin^2\theta^{\rm lept}_{\rm eff}$, and the mass of the $W$-boson, $m_W$. 
Thanks to its unique forward coverage, an upgraded LHCb can make important contributions to this programme.
The forward coverage of LHCb also allows a probe of electroweak boson production in a different regime from that of ATLAS and CMS, and the range of accessible physics topics is not limited to electroweak bosons.
For example, $t\bar{t}$ production proceeds predominantly by gluon-gluon fusion in the central region, but has a significant contribution from quark-antiquark annihilation in the forward region, giving a similar production regime to that studied at the Tevatron.

\subsubsection{$\sin^2\theta^{\rm lept}_{\rm eff}$}

The value of $\sin^2\theta^{\rm lept}_{\rm eff}$ can be extracted from $A_{\rm FB}$, the forward-backward asymmetry
of leptons produced in $Z$ decays. 
The raw value of $A_{\rm FB}$ measured in dimuon final states at the LHC is about five times larger than 
at an $e^+e^-$ collider, due to the initial state couplings, and so, in principle, it can be measured with a better relative precision, given equal amounts of data.  
The measurement however requires knowledge of the direction of the quark and antiquark that created the $Z$ boson, and any uncertainty in this quantity results in a dilution of the observed value of $A_{\rm FB}$.
This dilution is very significant in the central region, as there is an
approximately equal probability for each proton to contain the quark or anti-quark that is involved in the creation of the $Z$,
leading to an ambiguity in the definition of the axis required in the measurement.
However, the more forward the $Z$ boson is produced, the more likely it is that it follows the quark direction; 
for rapidities $y > 3$, the $Z$ follows the quark direction in around 95\% of the cases.
Furthermore, in the forward region, the partonic collisions that produce the $Z$ are nearly always between 
$u$-valence and $\bar u$-sea quark or $d$-valence and $\bar d$-sea quark.
The $s \bar s$ contribution, with a less well-known parton density function, is smaller than in the central region.
Consequently, the forward region is the optimum environment in which to measure $A_{\rm FB}$ at the LHC.  
Preliminary studies~\cite{ronan_afb} have shown that with a $50 \invfb$ data sample collected by the LHCb upgrade, $A_{\rm FB}$ 
 could be measured with a statistical precision of around 0.0004.  
 This would give a statistical uncertainty on $\sin^2 \theta^{\rm lept}_{\rm eff}$  of better than 0.0001, which is
a significant improvement in precision on the current world average value.
It is also worth remarking that the two most precise values entering this
world average at present, the forward-backward $b\bar{b}$ asymmetry
measured at LEP ($\sin^2 \theta^{\rm lept}_{\rm eff} = 0.23221 \pm 0.00029$), and the left-right asymmetries measured at SLD with polarised
beams ($\sin^2 \theta^{\rm lept}_{\rm eff}=0.23098 \pm 0.00026$), are over $3\,\sigma$ discrepant with each other~\cite{ALEPH:2005ab}. 
LHCb will be able to bring clarity to this unsatisfactory situation.

\begin{figure}[hbt]
\centering
\subfigure{
\includegraphics[width=0.50\textwidth]{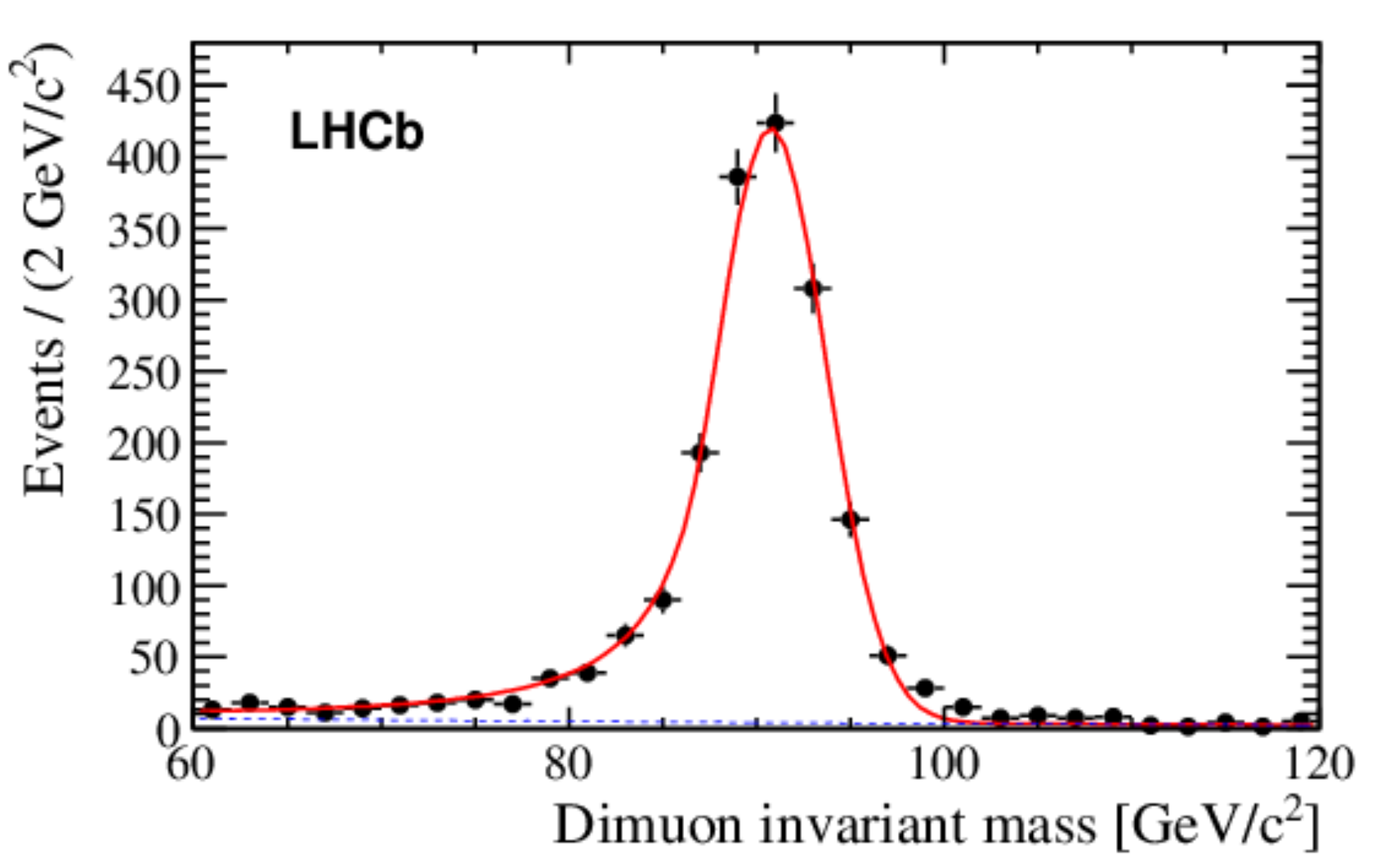}}
\subfigure{ 
\includegraphics[width=0.46\textwidth]{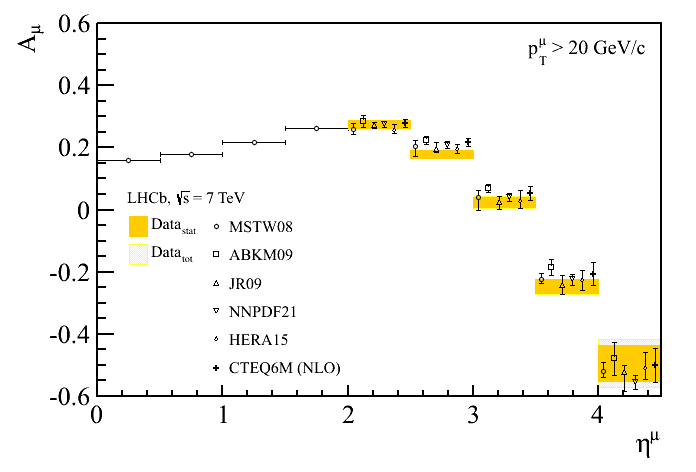}}
\caption{\small
  LHCb $Z$ and $W$ production results from $37 \invpb$ at $\sqrt{s}=7 \tev$~\cite{LHCb-PAPER-2012-008}. 
  Left: $Z \to \mu^+\mu^-$ peak.
  Right: $W^+ - W^-$ production asymmetry, where the bands correspond to the experimental uncertainties (only indicated within the LHCb acceptance), and the data points give predictions for various different parton density function sets. 
  Note that the kinematic range of the ATLAS and CMS experiments only extends up to lepton pseudorapidities of 2.5. 
} 
\label{fig:ewresults}
\end{figure}

More work is needed to identify the important systematic uncertainties on the $A_{\rm FB}$ measurement.
One source of error is the uncertainty in the parton density functions.  
With current knowledge this contribution would lead to an uncertainty of almost double the statistical precision estimate above, but this will reduce when the differential cross-section measurements from the LHC of the $W$ and $Z$ bosons, and those of Drell-Yan dimuon production at lower masses, are included in the global fits to the parton density functions.
LHCb has already embarked on this measurement programme. 
Figure~\ref{fig:ewresults}~(left) shows the $Z \to \mu^+\mu^-$ peak obtained with $37 \invpb$ of data~\cite{LHCb-PAPER-2012-008}.  
Figure~\ref{fig:ewresults}~(right) shows the measured asymmetry between $W^+$ and $W^-$ production as a function of lepton pseudorapidity. 
This measurement is already approaching the accuracy of the theoretical uncertainties. 
The $W$ and $Z$ measurements described in Ref.~\cite{LHCb-PAPER-2012-008} are being used to constrain parton density functions by some groups~\cite{nnpdf23}. 
A preliminary measurement of lower mass Drell-Yan production~\cite{LHCb-CONF-2012-013} will extend these constraints to lower $Q^2$ (masses above $5 \gevcc$ are currently considered) and Bjorken $x$.

\subsubsection{$m_W$}

Decreasing the uncertainty on $m_W$ from its present error of $15 \mevcc$ is one of the most challenging tasks for the LHC (it may also be reduced further at the Tevatron).  
Although no studies have yet been made of determining $m_W$ with LHCb itself, it is evident that the experiment can give important input to the measurements being made at ATLAS and CMS~\cite{Besson:2008zs}.  
A significant and potentially limiting external uncertainty on $m_W$ will again come from the knowledge of the parton density functions.
These are less constrained in the kinematic range accessible to LHCb, so that precise measurements of $W^+$, $W^-$, $Z$ and Drell-Yan production in this region can be used to improve the global picture.
Improved determinations of the shapes of the differential cross-sections are particularly important.
One specific area of concern arises from the knowledge of the heavy quarks in the proton. 
Around $20$--$30\,\%$ of $W$ production in the
central region is expected to involve $s$ and $c$ quarks, making
the understanding of this component very important for the $m_W$ measurement.  
LHCb can make a unique contribution to improving the knowledge of the heavy-quark parton density functions by exploiting its vertexing and particle identification capabilities to tag the relatively low-$p_T$ final-state quarks produced in processes such as $gs \to W c$, $gc \to Z c$, $gb \to Z b$, $g c \to \gamma c$ and  $g b \to \gamma b$.
These processes provide direct probes of the strange, charm and bottom partons, and can be probed at high and low values of Bjorken $x$ inside the LHCb acceptance.


\subsubsection{$t\bar{t}$ production}

Understanding the nature of top production, and in particular the asymmetry in $t \bar{t}$ events reported by Fermilab~\cite{Aaltonen:2011kc,Abazov:2011rq,CDFNote:10398,CDFNote:10584,Leone:Talk}, is of prime concern. 
As for the measurement of $\sin^2\theta^{\rm lept}_{\rm eff}$, identifying the forward direction of events is crucial. 
The LHCb acceptance for identifying both leptons from $t \bar{t}$ decays is far smaller than that of ATLAS and CMS (typically 2\,\% rather than 70\,\%, according to {\tt PYTHIA} generator level studies).
However, the higher $q \bar{q}$ production fraction and better determined direction in the LHCb forward acceptance combine to suggest that competitive measurements can be achieved. 
With the integrated luminosity offered by the upgrade, statistical precision will no longer be an issue, and LHCb measurements of the $t \bar{t}$ asymmetry will offer a competitive and complementary test of Tevatron observations~\cite{Kagan:2011yx}.

\subsection{Searches for exotic particles with displaced vertices}
\label{subsec:exotics}

Different theoretical paradigms have been proposed to 
solve the so-called ``hierarchy problem'', the most discussed being SUSY.
There are, however, many other ideas including various models involving extra dimensions, Technicolour and little Higgs models. 
These ideas approach the hierarchy problem from the direction of strong dynamics~\cite{Zurek:2010xf}. 

A growing subset of models features new massive long-lived particles with a macroscopic distance of flight. 
They can be produced by the decay of a single-produced resonance, such as a Higgs boson or a $Z^\prime$~\cite{Strassler:2006im,Carpenter:2007zz}, from the decay chain of SUSY particles~\cite{Strassler:2006ri}, or by a hadronisation-type mechanism in models where the long-lived particle is a bound state of quarks from a new confining gauge group, as discussed in Ref.~\cite{Strassler:2006im}. 
In the last case, the multiplicity of long-lived particles in an event can be large, while only one long-lived particle is expected to be produced in other models. 
The decay modes may also vary depending on the nature of the particle, from several jets in the final state~\cite{Carpenter:2007zz} to several leptons~\cite{FileviezPerez:2011kd} or lepton plus jets~\cite{deCampos:2008re}.
A comprehensive review of the experimental signatures is given in Ref.~\cite{Brooijmans:2010tn}. 

The common feature amongst these models is the presence of vertices displaced from the interaction region. 
Such signatures are well suited to LHCb, and in particular to
the upgraded experiment, which will be able to select events with displaced vertices 
at the earliest trigger level.

\begin{figure}[htb]
 \centering
\includegraphics[width=3.in]{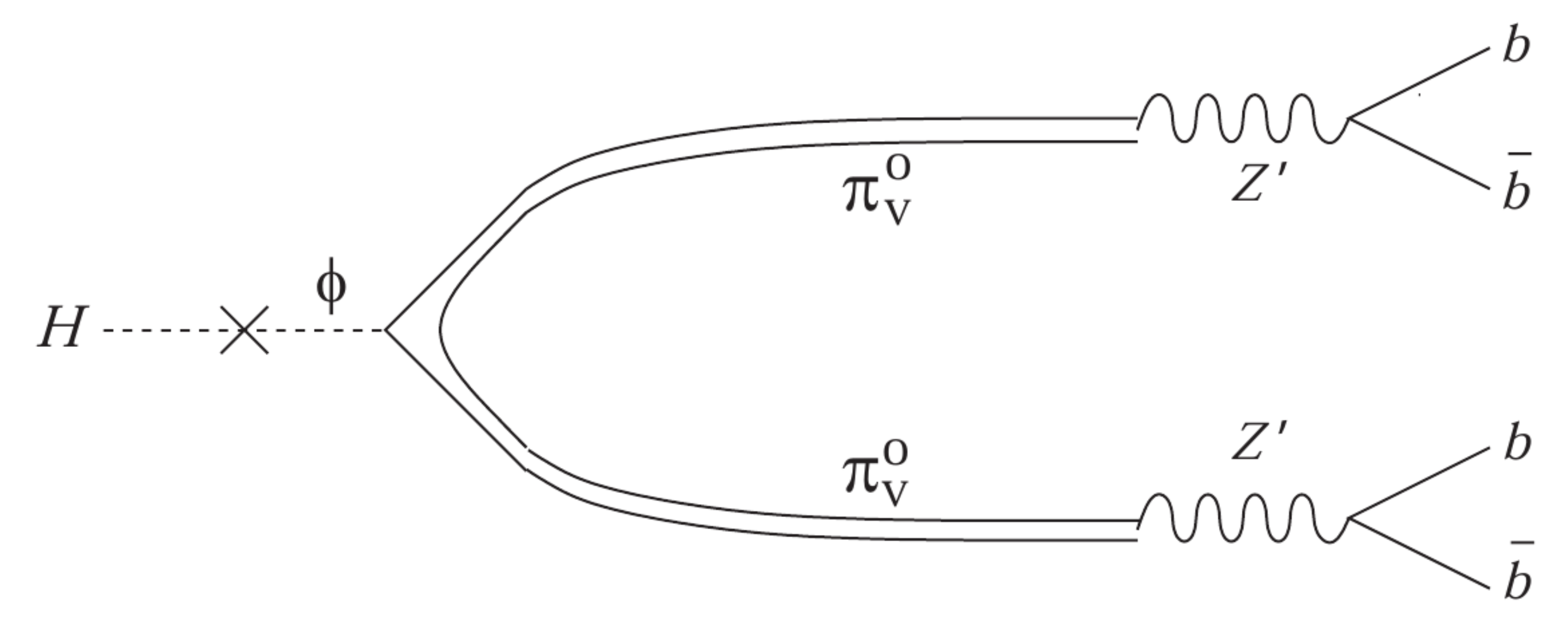}
\caption{\small
  Decay of a Higgs via a scalar field $\phi$ into two $\pi^0_v$ particles, with
$\pi^0_v$ charge equal to zero, which subsequently decay into $b\overline{b}$ jets. (Adapted from Ref.~\cite{Strassler:2006ri}.)}
  \label{matt-higgs-fig1}
\end{figure}

As an example, consider the hidden valley (HV) model already discussed in Ref.~\cite{CERN-LHCC-2011-001}.
In this model the hidden sector, or $v$-sector, contains two new heavy quarks: $U$ and $C$. 
Strassler and Zurek~\cite{Strassler:2006ri} suggest that an exotic Higgs boson could decay with a significant branching fraction to a pair of $\pi^0_v$ particles, where the $\pi^0_v$ is the 'neutral' member of the isotriplet of $v$-isospin 1 hadrons formed by $U$ and $C$ quarks.
The $\pi^0_v$ can decay in SM particles and if the mass of the spinless $\pi^0_v$ is below the $ZZ$ threshold it
will decay dominantly into $b\overline{b}$ pairs due to helicity conservation.
Here the $\pi^0_v$ widths are determined
by their lifetime which could be very long, resulting in narrow states. 
 The final state would consist of four $b$-jets, each pair being produced
from a displaced vertex corresponding to the $\pi^0_v$ decay as illustrated
in Fig.~\ref{matt-higgs-fig1}. 
If these decays exist, the lower limit on the Higgs mass set by LEP would be misleading,
as it assumes the prompt decay of the Higgs to $b\overline{b}$ to be dominant.

The potential of LHCb to search for such exotic Higgs decays at $\sqrt{s}=14 \tev$ has been discussed in Ref.~\cite{CERN-LHCC-2011-001}, and is briefly summarised here. 
The benchmark model uses $m_H = 120 \gevcc$, $m_{\pi^0_v} = 35 \gevcc$ and $\tau_{\pi^0_v} = 10 \ps$. 
By combining vertex and jet reconstruction, the capacity to reconstruct this final state is shown using full simulation of the detector, assuming 0.4 interactions per crossing. 
Backgrounds to this signal from other processes, such as the production of two pairs of $b\bar{b}$ quarks, have been considered and found to be negligible.

During 2010 and 2011 data taking, an inclusive displaced vertex trigger has been introduced in the second level of the software trigger. 
Preliminary studies~\cite{LHCb-CONF-2012-014} have demonstrated that for an output rate below $1\,\%$ of the overall trigger bandwidth, the efficiency of the whole trigger chain on
events with two offline reconstructible $\pi^0_v$ vertices with a minimum mass of $6 \gev$ and good vertex quality is of the order of $80\%$.  
This strategy has been tested up to on average two visible interactions per crossing which is what is expected for the upgraded experiment.

\begin{figure}[!hbt]
\centering
\includegraphics[width=3.in]{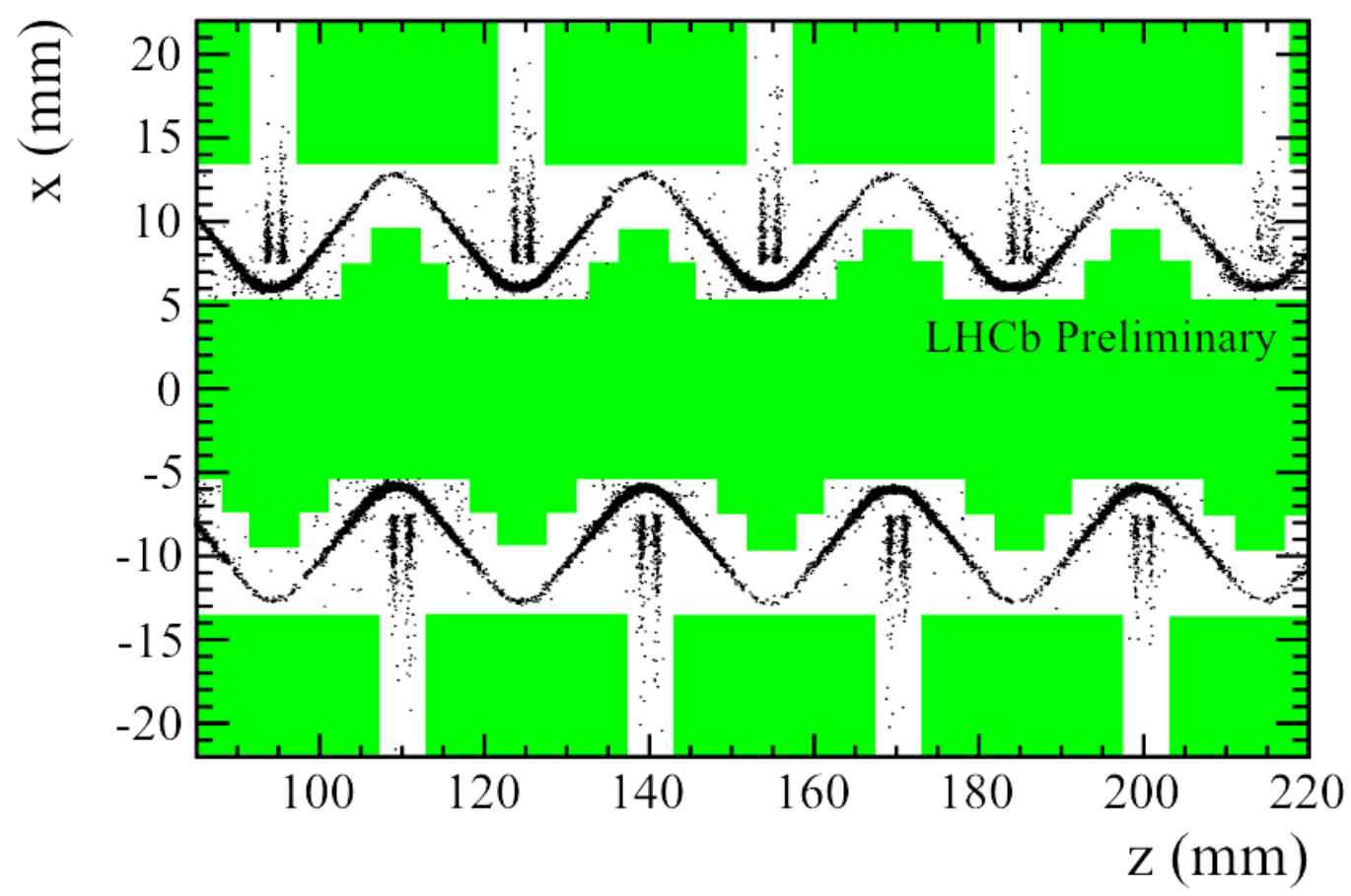}
\includegraphics[width=3.in]{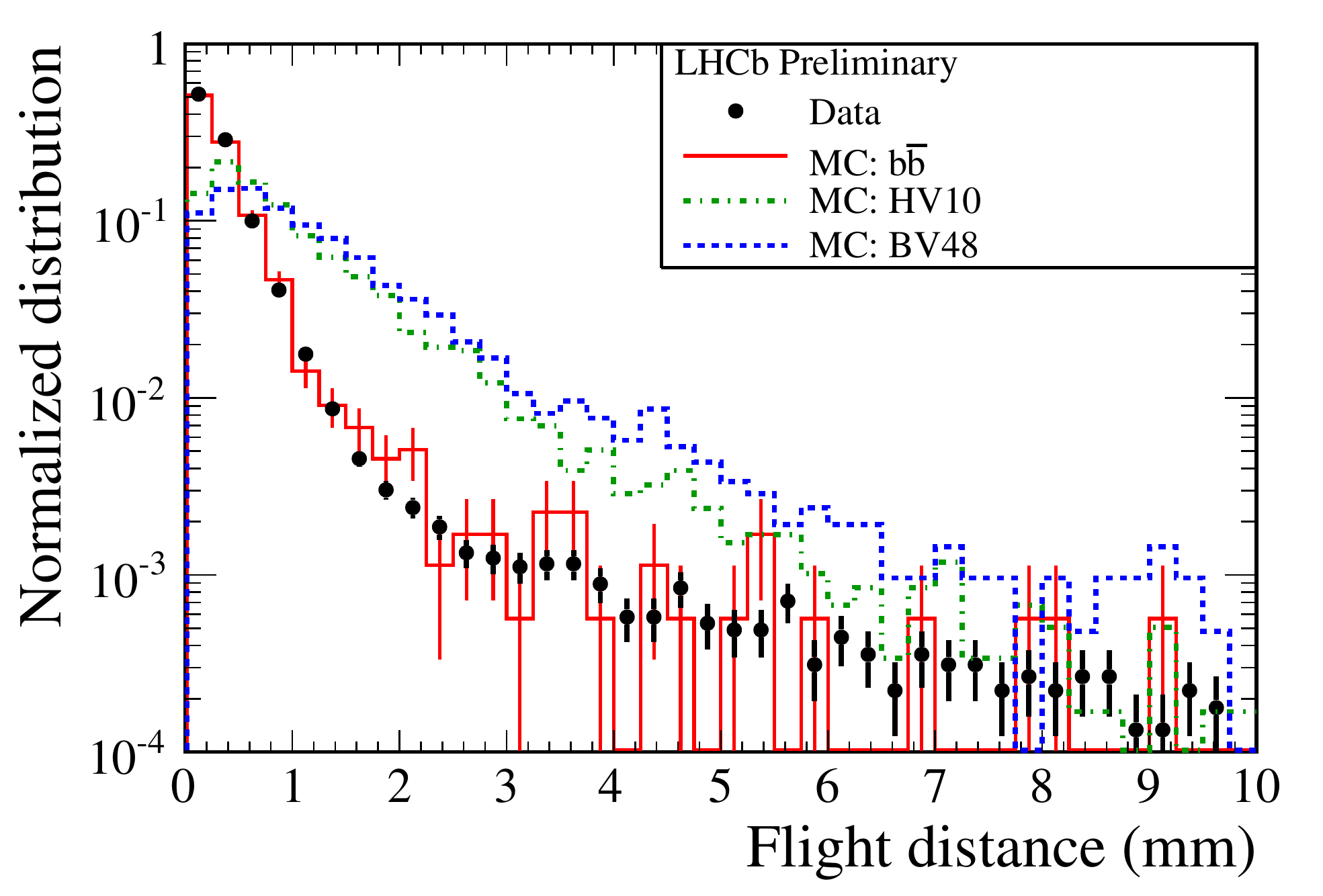}
\caption{\small
  Left: Distribution in $x$ and $z$, for $|y| < 1 \mm$, of the reconstructed vertices. 
  The visible structures reflect the geometry of the vertex detector, with the pairs of silicon sensors appearing as pairs of vertical bands and the corrugated (``RF'') foil as the two wave shapes. 
  The green shaded region represents the fiducial vacuum volume in which candidates are accepted. 
  Right: Flight distance of offline reconstructed vertices in events outside the matter region. 
  Data are compatible with $b\bar{b}$ background. 
  The black points are for data in $36 \invpb$~\cite{LHCb-CONF-2012-014}, the red line is a full simulation of $b\bar{b}$ production and the green dashed line is a full simulation of the HV benchmark channel. 
  The blue dashed line shows a simulation of a model with baryon number violating neutralino couplings.
} \label{HVPicture}
\end{figure}

The analysis of the trigger output showed that once vertices arising from hadronic interactions with material are rejected,
the dominant background is compatible with $b$ hadron decay vertices as shown in Fig.~\ref{HVPicture}.
Those $b$ hadron vertices are reconstructed with large masses because of the presence of fake or cloned tracks. 
With the present detector, it is difficult to keep the trigger rate down
for single candidate events without using tight cuts on the mass and the displacement of the candidates. 
In the previous model, the trigger efficiency for events with a single long-lived particle reconstructible in LHCb is only about $20\,\%$. 
This efficiency is expected to decrease for models where the mass of the long-lived particle is smaller. 
In addition, the number of events with at least one $\pi^0_v$ state in the acceptance is three times higher than the number of events with two $\pi^0_v$ particles. 
Improving the single candidate efficiency would increase sensitivity to this model. 
It would also give a better coverage for the models where only one long-lived particle is produced.

In the upgraded detector, the track fake rate in the vertex detector is expected to be below one percent~\cite{CERN-LHCC-2012-007}, compared to $6\,\%$ in the present detector. 
Other upgrades to the tracking detectors will also help to reduce the fake rate.
Moreover the use of an improved description for the complex RF foil shape will give a better control on the background arising from hadronic interactions.
It will enable the use of the true shape of the RF foil, rather than the loose fiducial volume cut used at present, 
which depending on the considered lifetime, rejects $10$--$30\,\%$ of the long-lived particles.
Those improvements would allow to decrease the thresholds on the single candidates trigger and therefore increase the reach of such searches.

As discussed in Ref.~\cite{CERN-LHCC-2011-001} the coupling of vertex information to jet reconstruction will
allow to reduce the physical backgrounds. Studies are on-going on this matter.
Assuming a Higgs production cross-section at $\sqrt{s} = 14 \tev$ of $50 \pb$, an integrated luminosity of $50 \invfb$ and a geometric efficiency of 10\,\%, 250\,000 Higgs bosons will be produced in LHCb. 
If $H^0\to\pi_v^0\pi_v^0$ is a dominant decay mode, then LHCb will be in an excellent position to observe this signal, taking advantage of  the software trigger's ability to select high-multiplicity events with good efficiency.

\subsection{Central exclusive production}
\label{subsec:cep}

Central exclusive production (CEP) processes provide a promising and novel way
to study QCD and the nature of new particles, from low mass glueball candidates 
up to the Higgs boson itself. The CEP of an object $X$ in a $pp$ collider may be written as follows
\begin{eqnarray}
  pp \to p \, + \, X \, + \, p \, ,
\nonumber
\end{eqnarray}
where the `$+$' signs denote the presence of large rapidity gaps. At high energies the $t$-channel
exchanges giving rise to these processes can only be zero-charge colour singlets.
Known exchanges include the photon and the pomeron.  Another possibility,
allowed in QCD, but not yet observed, is the odderon, a negative C-parity 
partner to the pomeron with at least three gluons. The most attractive aspect of
CEP reactions is that they offer a very clean environment in which to measure
the nature and quantum numbers of the centrally produced state $X$.

\begin{figure}[hbt]
\vspace{-.2cm}
\centering
\includegraphics[width=3in]{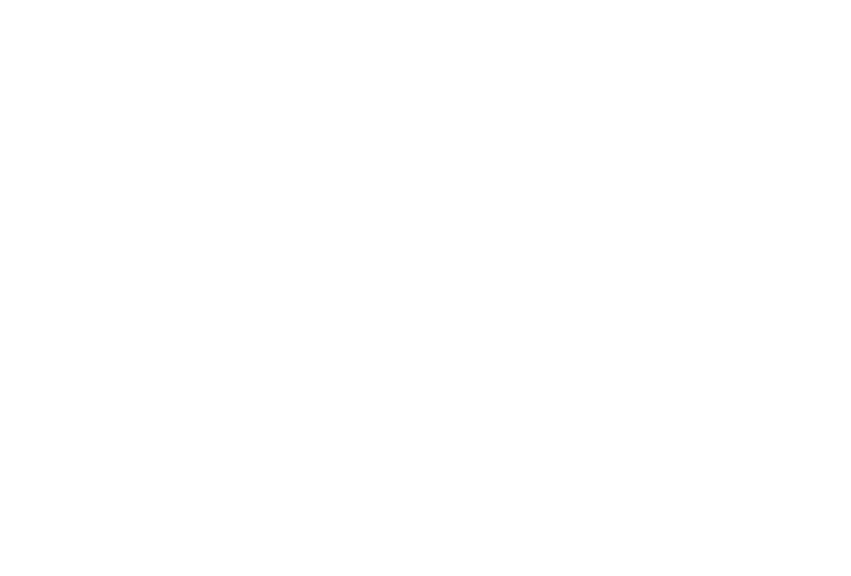}
\vspace{-0.4cm}
\caption{\small
  \label{fig:lhcbcep} 
  Preliminary LHCb results on central exclusive $\chi_c$ production~\cite{LHCb-CONF-2011-022}.  
  The $\jpsi\gamma$ invariant mass in data is compared to the expectation of the {\tt SuperCHIC} Monte Carlo generator~\cite{SUPERCHIC}, which has been normalised to the observed number of events.   
  The relative proportions of $\chi_{c0}$, $\chi_{c1}$ and $\chi_{c2}$ are 12\,\%, 36\,\% and 52\,\% respectively.
}
\end{figure}

Central exclusive $\gamma \gamma$~\cite{Aaltonen:2007am}, dijet~\cite{Aaltonen:2007hs,Abazov:2010bk} and $\chi_c$~\cite{Aaltonen:2009kg} production has been observed at the Tevatron.  
LHCb has presented preliminary results on candidate dimuon events compatible with CEP~\cite{LHCb-CONF-2011-022}.  
Figure~\ref{fig:lhcbcep} shows the invariant mass of CEP $\chi_c$ candidates.
These are events in which only a  $\jpsi \to \mu^+\mu^-$ decay and a $\gamma$ candidate are reconstructed, with no other activity (inconsistent with noise) seen elsewhere in the detector.
Important observables in CEP are the relative production rates of $\chi_{c0}$, $\chi_{c1}$ and $\chi_{c2}$.  
As is evident from Fig.~\ref{fig:lhcbcep}, the invariant mass resolution of LHCb is sufficient for this measurement.

Although not part of the baseline for the LHCb upgrade, additional instrumentation is being considered which could improve the potential of LHCb to study CEP processes.  
For example, the inclusion of forward shower counters (FSCs) on both sides of the interaction point, as proposed in Ref.~\cite{Lamsa:2009ej}, would be able to detect showers from very forward particles interacting in the beam pipe and surrounding material.  
The absence of a shower would indicate a rapidity gap and be helpful in increasing the purity of a CEP sample.  
More ambitiously, the deployment of semi-conductor detectors very close to the beam, within Roman pots, several hundred meters away from the interaction point, as proposed for other LHC experiments~\cite{Albrow:2008pn} would also be beneficial for LHCb.  
The ability to measure the directions of the deflected protons in the CEP interaction provides invaluable information in determining the quantum numbers of the centrally produced state. 

Several important physics goals have been identified for the LHCb CEP programme:
\begin{itemize}
\item{
  Accumulation and characterisation of large samples of exclusive $c\bar{c}$ and $b\bar{b}$ events.  
  A full measurement programme of these  `standard candles'
  will be essential to understand better the QCD mechanism of CEP~\cite{HarlandLang:2010ep},
  and may provide vital input if CEP is used for studies of Higgs and other new  particles~\cite{Heinemeyer:2010gs}.}
\item{
  Searches for structure in the mass spectra of decay states such
  as $K^+K^-$, $2 \pi^+ 2 \pi^-$, $K^+K^- \pi^+\pi^-$ and $p\bar{p}$.
  A particular interest of this study would be the hunt for glueballs, which
  are a key prediction of QCD.}
\item{
  Observation and study of exotic particles in CEP processes.
  For example, a detailed study of the CEP process $pp \to p \, + \, X(3872) \, + \, p$ would provide a valuable new tool to aid understanding of this state.  
  This and other states could be searched for in,
  for example, decays containing $D\bar{D}$, which if observed would
  shed light onto the nature of the parent particle~\cite{HarlandLang:2010ep}.}
\end{itemize}
There are several reasons which make LHCb a suitable detector for realising these goals,  particularly 
with the upgraded experiment:
\begin{itemize}
\item{Even when running at a luminosity of $10^{33} \cm^{-2} \sec^{-1}$ LHCb will have low pileup compared to ATLAS and CMS. This will be advantageous in triggering and reconstructing low mass CEP states.}
\item{The higher integrated luminosity that will be collected by the upgraded detector will allow
studies to be performed on states that are inaccessible with only a few $\invfb$.  This is true, 
for example, of central exclusive $\chi_b$ production, which is expected to be a factor of $\sim$~1000 less than that of $\chi_c$ mesons~\cite{HarlandLang:2010ep}.}  
\item{The particle identification capabilities of the LHCb ring-imaging Cherenkov detector system allow centrally produced states to be cleanly separated into decays involving pions, kaons and protons. }
\item{The low $\pt$ acceptance of LHCb, and high bandwidth trigger, will allow samples
of relatively low mass states to be collected and analysed.}
\end{itemize}

\clearpage

\section{Summary}
\label{sec:summary}

As described in the previous sections, LHCb has produced world-leading results across its physics programme, using the $1.0 \invfb$ data sample of $\sqrt{s} = 7 \tev$ $pp$ collisions collected in 2011.
The inclusion of the data collected at $\sqrt{s} = 8 \tev$ during 2012 will enable further improvements in precision in many key flavour physics observables.
However, an upgrade to the detector is needed to remove the bottleneck in the trigger chain that currently prevents even larger increases in the collected data sample.
The upgraded detector with trigger fully implemented in software is to be installed during the 2018 long shutdown of the LHC, and will allow a total data set of $50 \invfb$ to be collected.
With such a data sample, LHCb will not only reach unprecedented precision for a wide range of flavour physics observables, but the flexible trigger will allow it to exploit fully the potential of a forward physics experiment at a hadron collider.

In this section, some highlights of the LHCb physics output so far, and their implications on the theoretical landscape, are summarised. 
The sensitivity of the upgraded detector to key observables is then given, before a concluding statement on the importance of the LHCb upgrade to the global particle physics programme.

\subsection{Highlights of LHCb measurements and their implications}
\subsubsection{Rare decays}

Among rare decays, the LHCb limit on the rate of the decay $\Bs \to \mumu$~\cite{LHCb-PAPER-2012-007} places stringent limits on NP models that enhance the branching fraction.
The measurement
\begin{equation}
  \BRof\Bsmumu < 4.5 \times 10^{-9} \ (95\,\% \ {\rm confidence \ level})\,, \\
\end{equation}
can be compared to the SM prediction
$\BRof\Bsmumu_{\rm SM} = (3.1 \pm 0.2) \times 10^{-9}$~\cite{Buras:2012ts}.\footnote{
  It should be noted that the measured value is the time-integrated branching fraction, and the SM prediction should be increased by around 10\,\% to allow a direct comparison~\cite{deBruyn:2012wk}.
}
This result puts severe constraints --- far beyond the ATLAS and CMS search limits --- on supersymmetric models with large values of $\tan\beta$, \ie\ of the ratio of vacuum expectation values of the Higgs doublets (see, for example, Refs.~\cite{Buras:2012ts,Mahmoudi:2012uk,Buchmueller:2011sw}).

The measurement of the forward-backward asymmetry in $\Bd \to \Kstarz\mumu$~\cite{LHCb-PAPER-2011-020} has to be viewed as the start of a programme towards a full angular analysis of these decays.
The full analysis will allow determination of numerous NP-sensitive observables (see, for example, Refs.~\cite{Egede:2008uy,Egede:2010zc}).
The measurements that will be obtained from such an analysis, as well as similar studies of related channels, such as $\Bs\to \phi\mumu$~\cite{LHCb-CONF-2012-003}, allow model-independent constraints on NP, manifested as limits on the operators of the effective Hamiltonian (see, for example, Refs.~\cite{Altmannshofer:2012ir,Hurth:2012jn}).
Indeed, the first results already impose important constraints. 
Studies of radiative decays such as $\Bs \to \phi\gamma$~\cite{LHCb-PAPER-2011-042,LHCb-PAPER-2012-019} provide additional information since they allow to measure the polarisation of the emitted photon, and are therefore especially sensitive to models that predict new right-handed currents.
Similarly, studies of observables such as isospin asymmetries~\cite{LHCb-PAPER-2012-011} are important since they allow to pin down in which operators the NP effects occur.

Several new opportunities with rare decays at LHCb are becoming apparent.  The observation of $\Bp \to \pip\mumu$~\cite{LHCb-PAPER-2012-020}, the rarest $B$ decay yet discovered, enables a new approach to measure the ratio of CKM matrix elements $\left| V_{td}/V_{ts} \right|$.  Decays to final states containing same-sign leptons~\cite{LHCb-PAPER-2011-038} allow searches for Majorana neutrinos complementary to those based on neutrinoless double beta decay.  LHCb can also reach competitive sensitivity for some lepton flavour violating decays such as $\taup \to \mumu\mup$~\cite{LHCb-CONF-2012-015}.

\subsubsection{\CP violation in the \B sector}

Measurements of the neutral \B meson mixing parameters provide an excellent method to search for NP effects, due to the low theoretical uncertainties associated to several observables.  The LHCb measurements of the \CP-violating phase, $\phi_s$, and the width difference, $\Delta \Gamma_s$, in the \Bs system~\cite{LHCb-CONF-2012-002,LHCb-PAPER-2011-028,LHCb-PAPER-2011-021,LHCb-PAPER-2012-006} significantly reduce the phase space for NP:
\begin{equation}
  \phi_s = -0.002 \pm 0.083 \pm 0.027 \ {\rm rad} \, ,
  \hspace{0.3cm}
  \Delta \Gamma_s = 0.116 \pm 0.018 \,{\rm (stat)} \pm 0.006 {\,\rm (syst)} \invps \, .
\end{equation}
However 
deviations from the SM predictions~\cite{Lenz:2006hd,Charles:2011va} are still possible.  
Effects of ${\cal O}(0.1)$ are typical of some well-motivated NP models that survive the present ATLAS and CMS bounds (such as in Ref.~\cite{Barbieri:2011ci}).
The experimental uncertainty on $\phi_s$ is still a factor of 40 larger than that on the prediction, therefore improved measurements are needed to reach the level of sensitivity demanded by theory.
It should also be noted that compared to the \CP-violating phase in the \Bd system ($2\beta$), $\phi_s$ is much more precisely predicted, and therefore presents stronger opportunities for NP searches.

In addition, to understand the origin of the anomalous dimuon asymmetry seen by \dzero~\cite{Abazov:2011yk}, improved measurements of semileptonic asymmetries in both \Bs and \Bd systems are needed.
LHCb has just released its first results on the \Bs asymmetry~\cite{LHCb-CONF-2012-022}, demonstrating the potential to search for NP effects with more precise measurements.
Moreover, a constraint on, or a measurement of, the rate of the decay $\Bs\to\tau^+\tau^-$ is important to provide knowledge of possible NP contributions to $\Gamma_{12}$ (see, for example, Refs.~\cite{Dighe:2010nj,Bobeth:2011st}).

Among the \Bd mixing parameters, improved measurements of both $\phi_d$ (\ie, $\sin2\beta$) and $\Delta \Gamma_d$ are needed.  Reducing the uncertainty on the former will help to improve the global fits to the CKM matrix~\cite{Charles:2004jd,Bona:2005vz}, and may clarify the current situation regarding the tension between various inputs to the fits (see, for example, Ref.~\cite{Lunghi:2010gv}).
Another crucial observable is the angle $\gamma$, which, when measured in the tree-dominated $B \to DK$ processes, provides a benchmark measurement of \CP violation.
The first measurements from LHCb already help to improve the uncertainty on $\gamma$~\cite{LHCb-PAPER-2012-001,LHCb-PAPER-2012-027}: further improvements are both anticipated and needed.

Comparisons of values of $\gamma$ from loop-dominated processes with the SM benchmark from tree-dominated processes provide important ways to search for new sources of \CP violation.
In particular, the study of $\Bs \to \Kp\Km$ and $\Bd \to \pip\pim$ decays~\cite{LHCb-CONF-2012-007}, which are related by U-spin, allows a powerful test of the consistency of the observables with the SM~\cite{Fleischer:2010ib,Ciuchini:2012gd}.  Similarly, the U-spin partners $\Bs \to \Kstarz\Kstarzb$~\cite{LHCb-PAPER-2011-012} and $\Bd \to \Kstarz\Kstarzb$ are among the golden channels to search for NP contributions in $b \to s q\bar{q}$ penguin amplitudes~\cite{Ciuchini:2007hx}.  Another important channel in this respect is $\Bs \to \phi\phi$~\cite{LHCb-PAPER-2012-004}, for which the \CP-violating observables are predicted with low theoretical uncertainty in the SM.  
Studies of \CP violation in multibody $b$ hadron decays~\cite{LHCb-CONF-2012-018,LHCb-CONF-2012-028} offer additional possibilities to search for both the existence and features of NP.

\subsubsection{Charm mixing and \CP violation}

In the charm sector, the evidence for \CP violation in the observable $\dacp$ has prompted a large amount of theoretical work.  
The measurement
\begin{equation}
  \dacp  = \ACP(\Kp\Km) - \ACP(\pip\pim) = (-0.82 \pm 0.21 \pm 0.11)\,\% \, ,
\end{equation}
is different from zero by $3.5$ standard deviations~\cite{LHCb-PAPER-2011-023}.
While $\ACP$ represents a time-integrated \CP asymmetry, $\Delta \ACP$ originates predominantly from direct \CP violation.
The emergent consensus 
is that while an asymmetry of the order of 1\,\% is rather unlikely in the SM, it cannot be ruled out that QCD effects cause enhancements of that size.  Further measurements are needed in order to establish if NP effects are present in the charm sector.  Among the anticipated results are updates of the $\Delta \ACP$ measurement as well as of the individual \CP asymmetries in $\Dz \to \Kp\Km$ and $\Dz \to \pip\pim$.  It is of great interest to look for direct \CP violation in decays to other final states, and in decays of other charmed hadrons ($\Dp$, $\Dsp$ and $\Lc$).  

The SM predictions are somewhat cleaner for indirect \CP violation effects, and therefore it is also essential to search for \CP violation in charm mixing.  New results from time-dependent analyses of $\Dz \to \Kp\Km$~\cite{LHCb-PAPER-2011-032} and $\Dz \to \KS \pip\pim$ will improve the current knowledge, and additional channels will also be important with high statistics.

Several authors have noted correlations between \CP violation in charm and various other observables (for example, Refs.~\cite{Isidori:2011qw,Hochberg:2011ru}).  These correlations appear in, and differ between, certain theoretical models, and can therefore be used to help identify the origin of the effects.  Observables of interest in this context include those that can be measured at high-\pt experiments, such as $t\bar{t}$ asymmetries, as well as rare charm decays.  Among the latter, it has been noted that \CP asymmetries are possible in radiative decays such as $\Dz \to \phi \gamma$~\cite{Isidori:2012yx}, and that searches for decays involving dimuons, such as $\Dz \to \mumu$~\cite{LHCb-CONF-2012-005} and $\Dp \to \pip\mumu$ are well motivated.

\subsubsection{Measurements exploiting the unique kinematic acceptance of LHCb}

The unique kinematic region covered by the LHCb acceptance enables measurements that cannot be performed at other experiments, and that will continue to be important in the upgrade era.  
These include probes of QCD both in production, such as studies of multi-parton scattering~\cite{LHCb-PAPER-2011-013,LHCb-PAPER-2012-003}, and in decay, such as studies of exotic hadrons like the $X(3872)$~\cite{LHCb-PAPER-2011-034} and the putative $Z(4430)^+$ state.  Conventional hadrons can also be studied with high precision: one important goal will be to establish the existence of doubly heavy baryons.  Central exclusive production of conventional and exotic hadrons can also be studied; the sensitivity of the upgraded experiment will be significantly enhanced due to the software trigger.

Measurements of production rates and asymmetries of electroweak gauge bosons in the LHCb acceptance are important to constrain parton density functions~\cite{LHCb-PAPER-2012-008}.  With high statistics, LHCb will be well placed to make a precision measurement of the sine of the effective electroweak mixing angle for leptons, $\sin^2\theta^{\rm lept}_{\rm eff}$, from the forward-backward asymmetry of leptons produced in the $Z \to \mumu$ decay.  Improved knowledge of parton density functions, as can be obtained from studies of production of gauge bosons in association with jets~\cite{LHCb-CONF-2012-016}, will help to reduce limiting uncertainties on the measurement of the $W$ boson.  These studies are also an important step towards a top physics programme at LHCb, which will become possible once the LHC energy approaches the nominal $14 \tev$.  

The importance of having a detector in the forward region can be illustrated with the recent discovery by ATLAS and CMS of a new particle that may be the Higgs boson.  It is now essential to determine if this particle has the couplings to bosons, leptons and quarks expected in the SM.  In particular, at the observed mass the highest branching ratio is expected to be for $H \to b\bar{b}$ --- however this is a difficult channel for ATLAS and CMS due to the large SM background. LHCb with its excellent $b$-hadron sensitivity will be able to search for such decays. The forward geometry of LHCb is also advantageous to observe new long-lived particles that are predicted in certain NP models, including some with extended Higgs sectors.  Although limits can be set with the current detector~\cite{LHCb-CONF-2012-014}, this is an area that benefits significantly from the flexible software trigger of the upgraded experiment.  Models with extended Higgs sectors also produce characteristic signals in flavour physics observables, which emphasises the need for the LHCb upgrade as part of the full exploitation of the LHC.

\subsection{Sensitivity of the upgraded LHCb experiment to key observables}

As mentioned in Sec.~\ref{sec:Introduction}, the LHCb upgrade is necessary to progress beyond the limitations imposed by the current hardware trigger that, due to its maximum output rate of $1 \mhz$, restricts the instantaneous luminosity at which data can most effectively be collected.  
To overcome this, the upgraded detector will be read out at the maximum LHC bunch-crossing frequency of $40 \mhz$ so that the trigger can be fully implemented in software.
The upgraded detector will be installed during the long shutdown of the LHC planned for 2018.
A detailed description of the upgraded LHCb experiment can be found in the Letter of Intent (LoI)~\cite{CERN-LHCC-2011-001}, complemented by the recent framework technical design report (FTDR)~\cite{CERN-LHCC-2012-007}, which sets out the timeline and costing for the project.
 A summary has been prepared for the European Strategy Preparatory Group~\cite{LHCb-PUB-2012-010}.

\begin{sidewaystable}
  \begin{center}
    \caption{\small
      Statistical sensitivities of the LHCb upgrade to key observables. For each observable the current sensitivity is compared to that which will be achieved by LHCb before the upgrade, and that which will be achieved with $50 \invfb$ by the upgraded experiment.
      Systematic uncertainties are expected to be non-negligible for the most precisely measured quantities.
      Note that the current sensitivities do not include new results presented at ICHEP 2012 or CKM2012.
    }
    \ifthenelse{\boolean{firstbigtable}}{\label{tab:upgrade:sensitivities}\setboolean{firstbigtable}{false}}{\label{tab:upgrade:sensitivities2}}
    \begin{tabular}{cccccc}
\hline
\hline
Type & Observable & Current   & LHCb & {\bf Upgrade} & Theory    \\
     &            & precision & 2018 & ($50 \invfb$) & uncertainty \\
\hline
$\Bs$ mixing & $2\beta_s$ ($\Bs\to \jpsi\,\phi$)     & $0.10$~\cite{LHCb-CONF-2012-002} & $0.025$ & $0.008$ & $\sim 0.003$ \\
             & $2\beta_s$ ($\Bs\to \jpsi\,f_0(980)$) & $0.17$~\cite{LHCb-PAPER-2012-006} & $0.045$ & $0.014$ & $\sim 0.01$ \\
             & $a_{\rm sl}^s$ & $6.4 \times 10^{-3}$~\cite{HFAG} & $0.6 \times 10^{-3}$ & $0.2 \times 10^{-3}$ & $0.03 \times 10^{-3}$ \\
\hline
Gluonic & $2\beta_s^{\rm eff}(\Bs \to \phi\phi)$          & --      & $0.17$ & $0.03$ & $0.02$ \\
penguins& $2\beta_s^{\rm eff}(\Bs \to K^{*0} \bar{K}^{*0})$ & --      & $0.13$ & $0.02$ & $ <0.02$ \\
	& $2\beta^{\rm eff}(\Bd \to \phi K^0_S)$         & $0.17$~\cite{HFAG} & $0.30$ & $0.05$ & $0.02$ \\
\hline
Right-handed & $2\beta_s^{\rm eff}(\Bs \to \phi \gamma)$              & -- & $0.09$ & $0.02$ & $<0.01$ \\
currents     & $\tau^{\rm eff}(\Bs \to \phi \gamma)/\tau_{\Bs}$ & -- & $5\,\%$ & $1\,\%$ & $0.2\,\%$ \\
\hline
Electroweak & $S_3(\Bd \to K^{*0} \mu^+ \mu^-; 1 < q^2 < 6 \gevgevcccc)$ & $0.08$~\cite{LHCb-CONF-2012-008} & $0.025$ & $0.008$ & $0.02$ \\
penguins    & $s_0 \, A_{\rm FB}(\Bd \to K^{*0}\mu^+ \mu^-)$ & $25\,\%$~\cite{LHCb-CONF-2012-008} & $6\,\%$ & $2\,\%$ & $7\,\%$ \\
         & $A_{\rm I}(K\mu^+\mu^-; 1 < q^2 < 6 \gevgevcccc)$ & $0.25$~\cite{LHCb-PAPER-2012-011} & $0.08$ & $0.025$ & $\sim 0.02$ \\
         & ${\cal B}(\Bp \to \pip\mup\mun)/{\cal B}(\Bp \to \Kp\mup\mun)$ & $25\,\%$~\cite{LHCb-PAPER-2012-020} & $8\,\%$ & $2.5\,\%$ & $\sim 10\,\%$ \\
\hline
Higgs   & ${\cal B}(\Bs \to \mu^+ \mu^-)$ & $1.5 \times 10^{-9}$~\cite{LHCb-PAPER-2012-007} & $0.5 \times 10^{-9}$ & $0.15 \times 10^{-9}$ & $0.3 \times 10^{-9}$ \\
penguins& ${\cal B}(\Bd \to \mu^+ \mu^-)/{\cal B}(\Bs \to \mu^+ \mu^-)$ & -- & $\sim 100\,\%$ & $\sim 35\,\%$ & $\sim 5\,\%$ \\
\hline
Unitarity & $\gamma$ ($B \to D^{(*)}K^{(*)}$) & $\sim 10$--$12^\circ$~\cite{Charles:2004jd,Bona:2005vz} & $4^\circ$ & $0.9^\circ$ & negligible \\
triangle  & $\gamma$ ($\Bs \to D_s K$)      & --              & $11^\circ$ & $2.0^\circ$ & negligible \\
angles    & $\beta$ ($\Bd \to \jpsi\,\KS$) & $0.8^\circ$~\cite{HFAG} & $0.6^\circ$ & $0.2^\circ$ & negligible \\
\hline
Charm & $A_\Gamma$       & $2.3 \times 10^{-3}$~\cite{HFAG} & $0.40 \times 10^{-3}$ & $0.07 \times 10^{-3}$ & -- \\
\CP violation   & $\Delta \ACP$ & $2.1 \times 10^{-3}$~\cite{LHCb-PAPER-2011-023} & $0.65 \times 10^{-3}$ & $0.12 \times 10^{-3}$ & -- \\
      \hline
      \hline
    \end{tabular}
  \end{center}
\end{sidewaystable}

The sensitivity to various flavour observables is summarised in Table~\ref{tab:upgrade:sensitivities}, which is taken from the FTDR~\cite{CERN-LHCC-2012-007}.
This is an updated version of a similar summary that appears as Table 2.1 in the LoI~\cite{CERN-LHCC-2011-001}.
The measurements considered include \CP-violating observables, rare decays and fundamental parameters of the CKM unitarity triangle.  
More details about these observables are given below.
The current precision, either from LHCb measurements or averaging groups~\cite{HFAG,Charles:2004jd,Bona:2005vz}, is given and compared to the estimated sensitivity with the upgrade.
As an intermediate step, the estimated precision that can be achieved prior to the upgrade is also given for each observable.
For this, a total integrated luminosity of $1.0\ (1.5, 4.0) \invfb$ at $pp$ centre-of-mass collision energy $\sqrt{s} = 7\ (8, 13) \tev$ recorded in 2011 (2012, 2015--17) is assumed.
Another assumption is that the current efficiency of the muon hardware trigger can be maintained at higher $\sqrt{s}$, but that higher thresholds will be necessary for other triggers, reducing the efficiency for the relevant channels by a factor of 2 at $\sqrt{s} = 14 \tev$.

In LHCb measurements to date, the \CP-violating phase in \Bs mixing, measured in both $\jpsi\,\phi$ and $\jpsi\,f_0(980)$ final states, has been denoted $\phi_s$.  In the upgrade era it will be necessary to remove some of the assumptions that have been made in the analyses to date, related to possible penguin amplitude contributions, and therefore the observables in $b \to c\bar{c}s$ transitions are denoted by $2\beta_s = - \phi_s$, while in $b \to q\bar{q}s \ (q = u,d,s)$ transitions the notation $2\beta^{\rm eff}_s$ is used.  
This parallels the established notation used in the $\Bd$ system (the $\alpha, \beta, \gamma$ convention for the CKM unitarity triangle angles is used).
The penguin contributions are expected to be small, and therefore a theory uncertainty on $2\beta_s\,(\Bs\to \jpsi\,\phi) \sim 0.003$ is quoted, comparable to the theory uncertainty on $2\beta\,(\Bd\to \jpsi\,\KS)$.  
However, larger effects cannot be ruled out at present.  
Data-driven methods to determine the penguin amplitudes are also possible~\cite{Faller:2008gt,Chiang:2009ev,Bhattacharya:2012ph}: at present these given much larger estimates of the uncertainty, but improvement can be anticipated with increasing data samples.
The flavour-specific asymmetry in the \Bs system, $a_{\rm sl}^s$ in Table~\ref{tab:upgrade:sensitivities}, probes \CP violation in mixing.  The ``sl'' subscript is used because the measurement uses semileptonic decays.

Sensitivity to the emitted photon polarisation is encoded in the effective lifetime, $\tau^{\rm eff}$ of $\Bs \to \phi \gamma$ decays, together with the effective \CP-violation parameter $2\beta_s^{\rm eff}$.
Two of the most interesting of the full set of angular observables in $\Bd \to \Kstarz \mumu$ decays~\cite{Altmannshofer:2008dz}, are $S_3$, which is related to the transverse polarisation asymmetry~\cite{Kruger:2005ep}, and the zero-crossing point ($s_0$) of the forward-backward asymmetry.
As discussed above, isospin asymmetries, denoted $A_I$, are also of great interest.

In the charm sector, it is important to improve the precision of $\Delta \ACP$, described above, and related measurements of direct \CP violation.  One of the key observables related to indirect \CP violation is the difference in inverse effective lifetimes of $\Dz \to \Kp\Km$ and $\Dzb \to \Kp\Km$ decays, $A_{\Gamma}$. 

The extrapolations in Table~\ref{tab:upgrade:sensitivities} assume the central values of the current measurements, or the SM where no measurement is available.
While the sensitivities given include statistical uncertainties only, preliminary studies of systematic effects suggest that these will not affect the conclusions significantly, except in the most precise measurements, such as those of $a_{\rm sl}^s$, $A_\Gamma$ and $\Delta \ACP$.
Branching fraction measurements of $\Bs$ mesons require knowledge of the ratio of fragmentation fractions $f_s/f_d$ for normalisation~\cite{LHCb-PAPER-2011-018}.
The uncertainty on this quantity is limited by knowledge of the branching fraction of $\Ds\to\Kp\Km\pip$, and improved measurements of this quantity will be necessary to avoid a limiting uncertainty on, for example, ${\cal B}(\Bs \to \mu^+ \mu^-)$.
The determination of $2\beta_s$ from $\Bs\to \jpsi\,\phi$ provides an example of how systematic uncertainties can be controlled for measurements at the LHCb upgrade.
In the most recent measurement~\cite{LHCb-CONF-2012-002}, the largest source of systematic uncertainty arises due to the constraint of no direct \CP violation that is imposed in the fit.  With larger statistics, this constraint can be removed, eliminating this source of uncertainty.
Other sources, such as the background description and angular acceptance, are already at the 0.01~\rad level, and can be reduced with more detailed studies.

Experiments at upgraded $e^+e^-$ \B factories and elsewhere will study flavour-physics observables in a similar timeframe to the LHCb upgrade.  
However, the LHCb sample sizes in most exclusive \B and \D final states will be far larger than those that will be collected elsewhere, and the LHCb upgrade will have no serious competition in its study of \Bs decays, $b$-baryon decays, mixing and \CP violation.
Similarly the yields in charmed-particle decays to final states consisting of only charged tracks cannot be matched by any other experiment.
On the other hand, the $e^+e^-$ environment is advantageous for inclusive studies and for measurements of decay modes including multiple neutral particles~\cite{Bona:2007qt,Browder:2007gg,Browder:2008em,Aushev:2010bq,Ciuchini:2011ca}, and therefore enables complementary measurements to those that will be made with the upgraded LHCb experiment.

\subsection{Importance of the LHCb upgrade}

The study of deviations from the SM in quark flavour physics provides key information about any extension of the SM.
It is already known that the NP needed to stabilize the electroweak sector must have a non-generic flavour structure in order to be compatible with the tight constraints of flavour-changing processes, even if the precise form of this structure is still unknown.
Hopefully, ATLAS and CMS will detect new particles belonging to these models, but the couplings of the theory and, in particular, its flavour structure, cannot be determined only using high-\pt data.

Therefore, the LHCb upgrade will play a vital role in any scenario.
It allows the exploration of NP phase space that {\it a priori} cannot be studied by high energy searches.
Future plans for full exploitation of the LHC should be consistent with a co-extensive LHCb programme.


\clearpage

\section*{Acknowledgements}

\noindent 
The LHCb collaboration expresses its gratitude to its colleagues in the CERN
accelerator departments for the excellent performance of the LHC. 
LHCb thanks the technical and administrative staff at the LHCb institutes,
and acknowledges support from CERN and from the national
agencies: CAPES, CNPq, FAPERJ and FINEP (Brazil); NSFC (China);
CNRS/IN2P3 and Region Auvergne (France); BMBF, DFG, HGF and MPG
(Germany); SFI (Ireland); INFN (Italy); FOM and NWO (The Netherlands);
SCSR (Poland); ANCS/IFA (Romania); MinES, Rosatom, RFBR and NRC
``Kurchatov Institute'' (Russia); MinECo, XuntaGal and GENCAT (Spain);
SNSF and SER (Switzerland); NAS Ukraine (Ukraine); STFC (United
Kingdom); NSF (USA). 
LHCb also acknowledges the support received from the ERC under FP7. 
The Tier1 computing centres are supported by IN2P3
(France), KIT and BMBF (Germany), INFN (Italy), NWO and SURF (The
Netherlands), PIC (Spain), GridPP (United Kingdom). 
LHCb is thankful for the computing resources put at its
disposal by Yandex LLC (Russia), as well as to the communities behind
the multiple open source software packages that are depended upon.

The work of A.~Datta was supported by the National Science Foundation under Grant No. NSF PHY-1068052.
D.M.~Straub was supported by the EU ITN ``Unification in the LHC Era'', contract PITN-GA-2009-237920 (UNILHC).
We thank D.~Gorbunov for useful comments.

\clearpage

\addcontentsline{toc}{section}{References}
\bibliographystyle{LHCb}
\bibliography{main,LHCb-PAPER,LHCb-CONF,LHCb-DP,bmixing,bspenguin,quarkonia,rare,gammafromloops,gammafromtrees,charm,ewexotic,ckm}

\end{document}